\documentclass[12pt,english,american]{article}
\usepackage[T1]{fontenc}
\usepackage[latin9]{inputenc}
\usepackage{geometry}
\usepackage{color}
\usepackage{yfonts}
\usepackage{slashed}
\usepackage{babel}
\usepackage{amsthm}
\usepackage{mathrsfs}
\usepackage{amstext}
\usepackage{amssymb}
\usepackage[unicode=true,pdfusetitle,
 bookmarks=true,bookmarksnumbered=false,bookmarksopen=false,
 breaklinks=true,pdfborder={0 0 0},backref=false,colorlinks=true]
 {hyperref}
\hypersetup{
 linkcolor=blue,citecolor=blue, urlcolor=blue}

\makeatletter

\def\qed{$\Box$\medskip}

\newtheorem{theoreme}{Theorem } [section]
\newtheorem{proposition}[theoreme]{Proposition}
\newtheorem{lemma}[theoreme]{Lemma}
\newtheorem{definition}[theoreme]{Definition}
\newtheorem{corollary}[theoreme]{Corollary}
\newtheorem{remark}[theoreme]{Remark}
\newtheorem{example}[theoreme]{Example}
\newtheorem{criterion}[theoreme]{Criterion}

\newcommand{\beq}{\begin{equation}}
\newcommand{\eeq}{\end{equation}}
\newcommand{\beqa}{\begin{eqnarray}}
\newcommand{\eeqa}{\end{eqnarray}}
\newcommand{\ben}{\begin{arabicenumerate}}
\newcommand{\een}{\end{arabicenumerate}}
\newcommand{\bex}{\begin{example}}
\newcommand{\eex}{\end{example}}
\newcommand{\ber}{\begin{remark}}
\newcommand{\eer}{\end{remark}}
\newcommand{\bec}{\begin{corollary}}
\newcommand{\eec}{\end{corollary}}
\newcommand{\bep}{\begin{proposition}}
\newcommand{\eep}{\end{proposition}}
\newcommand{\becr}{\begin{criterion}}
\newcommand{\eecr}{\end{criterion}}

\def\bel{\begin{lem} } 
\def\eel{\end{lem} }
\def\bet{\begin{theoreme}}
\def\eet{\end{theoreme}}
\def\bed{\begin{defn}}
\def\eed{\end{defn} }

\usepackage{enumitem}		% customizable list environments
\theoremstyle{plain}
\newtheorem{thm}{\protect\theoremname}[section]
\theoremstyle{definition}
\newtheorem{defn}[thm]{\protect\definitionname}
\theoremstyle{plain}

\theoremstyle{remark}

\theoremstyle{plain}
\newtheorem{lem}[thm]{\protect\lemmaname}
\theoremstyle{plain}

\usepackage{txfonts,refstyle,xcolor}

\newref{con}{name = Conjecture\ }
\newref{prop}{name = Proposition\ }
\newref{def}{name = Definition\ }
\newref{sec}{name = Section\ }
\newref{sub}{name = Section\ }
\newref{thm}{name = Theorem\ }
\newref{lem}{name = Lemma\ }
\newref{cor}{name = Corollary\ }
\newref{fig}{name = Figure\ }

\usepackage{bbm}
\newcommand{\charf}{\mathbbm{1}}

\usepackage[all]{xy}

\newcommand{\xyR}[1]{%
     \makeatletter
     \xydef@\xymatrixrowsep@{#1}
     \makeatother
}

\newcommand{\xyC}[1]{%
     \makeatletter
     \xydef@\xymatrixcolsep@{#1}
     \makeatother
}

\newcommand{\ncol}[1]{\color{normalcolor}}

\makeatother

\addto\captionsamerican{\renewcommand{\corollaryname}{Corollary}}
\addto\captionsamerican{\renewcommand{\definitionname}{Definition}}
\addto\captionsamerican{\renewcommand{\lemmaname}{Lemma}}
\addto\captionsamerican{\renewcommand{\propositionname}{Proposition}}
\addto\captionsamerican{\renewcommand{\remarkname}{Remark}}
\addto\captionsamerican{\renewcommand{\theoremname}{Theorem}}
\addto\captionsenglish{\renewcommand{\corollaryname}{Corollary}}
\addto\captionsenglish{\renewcommand{\definitionname}{Definition}}
\addto\captionsenglish{\renewcommand{\lemmaname}{Lemma}}
\addto\captionsenglish{\renewcommand{\propositionname}{Proposition}}
\addto\captionsenglish{\renewcommand{\remarkname}{Remark}}
\addto\captionsenglish{\renewcommand{\theoremname}{Theorem}}
\providecommand{\corollaryname}{Corollary}
\providecommand{\definitionname}{Definition}
\providecommand{\lemmaname}{Lemma}
\providecommand{\propositionname}{Proposition}
\providecommand{\remarkname}{Remark}
\providecommand{\theoremname}{Theorem}
\begin{document}
\title{Bose particles in a box I.  A convergent expansion of  the ground state of a three-modes Bogoliubov Hamiltonian.} 
  \author{A. Pizzo \footnote{email: pizzo@mat.uniroma2.it}\\
 Dipartimento di Matematica, Universit\`a di Roma ``Tor Vergata",\\
 Via della Ricerca Scientifica 1, I-00133 Roma, Italy}

 % \author{ A. Pizzo}

\date{15/01/2017}

\maketitle

\abstract{In this paper we introduce a novel multi-scale technique to study many-body quantum systems where the total number of particles is kept fixed. The method is based on Feshbach-Schur map and the scales are represented by occupation numbers of particle states. Here, we consider a \emph{three-modes} (including the zero mode) Bogoliubov Hamiltonian  for a sufficiently small ratio between the kinetic energy and the Fourier component of the (positive type) potential corresponding to the two nonzero modes.
   For any space dimension $d\geq 1$ and in the mean field limiting regime (i.e., at fixed box volume $|\Lambda|$ and for a number of particles, $N$, sufficiently large) this method provides the construction  of the ground state and its expansion in terms of the bare operators that in the limit $N\to \infty$ is  up to any desired precision.  In space dimension $d\geq 3$  the method provides similar results  for an arbitrarily large (finite) box and a \emph{large but fixed particle density $\rho$}, i.e., $\rho$ is independent of the size of the box. 
   %{\color{red}Nevertheless, there is some evidence that also for space dimension $d=3$ the same procedure should work at \emph{large but fixed particle density $\rho$}.
}
\\

\noindent
{\bf{Summary of contents}}
\begin{itemize}
\item In Sections \ref{introduction} and \ref{hamiltonians} a model of a  gas of Bose particles in a box is defined along with the notation used throughout the paper. After introducing the \emph{particle number preserving} Bogoliubov Hamiltonian (from now on Bogoliubov Hamiltonian), the main ideas of the multi-scale technique  are presented.
\item In Section \ref{multiscale-HBog} the multi-scale analysis in the particle states occupation numbers is implemented for the Bogoliubov Hamiltonian of a model where only three modes (including the zero mode) interact. In fact, the treatment of the full Bogoliubov Hamiltonian can be thought of as a repeated application of the multi-scale analysis to a collection of three-modes systems (see \cite{Pi2}). The Feshbach-Schur flow is described informally in Section \ref{Feshbach} and the main results are stated in Section \ref{statement}.
%The motivations of the strategy are explained a-posteriori in Remark \ref{motivations}.
\item In Section \ref{groundstate}  the ground state of the "three-modes Bogoliubov Hamiltonian" is constructed as a byproduct of the Feshbach-Schur flow. In the mean field limit, this also provides a convergent expansion of the vector in terms of the bare operators up to any desired precision.
\item
Section \ref{appendix} is an Appendix where some of the proofs are deferred.
\end{itemize}
\setcounter{equation}{0}

\section{Introduction: interacting Bose gas in a box}\label{introduction}

%\[
%a(x)=\frac{\sum_{\boldsymbol{n}}a_{\boldsymbol{n}}e^{ik_{\boldsymbol{n}}\cdot x}}{|\Lambda_{L'}|^{\frac{1}{2}}}
%\]
%where $|\Lambda_{L'}|=L'^{3}$and $k_{\boldsymbol{n}}=\frac{2\pi\boldsymbol{n}}{L'}$
%and $\boldsymbol{n}\in Z^{3}$. Commutation relations:

%\[
%[a_{\boldsymbol{n}},a_{\boldsymbol{n'}}^{*}]=\delta_{\boldsymbol{n},\boldsymbol{n}'}
%\]
%Hamiltonian
%\[
%H'=\int\frac{1}{2m}(\nabla a^{*})(\nabla a)(x)dx+\frac{\lambda}{2}\int\int(a^{*}(x)a(x)-\rho)\phi(x-y)(a^{*}(x)a(x)-\rho)dxdy
%\]
%where the integration is over $\Lambda_{L'}$ , and 
%\[
%\phi(x-y)=(2\pi)^{-\frac{3}{2}}\sum_{\boldsymbol{n}}\tilde{\phi}(k_{\boldsymbol{n}})\frac{e^{ik_{\boldsymbol{n}}\cdot(x-y)}}{|\Lambda_{L'}|^{\frac{1}{2}}}
%\]
%\\
We study the Hamiltonian describing a gas of (spinless) nonrelativistic Bose particles that, at zero temperature, are constrained to a $d-dimensional$ box of side $L$ with $d\geq 1$. The particles interact through a pair potential with a coupling constant proportional to the inverse of the particle density  $\rho$.  The rigorous description of this system has many intriguing mathematical aspects not completely clarified yet. In spite of remarkable contributions also in recent years, some important problems are still open to date, in particular in connection to the thermodynamic limit and the exact structure of the ground state vector. We shall briefly mention the results closer to our present work  and give references to the reader for the details. 

Some of the results have been concerned with the low energy spectrum of the Hamiltonian that in the mean field limit was predicted by Bogoliubov \cite{Bo1}, \cite{Bo2}. The expression predicted by Bogoliubov for the ground state energy has been rigorously proven for certain systems in \cite{LS1}, \cite{LS2}, \cite{ESY}, \cite{YY}. Concerning  the excitation spectrum, in Bogoliubov theory it consists of elementary excitations whose energy is linear in the momentum for small momenta. 
After some important results restricted to one-dimensional models (see \cite{G}, \cite{LL}, \cite{L}), this conjecture was proven by Seiringer in \cite{Se1} (see also \cite{GS}) for the low-energy spectrum of an interacting Bose gas in a finite box and in the mean field limiting regime,  where  the pair potential is of positive type. In  \cite{LNSS}  it has been extended to a more general class of potentials and the limiting behavior of the low energy eigenstates has been studied.  Later,  the result of \cite{Se1} has been proven to be valid in a sort of diagonal limit where the particle density and the box volume diverge according to a prescribed asymptotics; see  \cite{DN}. Recently, Bogoliubov's prediction of the energy spectrum in the mean field limit has been shown to be valid also for the high energy eigenvalues (see \cite{NS}).

\noindent
These results are based on clever energy estimates starting from the spectrum of the corresponding Bogoliubov Hamiltonian. 

A different approach to studying a gas of Bose particles is based on renormalization group.  In this respect, we mention the paper by Benfatto, \cite{Be}, where he has provided \emph{an order by order control} of the Schwinger  functions of this system in three dimensions and with an ultraviolet cut-off. His analysis holds at zero temperature in the infinite volume limit  and at finite particle density. Thus, it contains  a fully consistent treatment of the infrared divergences at a perturbative level. This program has been later developed in  \cite{CDPS1}, \cite{CDPS2}, and, more recently, in  \cite{C} and \cite{CG} by making use of \emph{Ward identities} to deal also with two-dimensional systems where some partial control of the renormalization flow has been provided; see \cite{C} for a detailed review of previous related results.

\noindent
Within the renormalization group approach, we also mention some results towards a rigorous construction of the functional integral for this system contained in \cite{BFKT1}, \cite{BFKT2}, and \cite{BFKT}.

Both in the grand canonical and in the canonical ensemble  approach (see \cite{Se1}), starting from the Hamiltonian of the system one can define an approximated one, the Bogoliubov Hamiltonian. For a finite box and a large class of pair potentials, upon a  unitary transformation the Bogoliubov Hamiltonian describes\footnote{In the canonical ensemble approach the spectrum of the (particle preserving) Bogoliubov Hamiltonian coincides with the \emph{Bogoliubov spectrum} only in the limit $N\to \infty$ (see \cite{Se1}).}  a system of non-interacting bosons with a new energy dispersion law, which in fact provides the correct description of the energy spectrum of the Bose particles system in the mean field limit.

%Somehow related to the content of this paper and of the companion papers,
%The literature on the Bose gas in general is vast and goes beyond the problems and the model addressed in this paper. However, it is surely .... 

With regard to Bose-Einstein condensation, we recall the breakthrough results obtained for a system of trapped Bose particles in the so called Gross-Pitaeveskii limiting regime: see \cite{LSY}, \cite{LS}, \cite{LSSY}, and \cite{NRS}. More recently, Bose-Einstein condensation has been proven for particles interacting with a delta potential and in the mean field limit; see \cite{LNR}.  We also mention  the progress in the control of the dynamical properties of Bose gases. For references and for an update of the state of the art, we refer the reader to the introduction of \cite{DFPP}.
\\

In our paper, we consider the number of particles fixed but we use the formalism of second quantization. The Hamiltonian corresponding to the pair potential $\phi(x-y)$ and to the coupling constant $\lambda>0$ is

\begin{equation}\label{initial-ham}
\mathscr{H}:=\int\frac{1}{2m}(\nabla a^{*})(\nabla a)(x)dx+\frac{\lambda}{2}\int\int a^{*}(x)a^*(y)\phi(x-y)a(x)a(y)dxdy\,,
\end{equation}
where 
reference to the integration domain $\Lambda:=\{x\in \mathbb{R}^d\,|\,|x_i|\leq \frac{L}{2}\,,\,i=1,2,\dots, d\}$ is omitted, periodic boundary conditions are assumed, and $dx$ is Lebesgue measure in $d$ dimensions. Concerning units, we have set $\hbar$ equal to $1$. Here, the operators $a^{*}(x)\:,a(x)$ are the usual operator-valued distributions on the bosonic Fock space $$\mathcal{F}:= \Gamma\left(L^{2}\left(\Lambda,\mathbb{C};dx\right)\right)\,$$  that satisfy the canonical
commutation relations (CCR)
\[
[a^{\#}(x),a^{\#}(y)]=0,\quad\quad[a(x),a^{*}(y)]=\delta(x-y)\charf_{\mathcal{F}},
\]
with $a^{\#}:=a$ or $a^{*}$. In terms of the field modes they read
\[
a(x)=\sum_{\bold{j}\in \mathbb{Z}^d}\frac{a_{\bold{j}}e^{ik_{\bold{j}}\cdot x}}{|\Lambda|^{\frac{1}{2}}},\quad a^{*}(x)=\sum_{\bold{j}\in \mathbb{Z}^d}\frac{a_{\bold{j}}^*e^{-ik_{\bold{j}}\cdot x}}{|\Lambda|^{\frac{1}{2}}},\]
where $k_{\bold{j}}:=\frac{2\pi}{L} \bold{j}$, $\bold{j}=( j_1,  j_2,\dots,  j_d)$, $j_1, j_2,\dots,  j_d \in \mathbb{Z}$, and $|\Lambda|=L^d$,  with CCR

\begin{equation}
[a^{\#}_{\bold{j}},a^{\#}_{\bold{j}'}]=0,\quad\quad[a_{\bold{j}},a^{*}_{\bold{j}'}]=\delta_{\bold{j}\,,\,\bold{j}'}\,.
\end{equation}
The unique (up to a phase) vacuum vector of $\mathcal{F}$ is denoted by $\Omega$ ($\|\Omega\|=1$). In $\mathcal{F}$,  $a_{\bold{j}}\,/\,a^{*}_{\bold{j}}$ are the annihilation / creation operators  of a particle of momentum $k_{\bold{j}}$.
\\

Given any function $\varphi\in L^{2}\left(\Lambda,\mathbb{C};dz\right)$, we express it in terms of its Fourier components $\varphi_{\bold{j}}$, i.e.,
\begin{equation}
\varphi(z)=\frac{1}{|\Lambda|}\sum_{\bold{j}\in \mathbb{Z}^d}\varphi_{\bold{j}}e^{ik_{\bold{j}}z}\,.
\end{equation}
%and the Parseval identity reads
%\begin{equation}
%\int dz |\varphi|^2(z)=\frac{1}{|\Lambda|}\sum_{\bold{j}\in \mathbb{Z}^d}|\varphi_{\bold{j}}|^2<\infty\,.
%\end{equation}

\begin{definition}\label{def-pot}
The pair potential $\phi(x-y)$ is a bounded, real-valued function that is periodic, i.e., $\phi(z)=\phi(z+\bold{j}L)$  for $\bold{j}\in \mathbb{Z}^{d}$, and satisfies  the following conditions:
\begin{enumerate}
%\item
%$\phi(x-y)\geq 0$
%\item
%$\phi(x-y)=\phi(y-x)$, hence we write $\phi(x-y)=\frac{1}{|\Lambda|}\sum_{\bold{j}\in \mathbb{Z}^3}\phi_{\bold{j}}e^{ik_{\bold{j}}(x-y)}$
\item  $\phi(z)$ is an even function, in consequence $\phi_{\bold{j}}=\phi_{-\bold{j}}$. 
\item 
$\phi(z)$ is of positive type, i.e., the Fourier components $\phi_{\bold{j}}$ are nonnegative.
\item
The pair interaction has a fixed but arbitrarily large ultraviolet cutoff (i.e., the nonzero Fourier components $\phi_{\bold{j}}$  form a finite set)  with the requirements below to be satisfied:
%(see also point 3) in Remark \ref{motivations} and Remark \ref{high regime}).

3.1) (\underline{Strong Interaction Potential Assumption})
 The ratio $\epsilon_{\bold{j}}$ between the kinetic energy of the modes $\pm\bold{j}\neq \bold{0}=(0,\dots,0)$ and the corresponding Fourier component $\phi_{\bold{j}}(\neq 0)$  of the potential  is sufficiently  small. 

\noindent
3.2) For all nonzero $\phi_{\bold{j}}$ there exist some $1>\mu>0$ and $\theta >0$ such that
\begin{equation}\frac{\phi_{\bold{j}}}{\Delta_0}\frac{N^{\mu}}{N(N-N^{\mu})}<\frac{1}{2}\quad,\quad  \frac{1}{N^{\mu}}\leq \mathcal{O}((\sqrt{\epsilon_{\bold{j}}})^{1+\theta})\,,\label{mu-cond}\end{equation}
where $\Delta_0=\min\, \Big\{k_{\bold{j}}^2\,|\,\bold{j}\in \mathbb{Z}^d\setminus\{\bold{0}\}\Big\}\,$ and $N$ is the number of particles in the box.

\end{enumerate}

\end{definition}
\begin{remark}
Notice that $\epsilon_{\bold{j}}$ small corresponds either to a low energy mode  $\frac{2\pi\bold{j}}{L}$  or/and to a large potential $\phi_{\bold{j}}$.
\end{remark}
\begin{remark}
Different regimes for the ratio $\epsilon_{\bold{j}}$ can be explored with the same method by suitable modifications of some estimates (see \cite{CP}). In the present paper we are interested in the behavior of the (three-modes) system in the thermodynamic limit, therefore the   ``Strong Interaction Potential Assumption" is the right regime to study.
\end{remark}
%\begin{remark}
%The class of potentials chosen in Definition 1.1 is just to use only one method, i.e., the multi-scale analysis in the number of zero-mode particles using the Feshbach-Schur map. This method is quite effective with high potentials, because of the artificial gap induced by the chosen projection in the resolvents of the expansion: the higher is the potential with respect to the kinetic energy,  the higher is the gap. In fact,  the gap is the difference between the $\inf$ of the free Hamiltonian and the value of the spectral parameter $z\in \mathbb{R}, z<0$,  entering the Feshbach-Schur map. The procedure is expected to work for $z$ less or close to the Bogoliubov energy.
%\end{remark}
%\begin{remark}
%As it will be explained in the last section, in order to have an explicit expression (i.e., in terms of bare quantities) for the ground state --  we call this expression main term -- up to a remainder which is estimated in norm less than some fraction, say $1/16$, of the main term, we have to decompose the expansion into two steps (at least), by considering ``high" and ``low" components of the potential separately. Hopefully, at least in the case in which we have a bunch of high frequencies and another bunch of very low frequencies, the two sets of frequencies being well separated, this method should work.
%\end{remark}

We restrict $\mathscr{H}$ to the Fock subspace $\mathcal{F}^{N}$ of vectors with $N$ particles 
\begin{equation}
\mathscr{H}\upharpoonright_{\mathcal{F}^N}=\Big(\int\frac{1}{2m}(\nabla a^{*})(\nabla a)(x)dx+\frac{\lambda}{2}\int\int a^{*}(x)a^{*}(y) \phi(x-y)a(y)a(x)dxdy\Big)\upharpoonright_{\mathcal{F}^N}\,.
\end{equation}
From now on, we study the Hamiltonian
\begin{equation}\label{complete-ham}
H:=\int\frac{1}{2m}(\nabla a^{*})(\nabla a)(x)dx+\frac{\lambda}{2}\int\int a^{*}(x)a^{*}(y)\phi(x-y)a(y)a(x)dxdy+c_N\charf
\end{equation}
where $c_N=\frac{\lambda\phi_{\bold{0}}}{2|\Lambda|}N-\frac{\lambda\phi_{\bold{0}}}{2|\Lambda|}N^2$ with $\bold{0}=\{0,\dots,0\}$. The operator $H$ is meant to be restricted to the subspace $\mathcal{F}^N$, and $\lambda$ will be eventually chosen equal to $\frac{|\Lambda|}{N}$. 
%{\color{red} The value $-c_N+\frac{\lambda\phi_{\bold{0}}}{2|\Lambda|}N$ is in fact the leading order term of the ground state energy of $\mathscr{H}\upharpoonright_{\mathcal{F}^N}$. 
We subtract $-c_N$ as clarified in (\ref{eq-c_n-1})-(\ref{eq-c_n-2}).  Notice that
\begin{equation}
\mathscr{H}\upharpoonright_{\mathcal{F}^N}=(H-c_N\charf)\upharpoonright_{\mathcal{F}^N}\,.
\end{equation}

The main technical features of the scheme introduced in this paper are highlighted in Section \ref{motivations} after the outline of the procedure in Section \ref{Nonrigorous derivation}. In Section \ref{statement} we summarize the obtained results. Here, we present some motivations that can help the reader understand the scheme. 
\\

We know that, at  fixed volume $|\Lambda|$, the expectation value of the number operator\footnote{The operator  $\sum_{\bold{j}\in \mathbb{Z}^d\setminus \{\bold{0}\}}a^*_{\bold{j}}a_{\bold{j}}$ counts the number  of particles in the nonzero modes states.} $\sum_{\bold{j}\in \mathbb{Z}^d\setminus \{\bold{0}\}}a^*_{\bold{j}}a_{\bold{j}}$  in the ground state of the Hamiltonian (\ref{complete-ham})   remains bounded in the mean field limit (i.e.,  $\lambda= \frac{|\Lambda|}{N}$ and $N\to \infty$); see \cite{Se1} and \cite{LNSS}. Starting from this fact, one might think of a multi-scale procedure leading to an effective Hamiltonian for spectral values in a neighborhood of the ground state energy.  An obvious  candidate for such an effective Hamiltonian  is (a multiple of) the orthogonal projection onto the state where all the particles are in the zero mode.
\\

The Feshbach-Schur map is a very useful tool to construct effective Hamiltonians. We recall that given the (separable) Hilbert space $\mathcal{H}$,  the projections $\mathscr{P}$, $\overline{\mathscr{P}}$ ($\mathscr{P}=\mathscr{P}^2$, $\overline{\mathscr{P}}=\overline{\mathscr{P}}^2$) where $\mathscr{P}+\overline{\mathscr{P}}=\charf_{\mathcal{H}}$, and a closed operator $K-z\charf $ acting on $\mathcal{H}$ ($z$ in a subset of $ \mathbb{C}$) the Feshbach-Schur map associated with the couple $\mathscr{P}\,,\,\overline{\mathscr{P}}$ maps $K-z\charf$ to the operator $\mathscr{F}(K-z\charf)$ acting on $\mathscr{P}\mathcal{H}$ where  (formally)
\begin{equation}\label{feshbach-general}
\mathscr{F}(K-z\charf):=\mathscr{P}(K-z\charf )\mathscr{P}-\mathscr{P}K\overline{\mathscr{P}}\frac{1}{\overline{\mathscr{P}}(K-z\charf)\overline{\mathscr{P}}}\overline{\mathscr{P}}K\mathscr{P}\,.
\end{equation}
%We recall that given the (separable) Hilbert space $\mathcal{H}$ and the projections $\mathscr{P}$, $\overline{\mathscr{P}}$ ($\mathscr{P}=\mathscr{P}^2$, $\overline{\mathscr{P}}=\overline{\mathscr{P}}^2$) where $\mathscr{P}+\overline{\mathscr{P}}=\charf_{\mathcal{H}}$, the Feshbach map associated with $\mathscr{P}$ and $\overline{\mathscr{P}}$ maps a normal operator $K-z$ acting on $\mathcal{H}$ (here $z$ belongs to a subset of $ \mathbb{C}$)  to the operator $\mathscr{F}(K-z)$ acting on $\mathscr{P}\mathcal{H}$  and which is formally defined as
%\begin{equation}
%\mathscr{F}(K-z):=\mathscr{P}(K-z)\mathscr{P}-\mathscr{P}K\overline{\mathscr{P}}\frac{1}{\overline{\mathscr{P}}(K-z)\overline{\mathscr{P}}}\overline{\mathscr{P}}K\mathscr{P}\,.\label{feshbach-general}
%\end{equation}
The Feshbach-Schur map is ``isospectral" (see  \cite{BFS}), i.e.,  assuming that $\mathscr{F}(K-z\charf )$ is a well defined closed operator on $\mathscr{P}\mathcal{H}$ then: 1) $\mathscr{F}(K-z\charf )$ is bounded invertible if and only if $z$ is in the resolvent set of $K$; 2)  $z$ is an eigenvalue of  $K$ if and only if $0$ is an eigenvalue of $\mathscr{F}(K-z\charf )$. Moreover, the map provides an algorithm to reconstruct the eigenspace corresponding to the eigenvalue $z$ from the kernel of the operator $\mathscr{F}(K-z\charf)$, and their dimensions coincide.
\\

The use of the Feshbach-Schur map for the spectral  analysis of quantum field theory systems started with the seminal work by V. Bach, J. Fr\"ohlich, and I.M. Sigal, \cite{BFS},  followed by refinements of the technique and variants (see \cite{BCFS} and  \cite{GH}). In those papers, the use of the Feshbach-Schur map is in the spirit of the functional renormalization group, and the  projections ($\mathscr{P}$, $\overline{\mathscr{P}}$) are directly related to energy subspaces of the free Hamiltonian. However, as a mathematical tool the Feshbach-Schur map  enjoys an enormous flexibility due to the freedom in the choice of the couple of  projections $\mathscr{P}$, $\overline{\mathscr{P}}$ . The effectiveness of the choice depends on the features of the Hamiltonian.
\\

In the system that we study  the total number of particles is conserved under time evolution. The effective Hamiltonian that we want to construct suggests to relate the Feshbach-Schur projections $(\mathscr{P}$, $\overline{\mathscr{P}})$ to subspaces of states with definite number of particles in the modes labeled by $\Big\{\frac{2\pi}{L}\bold{j}\,;\,\bold{j}\in \mathbb{Z}^d \Big\}$.  More precisely, consider the eigenspace  of  $\sum_{\bold{j}=\pm \bold{j}_*}a^*_{\bold{j}}a_{\bold{j}}$ corresponding to the eigenvalue $i$, i.e., the subspace of states containing $i$ particles in the modes associated with $\pm \frac{2\pi}{L}\bold{j}_*$. Observe that the interaction part in the second quantized Hamiltonian in (\ref{complete-ham}) can connect  two eigenspaces  corresponding to distinct eigenvalues, $i$ and $i'$,  only  if $i-i'=\pm 1, \pm 2$.  The selection rules of the interaction Hamiltonian with respect to the occupation numbers of the particle states associated with the modes  $\Big\{\frac{2\pi}{L}\bold{j}\,;\,\bold{j}\in \mathbb{Z}^d \Big\}$ suggest to construct a flow of Feshbach-Schur maps associated with projections onto such eigenspaces with decreasing eigenvalue $i$.
% and to exploit the selection rules of the Hamiltonian $H$ for the matrix elements of $\mathscr{P}H\overline{\mathscr{P}}$.
\\

The (formal) Rayleigh-Schr\"odinger series of the ground state of the Hamiltonian (\ref{complete-ham}) of the system calls for the use of the Feshbach-Schur map. Indeed, one can observe that the series is not under control for interaction potentials that are strong with respect to the minimum (nonzero) kinetic energy. 
%regime that we are interested in (see Remark \ref{high regime}). 
%But replacing the free Hamiltonian ground state energy with the expected nonperturbative value of the ground state energy of the interacting system, the individual terms of the  series make sense. 
Then, one might wonder whether it is possible to organize an expansion (up to any desired precision) of the ground state in terms of the ground state of the free Hamiltonian and in terms of \emph{bare operators}, around a reference energy close to the expected value of the ground state energy of the (interacting) system.  The expansion provided in Section \ref{conv-exp}  (starting from the formula in (\ref{gs-1})-(\ref{gs-2})) answers this question  into affirmative for a three-modes Bogoliubov Hamiltonian. In this expansion the flow of Feshbach-Schur maps plays a crucial role thanks to the choice of the perpendicular projections, $\overline{\mathscr{P}}$,  entering the Feshbach-Schur map at each step of the flow. These projections prevent \emph{small denominator} problems in the expansion,  even for an arbitrarily small (positive) ratio between the  kinetic energy $k_{\bold{j}}^2$ and the Fourier component $\phi_{\bold{j}}$, provided the number of particles $N$ is sufficiently large.
\\

 Indeed, the method presented in the next sections and developed in \cite{Pi2} and \cite{Pi3} is applied to a potential $\phi$ with an ultraviolet cut-off and in the \emph{strong interaction potential regime}: by this we mean that the ratio between each nonzero Fourier component of the potential, $\phi_{\bold{j}}$, and the corresponding kinetic energy, $k_{\bold{j}}^2$, must be sufficiently large. For a (positive definite) potential $\phi \in L^1$ such that $\int \phi(z)dz>0$, this is precisely the regime that is relevant in the thermodynamic limit because at fixed $\bold{j}$ the ratio $\phi_{\bold{j}}/(k_{\bold{j}})^2$ diverges like $L^2$, being $k_{\bold{j}}:=\frac{2\pi}{L} \bold{j}$ and $L$ the side of the box.
%Indeed, controlling the expansion of the ground state of the three-modes system  associated with the modes $\pm\bold{j}$ as the ratio $\phi_{\bold{j}}/(k_{\bold{j}})^2$ diverges  -- at fixed potential this ratio diverges as $L^2$, $L$ being the side of the box -- is a problem independent of the procedure to sum up the contributions of all the modes.
\\

In this scheme we never implement a Bogoliubov transformation yielding a new Hamiltonian in terms of quasi-particles degrees of freedom. The occupation numbers are always referred to the real particles.
% and do not involve the so called condensate degrees of freedom. 
In this respect, the method might be robust enough to deal with systems and regimes where the features of the Bogoliubov diagonalization is not clear \emph{a priori}.  
Furthermore,  if  the range of the spectral parameter $z$ (see  (\ref{feshbach-general})) extends to the first $q$ eigenvalues (with multiplicity) above the ground state energy the same method should also provide an effective Hamiltonian acting on a $q-$dependent, finite-dimensional  subspace. Some numerical simulations for a three-modes system seem to confirm this scenario. 
%situation the Feshbach flow is well defined up to the step such that the corresponding Feshbach Hamiltonian is a $q$-dimensional matrix.}
\\

The three-modes system analyzed in this paper represents the main building block in the construction of the ground state of the Bogoliubov Hamiltonian (see (\ref{Bog})) and of the complete Hamiltonian (see (\ref{complete-ham})) in the mean field limiting regime, provided the potential fulfills the requirements in Definition \ref{def-pot}; see \cite{Pi2} and \cite{Pi3}, respectively. Within this technique the three-modes system is like a solvable model, in the sense that the interaction is so constrained that the Feshbach-Schur flow can be closely followed. 

The Bogoliubov Hamiltonian is analyzed as a collection of three-modes (i.e., $\{\bold{j},-\bold{j},\bold{0}\}$)  systems. Consequently, the technical challenge consists in showing that in the mean field limit they can be treated as independent couples of modes that interact only within each couple through the zero-mode. In \cite{Pi2}, we show this result and control deviations from the mean field limit. 

In the third paper \cite{Pi3}, because of the interaction terms that are neglected in the Bogoliubov Hamiltonian (the so called ``cubic" and ``quartic" terms in the nonzero modes) a refined choice of the Feshbach-Schur projections is required.
%The new interaction terms enhance more connection between the states and therefore the Feshbach projections have to turn off these connections as long as they produce 

\section{The Hamiltonian $H$ and the Hamiltonian $H^{Bog}$}\label{hamiltonians}
\setcounter{equation}{0}
%In the next formulae of this section, we use the notation that will be necessary for a general potential, that means for a potential where more than three modes have corresponding components different from zero (and positive).

For later convenience, we define
\begin{equation}
\label{b in the strip}
a_{+}(x):=\sum_{\bold{j}\in\mathbb{Z}^d\setminus \{\bold{0}\} }\frac{a_{\bold{j}}}{|\Lambda|^{\frac{1}{2}}}e^{ik_{\bold{j}}\cdot x}\,\quad,\quad a_{\bold{0}}(x):=\frac{a_{\bold{0}}}{|\Lambda|^{\frac{1}{2}}}\,
\end{equation}
where $\bold{0}:=(0,\dots,0)$.
Then, the Hamiltonian $H$ reads
\begin{eqnarray}
H& = & \sum_{\bold{j}\in  \mathbb{Z}^d}\frac{k^2_{\bold{j}}}{2m}a_{\bold{j}}^{*}a_{\bold{j}}\\
% &  & +\frac{\lambda}{2}\int \int b^*|_{n,l+1}^{n,l}(x)b^*|_{n,l}^{0,0}(y)\phi(x-y)b|_{n,l+1}^{n,l}(x)b|_{n,l}^{0,0}(y)dxdy\\
& &+\frac{\lambda}{2}\int\int a^{*}_{+}(x)a^*_{+}(y)\phi(x-y)a_{+}(x)a_{+}(y)dxdy \\
& &+\lambda\int \int \{ a^*_{+}(x)a^*_{+}(y)\phi(x-y)a_{+}(x)a_{\bold{0}}(y)
 %&  & +\frac{\lambda}{2}\int \int a^*_{+}(x)a^*_{+}(y)\phi(x-y)a_{\bold{0}}(x)a_{+}(y)dxdy
  +h.c.\}dxdy\\
% &  & +\frac{\lambda}{2}\int \int a^*_{\bold{0}}(x)a^*_{+}(y)\phi(x-y)a_{+}(x)a_{+}(y)dxdy\\
 &  & +\frac{\lambda}{2}\int \int \{ a^*_{\bold{0}}(x)a^*_{\bold{0}}(y)\phi(x-y)a_{+}(x)a_{+}(y)
 + h.c.\}dxdy\\
 &  & +\lambda\int \int a^*_{\bold{0}}(x)a^*_{+}(y)\phi(x-y)a_{\bold{0}}(x)a_{+}(y)dxdy\\
 &  & +\lambda\int \int a^*_{\bold{0}}(x)a^*_{+}(y)\phi(x-y)a_{\bold{0}}(y)a_{+}(x)dxdy\\
& & +\frac{\lambda}{2}\int\int a^{*}_{\bold{0}}(x)a^*_{\bold{0}}(y)\phi(x-y)a_{\bold{0}}(x)a_{\bold{0}}(y)dxdy  \\
& &+c_N\charf\,.
\end{eqnarray}
Given the (number) operators
\begin{equation}
\mathscr{N}_{\bold{0}}:= \int a^*_{\bold{0}}(x)a_{\bold{0}}(x)dx\quad , \quad\mathscr{N}_{+}:= \int a^*_{+}(x)a_{+}(x)dx\,,
\end{equation}
we observe that
\begin{equation}
\lambda\int \int a^*_{\bold{0}}(x)a^*_{+}(y)\phi(x-y)a_{\bold{0}}(x)a_{+}(y)dxdy= \frac{\lambda \phi_{\bold{0}}}{|\Lambda|} \mathscr{N}_{+}\mathscr{N}_{\bold{0}}\,,
\end{equation}
\begin{equation}
\frac{\lambda}{2}\int\int a^{*}_{\bold{0}}(x)a^*_{\bold{0}}(y)\phi(x-y)a_{\bold{0}}(x)a_{\bold{0}}(y)dxdy=\frac{\lambda\phi_{\bold{0}}}{2|\Lambda|}(\mathscr{N}_{\bold{0}})^2- \frac{\lambda\phi_{\bold{0}}}{2|\Lambda|}\mathscr{N}_{\bold{0}}\,,
\end{equation}
and
\begin{equation}
\frac{\lambda}{2}\int\int a^{*}_{+}(x)a^*_{+}(y)\phi_{(0)}(x-y)a_{+}(x)a_{+}(y)dxdy\\
=  \frac{\lambda\phi_{\bold{0}}}{2|\Lambda|}(\mathscr{N}_{+})^2- \frac{\lambda\phi_{\bold{0}}}{2|\Lambda|}\mathscr{N}_{+}
\end{equation}
where $\phi_{(0)}(x-y):=\frac{\phi_{\bold{0}}}{|\Lambda|}$. Hence, because of the implicit restriction to $\mathcal{F}^N$, we conclude that
\begin{eqnarray}
&  &\frac{\lambda}{2}\int\int a^{*}_{+}(x)a^*_{+}(y)\phi_{(0)}(x-y)a_{+}(x)a_{+}(y)dxdy  \label{eq-c_n-1}\\
 &  & +\lambda\int \int a^*_{\bold{0}}(x)a^*_{+}(y)\phi(x-y)a_{\bold{0}}(x)a_{+}(y)dxdy\\
 & &+\frac{\lambda}{2}\int\int a^{*}_{\bold{0}}(x)a^*_{\bold{0}}(y)\phi(x-y)a_{\bold{0}}(x)a_{\bold{0}}(y)dxdy   \label{eq-c_n-2} \\
 & &+c_N \charf \\
 &=&0\,.
\end{eqnarray}
Therefore, we can write
\begin{eqnarray}
H& = & \sum_{\bold{j}\in  \mathbb{Z}^d}\frac{k^2_{\bold{j}}}{2m}a_{\bold{j}}^{*}a_{\bold{j}}\\
% &  & +\frac{\lambda}{2}\int \int b^*|_{n,l+1}^{n,l}(x)b^*|_{n,l}^{0,0}(y)\phi(x-y)b|_{n,l+1}^{n,l}(x)b|_{n,l}^{0,0}(y)dxdy\\
%& &+ \frac{\lambda\phi_{\bold{0}}}{2|\Lambda|^2}\int\int a^{*}_{+}(x)a_{+}(y)dxdy
& &+\frac{\lambda}{2}\int\int a^{*}_{+}(x)a^*_{+}(y)\phi_{(\neq 0)}(x-y)a_{+}(x)a_{+}(y)dxdy  \\
& &+\lambda\int \int \{a^*_{+}(x)a^*_{+}(y)\phi_{(\neq 0)}(x-y)a_{+}(x)a_{\bold{0}}(y)
 %&  & +\frac{\lambda}{2}\int \int a^*_{+}(x)a^*_{+}(y)\phi(x-y)a_{\bold{0}}(x)a_{+}(y)dxdy\\
 + a^*_{+}(x)a^*_{\bold{0}}(y)\phi_{(\neq 0)}(x-y)a_{+}(x)a_{+}(y)\}dxdy\\
% &  & +\frac{\lambda}{2}\int \int a^*_{\bold{0}}(x)a^*_{+}(y)\phi(x-y)a_{+}(x)a_{+}(y)dxdy\\
 &  & +\frac{\lambda}{2}\int \int \{a^*_{\bold{0}}(x)a^*_{\bold{0}}(y)\phi_{(\neq 0)}(x-y)a_{+}(x)a_{+}(y)
 + a^*_{+}(x)a^*_{+}(y)\phi_{(\neq 0)}(x-y)a_{\bold{0}}(x)a_{\bold{0}}(y)\}dxdy\quad\quad\quad\quad\\
% &  & +\lambda\int \int a^*_{\bold{0}}(x)a^*_{+}(y)\phi(x-y)a_{\bold{0}}(x)a_{+}(y)dxdy\\
 &  & +\lambda\int \int a^*_{\bold{0}}(x)a^*_{+}(y)\phi_{(\neq 0)}(x-y)a_{\bold{0}}(y)a_{+}(x)dxdy
%& & +\frac{\lambda}{2}\int\int a^{*}_{\bold{0}}(x)a^*_{\bold{0}}(y)\phi(x-y)a_{\bold{0}}(x)a_{\bold{0}}(y)dxdy  \\
%& &+c_N
\end{eqnarray}
where $\phi_{(\neq 0)}(x-y):=\phi (x-y)-\phi_{(0)}(x-y)$.

Next, we define the \emph{particles number preserving} Bogoliubov Hamiltonian
 \begin{eqnarray}
H^{Bog} &:=  &\sum_{\bold{j}\in  \mathbb{Z}^d}\frac{k^2_{\bold{j}}}{2m}a_{\bold{j}}^{*}a_{\bold{j}}\\
& &+ \frac{\lambda}{2}\int \int a^*_{\bold{0}}(x)a^*_{\bold{0}}(y)\phi_{(\neq 0)}(x-y)a_{+}(x)a_{+}(y)dxdy\label{quartic-two-one}\\
& &+\frac{\lambda}{2}\int \int a^*_{+}(x)a^*_{+}(y)\phi_{(\neq 0)}(x-y)a_{\bold{0}}(x)a_{\bold{0}}(y)dxdy\label{quartic-two-two}\\
  &  & +\lambda\int \int a^*_{\bold{0}}(x)a^*_{+}(y)\phi_{(\neq 0)}(x-y)a_{\bold{0}}(y)a_{+}(x)dxdy\,,
 \end{eqnarray}
that in terms of the field modes reads
\begin{eqnarray}
H^{Bog}
&=&\sum_{\bold{j}\in\mathbb{Z}^d\setminus \{\bold{0}\}} (\frac{k^2_{\bold{j}}}{2m}+\lambda\frac{\phi_{\bold{j}}}{|\Lambda|}a^*_{\bold{0}}a_{\bold{0}})a_{\bold{j}}^{*}a_{\bold{j}}+\frac{\lambda}{2}\sum_{\bold{j}\in \mathbb{Z}^d\setminus \{\bold{0}\}}\frac{\phi_{\bold{j}}}{|\Lambda|}\,\Big\{a^*_{\bold{0}}a^*_{\bold{0}}a_{\bold{j}}a_{-\bold{j}}+a^*_{\bold{j}}a^*_{-\bold{j}}a_{\bold{0}}a_{\bold{0}}\Big\}\,.\label{Bog}
\end{eqnarray}
We also define
\begin{eqnarray}
V& :=&\lambda\int \int a^*_{+}(x)a^*_{\bold{0}}(y)\phi_{(\neq 0)}(x-y)a_{+}(x)a_{+}(y)dxdy \\
& &+\lambda\int \int a^*_{+}(x)a^*_{+}(y)\phi_{(\neq 0)}(x-y)a_{+}(x)a_{\bold{0}}(y)dxdy \\
& & +\frac{\lambda}{2}\int\int a^{*}_{+}(x)a^*_{+}(y)\phi_{(\neq 0)}(x-y)a_{+}(x)a_{+}(y)dxdy
\end{eqnarray}
so that
\begin{equation}\label{complete-H}
H=H^{Bog}+V\,.
\end{equation}
From now on, we set 
\begin{equation}\label{definitions}
\lambda=\frac{1}{\rho}\quad\text{with}\quad\rho>0,\quad m=\frac{1}{2}\quad,\quad N=\rho |\Lambda|\,\,\, \text{and}\,\,\text{even}.
\end{equation}
With obvious modifications the analysis can be done also for $N\gg 1$ being an odd number.
\\

%\begin{remark}
%The Hamiltonian $\mathscr{H}\upharpoonright_{\mathcal{F}^N}$ coincides with the Hamiltonian in Eq. (1) of Seiringer's paper for 
%\begin{itemize}
%\item
%$\lambda=\frac{1}{N-1}$ and $m=\frac{1}{2}$
%\item
% $\Lambda\equiv$ the box of side $L=1$
% \item
%$ \phi(x-y)\equiv v(x-y)$.
%\end{itemize}
%Notice also that for $|\Lambda|=1$ and $\lambda=\frac{1}{N-1}$
%\begin{equation}
%-c_N=\frac{N^2\phi_{\bold{0}}}{2(N-1)}-\frac{N\phi_{\bold{0}}}{2(N-1)}
%\end{equation}
%\end{remark}
{\bf{Notation}}

\begin{enumerate}
\item
The symbol $\charf$ stands for the identity operator. If helpful we specify the Hilbert space where it acts, e.g., $\charf_{\mathcal{F}^N}$. For $c-$number operators, e.g., $z\charf$, we may omit the symbol $\charf$.
\item
The symbol $\langle\,\,,\,\,\rangle$ stands for the scalar product in $\mathcal{F}^N$.
\item
The symbol $o(\alpha)$ stands for a quantity such that $o(\alpha)/\alpha \to 0$ as $\alpha \to 0$. The symbol $\mathcal{O}(\alpha)$ stands for a quantity bounded in absolute value by a constant times $\alpha$ ($\alpha>0$). Throughout the paper the related implicit multiplicative constants are always independent of $\rho$, $L$, and $d$.
%\item
% The implicit multiplicative constant is always independent of $\rho$, $L$, and $d$.
\item
The symbol $|\psi \rangle \langle \psi|$, with $\|\psi\|=1$,  stands for the one-dimensional projection onto the state $\psi$.
\item
The word mode is used for the wavelength $\frac{2\pi}{L}\bold{j}$ (or simply for $\bold{j}$) when we refer to the field mode associated with it.
\end{enumerate}
\section{Multi-scale analysis in the particle states occupation numbers for the Hamiltonian $H^{Bog}_{\bold{j}_{*}}$}\label{multiscale-HBog}
\setcounter{equation}{0}
%Since the momentum operator $ \sum_{\bold{j}\in  \mathbb{Z}^3}k_{\bold{j}}a_{\bold{j}}^{*}a_{\bold{j}}$ commutes with $H$ and $H^{Bog}$, it is convenient to consider the decomposition of $\mathcal{F}^N$ into sectors $\mathcal{F}^N_{P}$ where $P$ is the sum of a collection of $k_{\bold{j}}$. Then, we consider the Hamiltonians $H_P$, $H^{Bog}_P$, and the interaction $\Delta H_P$ at any fixed total momentum $P$. 
The terms in $H^{Bog}$ that do not conserve the number of zero-mode particles are
\begin{equation}
\phi_{\bold{j}}\frac{a^*_{\bold{0}}a^*_{\bold{0}}a_{\bold{j}}a_{-\bold{j}}}{N}=:W_{\bold{j}}\,\quad,\quad
\phi_{\bold{j}}\frac{a_{\bold{0}}a_{\bold{0}}a^*_{\bold{j}}a^*_{-\bold{j}}}{N}=:W^*_{\bold{j}}\,.
\end{equation} 
For later convenience, we define
\begin{equation}\label{H0j}
\hat{H}^0_{\bold{j}}:=(k^2_{\bold{j}}+\phi_{\bold{j}}\frac{a^*_{\bold{0}}a_{\bold{0}}}{N})a_{\bold{j}}^{*}a_{\bold{j}}+(k^2_{\bold{j}}+\phi_{\bold{j}}\frac{a^*_{\bold{0}}a_{\bold{0}}}{N})a_{-\bold{j}}^{*}a_{-\bold{j}}\quad,\quad H_0:=\frac{1}{2}\sum_{\bold{j}\in\mathbb{Z}^d\setminus \{\bold{0}\}}\hat{H}^{0}_{\bold{j}}\,,
\end{equation}
and
\begin{equation}\label{check-HBogj}
\hat{H}^{Bog}_{\bold{j}}:=\hat{H}^0_{\bold{j}} +W_{\bold{j}}+W^*_{\bold{j}}
\end{equation}
so that
\begin{equation}
H^{Bog}=\frac{1}{2}\sum_{\bold{j}\in\mathbb{Z}^d\setminus \{\bold{0}\}}\hat{H}^{Bog}_{\bold{j}}\,.
\end{equation}

\noindent
The Bogoliubov energy is, by definition,
\begin{equation}
E^{Bog}:=\frac{1}{2}\sum_{\bold{j}\in\mathbb{Z}^d\setminus \{\bold{0}\}}E^{Bog}_{\bold{j}}\,
\end{equation}
where
%Therefore, for the three-modes system we have
\begin{equation}
E^{Bog}_{\bold{j}}:=-\Big[k^2_{\bold{j}}+\phi_{\bold{j}}-\sqrt{(k^2_{\bold{j}})^2+2\phi_{\bold{j}}k^2_{\bold{j}}}\Big]\,.
\end{equation}

Now, we focus on a three-modes system, i.e.,  $\phi_{\bold{j}}\neq 0$ only for $\bold{j}= \pm \bold{j}_{*}\neq \bold{0}$, and we construct the ground state of the corresponding Bogoliubov Hamiltonian:
%Notice however that because of the selection rules on the particle momentum, the operators $V$, $V^*$, and $U$ are identically zero if the Hilbert space contains only the modes $ \{\bold{0};\bold{j};-\bold{j} \}$.  
%The treatment of the Hamiltonian $H$ in Section \ref{multiscale-H} is very similar but the algebra is slightly more involved. 
%\begin{eqnarray}
%H^{Bog}
%&=&\sum_{\bold{j}\in\mathbb{Z}^3\setminus \{\bold{0}\}} (\frac{k^2_{\bold{j}}}{2m}+\lambda\frac{\phi_{\bold{j}}}{|\Lambda|}a^*_{\bold{0}}a_{\bold{0}})a_{\bold{j}}^{*}a_{\bold{j}}+\frac{\lambda}{2}\sum_{\bold{j}\in \mathbb{Z}^3\setminus \{\bold{0}\}}\frac{\phi_{\bold{j}}}{|\Lambda|}\,\Big\{a^*_{\bold{0}}a^*_{\bold{0}}a_{\bold{j}}a_{-\bold{j}}+a^*_{\bold{j}}a^*_{-\bold{j}}a_{\bold{0}}a_{\bold{0}}\Big\}\,.
%\end{eqnarray}
\begin{eqnarray}
H^{Bog}_{\bold{j}_{*}}
&:=&\sum_{\bold{j}\in\mathbb{Z}^d\setminus \{\pm\bold{j}_{*} \}} k^2_{\bold{j}}a_{\bold{j}}^{*}a_{\bold{j}}+\hat{H}^{Bog}_{\bold{j}_{*}}\,.
%+\phi_{\bold{j}_{*}}\frac{a^*_{\bold{0}}a_{\bold{0}}}{N}a_{\bold{j}_{*}}^{*}a_{\bold{j}_{*}}+\phi_{\bold{j}_{*}}\frac{a^*_{\bold{0}}a_{\bold{0}}}{N}a_{-\bold{j}_{*}}^{*}a_{-\bold{j}_{*}}+\frac{\phi_{\bold{j}}}{N}\,\Big\{a^*_{\bold{0}}a^*_{\bold{0}}a_{\bold{j}_{*}}a_{-\bold{j}_{*}}+a^*_{\bold{j}_{*}}a^*_{-\bold{j}_{*}}a_{\bold{0}}a_{\bold{0}}\Big\}\,\quad
\end{eqnarray}
\begin{remark}
Notice that $H^{Bog}_{\bold{j}_{*}}$ contains the kinetic energy corresponding to all the modes whereas $\hat{H}^{Bog}_{\bold{j}_{*}}$ contains the kinetic energy associated with the interacting modes only.
\end{remark}
\subsection{Feshbach-Schur projections and Feshbach-Schur Hamiltonians for $H^{Bog}_{\bold{j}_*}$}\label{Feshbach}
In the following, we describe the construction of the Feshbach-Schur Hamiltonians starting from the definition of the Feshbach-Schur projections. In Remark \ref{motivations} we highlight some important features of the strategy. In Section \ref{Control of the Feshbach-Schur flow} after Theorem \ref {theorem-Bog}
 we explain why the Feshbach-Schur maps defined below fulfill the isospectrality property.
\\

\noindent
We consider $H^{Bog}_{\bold{j}_*}$ applied\footnote{Notice that $W_{\bold{j}_*}$, $W^*_{\bold{j}_*}$ are  bounded operators when restricted to $\mathcal{F}^N$. Then, $H^{Bog}_{\bold{j}_*}$ is essentially selfadjoint on any core of $H^0_{\bold{j}_*}$.} to vectors in $\mathcal{F}^{N}$, and we define
\begin{itemize}
\item
$Q^{(0,1)}_{\bold{j}_*}:=$
the projection (in $\mathcal{F}^{N}$) onto the subspace generated by vectors with $N-0=N$ or $N-1$ particles in the modes $\bold{j}_{*}$ and $-\bold{j}_{*}$, i.e., the operator $a^*_{\bold{j}_*}a_{\bold{j}_*}+a^*_{-\bold{j}_*}a_{-\bold{j}_*}$ has eigenvalues $N$ and $N-1$ when restricted to $Q^{(0,1)}_{\bold{j}_*}\mathcal{F}^N$.
\item
$Q^{(>1)}_{\bold{j}_*}:=$ the projection onto the orthogonal complement of $Q^{(0,1)}_{\bold{j}_*}\mathcal{F}^{N}$ in $\mathcal{F}^{N}$.
\end{itemize}
Hence, we can write $$Q^{(0,1)}_{\bold{j}_*}+Q^{(>1)}_{\bold{j}_*}=\charf_{\mathcal{F}^{N}}\,.$$ 
 
 \noindent
Analogously, starting from $i=2$ up to $i=N-2$ with $i$ even, we define:
\begin{itemize}
\item
 $Q^{(i, i+1)}_{\bold{j}_*}$
the projection onto the subspace of $Q^{(>i-1)}_{\bold{j}_*}\mathcal{F}^{N}$ spanned by the vectors with $N-i$ or $N-i-1$ particles in the modes $\bold{j}_{*}$ and $-\bold{j}_{*}$; 
\item$Q^{(>i+1)}_{\bold{j}_*}$ the projection onto the orthogonal complement of $Q^{(i, i+1)}_{\bold{j}_*}Q^{(>i-1)}_{\bold{j}_*}\mathcal{F}^{N}$ in $Q^{(>i-1)}_{\bold{j}_*}\mathcal{F}^{N}$\,.
\end{itemize}
Hence, we can write
\begin{equation}
Q^{(>i+1)}_{\bold{j}_*}+Q^{(i,i+1)}_{\bold{j}_*}=Q^{(>i-1)}_{\bold{j}_*}\,.
\end{equation}
%\begin{remark}
%Notice that if we denote by $\mathcal{F}^{N}_{\{\bold{0};\bold{j}_{*};-\bold{j}_{*}\}}$ the subspace of vectors that contain only particles in the modes $\{\bold{0};\bold{j}_{*};-\bold{j}_{*}\}$, then $Q^{(>N-1)}\mathcal{F}^{N}_{\{\bold{0};\bold{j}_{*};-\bold{j}_{*}\}}\equiv Q^{(N)}\mathcal{F}^{N}_{\{\bold{0};\bold{j}_{*};-\bold{j}_{*}\}}$ where the R-H-S is the one-dimensional subspace generated by the state $$\eta:=\frac{1}{\sqrt{N!}}a_{\bold{0}}^*\dots a_{\bold{0}}^*\Omega$$ with all the $N$ particles in the zero-mode state. 
%\end{remark}

We recall that given the (separable) Hilbert space $\mathcal{H}$ and the projections $\mathscr{P}$, $\overline{\mathscr{P}}$ where $\mathscr{P}+\overline{\mathscr{P}}=\charf_{\mathcal{H}}$, the Feshbach-Schur map associated with $\mathscr{P}$ and $\overline{\mathscr{P}}$ maps the (closed) operator $K-z$, $z$ in a subset of $ \mathbb{C}$,  acting on $\mathcal{H}$ to the operator $\mathscr{F}(K-z)$ acting on $\mathscr{P}\mathcal{H}$ where (formally)
\begin{equation}
\mathscr{F}(K-z):=\mathscr{P}(K-z)\mathscr{P}-\mathscr{P}K\overline{\mathscr{P}}\frac{1}{\overline{\mathscr{P}}(K-z)\overline{\mathscr{P}}}\overline{\mathscr{P}}K\mathscr{P}\,.
\end{equation}

In Section \ref{Nonrigorous derivation} we provide an informal derivation of the Feshbach-Schur Hamiltonians. The rigorous control of the Feshbach-Schur flow up to $i=N-2$ is the content of  Section \ref{Control of the Feshbach-Schur flow}. From now on, we consider $z\in \mathbb{R}$.

\subsubsection{Outline of the Feshbach-Schur flow}\label{Nonrigorous derivation}

We shall iterate the Feshbach-Schur map starting from $i=0$ up to $i=N-2$ with $i$ even, using the projections $\mathscr{P}^{(i)}$ and $\overline{{\mathscr{P}}^{(i)}}$  for the i-th  step\footnote{We use this notation though the number of steps is in fact $1+i/2$ being $i$ an even number.}  of the iteration where
\begin{equation}\label{projections}
\mathscr{P}^{(i)}:= Q^{(>i+1)}_{\bold{j}_*}\quad,\quad \overline{{\mathscr{P}}^{(i)}}:= Q^{(i,i+1)}_{\bold{j}_*}\,.
\end{equation}
We denote by  $\mathscr{F}^{(i)}$ the Feshbach-Schur map at the i-th step. We start applying $\mathscr{F}^{(0)}$ to $H^{Bog}_{\bold{j}_*}-z$ and compute
\begin{eqnarray}
& &\mathscr{K}^{Bog\,(0)}_{\bold{j}_*}(z)\\
&:=&\mathscr{F}^{(0)}(H^{Bog}_{\bold{j}_*}-z)\\
&=&Q^{(>1)}_{\bold{j}_*}(H^{Bog}_{\bold{j}_*}-z)Q^{(>1)}_{\bold{j}_*}-Q^{(>1)}_{\bold{j}_*}H^{Bog}_{\bold{j}_*}Q^{(0,1)}_{\bold{j}_*}\frac{1}{Q^{(0,1)}_{\bold{j}_*}(H^{Bog}_{\bold{j}_*}-z)Q^{(0,1)}_{\bold{j}_*}}Q^{(0,1)}_{\bold{j}_*}H^{Bog}_{\bold{j}_*}Q^{(>1)}_{\bold{j}_*}\quad\quad\\
&=&Q^{(>1)}_{\bold{j}_*}(H^{Bog}_{\bold{j}_*}-z)Q^{(>1)}_{\bold{j}_*}-Q^{(>1)}_{\bold{j}_*}W_{\bold{j}_*}Q^{(0,1)}_{\bold{j}_*}\frac{1}{Q^{(0,1)}_{\bold{j}_*}(H^{Bog}_{\bold{j}_*}-z)Q^{(0,1)}_{\bold{j}_*}}Q^{(0,1)}_{\bold{j}_*}W^*_{\bold{j}_*}Q^{(>1)}_{\bold{j}_*}\,. \label{KappaBog0}
\end{eqnarray}
Then,  we iteratively define
\begin{equation}\label{definition-kappa-operators}
\mathscr{K}^{Bog\,(i)}_{\bold{j}_*}(z):=\mathscr{F}^{(i)}(\mathscr{K}^{Bog\,(i-2)}_{\bold{j}_*}(z))\,,\quad i=0,\dots,N-2\quad \text{with}\,\, i\,\, \text{even},
\end{equation}
where $\mathscr{K}^{Bog\,(-2)}_{\bold{j}_*}(z)\equiv H^{Bog}_{\bold{j}_*}-z$.

%We stress that the operator $\mathscr{K}^{(i)}_{\bold{j}_*}(z)$ acts on $Q^{(>i+1)}_{\bold{j}_*}\mathcal{F}^N$.

Notice that, for $l$ and $l'$ even numbers, $Q^{(l,l+1)}_{\bold{j}_*}W_{\bold{j}_*}Q^{(l',l'+1)}_{\bold{j}_*}\neq 0$ only if $l-l'=2$ and $Q^{(l,l+1)}_{\bold{j}_*}W^*_{\bold{j}_*}Q^{(l',l'+1)}_{\bold{j}_*}\neq 0$ only if $l-l'=-2$. This implies
%\begin{equation}
%Q^{(2,3)}_{\bold{j}_*}\mathscr{K}_{\bold{j}_*}^{Bog\,(0)}(z)Q^{(2,3)}_{\bold{j}_*}=Q^{(2,3)}_{\bold{j}_*}(H^{Bog}_{\bold{j}_*}-z)Q^{(2,3)}_{\bold{j}_*}-Q^{(2,3)}_{\bold{j}_*}W_{\bold{j}_*}Q^{(0,1)}_{\bold{j}_*}\frac{1}{Q^{(0,1)}_{\bold{j}_*}(H^{Bog}_{\bold{j}_*}-z)Q^{(0,1)}_{\bold{j}_*}}Q^{(0,1)}_{\bold{j}_*}W^*_{\bold{j}_*}Q^{(2,3)}_{\bold{j}_*}
%\end{equation}
\begin{equation}\label{QWQ-Bog}
Q^{(>3)}_{\bold{j}_*}\mathscr{K}^{Bog\,(0)}_{\bold{j}_*}(z)\,Q^{(2,3)}_{\bold{j}_*}=Q^{(>3)}_{\bold{j}_*}W_{\bold{j}_*}\,Q^{(2,3)}_{\bold{j}_*}\,
\end{equation}
and
\begin{equation}\label{QWQ-Bog-bis}
Q^{(>3)}_{\bold{j}_*}\mathscr{K}^{Bog\,(0)}_{\bold{j}_*}(z)\,Q^{(>3)}_{\bold{j}_*}=Q^{(>3)}_{\bold{j}_*}H^{Bog}_{\bold{j}_*}\,Q^{(>3)}_{\bold{j}_*}\,.
\end{equation}
Hence,  a straightforward calculation shows that
%\begin{eqnarray}
%\mathscr{K}^{(0)}(z)&=&Q^{(>0)}(H^{Bog}-z)Q^{(>0)}-\frac{\lambda^2}{4}Q^{(>0)} W\,Q^{(0)} \frac{1}{Q^{(0)}(H^{Bog}-z)Q^{(0)}}Q^{(0)}W^*Q^{(>0)}\quad\quad\quad 
%\end{eqnarray}
\begin{eqnarray}
& &\mathscr{K}^{Bog\,(2)}_{\bold{j}_*}(z)\\
&=&Q^{(>3)}_{\bold{j}_*}(H^{Bog}_{\bold{j}_*}-z)Q^{(>3)}_{\bold{j}_*}\\
& &-Q^{(>3)}_{\bold{j}_*}W_{\bold{j}_*}\,Q^{(2,3)}_{\bold{j}_*}\frac{1}{Q^{(2,3)}_{\bold{j}_*}(H^{Bog}_{\bold{j}_*}-W_{\bold{j}_*}Q^{(0,1)}_{\bold{j}_*}\frac{1}{Q^{(0,1)}_{\bold{j}_*}(H^{Bog}_{\bold{j}_*}-z)Q^{(0,1)}_{\bold{j}_*}}Q^{(0,1)}_{\bold{j}_*}W^*_{\bold{j}_*}-z)Q^{(2,3)}_{\bold{j}_*}}Q^{(2,3)}_{\bold{j}_*}W^*_{\bold{j}_*}Q^{(>3)}_{\bold{j}_*}\,.\quad\quad\quad  \label{KappaBog1}
%& &-Q^{(>1)}_{\bold{j}_*}W_{\bold{j}_*}\,Q^{(1)}_{\bold{j}_*}\frac{1}{Q^{(1)}_{\bold{j}_*}(H^{Bog}_{\bold{j}_*}-z)Q^{(1)}_{\bold{j}_*}}Q^{(1)}_{\bold{j}_*}W^*_{\bold{j}_*}Q^{(>1)}_{\bold{j}_*}\quad\quad\quad  \nonumber
\end{eqnarray}
Assuming that the  expansion
\begin{eqnarray}
& &Q^{(2,3)}_{\bold{j}_*}\frac{1}{Q^{(2,3)}_{\bold{j}_*}(H^{Bog}_{\bold{j}_*}-W_{\bold{j}_*}\,Q^{(0,1)}_{\bold{j}_*}\frac{1}{Q^{(0,1)}_{\bold{j}_*}(H^{Bog}_{\bold{j}_*}-z)Q^{(0,1)}_{\bold{j}_*}}Q^{(0,1)}_{\bold{j}_*}W^*_{\bold{j}_*}-z)Q^{(2,3)}_{\bold{j}_*}}Q^{(2,3)}_{\bold{j}_*} \\
&=& Q^{(2,3)}_{\bold{j}_*}\sum_{l_2=0}^{\infty}\frac{1}{Q^{(2,3)}_{\bold{j}_*}(H^{Bog}_{\bold{j}_*}-z)Q^{(2,3)}_{\bold{j}_*}}\times \\
&  &\quad\quad\quad \times\Big[Q^{(2,3)}_{\bold{j}_*}W_{\bold{j}_*}\,Q^{(0,1)}_{\bold{j}_*}\frac{1}{Q^{(0,1)}_{\bold{j}_*}(H^{Bog}_{\bold{j}_*}-z)Q^{(0,1)}_{\bold{j}_*}}Q^{(0,1)}_{\bold{j}_*}W^*_{\bold{j}_*}Q^{(2,3)}_{\bold{j}_*}\frac{1}{Q^{(2,3)}_{\bold{j}_*}(H^{Bog}_{\bold{j}_*}-z)Q^{(2,3)}_{\bold{j}_*}}\Big]^{l_2} Q^{(2,3)}_{\bold{j}_*} \,\nonumber
\end{eqnarray}
 is well defined,
and using the notation $$W_{\bold{j}_*\,;\,i,i'}:=Q^{(i,i+1)}_{\bold{j}_*}W_{\bold{j}_*}Q^{(i',i'+1)}_{\bold{j}_*}\quad,\quad W^*_{\bold{j}_*\,;\,i,i'}:=Q^{(i,i+1)}_{\bold{j}_*}W^*_{\bold{j}_*}Q^{(i',i'+1)}_{\bold{j}_*} \,,$$ we can write
%\begin{eqnarray}
%\mathscr{K}^{(2)}(z)&=&Q^{(>2)}(H^{Bog}-z)Q^{(>2)}-\frac{\lambda^2}{4}Q^{(>2)}W\,Q^{(1)}\frac{1}{Q^{(1)}(H^{Bog}-z)Q^{(1)}}Q^{(1)}W^*Q^{(>2)}\quad\quad\quad \\
%& &-\frac{\lambda^2}{4}\sum_{l=0}^{\infty}Q^{(>2)}W\,Q^{(2)}\frac{1}{Q^{(2)}(H^{Bog}-z)Q^{(2)}}\Big[\frac{\lambda^2}{4}Q^{(2)}W\,Q^{(0)}\frac{1}{Q^{(0)}(H^{Bog}-z)Q^{(0)}}Q^{(0)}W^*Q^{(2)}\frac{1}{Q^{(2)}(H^{Bog}-z)Q^{(2)}}\Big]^lQ^{(2)}W^*Q^{(>2)}\quad\quad\quad 
%\end{eqnarray}
%We use the notation $W_{i,i'}=Q^{(i)}WQ^{(i')}$ and $W^*_{i,i'}=Q^{(i)}W^*Q^{(i')}$
\begin{eqnarray}
& &\mathscr{K}^{Bog\,(2)}_{\bold{j}_*}(z)\\
&=&Q^{(>3)}_{\bold{j}_*}(H^{Bog}_{\bold{j}_*}-z)Q^{(>3)}_{\bold{j}_*}\quad\quad\quad  \label{Kappa2}\\
& &-\sum_{l_2=0}^{\infty}Q^{(>3)}_{\bold{j}_*}W_{\bold{j}_*}Q^{(2,3)}_{\bold{j}_*}\frac{1}{Q^{(2,3)}_{\bold{j}_*}(H^{Bog}_{\bold{j}_*}-z)Q^{(2,3)}_{\bold{j}_*}}\times\\
& &\quad\quad \times\Big[W_{\bold{j}_*\,;\,2,0}\,\frac{1}{Q^{(0,1)}_{\bold{j}_*}(H_{\bold{j}_*}^{Bog}-z)Q^{(0,1)}_{\bold{j}_*}}W_{\bold{j}_*\,;\,0,2}^*\frac{1}{Q^{(2,3)}_{\bold{j}_*}(H^{Bog}_{\bold{j}_*}-z)Q^{(2,3)}_{\bold{j}_*}}\Big]^{l_2}Q^{(2,3)}_{\bold{j}_*}W^*_{\bold{j}_*}Q_{\bold{j}_*}^{(>3)}\,.\quad\quad\quad \nonumber
\end{eqnarray}

\noindent
With the definition
\begin{equation}
R^{Bog}_{\bold{j}_*\,;\,i,i}(z):=Q^{(i,i+1)}_{\bold{j}_*}\frac{1}{Q^{(i,i+1)}_{\bold{j}_*}(H^{Bog}_{\bold{j}_*}-z)Q^{(i,i+1)}_{\bold{j}_*}}Q^{(i,i+1)}_{\bold{j}_*}\,,
\end{equation}
for $4\leq i \leq N-2$ we get
\begin{eqnarray}
& &\mathscr{K}^{Bog\,(i)}_{\bold{j}_*}(z)\\
&=&Q^{(>i+1)}_{\bold{j}_*}(H^{Bog}_{\bold{j}_*}-z)Q^{(>i+1)}_{\bold{j}_*}\\
& &-\sum_{l_i=0}^{\infty}Q^{(>i+1)}_{\bold{j}_*}W_{\bold{j}_*}R^{Bog}_{\bold{j}_*\,;\,i,i}(z)\Big[W_{i,i-2}\,R^{Bog}_{\bold{j}_*\,;\,i-2,i-2}(z)\times \label{exp-1}\\
& &\quad\quad\quad\times \sum_{l_{i-2}=0}^{\infty}\Big[W_{\bold{j}_*\,;\,i-2,i-4}\,\dots W^*_{\bold{j}_*\,;\,i-4,i-2}R^{Bog}_{\bold{j}_*\,;\,i-2,i-2}(z)\Big]^{l_{i-2}}W^*_{\bold{j}_*\,;\,i-2,i}R^{Bog}_{\bold{j}_*\,;\,i,i}(z)\Big]^{l_i} W^*_{\bold{j}_*}Q^{(>i+1)}_{\bold{j}_*}\quad \label{dots-2}\quad\quad
%& &\quad\quad\quad\quad\quad\quad\quad\quad\quad \times W^*_{\bold{j}_*}Q^{(>i+1)}_{\bold{j}_*}\quad\quad\quad \nonumber
\end{eqnarray}
where $i$ is an even number and the expression corresponding to $\dots$ in (\ref{dots-2}) is made precise in Theorem \ref{theorem-Bog}.

\begin{definition}
%We define
%\begin{equation}
%E^{Bog}_{\mathcal{G}_{\bold{j}}}:=\sum_{\bold{j}'\in \mathcal{G}_{\bold{j}}}E^{Bog}_{\bold{j}'}\,
%\end{equation}
%where
%\begin{equation}
%E^{Bog}_{\bold{j}}:=-\frac{1}{2}\Big[k^2_{\bold{j}}+\phi_{\bold{j}}-\sqrt{(k^2_{\bold{j}})^2+2\phi_{\bold{j}}k^2_{\bold{j}}}\Big].
%\end{equation}
%therefore
%\begin{equation}
%E^{Bog}:=\sum_{\bold{l}=\bold{j}_{*}, -\bold{j}_{*}}E^{Bog}_{\bold{l}}\,
%\end{equation}
We  define
\begin{equation}
\frac{k_{\bold{j}_{*}}^2}{\phi_{\bold{j}_{*}}}=:\epsilon_{\bold{j}_*}\,,
\end{equation}
thus 
\begin{equation}
\frac{E^{Bog}_{\bold{j}_{*}}}{\phi_{\bold{j}_{*}}}=-\Big[\epsilon_{\bold{j}_*}+1-\sqrt{\epsilon_{\bold{j}_*}^2+2\epsilon_{\bold{j}_*}}\,\Big]\,.
\end{equation}
\end{definition}
%\begin{remark}\label{high regime}
%{\color{red} Notice that $\epsilon_{\bold{j}_*}\to 0$ either as $\phi_{\bold{j}_{*}}\to \infty$ or as $L\to \infty$ at fixed $\phi_{\bold{j}_{*}}$.}
%\end{remark}

\subsubsection{Motivations and features of the strategy} \label{motivations}
After the heuristic implementation of the Feshbach-Schur flow, we can better explain the main motivations and features of this strategy.
\\

\begin{enumerate}
\item
In the case of a finite box, as $N\to \infty$ the ground state of $H^{Bog}_{\bold{j}_*}$ (restricted to $\mathcal{F}^N$) converges in the sense of (rescaled) one-particle reduced density matrix  to the state $\eta$ $$\eta:=\frac{1}{\sqrt{N!}}a_{\bold{0}}^*\dots a_{\bold{0}}^*\Omega$$ where all the $N$ particles are in the zero-mode state, $\Omega$ being the vacuum vector. Therefore, the contribution of the components with a macroscopic  number of particles in the nonzero modes states has to be irrelevant in the limit $N\to \infty$. 
\\

\item In connection to the previous remark, we define $(\mathcal{F}^N)_{\{\bold{0}; \pm\bold{j}_*\}}\subset \mathcal{F}^N$ the subspace spanned by vectors containing particles only in the modes $\bold{0}, \pm\bold{j}_*$. Consider for the moment the Hamiltonian $\hat{H}^{Bog}_{\bold{j}_*}$ instead of $H^{Bog}_{\bold{j}_*}$. We observe that, if the Feshbach-Schur flow associated with the operator  $(\hat{H}^{Bog}_{\bold{j}_*}-z)\upharpoonright_{(\mathcal{F}^N)_{\{\bold{0}; \pm\bold{j}_*\}}}$ is well defined, the Feshbach-Schur Hamiltonian at the $N-2-th$ step is an operator proportional to the projection $|\eta\rangle\langle \eta |$, where the multiplicative factor is a function $f(z)$.
\\

\item
Starting from the previous observation (see point 2.) we recall that if $f(z_*)=0$ for some $z_*$ then $z_*$ is an eigenvalue of the original Hamiltonian $\hat{H}^{Bog}_{\bold{j}_*}$ due to the isospectrality that holds at each step of the Feshbach-Schur flow. Feshbach-Schur theory provides also an algorithm to reconstruct  the eigenvector of the original Hamiltonian $\hat{H}^{Bog}_{\bold{j}_*}$ associated with the eigenvalue $z_*$ from the eigenvector ($\eta$) with eigenvalue zero of the Feshbach-Schur Hamiltonian,  $f(z_*)|\eta\rangle\langle \eta |$,  at the $N-2-th$ step. 
%The isospectrality property of the Feshbach-Schur map is used in Section \ref{groundstate} to provide the expression of the ground state vector of the Hamiltonian $H^{Bog}_{\bold{j}_*}$ (acting on $\mathcal{F}^N$) in (\ref{gs-1})-(\ref{gs-2}).
\\

\item
With regard to the estimates that are needed to control the series expansions in (\ref{exp-1})-(\ref{dots-2}), we explain the role of the projections in (\ref{projections}). Note that in the resolvent
\begin{eqnarray}
& &R^{Bog}_{\bold{j}_*\,;\,i,i}(z)\\
&:=&Q^{(i,i+1)}_{\bold{j}_*}\frac{1}{Q^{(i,i+1)}_{\bold{j}_*}(H^{Bog}_{\bold{j}_*}-z)Q^{(i,i+1)}_{\bold{j}_*}}Q^{(i,i+1)}_{\bold{j}_*}\\
&=&Q^{(i,i+1)}_{\bold{j}_*}\frac{1}{Q^{(i,i+1)}_{\bold{j}_*}(\hat{H}^0_{\bold{j}_*}+\sum_{\bold{j}\in\mathbb{Z}^d\setminus \{\pm\bold{j}_{*} \}} k^2_{\bold{j}}a_{\bold{j}}^{*}a_{\bold{j}}-z)Q^{(i,i+1)}_{\bold{j}_*}}Q^{(i,i+1)}_{\bold{j}_*}\,
\end{eqnarray}
the interaction terms $W_{\bold{j}_*}$ and $W^*_{\bold{j}_*}$ disappear due to the perpendicular projection $Q^{(i,i+1)}_{\bold{j}_*}$. This mechanism yields an artificial gap because $z$ will be chosen close to the Bogoliubov energy. The expansion in (\ref{exp-1})-(\ref{dots-2}) turns out to be well defined when the ratio $\epsilon_{\bold{j}_*}$ between the kinetic energy $k_{\bold{j}_{*}}^2$ and the Fourier component $\phi_{\bold{j}_{*}}$ is sufficiently small. In fact, it can be arbitrarily small (but positive) provided $N$ is sufficiently large; see Section \ref{statement} below. However, it is important to stress that there is no \emph{small} parameter in the expansions that we use to define the Feshbach flow, in the sense that the operators are not defined as $\epsilon_{\bold{j}_*}$ tends to zero.
\end{enumerate}

\subsubsection{Statement of the results and role of the assumptions}\label{statement}
In the list of remarks below we specify the results that are obtained for the Hamiltonian $H^{Bog}_{\bold{j}_*}$ acting on $\mathcal{F}^N$, commenting on the role of the \emph{Strong interaction potential assumption} and of Condition 3.2 in Definition \ref{def-pot}. \emph{We stress that $\bold{j}_*$ is kept fixed in the sequel.}
\begin{enumerate}

\item
For the implementation of the Feshbach-Schur map up to the $N-2-th$ step we shall require   $\frac{1}{N}\leq \epsilon^{\nu}_{\bold{j}_*}$ for some $\nu >\frac{11}{8}$ and $\epsilon_{\bold{j}_*}$  sufficiently small (that we always assume in the sequel). The bound $\frac{1}{N}\leq \epsilon^{\nu}_{\bold{j}_*}$ holds in the mean field limiting regime where the box is kept fixed and the number of particles, $N$, can be chosen sufficiently large irrespective of the box size. For space dimension $d\geq 3$, at fixed particle density,  the bound $\frac{1}{N}\leq \epsilon^{\nu}_{\bold{j}_*}$  is fulfilled (for $\nu<\frac{3}{2}$) if the box is sufficiently large. For $d=1,2$, if at fixed $\phi_{\bold{j}_*}$ (and  $\bold{j}_*$) the box size tends to infinity the particle density $\rho$ must be suitably divergent to ensure the bound $\frac{1}{N}\leq \epsilon^{\nu}_{\bold{j}_*}$.
\\

\item
For the last step of the Feshbach-Schur flow (see Section \ref{groundstate}),  Condition 3.2) in Definition \ref{def-pot} is also necessary to include values of the spectral parameter $z$ belonging to a neighborhood of the ground state energy of $H^{Bog}_{\bold{j}_*}$. This condition is fulfilled for any dimension $d$  in the mean field limiting regime.  At fixed particle density and  for $d\geq 2$, Condition 3.2) is fulfilled  if $L$ is sufficiently large.  The final Feshbach-Schur Hamiltonian has the form
\begin{equation}\label{final-H-0}
\mathscr{K}^{Bog\,(N)}_{\bold{j}_*}(z)=f_{\bold{j}_*}(z)|\eta \rangle \langle \eta |\,.
\end{equation}
%(Of course it is also fulfilled if $\rho$ is sufficiently large.)

\item
The existence of the point $z_*$ such that $f_{\bold{j}_*}(z_*)=0$, i.e., the ground state energy of $H^{Bog}_{\bold{j}_*}$ (due to the isospectrality property of Feshbach-Schur map),  is established  for any space dimension $d\geq 1$ in the mean field limiting regime. 

\noindent
With regard to a  box of arbitrarily large side $1<L<\infty$, the existence of $z_*$ (see  Remark \ref{cond-gamma})  is established if $\rho \geq \rho_0(L/L_0)^{3-d}$ where $\rho_0$ is sufficiently large and $L_0=1$. Hence, for $d\geq 3$ it is enough to require $\rho$ be sufficiently large but independent of $L(>1)$ and the result  holds for a finite box of arbitrarily large (finite) volume $|\Lambda|$.  
%{\color{red}We believe that some estimates can be improved and that the same result can be shown in space dimension $d= 3$.}
\\

\item
In all cases where the existence of $z_*$ is proven, we can construct  the ground state of the Hamiltonian $H^{Bog}_{\bold{j}_*}$; see (\ref{gs-1})-(\ref{gs-2}) in Corollary \ref{col-hbog}, Section \ref{groundstate}. Next, in Lemma  \ref{eigenvalue} we show that  in the mean field limiting regime the inequality $|z_*-E^{Bog}_{\bold{j}_*}|\leq \mathcal{O}(\frac{1}{N^{\beta}})$ holds  for any $0<\beta <1$. In addition, Lemma  \ref{eigenvalue} shows that, for any scaling $\rho=\rho_0 (\frac{L}{L_0})^{s}$ with $s>0$, in space dimension $d=3$ the ground state energy of $H^{Bog}_{\bold{j}_*}$ tends to $E^{Bog}_{\bold{j}_*}$ as $L\to \infty$. This readily implies that  in space dimension $d\geq 4$ the ground state energy of $H^{Bog}_{\bold{j}_*}$ tends to $E^{Bog}_{\bold{j}_*}$ in the thermodynamic limit, i.e., in the limit $L\to \infty$ at fixed $\bold{j}_*$, $\phi_{\bold{j}_*}$ and $\rho$. Concerning $d=3$, in Remark  \ref{EBoglimit} we outline the argument of the convergence $z_*\to E^{Bog}_{\bold{j}_*}$ in the thermodynamic limit  at large (but fixed) density $\rho$.

\item
We provide an expansion (with controlled remainder ) of the ground state vector in terms of the vector $\eta$ and of a finite sum of products of the interaction terms $W^*_{\bold{j}_*}\,,\,W_{\bold{j}_*}$, and of the resolvent (see (\ref{H0j})) $\frac{1}{\hat{H}^0_{\bold{j}_*}-z_*}$; see Section \ref{expansion-gs}. In the mean field limit (i.e., for a fixed box and $N\to \infty$) the expansion in terms of the bare quantities, i.e., in terms of the vector $\eta$ and of a finite sum of products of the interaction terms $W^*_{\bold{j}_*}\,,\,W_{\bold{j}_*}$, and of the resolvent $\frac{1}{\hat{H}^0_{\bold{j}_*}-E^{Bog}_{\bold{j}_*}}$,  is up to any desired precision. 
\end{enumerate}

%\begin{remark}
%For our final purpose, i.e., the construction of the ground state of the Hamiltonian $H$, we need to have the result in the lemma below for values $z$ strictly above $E^{Bog}$. This is not a problem thanks to the gap of the unperturbed Hamiltonian. In other words, for fixed potential one can find a $\Delta$ sufficiently small such that the proof works also for $z<E^{Bog}+\Delta$ with a $\delta^{Bog}_N$ still larger than $1$.
%\end{remark}

%\begin{remark}
%.....TO BE CHANGED OR ERASED.....In order to have an explicit expression of the ground state,  and a self contained paper where we prove that the ground state energy is $E^{Bog}$ plus corrections that vanish as $N\to \infty$, we have to proceed differently as follows. By splitting the potential in high and low components, at first one construct the ground state for the high components according to the multi-scale analysis in the number of particles. Then, one uses ordinary perturbation theory for the low components. This way, without invoking the result obtained with different methods,  it should be possible to determine the ground state energy and then to have an explicit expression of the ground state energy up to small corrections (but not arbitrarily small!)
%\end{remark}

\subsection{Control of the Feshbach-Schur flow}\label{Control of the Feshbach-Schur flow}
In Theorem \ref{theorem-Bog} we prove that the flow of  Feshbach-Schur Hamiltonians is well defined up to step $i=N-2$ for spectral values $z$ up to $  E^{Bog}_{\bold{j}_*}+ (\delta-1)\phi_{\bold{j}^*}\sqrt{\epsilon_{\bold{j}_*}^2+2\epsilon_{\bold{j}_*}}$ with $\delta>1$ but very close to $1$. We recall that in the mean field limit the first excited energy level of the Hamiltonian $H^{Bog}_{\bold{j}_*}$ is  expected to be located  at $$E^{Bog}_{\bold{j}_*}+\min \Big\{\phi_{\bold{j}_*}\sqrt{\epsilon_{\bold{j}_*}^2+2\epsilon_{\bold{j}_*}}\,;\, \min\{k_{\bold{j}}^2\,:\, \bold{j}\in \mathbb{Z}^d \setminus \{\bold{0},\pm \bold{j}_*\}\}\Big\}\,.$$
The proof of Theorem \ref{theorem-Bog} requires a key estimate which is the content of the next lemma. 

%\begin{remark}
%In the next lemma, we assume that $\epsilon_{\bold{j}_{*}}/N$ can be made arbitrarily small. This can be obtained for $\rho$ sufficiently large.
%\end{remark}
\begin{lemma}\label{main-lemma-Bog}
Let 
\begin{equation}
z\leq  E^{Bog}_{\bold{j}_*}+ (\delta-1)\phi_{\bold{j}_*}\sqrt{\epsilon_{\bold{j}_*}^2+2\epsilon_{\bold{j}_*}}(<0)
\end{equation} with\footnote{We set this upper bound for $\delta$ because the last step of the Feshbach-Schur flow (implemented in Section \ref{groundstate}) is defined for values of $z$ strictly smaller than the first excited eigenvalue.} $\delta< 2$,  $\frac{1}{N}\leq \epsilon^{\nu}_{\bold{j}_*}$ for some $\nu >1$, and $\epsilon_{\bold{j}_*}$ be sufficiently small. Then
\begin{eqnarray}\label{estimate-main-lemma-Bog}
%& &\frac{\sqrt{\frac{(N-l-1)}{(N-l-2)}}}{\Big[1+\frac{N}{N-l-2}(\epsilon+\frac{1+\epsilon-\sqrt{\epsilon^2+2\epsilon}}{l})\Big]^{\frac{1}{2}} \Big[1+\frac{N}{N-l}(\epsilon-\frac{1+\epsilon+\sqrt{\epsilon^2+2\epsilon}}{l})\Big]^{\frac{1}{2}}}\\
& &\| \Big[R^{Bog}_{\bold{j}_*\,;\,i,i}(z)\Big]^{\frac{1}{2}}\,W_{\bold{j}_*\,;\,i,i-2}\,\Big[R^{Bog}_{\bold{j}_*\,;\,i-2,i-2}(z)\Big]^{\frac{1}{2}}\|\,\|\Big[R^{Bog}_{\bold{j}_*\,;\,i-2,i-2}(z)\Big]^{\frac{1}{2}}\,W^*_{\bold{j}_*\,;\,i-2,i}\,\Big[R^{Bog}_{\bold{j}_*\,;\,i,i}(z)\Big]^{\frac{1}{2}}\|\quad \\
%%%%
%& &\|\Big[R^{Bog}_{i-2,i-2}(z)\Big]^{\frac{1}{2}}\,W_{i-2,i}\Big[R^{Bog}_{i,i}(z)\Big]^{\frac{1}{2}}\|\,\|\Big[R^{Bog}_{i,i}(z)\Big]^{\frac{1}{2}}W^*_{i,i-2}\,\Big[R^{Bog}_{i-2,i-2}(z)\Big]^{\frac{1}{2}}\|\\
%&\leq&\frac{1}{4\Big[1+\epsilon+\frac{\epsilon+1-\sqrt{\epsilon^2+2\epsilon}}{(N-i+2)}\Big]}\frac{1}{ \Big[1+\epsilon-\frac{\epsilon+1+\sqrt{\epsilon^2+2\epsilon}}{(N-i+2)}\Big]}+CN^{-\eta}\\
%%%%
%&\leq&\frac{1}{4\Big[1+\epsilon_{\bold{j}_*}+o(\epsilon_{\bold{j}_*})+\frac{\epsilon_{\bold{j}_*}+1-\delta\sqrt{\epsilon_{\bold{j}_*}^2+2\epsilon_{\bold{j}_*}}}{(N-i+1)}\Big]}\frac{1}{ \Big[1+\epsilon_{\bold{j}_*}+o(\epsilon_{\bold{j}_*})-\frac{\epsilon_{\bold{j}_*}+1+\delta\sqrt{\epsilon_{\bold{j}_*}^2+2\epsilon_{\bold{j}_*}}}{(N-i+1)}\Big]}\label{main-estimate-intermediate-a}\\
%%%
%\frac{1}{2\Big[1+(\epsilon+\frac{1+\epsilon-\sqrt{\epsilon^2+2\epsilon}}{N-i})-\mathcal{O}(\epsilon^2)\Big]^{\frac{1}{2}} \Big[1+(\epsilon-\frac{1+\epsilon+\sqrt{\epsilon^2+2\epsilon}}{N-i})-\mathcal{O}(\epsilon^2)\Big]^{\frac{1}{2}}}\\
%%%
&\leq &\frac{1}{4(1+a_{\epsilon_{\bold{j}_*}}-\frac{2b^{(\delta)}_{\epsilon_{\bold{j}_*}}}{N-i+1}-\frac{1-c^{(\delta)}_{\epsilon_{\bold{j}_*}}}{(N-i+1)^2})}\quad\label{def-deltabog}
\end{eqnarray}
holds for all $2\leq i\leq N-2$ where $i$ is even. Here,  \begin{equation}\label{a}
a_{\epsilon_{\bold{j}_{*}}}:=2\epsilon_{\bold{j}_{*}}+\mathcal{O}(\epsilon^{\nu}_{\bold{j}_{*}})\,,
\end{equation}
\begin{equation}\label{b}
b^{(\delta)}_{\epsilon_{\bold{j}_{*}}}:=(1+\epsilon_{\bold{j}_{*}})\delta \, \chi_{[0,2)}(\delta)\sqrt{\epsilon_{\bold{j}_{*}}^2+2\epsilon_{\bold{j}_{*}}}
\end{equation}
and
\begin{equation}\label{c}
c^{(\delta)}_{\epsilon_{\bold{j}_{*}}}:=-(1-\delta^2\, \chi_{[0,2)}(\delta))(\epsilon_{\bold{j}_{*}}^2+2\epsilon_{\bold{j}_{*}})\,
\end{equation}
with $\chi_{[0,2)}$ the characteristic function of the interval $[0,2)$.
\end{lemma}

\noindent
\emph{Proof}

\noindent
%We start with our main estimate
%\begin{equation}
%\|\Big[R^{Bog}_n_{\bold{j}_0}n_{\bold{j}_0}{i-2,i-2}(z)\Big]^{\frac{1}{2}}\,W^*_{i-2,i}\,\Big[R^{Bog}_{i,i}(z)\Big]^{\frac{1}{2}}\|\leq \mathcal{O}(\rho^{\frac{1}{2}})
%\end{equation}
\noindent
%We show the result for $z\leq  E^{Bog}$. Then, it will be clear that an analogous result holds for $z\leq \xi^{Bog} E^{Bog}$ for some $\xi^{Bog}(<1)$ sufficiently close to $1$.
We observe that
\begin{eqnarray}
& &\| \Big[R^{Bog}_{\bold{j}_*\,;\,i,i}(z)\Big]^{\frac{1}{2}}\,W_{\bold{j}_*\,;\,i,i-2}\,\Big[R^{Bog}_{\bold{j}_*\,;\,i-2,i-2}(z)\Big]^{\frac{1}{2}}\|\,\|\Big[R^{Bog}_{\bold{j}_*\,;\,i-2,i-2}(z)\Big]^{\frac{1}{2}}\,W^*_{\bold{j}_*\,;\,i-2,i}\,\Big[R^{Bog}_{\bold{j}_*\,;\,i,i}(z)\Big]^{\frac{1}{2}}\|\\
&=&\|\Big[R^{Bog}_{\bold{j}_*\,;\,i,i}(z)\Big]^{\frac{1}{2}}\,W_{\bold{j}_*\,;\,i,i-2}\,\Big[R^{Bog}_{\bold{j}_*\,;\,i-2,i-2}(z)\Big]^{\frac{1}{2}}\,\,\Big[R^{Bog}_{\bold{j}_*\,;\,i-2,i-2}(z)\Big]^{\frac{1}{2}}\,W^*_{\bold{j}_*\,;\,i-2,i}\Big[R^{Bog}_{\bold{j}_*\,;\,i,i}(z)\Big]^{\frac{1}{2}}\|\\
&= & \sup_{\psi \in Q^{(i,i+1)}_{\bold{j}_*}\mathcal{D}\,,\,\|\psi\|=1}\langle \psi \,,\,\Big[R^{Bog}_{\bold{j}_*\,;\,i,i}(z)\Big]^{\frac{1}{2}}\,W_{\bold{j}_*\,;\,i,i-2}\,\Big[R^{Bog}_{\bold{j}_*\,;\,i-2,i-2}(z)\Big]\,W^*_{\bold{j}_*\,;\,i-2,i}\Big[R^{Bog}_{\bold{j}_*\,;\,i,i}(z)\Big]^{\frac{1}{2}}\,\psi \rangle \quad\quad\quad \label{main-lemma-eq-1}
%Q^{(i-2)}W^*Q^{(i)}\varphi\rangle\Big|}{\|(Q^{(i-2)}(H^{Bog}-z)Q^{(i-2)})^{\frac{1}{2}}\psi\|\,\|(Q^{(i)}(H^{Bog}-z)Q^{(i)})^{\frac{1}{2}}\varphi\|}
\end{eqnarray}
where $\mathcal{D}$ is 
%a core of the operator $\sum_{\bold{j}\in\mathbb{Z}^3\setminus \{\bold{0}\}} \frac{k^2_{\bold{j}}}{2m}a_{\bold{j}}^{*}a_{\bold{j}}$ and it is 
the span of product state vectors of eigenstates of the one-particle momentum operators. The operator in (\ref{main-lemma-eq-1}) preserves the number of particles for any mode. Therefore, we can consider  the two subspaces $ Q_{\bold{j}_*}^{(i)}\mathcal{F}^{N}$ and $Q^{(i+1)}_{\bold{j}_*}\mathcal{F}^{N}$ separately, where $Q^{(r)}_{\bold{j}_*}$ is the projection onto the subspace of vectors with exactly $N-r$ particles in the modes $\pm \bold{j}_{*}$. It is enough to discuss the subspace $Q^{(i+1)}_{\bold{j}_*}\mathcal{F}^{N}$ because the estimate that we shall derive holds for vectors in $ Q^{(i)}_{\bold{j}_*}\mathcal{F}^{N}$ as well.
Next,  we write the state $\psi=Q^{(i+1)}_{\bold{j}_*}\psi$ as a linear superposition of product state vectors, i.e., vectors with definite occupation numbers in the modes $\Big\{\frac{2\pi}{L}\bold{j}\,;\,\bold{j}\in \mathbb{Z}^d \Big\}$.

\noindent
For a chosen labeling of the modes, $\{ \bold{j}_l\in \mathbb{Z}^d\,,\, l\in \mathbb{N}_0\}$,
  with each product state vector we can associate a sequence 
 \begin{equation}
 \{n_{\bold{j}_0},n_{\bold{j}_1},n_{\bold{j}_2}, \dots \}
 \end{equation}
  that encodes the occupation numbers, $n_{\bold{j}}$,  of the modes $\bold{j}$. As the vectors are in $\mathcal{F}^{N}$ by hypothesis, the sum of the occupation numbers $\sum_{l=0}^{\infty}n_{\bold{j}_l}$ must equal $N$. Hence, for each sequence there is a value $\bar{l}$ such that $n_{\bold{j}_l}\equiv 0$ for $l\geq \bar{l}$.
 Then, we can write
 \begin{eqnarray}
 Q^{(i+1)}_{\bold{j}_*}\psi&=&\sum_{ \{n_{\bold{j}_0},n_{\bold{j}_1},n_{\bold{j}_2}, \dots \}}C^{Q^{(i+1)}_{\bold{j}_*}\psi}_{ \{n_{\bold{j}_0},n_{\bold{j}_1},n_{\bold{j}_2}, \dots \}}\frac{1}{\sqrt{n_{\bold{j}_0}!n_{\bold{j}_1}!n_{\bold{j}_2}! \dots}}(a_{\bold{j}_0}^*)^{n_{\bold{j}_0}}(a_{\bold{j}_1}^*)^{n_{\bold{j}_1}}(a_{\bold{j}_2}^*)^{n_{\bold{j}_2}}\dots\Omega \quad\quad \label{sequence-phi}\\
 &=: &\sum_{ \{n_{\bold{j}_0},n_{\bold{j}_1},n_{\bold{j}_2}, \dots \}}C^{Q^{(i+1)}_{\bold{j}_*}\psi}_{ \{n_{\bold{j}_0},n_{\bold{j}_1},n_{\bold{j}_2}, \dots \}}\varphi_{\{n_{\bold{j}_0},n_{\bold{j}_1},n_{\bold{j}_2}, \dots \}}\quad
 \end{eqnarray} 
%   \begin{equation}
% Q^{(i-2)}\psi=\sum_{ \{n'_{\bold{j}_0},n'_{\bold{j}_1},n'_{\bold{j}_2}, \dots \}}C^{Q^{(i-2)}\psi}_{ \{n'_{\bold{j}_0},n'_{\bold{j}_1},n'_{\bold{j}_2}, \dots \}}\frac{1}{\sqrt{n'_{\bold{j}_0}!n'_{\bold{j}_1}!n'_{\bold{j}_2}! \dots}}a_{\bold{j}_0}^{n'_{\bold{j}_0}}a_{\bold{j}_1}^{n'_{\bold{j}_1}}a_{\bold{j}_2}^{n'_{\bold{j}_2}}\dots\Omega \label{sequence-psi}
% \end{equation} 
 where the sum is over all possible sequences, and the coefficients $C^{Q^{(i+1)}_{\bold{j}_*}\psi}_{ \{n_{\bold{j}_0},n_{\bold{j}_1},n_{\bold{j}_2}, \dots \}}$ are complex numbers such that
\begin{equation}
\sum_{ \{n_{\bold{j}_0},n_{\bold{j}_1},n_{\bold{j}_2}, \dots \}}|C^{Q^{(i+1)}_{\bold{j}_*}\psi}_{ \{n_{\bold{j}_0},n_{\bold{j}_1},n_{\bold{j}_2}, \dots \}}|^2=1\,.
\end{equation} 
 Moreover, if we set $\bold{j}_0\equiv \bold{0}$, for any vector of the type $ Q^{(i+1)}_{\bold{j}_*}\psi$ we have the constraint $n_{\bold{j}_0}\leq i+1$. 
 
\noindent 
%We write the operator
%\begin{eqnarray}
%& &\Big[R^{Bog}_{i-2,i-2}(z)\Big]^{\frac{1}{2}}\,\sum_{\bold{l}\in\mathcal{J}_{*}}\phi_{\bold{l}}\frac{a_{\bold{0}}a_{\bold{0}}a^*_{\bold{l}}a^*_{-\bold{l}}}{2N}\,\Big[R^{Bog}_{i,i}(z)\Big]^{\frac{1}{2}}\Big[R^{Bog}_{i,i}(z)\Big]^{\frac{1}{2}}\,\sum_{\bold{j} \in\mathcal{J}_{*}}\phi_{\bold{j}}\frac{a^*_{\bold{0}}a^*_{\bold{0}}a_{\bold{j}}a_{-\bold{j}}}{2N}\,\Big[R^{Bog}_{i-2,i-2}(z)\Big]^{\frac{1}{2}}\\
%&= &\sum_{\bold{m}\,,\,\bold{j}\in\mathcal{J}_{*}}\phi^2_{\bold{j}_{*}}\Big[R^{Bog}_{i-2,i-2}(z)\Big]^{\frac{1}{2}}\,\frac{a_{\bold{0}}a_{\bold{0}}a^*_{\bold{m}}a^*_{-\bold{m}}}{2N}\,\Big[R^{Bog}_{i,i}(z)\Big]^{\frac{1}{2}}\Big[R^{Bog}_{i,i}(z)\Big]^{\frac{1}{2}}\,\frac{a^*_{\bold{0}}a^*_{\bold{0}}a_{\bold{j}}a_{-\bold{j}}}{2N}\,\Big[R^{Bog}_{i-2,i-2}(z)\Big]^{\frac{1}{2}}
%\end{eqnarray}
With the new definitions, in expression (\ref{main-lemma-eq-1}) we replace
\begin{eqnarray}\label{term-inside}
& &\langle \psi \,,\, \Big[R^{Bog}_{\bold{j}_*\,;\,i,i}(z)\Big]^{\frac{1}{2}}\,W_{\bold{j}_*\,;\,i,i-2}\,R^{Bog}_{\bold{j}_*\,;\,i-2,i-2}(z)\,W^*_{\bold{j}_*\,;\,i-2,i}\Big[R^{Bog}_{\bold{j}_*\,;\,i,i}(z)\Big]^{\frac{1}{2}} \psi
\rangle \quad 
\end{eqnarray}
with
\begin{eqnarray}
&   &\sum_{ \{n'_{\bold{j}_0},n'_{\bold{j}_1},n'_{\bold{j}_2}, \dots \}} \sum_{ \{n_{\bold{j}_0},n_{\bold{j}_1},n_{\bold{j}_2}, \dots \}} \overline{C^{Q^{(i+1)}_{\bold{j}_*}\psi}_{ \{n'_{\bold{j}_0},n'_{\bold{j}_1},n'_{\bold{j}_2}, \dots \}}}C^{Q^{(i+1)}_{\bold{j}_*}\psi}_{ \{n_{\bold{j}_0},n_{\bold{j}_1},n_{\bold{j}_2}, \dots \}}\times \\
& &\times \langle \varphi_{\{n'_{\bold{j}_0},n'_{\bold{j}_1},n'_{\bold{j}_2}, \dots \}} \,,\,\Big[R^{Bog}_{\bold{j}_*\,;\,i,i}(z)\Big]^{\frac{1}{2}}\,\,\phi_{\bold{j}_{*}}\frac{a^*_{\bold{0}}a^*_{\bold{0}}a_{\bold{j}_{*}}a_{-\bold{j}_{*}}}{N}\,R^{Bog}_{\bold{j}_*\,;\,i-2,i-2}(z)\times \label{scalar-product-occupation-0}\\
& &\quad\quad\quad \times \phi_{\bold{j}_{*}}\frac{a_{\bold{0}}a_{\bold{0}}a^*_{\bold{j}_{*}}a^*_{-\bold{j}_{*}}}{N}\,\Big[R^{Bog}_{\bold{j}_*\,;\,i,i}(z)\Big]^{\frac{1}{2}}\,\,\varphi_{\{n_{\bold{j}_0},n_{\bold{j}_1},n_{\bold{j}_2}, \dots \}} \rangle\,. \quad\quad \quad\quad \nonumber
\end{eqnarray}
 The scalar product in (\ref{scalar-product-occupation-0}) is nonzero only if $n_{\bold{j}_l}=n'_{\bold{j}_l}$ for all $l$. % \item [C2)]
%$n_{\bold{j}_{*}}=n'_{\bold{j}_{*}}+1$, $n_{-\bold{j}_{*}}=n'_{-\bold{j}_{*}}+1$
%\item [C-2)]
%%\end{itemize} 
Therefore, we can write
%actually sum  over the sequences $ \{n_{\bold{j}_0},n_{\bold{j}_1},n_{\bold{j}_2}, \dots \}$ only,
\begin{eqnarray}
%& &\langle \psi\,,\,\Big[R^{Bog}_{\bold{j}_*\,;\,i-2,i-2}(z)\Big]^{\frac{1}{2}}\,W^*_{\bold{j}_*\,;\,i-2,i}\,\Big[R^{Bog}_{\bold{j}_*\,;\,i,i}(z)\Big]^{\frac{1}{2}}\Big[R^{Bog}_{\bold{j}_*\,;\,i,i}(z)\Big]^{\frac{1}{2}}\,W_{\bold{j}_*\,;\,i,i-2}\,\Big[R^{Bog}_{\bold{j}_*\,;\,i-2,i-2}(z)\Big]^{\frac{1}{2}} \psi
%\rangle \quad \\
& &(\ref{term-inside})\\
&=   &\sum_{ \{n_{\bold{j}_0},n_{\bold{j}_1},n_{\bold{j}_2}, \dots \}} \overline{C^{Q^{(i+1)}_{\bold{j}_*}\psi}_{ \{n_{\bold{j}_0},n_{\bold{j}_1},n_{\bold{j}_2},\dots\}}}C^{Q^{(i+1)}_{\bold{j}_*}\psi}_{ \{n_{\bold{j}_0},n_{\bold{j}_1},n_{\bold{j}_2}, \dots \}}\times \\
& &\times \langle \varphi_{\{n_{\bold{j}_0},n_{\bold{j}_1},n_{\bold{j}_2}, \dots \}} \,,\,\Big[R^{Bog}_{\bold{j}_*\,;\,i,i}(z)\Big]^{\frac{1}{2}}\,\,\phi_{\bold{j}_{*}}\frac{a^*_{\bold{0}}a^*_{\bold{0}}a_{\bold{j}_{*}}a_{-\bold{j}_{*}}}{N}\,R^{Bog}_{\bold{j}_*\,;\,i-2,i-2}(z)\times \label{scalar-product-occupation-bis}\\
& &\quad\quad\quad \times \phi_{\bold{j}_{*}}\frac{a_{\bold{0}}a_{\bold{0}}a^*_{\bold{j}_{*}}a^*_{-\bold{j}_{*}}}{N}\,\Big[R^{Bog}_{\bold{j}_*\,;\,i,i}(z)\Big]^{\frac{1}{2}}\,\,\varphi_{\{n_{\bold{j}_0},n_{\bold{j}_1},n_{\bold{j}_2}, \dots \}} \rangle\,. \quad\quad \quad\quad\nonumber
\end{eqnarray}
We observe that in expression (\ref{scalar-product-occupation-bis}) the operator
\begin{eqnarray}
& &(R^{Bog}_{\bold{j}_*\,;\,i,i}(z))^{\frac{1}{2}}W_{\bold{j}_*\,;\,i,i-2}\,(R^{Bog}_{\bold{j}_*\,;\,i-2,i-2}(z))^{\frac{1}{2}}(R^{Bog}_{\bold{j}_*\,;\,i-2,i-2}(z))^{\frac{1}{2}}W^*_{\bold{j}_*\,;\,i-2,i}(R^{Bog}_{\bold{j}_*\,;\,i,i}(z))^{\frac{1}{2}}\\
&=&(R^{Bog}_{\bold{j}_*\,;\,i,i}(z))^{\frac{1}{2}}\phi_{\bold{j}}\frac{a^*_{\bold{0}}a^*_{\bold{0}}a_{\bold{j}_{*}}a_{-\bold{j}_{*}}}{N}\,R^{Bog}_{\bold{j}_*\,;\,i-2,i-2}(z)\,\phi_{\bold{j}}\frac{a_{\bold{0}}a_{\bold{0}}a^*_{\bold{j}_{*}}a^*_{-\bold{j}_{*}}}{N}(R^{Bog}_{\bold{j}_*\,;\,i,i}(z))^{\frac{1}{2}}\quad \label{operator-cnumbers}
\end{eqnarray}
can be replaced with a function of the number operators $a^*_{\bold{j}}a_{\bold{j}}$. Indeed, it is enough to pull the operator $a_{\bold{j}_*}a_{-\bold{j}_*}$ contained in $W_{\bold{j}_*}$ through $R^{Bog}_{\bold{j}_*\,;\,i-2,i-2}(z)$
% and next to the operator $a^*_{\bold{j}_*}a^*_{-\bold{j}_*}$ contained in $W^*_{\bold{j}_*}$. 
and observe that
\begin{equation}
a_{\bold{j}_*}a_{-\bold{j}_*}a^*_{\bold{j}_*}a^*_{-\bold{j}_*}=(a^*_{\bold{j}_*}a_{\bold{j}_*}+1)(a^*_{-\bold{j}_*}a_{-\bold{j}_*}+1)\,.
\end{equation}
Analogously, we pull the operator $a_{\bold{0}}a_{\bold{0}}$ to the left next to $a^*_{\bold{0}}a^*_{\bold{0}}$ and observe that
\begin{equation}
a^*_{\bold{0}}a^*_{\bold{0}}a_{\bold{0}}a_{\bold{0}}=a_{\bold{0}}^*a_{\bold{0}}a_{\bold{0}}^*a_{\bold{0}}-a_{\bold{0}}^*a_{\bold{0}}\,.
\end{equation}
Finally, we can write
 \begin{eqnarray}
 & &\langle \varphi_{ \{n_{\bold{j}_0},n_{\bold{j}_1},\dots\}} \,,\,\Big[R^{Bog}_{\bold{j}_*\,;\,i,i}(z)\Big]^{\frac{1}{2}}\,\phi_{\bold{j}_{*}}\frac{a^*_{\bold{0}}a^*_{\bold{0}}a_{\bold{j}_{*}}a_{-\bold{j}_{*}}}{N}\,R^{Bog}_{\bold{j}_*\,;\,i-2,i-2}(z)\,\phi_{\bold{j}_{*}}\frac{a_{\bold{0}}a_{\bold{0}}a^*_{\bold{j}_{*}}a^*_{-\bold{j}_{*}}}{N}\,\Big[R^{Bog}_{\bold{j}_*\,;\,i,i}(z)\Big]^{\frac{1}{2}}\,\varphi_{\{n_{\bold{j}_0},n_{\bold{j}_1}, \dots \}} \rangle \quad\quad\quad \label{scalar-product}\\
%& &\|\Big[R^{Bog}_{i-2,i-2}(z)\Big]^{\frac{1}{2}}\,W^*_{i-2,i}\,\Big[R^{Bog}_{i,i}(z)\Big]^{\frac{1}{2}}\Big[R^%{Bog}_{i,i}(z)\Big]^{\frac{1}{2}}\,W_{i,i-2}\,\Big[R^{Bog}_{i-2,i-2}(z)\Big]^{\frac{1}{2}}\|\\
%& &  \langle \varphi_{\{n_{\bold{j}_0},n_{\bold{j}_1},n_{\bold{j}_2}, \dots \}} \,,\,\Big[R^{Bog}_{i-2,i-2}(z)\Big]^{\frac{1}{2}}\,W^*_{i-2,i}\,\Big[R^{Bog}_{i,i}(z)\Big]^{\frac{1}{2}}\Big[R^{Bog}_{i,i}(z)\Big]^{\frac{1}{2}}\,W_{i,i-2}\,\Big[R^{Bog}_{i-2,i-2}(z)\Big]^{\frac{1}{2}}\,\varphi_{\{n_{\bold{j}_0},n_{\bold{j}_1},n_{\bold{j}_2}, \dots \}} \rangle\\
&= & \frac{(n_{\bold{j}_0}-1)n_{\bold{j}_0}}{N^2}\,\phi^2_{\bold{j}_{*}}\,\frac{ (n_{\bold{j}_{*}}+1)(n_{-\bold{j}_{*}}+1)}{\Big[\sum_{\bold{j}\notin \{ \pm\bold{j}_*\}}(n_{\bold{j}}+n_{-\bold{j}})(k_{\bold{j}})^2+(\frac{n_{\bold{j}_0}}{N}\phi_{\bold{j}_{*}}+k_{\bold{j}_{*}}^2)(n_{\bold{j}_{*}}+n_{-\bold{j}_{*}})-z\Big]}\times\\
& &\quad\quad\quad \times\frac{1}{\Big[\sum_{\bold{j}\notin \{\pm\bold{j}_*\}}(n_{\bold{j}}+n_{-\bold{j}})(k_{\bold{j}})^2+(\frac{(n_{\bold{j}_0}-2)}{N}\phi_{\bold{j}_{*}}+k_{\bold{j}_{*}}^2)(n_{\bold{j}_{*}}+n_{-\bold{j}_{*}})+2(\frac{(n_{\bold{j}_0}-2)}{N}\phi_{\bold{j}_{*}}+k_{\bold{j}_{*}}^2)-z\Big]}\quad\quad\quad\nonumber\\
%& &\times \,\frac{ \sqrt{n'_{\bold{m}}n'_{-\bold{m}}}}{\Big[\sum_{\bold{l}\in\mathbb{Z}^3\setminus \{\bold{0}\}}(\frac{n_{\bold{l}_0}}{N}\phi_{\bold{l}}+(k_{\bold{l}}^2))n'_{\bold{l}}-E^{Bog}\Big]^{\frac{1}{2}}\Big[\sum_{\bold{h}\in\mathbb{Z}^3\setminus \{\bold{0}\}}(\frac{(n_{\bold{j}_0}+2)}{N}\phi_{\bold{h}}+(k_{\bold{h}}^2))n'_{\bold{h}}-2(\frac{(n_{\bold{j}_0}+2)}{N}\phi_{\bold{m}}+(k_{\bold{m}}^2))-E^{Bog}\Big]^{\frac{1}{2}}}\nonumber
%&= & \frac{(n_{\bold{j}_0}+2)(n_{\bold{j}_0}+1)}{N^2}\,\phi^2_{\bold{j}_{*}}\\
%&  &\times\,\frac{ n_{\bold{j}_{*}}n_{-\bold{j}_{*}}}{\Big[\sum_{\bold{j}\neq \pm\bold{j}_*}(n_{\bold{j}}+n_{-\bold{j}})(k_{\bold{j}})^2+(\frac{n_{\bold{j}_0}}{N}\phi_{\bold{j}_{*}}+(k_{\bold{j}_{*}}^2))(n_{\bold{j}_{*}}+n_{-\bold{j}_{*}})-z\Big]}\\
%& &\quad\quad\quad \times\frac{1}{\Big[\sum_{\bold{j}\neq \pm\bold{j}_*}(n_{\bold{j}}+n_{-\bold{j}})(k_{\bold{j}})^2+(\frac{(n_{\bold{j}_0}+2)}{N}\phi_{\bold{j}_{*}}+(k_{\bold{j}_{*}}^2))(n_{\bold{j}_{*}}+n_{-\bold{j}_{*}})-2(\frac{(n_{\bold{j}_0}+2)}{N}\phi_{\bold{j}_{*}}+(k_{\bold{j}_{*}}^2))-z\Big]}\quad\quad\quad
%\end{eqnarray}
%\begin{eqnarray}
&\leq& \frac{(n_{\bold{j}_0}-1)n_{\bold{j}_0}}{N^2}\,\phi^2_{\bold{j}_{*}}\,\frac{(n_{\bold{j}_{*}}+n_{-\bold{j}_{*}}+2)^2}{4\Big[(\frac{n_{\bold{j}_0}}{N}\phi_{\bold{j}_{*}}+k_{\bold{j}_{*}}^2)(n_{\bold{j}_{*}}+n_{-\bold{j}_{*}})-z\Big]\Big[(\frac{(n_{\bold{j}_0}-2)}{N}\phi_{\bold{j}_{*}}+k_{\bold{j}_{*}}^2)(n_{\bold{j}_{*}}+n_{-\bold{j}_{*}}+2)-z\Big]}\nonumber\end{eqnarray}
where we have used $\|\varphi_{ \{n_{\bold{j}_0},n_{\bold{j}_1},\dots\}}\|=1$.

\noindent
We recall that $n_{\bold{j}_{*}}+n_{-\bold{j}_{*}}=N-i-1 $ for a vector $\varphi_{\{n_{\bold{j}_0},n_{\bold{j}_1}, \dots \}}\in Q^{(i+1)}_{\bold{j}_*}\mathcal{F}^{N}$. Finally, we can estimate
\begin{eqnarray}
(\ref{scalar-product})&\leq &\frac{\,\frac{n_{\bold{j}_0}-1}{N}\phi_{\bold{j}_{*}}(N-i+1)}{4\Big[(N-i-1)\Big(\frac{n_{\bold{j}_0}}{N}\phi_{\bold{j}_{*}}+k_{\bold{j}_{*}}^2\Big)-z\Big]}\,\frac{\frac{n_{\bold{j}_0}}{N}\phi_{\bold{j}_{*}}(N-i+1)}{ \Big[(N-i+1)\Big(\frac{n_{\bold{j}_0}-2}{N}\phi_{\bold{j}_{*}}+k_{\bold{j}_{*}}^2\Big)-z\Big]}\\
&= &\frac{\,\frac{n_{\bold{j}_0}-1}{N}\phi_{\bold{j}_{*}}}{4\Big[\Big(\frac{n_{\bold{j}_0}}{N}\phi_{\bold{j}_{*}}+k_{\bold{j}_{*}}^2\Big)(1-\frac{2}{N-i+1})-\frac{z}{N-i+1}\Big]}\,\frac{\frac{n_{\bold{j}_0}}{N}\phi_{\bold{j}_{*}}}{ \Big[\Big(\frac{n_{\bold{j}_0}-2}{N}\phi_{\bold{j}_{*}}+k_{\bold{j}_{*}}^2\Big)-\frac{z}{N-i+1}\Big]} \label{step-a}\\
%&=&\frac{1}{4\Big[(1+\frac{N\epsilon_{\bold{j}_*}+1}{n_{\bold{j}_0}-1})(1-\frac{2}{N-i+1})-\frac{N}{(N-i+1)(n_{\bold{j}_{0}}-1)}\frac{z}{\phi_{\bold{j}_{*}}}\Big]}\, \frac{1}{ \Big[1+\frac{N\epsilon_{\bold{j}_{*}}-2}{n_{\bold{j}_0}}-\frac{N}{n_{\bold{j}_0}(N-i+1)}\frac{z}{\phi_{\bold{j}_{*}}}\Big]}\quad\quad\label{fin}\\
&=&\frac{1}{4\Big[(1+\frac{N\epsilon_{\bold{j}_*}}{n_{\bold{j}_0}})(1-\frac{2}{N-i+1})-\frac{N}{(N-i+1)n_{\bold{j}_{0}}}\frac{z}{\phi_{\bold{j}_{*}}}\Big]}\, \frac{1}{ \Big[1+\frac{N\epsilon_{\bold{j}_{*}}-1}{n_{\bold{j}_0}-1}-\frac{N}{(n_{\bold{j}_0}-1)(N-i+1)}\frac{z}{\phi_{\bold{j}_{*}}}\Big]}\quad\quad\label{fin}
%&=&\frac{1}{4\Big[1+\frac{N\epsilon_{\bold{j}_{*}}-1}{n_{\bold{j}_0}+1}-\frac{1}{(N-i-1)}\frac{z}{\phi_{\bold{j}_{*}}}\Big]\,\Big[1+\frac{N}{n_{\bold{j}_0}+1}\epsilon_{\bold{j}_{*}}(1-\frac{2}{N-i-1})-\frac{2}{N-i-1}-\frac{1}{N-i-1}\frac{z}{\phi_{\bold{j}_{*}}}\Big]}\quad\quad\quad \label{fin}
\end{eqnarray}
where $n_{\bold{j}_0}\geq 2$ otherwise $(\ref{scalar-product})=0$.
%We recall that $ n'_{\bold{j}_0}=n_{\bold{j}_0} \leq i+1 $.
We observe that $ -\frac{z}{\phi_{\bold{j}_{*}}}$ and $\epsilon_{\bold{j}_{*}}\equiv\frac{k_{\bold{j}_{*}}^2}{\phi_{\bold{j}_{*}}}$ are both positive in the considered ranges, i.e., for $\epsilon_{\bold{j}_{*}}$ sufficiently small. Furthermore, we notice that $N-i+1\geq 3$ for $i\leq N-2$, and, by hypothesis, $N \epsilon_{\bold{j}_{*}}> 1$. Hence, the maximum of (\ref{fin}) is attained at the maximum allowed value of $n_{\bold{j}_0}$ that is $n_{\bold{j}_0}\equiv i+1\leq N-1$.
%Q^{(i-2)}W^*Q^{(i)}\varphi\rangle\Big|}{\|(Q^{(i-2)}(H^{Bog}-z)Q^{(i-2)})^{\frac{1}{2}}\psi\|\,\|(Q^{(i)}(H^{Bog}-z)Q^{(i)})^{\frac{1}{2}}\varphi\|}
%where in the step from (\ref{step-a}) to (\ref{step-b}) we have used $n_{\bold{j}_0}+1<n_{\bold{j}_0}+2\leq N$ and $-\frac{z}{\phi_{\bold{j}_{*}}}>0$.
%We observe that
%\begin{equation}
%\frac{n_{\bold{j}_0}}{n_{\bold{j}_0}+1}+\frac{N}{n_{\bold{j}_0}+1}\epsilon\geq 1+\epsilon
%\end{equation}

%Since , it follows that if $( N-i \geq)n_{\bold{j}_0}<c_{\epsilon_{\bold{j}_{*}}}N$ with $c_{\epsilon_{\bold{j}_{*}}}=o(\epsilon_{\bold{j}_{*}})$ then the ratio in (\ref{fin}) can be made smaller than $\frac{1}{4}$ provided $\epsilon_{\bold{j}_{*}}$ is small enough, because .
\begin{remark}\label{dimensions}
The lower bound $\epsilon_{\bold{j}_{*}}\geq \frac{4\pi^2}{\phi_{\bold{j}_{*}}L^2}$ holds by construction. Therefore, at finite $\rho$ and at fixed $\bold{j}_{*}$, in space dimension larger or equal to three the product $N \epsilon_{\bold{j}_{*}}=\rho |\Lambda| \epsilon_{\bold{j}_{*}}$ is divergent as $L\to \infty$. In dimension two, at finite $\rho$ the product $N \epsilon_{\bold{j}_{*}}$ can be less than  $1$ uniformly in $\Lambda$.
\end{remark}
Therefore,  we can estimate the scalar product in (\ref{scalar-product}) from above by replacing $n_{\bold{j}_0}$ with $N$ and $n_{\bold{j}_0}-1$ with $N$ in the left factor and in the right factor of the denominator in (\ref{fin}),  respectively.
%{\color{red} and estimating with $o(\epsilon_{\bold{j}_{*}})$ the negative terms in the denominators in expression (\ref{fin}).} 
We recall that we have assumed \begin{equation}
z\leq  E^{Bog}_{\bold{j}_*}+ (\delta-1)\phi_{\bold{j}^*}\sqrt{\epsilon_{\bold{j}_*}^2+2\epsilon_{\bold{j}_*}}
\end{equation} 
where
\begin{equation}
\frac{E^{Bog} _{\bold{j}_*}}{\phi_{\bold{j}_{*}}}=-\Big[\epsilon _{\bold{j}_*} +1-\sqrt{\epsilon _{\bold{j}_*} ^2+2\epsilon _{\bold{j}_*}}\,\Big]
\end{equation}
by definition, hence
$$-\frac{z}{\phi_{\bold{j}_{*}}}\geq 1+\epsilon _{\bold{j}_*} -\delta \sqrt{\epsilon_{\bold{j}_*}^2+2\epsilon_{\bold{j}_*}}\,.$$
We observe that the expression in (\ref{fin}) is increasing in $z$ in the considered range.  Since $\frac{1}{N}\leq \epsilon^{\nu}_{\bold{j}_*}$ for some $\nu >1$, for  $\epsilon_{\bold{j}_*}$ sufficiently small and $\delta$ in the interval $[0,2)$ we get
%\footnote{Notice that $$\frac{\epsilon_{\bold{j}_{*}}+1+\delta\sqrt{\epsilon^2_{\bold{j}_{*}}+2\epsilon_{\bold{j}_{*}}}}{N-i+1}\leq\frac{1+\epsilon_{\bold{j}_{*}}+\mathcal{O}(\epsilon_{\bold{j}_{*}}^{\frac{1}{2}})}{3}$$ because $N-i\geq 2$ and $\epsilon_{\bold{j}_{*}}>0$. Therefore, the denominator in (\ref{meno-1}) is strictly positive for $\epsilon_{\bold{j}_{*}}$ sufficiently small.}
 \begin{eqnarray}
 & &(\ref{scalar-product})\\
&\leq &\frac{1}{ \Big[1+\epsilon_{\bold{j}_{*}}-\frac{2(1+\epsilon_{\bold{j}_{*}})}{(N-i+1)}+\frac{\Big[\epsilon_{\bold{j}_{*}}+1-\delta\sqrt{\epsilon_{\bold{j}_{*}}^2+2\epsilon_{\bold{j}_{*}}}\,\Big]}{(N-i+1)}\Big]}\frac{1}{4\Big[1+\epsilon_{\bold{j}_{*}}-\frac{1}{N}+\frac{\Big[\epsilon_{\bold{j}_{*}}+1-\delta\sqrt{\epsilon_{\bold{j}_{*}}^2+2\epsilon_{\bold{j}_{*}}}\,\Big]}{(N-i+1)}\Big]}\quad\quad\quad \label{meno-1}\\
%&= &\frac{1}{ \Big[1+\epsilon_{\bold{j}_{*}}+o(\epsilon_{\bold{j}_{*}})-\frac{\Big[\epsilon_{\bold{j}_{*}}+1+\delta\sqrt{\epsilon_{\bold{j}_{*}}^2+2\epsilon_{\bold{j}_{*}}}\,\Big]}{(N-i+1)}\Big]}\frac{1}{4\Big[1+\epsilon_{\bold{j}_{*}}+o(\epsilon_{\bold{j}_{*}})+\frac{\Big[\epsilon_{\bold{j}_{*}}+1-\delta\sqrt{\epsilon_{\bold{j}_{*}}^2+2\epsilon_{\bold{j}_{*}}}\,\Big]}{(N-i+1)}\Big]}\label{meno-1}\\
&\leq&\frac{1}{4\Big[1+a_{\epsilon_{\bold{j}_{*}}}-\frac{2b^{(\delta)}_{\epsilon_{\bold{j}_{*}}}}{N-i+1}-\frac{1-c^{(\delta)}_{\epsilon_{\bold{j}_{*}}}}{(N-i+1)^2}\label{meno-0}
\Big]}
\end{eqnarray}
for all $2\leq i \leq N-2$,  using the definitions in (\ref{a}), (\ref{b}), and (\ref{c}), and
where the step from (\ref{meno-1}) to (\ref{meno-0}) is explained in the Appendix, Lemma \ref{accessori}. Of course, if $\delta<0$ we can bound $(\ref{scalar-product})$ with the estimate provided  for $\delta=0$.

\noindent
This concludes the proof because $\sum_{ \{n_{\bold{j}_0},n_{\bold{j}_1},n_{\bold{j}_2}, \dots \}}|C^{Q_{\bold{j}_*}^{(i+1)}\psi}_{ \{n_{\bold{j}_0},n_{\bold{j}_1}, \dots \}}|^2=1$.
\qed

%%%%%% LEMMA LOWER BOUND %%%%%%
With the next lemma we prepare the ground for the result of Theorem \ref{theorem-Bog}.  The key tool is a sequence of real numbers constructed starting from the operator norm estimate established in Lemma \ref{main-lemma-Bog}.  For the use of this result in Theorem  \ref{theorem-Bog} we shall replace $\epsilon$ with $\epsilon_{\bold{j}_*}$. Notice also that  $\delta$ is set equal to $1+\sqrt{\epsilon}$ and a larger lower bound for $\nu$ is considered in Lemma \ref{lemma-sequence}. This will be however enough for our purposes.
\begin{lemma}\label{lemma-sequence}
Assume $\epsilon>0$ sufficiently small.  Consider  for $j\in \mathbb{N}_{0}$ the sequence defined iteratively according to
\begin{eqnarray}
X_{2j+2}&:=&1-\frac{1}{4(1+a_{\epsilon}-\frac{2b_{\epsilon}}{N-2j-1}-\frac{1-c_{\epsilon}}{(N-2j-1)^2})X_{2j}}
%x_{2j+3}&:=&1-\frac{1}{4(1+a_{\epsilon}-\frac{2b_{\epsilon}}{N-2j-1}-\frac{1-c_{\epsilon}}{(N-2j-1)^2})x_{2j+1}}
\end{eqnarray}
with initial condition $X_{0}=1$ up to $X_{2j=N-2}$ where $N(\geq 2)$ is even.  Here, 
\begin{equation}\label{adelta}
a_{\epsilon}:=2\epsilon+\mathcal{O}(\epsilon^{\nu})\,,\quad  \nu>\frac{11}{8}\,,
\end{equation}
\begin{equation}\label{bdelta}
b_{\epsilon}:=(1+\epsilon)\delta \, \chi_{[0,2)}(\delta)\sqrt{\epsilon^2+2\epsilon}\,\Big|_{\delta=1+\sqrt{\epsilon}}
\end{equation}
and
\begin{equation}\label{cdelta}
c_{\epsilon}:=-(1-\delta^2 \, \chi_{[0,2)}(\delta))(\epsilon^2+2\epsilon)\,\Big|_{\delta=1+\sqrt{\epsilon}}\,
\end{equation}
with $\chi_{[0,2)}(\delta)$ the characteristic function of the interval $[0,2)$.

\noindent
Then, the following estimate holds true for $2\leq N-2j\leq  N $,
\begin{equation}\label{bound-even}
X_{2j}\geq\frac{1}{2}\Big[1+\sqrt{\eta a_{\epsilon}}-\frac{b_{\epsilon}/\sqrt{\eta a_{\epsilon}}}{N-2j-\xi}\Big]\,
\end{equation}
with $\eta=1-\sqrt{\epsilon}$, $\xi=\epsilon^{\Theta}$ where $\Theta:=\min\{2(\nu-\frac{11}{8})\,;\,\frac{1}{4}\}$.
\end{lemma}

\noindent
\emph{Proof}

\noindent
See Lemma \ref{lemma-sequence-lower-bound} in the Appendix.
\qed
\\

We are now ready for the rigorous construction of the Feshbach-Schur Hamiltonians up to the value $i=N-2$ of the flow.

\begin{thm}\label{theorem-Bog}
For \begin{equation}
z\leq E^{Bog}_{\bold{j}_*}+ (\delta-1)\phi_{\bold{j}_*}\sqrt{\epsilon_{\bold{j}_*}^2+2\epsilon_{\bold{j}_*}}(<0)
\end{equation} with $\delta=  1+\sqrt{\epsilon_{\bold{j}_*}}$, $\frac{1}{N}\leq \epsilon^{\nu}_{\bold{j}_*}$ for some $\nu >\frac{11}{8}$, and $\epsilon_{\bold{j}_*}$ sufficiently small, the operators $\mathscr{K}^{Bog\,(i)}_{\bold{j}_*}(z)$, $0\leq i\leq N-2$ and even,  are well defined \footnote{$\mathscr{K}^{Bog\,(i)}_{\bold{j}_*}(z)$ is  self-adjoint on the domain of the Hamiltonian $Q^{(>i+1)}_{\bold{j}_*}(H^{Bog}_{\bold{j}_*}-z)Q^{(>i+1)}_{\bold{j}_*}$.}.  For $i=0$, it is given in (\ref{KappaBog0}).  For $i=2,4,6,\dots,N-2$ they correspond to
\begin{eqnarray}\label{KappaBog-i}
\mathscr{K}^{Bog\,(i)}_{\bold{j}_*}(z)&=&Q^{(>i+1)}_{\bold{j}_*}(H^{Bog}_{\bold{j}_*}-z)Q^{(>i+1)}_{\bold{j}_*}\\
%& &-\sum_{l_{i-1}=0}^{\infty}Q^{(>i)}_{\bold{j}_*}W_{\bold{j}_*}\,Q^{(i-1)}_{\bold{j}_*}R^{Bog}_{\bold{j}_*\,;\,i-1,i-1}(z)\,\Big[\Gamma^{Bog\,}_{\bold{j}_*\,;\,i-1,i-1}(z)R^{Bog}_{\bold{j}_*\,;\,i-1,i-1}(z)\Big]^{l_{i-1}}\,Q^{(i-1)}_{\bold{j}_*}W^*_{\bold{j}_*}Q^{(>i)}_{\bold{j}_*} \nonumber\\
& &-Q^{(>i+1)}_{\bold{j}_*}W_{\bold{j}_*}\,R^{Bog}_{\bold{j}_*\,;\,i,i}(z)\,\sum_{l_i=0}^{\infty}\,\Big[\Gamma^{Bog\,}_{\bold{j}_*\,;\,i,i}(z)R^{Bog}_{\bold{j}_*\,;\,i,i}(z)\Big]^{l_i}W^*_{\bold{j}_*}Q^{(>i+1)}_{\bold{j}_*} \nonumber
\end{eqnarray}
where:
\begin{itemize}
\item \begin{equation}\label{GammaBog-2}
\Gamma^{Bog\,}_{\bold{j}_*\,;\,2,2}(z):=W_{\bold{j}_*\,;\,2,0}\,R_{\bold{j}_*\,;\,0,0}^{Bog}(z)W_{\bold{j}_*\,;\,0,2}^*
\end{equation}
\item
 for $N-2\geq i\geq 4$,
%\begin{equation}
%\Gamma^{Bog\,}_{i-1,i-1}(z):=W_{i-1,i-3}\,R^{Bog}_{i-3,i-3}(z) \sum_{l_{i-3}=0}^{\infty}\Big[\Gamma^{Bog\,}_{i-3,i-3}(z)R^{Bog}_{i-3,i-3}(z)\Big]^{l_{i-3}}W^*_{i-3,i-1}
%\end{equation}
\begin{eqnarray}\label{GammaBog-i}
\Gamma^{Bog\,}_{\bold{j}_*\,;\,i,i}(z)&:=&W_{\bold{j}_*\,;\,i,i-2}\,R^{Bog}_{\bold{j}_*\,;\,i-2,i-2}(z) \sum_{l_{i-2}=0}^{\infty}\Big[\Gamma^{Bog}_{\bold{j}_*\,;\,i-2,i-2}(z)R^{Bog}_{\bold{j}_*\,;\,i-2,i-2}(z)\Big]^{l_{i-2}}W^*_{\bold{j}_*\,;\,i-2,i}\\
&=&W_{\bold{j}_*\,;\,i,i-2}\,(R^{Bog}_{\bold{j}_*\,;\,i-2,i-2}(z))^{\frac{1}{2}} \sum_{l_{i-2}=0}^{\infty}\Big[(R^{Bog}_{\bold{j}_*\,;\,i-2,i-2}(z))^{\frac{1}{2}}\Gamma^{Bog}_{\bold{j}_*\,;\,i-2,i-2}(z)(R^{Bog}_{\bold{j}_*\,;\,i-2,i-2}(z))^{\frac{1}{2}}\Big]^{l_{i-2}}\times \quad\quad\quad\\
& &\quad\quad\quad \times (R^{Bog}_{\bold{j}_*\,;\,i-2,i-2}(z))^{\frac{1}{2}}W^*_{\bold{j}_*\,;\,i-2,i}\,.\nonumber
\end{eqnarray}
\end{itemize}
%\begin{remark} We defer the last step corresponding $i=N-1$ to the next section.
%\end{remark}
%\begin{equation}\label{GammaBog-4}
%\Gamma^{Bog\,}_{4,4}(z):=W_{4,2}\,R^{Bog}_{2,2}(z)\sum_{l_2=0}^{\infty}\Big[W_{2,0}\,R_{0,0}^{Bog}(z)W_{0,2}^*R^{Bog}_{2,2}(z)
%\Big]^{l_2}W^*_{2,4}\,.
%\end{equation}
\end{thm}

\noindent
\emph{Proof}

\noindent
The expression in (\ref{KappaBog0}) is trivially well defined because $Q^{(0,1)}_{\bold{j}_*}H^{Bog}_{\bold{j}_*}Q^{(0,1)}_{\bold{j}_*}\geq 0$ and $z<0$ for $\epsilon_{\bold{j}_*}$ sufficiently small.
Thus, as it is also clear from the outline in Section \ref{Nonrigorous derivation}, the main task is showing that the Neumann expansion used at each successive step is  well defined. Therefore, we first show that the expression of $\mathscr{K}^{Bog\,(i)}_{\bold{j}_*}(z)$ for  $2\leq  i\leq N-2$ is formally correct and later we justify the Neumann expansions that have been used. 
\\

 We assume the given expression of $\mathscr{K}^{Bog\,(i)}_{\bold{j}_*}(z)$ for  $0 \leq i \leq N-4$ and derive $\mathscr{K}^{Bog\,(i+2)}_{\bold{j}_*}(z)$ according to the formula
\begin{eqnarray}
& &\mathscr{K}^{Bog\,(i+2)}_{\bold{j}_*}(z)\\
&:=&Q^{(>i+3)}_{\bold{j}_*}\mathscr{K}^{Bog\,(i)}_{\bold{j}_*}(z)Q^{(>i+3)}_{\bold{j}_*}\\
& &-Q^{(>i+3)}_{\bold{j}_*}\mathscr{K}^{Bog\,(i)}_{\bold{j}_*}(z)Q^{(i+2, i+3)}_{\bold{j}_*}\frac{1}{Q^{(i+2, i+3)}_{\bold{j}_*}\mathscr{K}^{Bog\,(i)}_{\bold{j}_*}(z)Q^{(i+2,i+3)}_{\bold{j}_*}}Q^{(i+2, i+3)}_{\bold{j}_*}\mathscr{K}^{Bog\,(i)}_{\bold{j}_*}(z)Q^{(>i+3)}_{\bold{j}_*}\,. \nonumber
\end{eqnarray}
Using $Q^{(>i+3)}_{\bold{j}_*}W_{\bold{j}_*} Q^{(i,i+1)}_{\bold{j}_*}=0$, we derive
\begin{eqnarray}
& &Q^{(>i+3)}_{\bold{j}_*}\mathscr{K}^{Bog\,(i)}_{\bold{j}_*}(z)Q^{(>i+3)}_{\bold{j}_*}\\
&=&Q^{(>i+3)}_{\bold{j}_*}Q^{(>i+1)}_{\bold{j}_*}(H^{Bog}-z)Q^{(>i+1)}_{\bold{j}_*}Q^{(>i+3)}_{\bold{j}_*}\\
%& &-Q^{(>i+3)}_{\bold{j}_*}Q^{(>i+1)}_{\bold{j}_*}W_{\bold{j}_*}\,Q^{(i-1)}_{\bold{j}_*}R^{Bog}_{\bold{j}_*\,;\,i-2,i-2}(z)\sum_{l_{i-2}=0}^{\infty}\Big[\Gamma_{\bold{j}_*\,;\,i-2,i-2}^{Bog\,}(z)R^{Bog}_{\bold{j}_*\,;\,i-2,i-2}(z)\Big]^{l_{i-2}}Q^{(i-1)}W_{\bold{j}_*}^*Q^{(>i+1)}_{\bold{j}_*} Q^{(>i+3)}_{\bold{j}_*}\quad\quad\quad\label{vanishing-term}\\
& &-Q^{(>i+3)}_{\bold{j}_*}Q^{(>i+1)}_{\bold{j}_*}W_{\bold{j}_*}\,R^{Bog}_{\bold{j}_*\,;\,i,i}(z)\sum_{l_i=0}^{\infty}\Big[\Gamma_{\bold{j}_*\,;\,i,i}^{Bog\,}(z)R^{Bog}_{\bold{j}_*\,;\,i,i}(z)\Big]^{l_i}W^*_{\bold{j}_*}Q^{(>i+1)}_{\bold{j}_*}Q^{(>i+3)}_{\bold{j}_*} \quad\quad\quad \label{vanishing-term}\\
&=&Q^{(>i+3)}_{\bold{j}_*}(H^{Bog}-z)Q^{(>i+3)}_{\bold{j}_*}
%& &-Q^{(>i+1)}_{\bold{j}_*}W_{\bold{j}_*}\,Q^{(i)}_{\bold{j}_*}R^{Bog}_{\bold{j}_*\,;\,i,i}(z)\sum_{l_i=0}^{\infty}\Big[\Gamma_{\bold{j}_*\,;\,i,i}^{Bog\,}(z)R^{Bog}_{\bold{j}_*\,;\,i,i}(z)\Big]^{l_i}Q^{(i)}_{\bold{j}_*}W^*_{\bold{j}_*}Q^{(>i+1)}_{\bold{j}_*}
\end{eqnarray}
where the term in (\ref{vanishing-term}) equals zero because 
\begin{equation}
Q^{(>i+3)}_{\bold{j}_*}W_{\bold{j}_*}R^{Bog}_{\bold{j}_*\,;\,i,i}(z)=Q^{(>i+3)}_{\bold{j}_*}W_{\bold{j}_*} Q^{(i,i+1)}_{\bold{j}_*}R^{Bog}_{\bold{j}_*\,;\,i,i}(z)=0.
\end{equation}
 Likewise, we get
\begin{eqnarray}
& &Q^{(>i+3)}_{\bold{j}_*}\mathscr{K}^{Bog\,(i)}_{\bold{j}_*}(z)Q^{(i+2,i+3)}_{\bold{j}_*}\\
&=&Q^{(>i+3)}_{\bold{j}_*}Q^{(>i+1)}_{\bold{j}_*}(H^{Bog}_{\bold{j}_*}-z)Q^{(>i+1)}_{\bold{j}_*}Q^{(i+1,i+3)}_{\bold{j}_*}\\
%& &-\sum_{l_i=0}^{\infty}Q^{(>i+1)}_{\bold{j}_*}Q^{(>i)}_{\bold{j}_*}W_{\bold{j}_*}\,Q^{(i-1)}_{\bold{j}_*}R^{Bog}_{\bold{j}_*\,;\,i-1,i-1}(z)\Big[\Gamma^{Bog\,}_{\bold{j}_*\,;\,i-1,i-1}(z)R^{Bog}_{\bold{j}_*\,;\,i-1,i-1}(z)\Big]^{l_{i-1}}Q^{(i-1)}_{\bold{j}_*}W^*_{\bold{j}_*}Q^{(>i)}_{\bold{j}_*} Q^{(i+1)}_{\bold{j}_*}\nonumber\\
& &-Q^{(>i+3)}_{\bold{j}_*}Q^{(>i+1)}_{\bold{j}_*}W_{\bold{j}_*}\,R^{Bog}_{\bold{j}_*\,;\,i,i}(z)\,\sum_{l_i=0}^{\infty}\Big[\Gamma^{Bog\,}_{\bold{j}_*\,;\,i,i}(z)R^{Bog}_{\bold{j}_*\,;\,i,i}(z)\Big]^{l_i}W^*_{\bold{j}_*}Q^{(>i+1)}_{\bold{j}_*}Q^{(i+2,i+3)}_{\bold{j}_*}\nonumber \\
&=&Q^{(>i+3)}_{\bold{j}_*}Q^{(>i+1)}_{\bold{j}_*}(H^{Bog}_{\bold{j}_*}-z)Q^{(>i+1)}_{\bold{j}_*}Q^{(i+2,i+3)}_{\bold{j}_*}\\
&=&Q^{(>i+3)}_{\bold{j}_*}W_{\bold{j}_*}Q^{(i+2,i+3)}_{\bold{j}_*}\,.
\end{eqnarray}
Combining these computations we obtain 
\begin{eqnarray}
& &\mathscr{K}^{Bog\,(i+2)}_{\bold{j}_*}(z)\\
&=&Q^{(>i+3)}_{\bold{j}_*}\mathscr{K}^{Bog\,(i)}_{\bold{j}_*}(z)Q^{(>i+3)}_{\bold{j}_*}\\
& &-Q^{(>i+3)}_{\bold{j}_*}\mathscr{K}^{Bog\,(i)}_{\bold{j}_*}(z)Q^{(i+2,i+3)}_{\bold{j}_*}\frac{1}{Q^{(i+2,i+3)}_{\bold{j}_*}\mathscr{K}^{Bog\,(i)}_{\bold{j}_*}(z)Q^{(i+2,i+3)}_{\bold{j}_*}}Q^{(i+1)}_{\bold{j}_*}\mathscr{K}^{Bog\,(i)}_{\bold{j}_*}(z)Q^{(>i+3)}_{\bold{j}_*} \nonumber\\
&=&Q^{(>i+3)}_{\bold{j}_*}(H^{Bog}_{\bold{j}_*}-z)Q^{(>i+3)}_{\bold{j}_*}\\
%& &-\sum_{l_i=0}^{\infty}Q^{(>i+1)}_{\bold{j}_*}W_{\bold{j}_*}\,Q^{(i)}_{\bold{j}_*}R^{Bog}_{\bold{j}_*\,;\,i,i}(z)\Big[\Gamma^{Bog\,}_{\bold{j}_*\,;\,i,i}(z)R^{Bog}_{\bold{j}_*\,;\,i,i}(z)\Big]^{l_i}Q^{(i)}_{\bold{j}_*}W^*_{\bold{j}_*}Q^{(>i+1)}_{\bold{j}_*} \\
& &-Q^{(>i+3)}_{\bold{j}_*}W_{\bold{j}_*}Q^{(i+2,i+3)}_{\bold{j}_*}\frac{1}{Q^{(i+2,i+3)}_{\bold{j}_*}\mathscr{K}_{\bold{j}_*}^{Bog\,(i)}(z)Q^{(i+2,i+3)}_{\bold{j}_*}}Q^{(i+2,i+3)}_{\bold{j}_*}W^*_{\bold{j}_*}Q^{(>i+3)}_{\bold{j}_*}\,.\label{third-term}
\end{eqnarray}
Now, we observe that
\begin{eqnarray}
& &Q^{(i+2,i+3)}_{\bold{j}_*}\mathscr{K}^{Bog\,(i)}_{\bold{j}_*}(z)Q^{(i+2,i+3)}_{\bold{j}_*}\label{sandwich-1}\\
&=&Q^{(i+2,i+3)}_{\bold{j}_*}Q^{(>i+1)}_{\bold{j}_*}(H^{Bog}_{\bold{j}_*}-z)Q^{(>i+1)}_{\bold{j}_*}Q^{(i+2,i+3)}_{\bold{j}_*}\\
%& &-\sum_{l_{i-1}=0}^{\infty}Q^{(i+1)}_{\bold{j}_*}Q^{(>i)}_{\bold{j}_*}W_{\bold{j}_*}\,Q^{(i-1)}_{\bold{j}_*}R^{Bog}_{\bold{j}_*\,;\,i-1,i-1}(z)\Big[\Gamma^{Bog\,}_{\bold{j}_*\,;\,i-1,i-1}(z)R^{Bog}_{\bold{j}_*\,;\,i-1,i-1}(z)\Big]^{l_{i-1}}Q^{(i-1)}_{\bold{j}_*}W^*_{\bold{j}_*}Q^{(>i)}_{\bold{j}_*} Q^{(i+1)}_{\bold{j}_*}\quad\quad\,\,\\
& &-Q^{(i+2,i+3)}_{\bold{j}_*}Q^{(>i+1)}_{\bold{j}_*}W_{\bold{j}_*}\,R^{Bog}_{\bold{j}_*\,;\,i,i}(z)\,\sum_{l_i=0}^{\infty}\Big[\Gamma^{Bog\,}_{\bold{j}_*\,;\,i,i}R^{Bog}_{\bold{j}_*\,;\,i,i}(z)\Big]^{l_i}W^*_{\bold{j}_*}Q^{(>i+1)}_{\bold{j}_*}Q^{(i+2,i+3)}_{\bold{j}_*} \label{vanishing-term2}\\
&=&Q^{(i+2,i+3)}_{\bold{j}_*}(H^{Bog}_{\bold{j}_*}-z)Q^{(i+2,i+3)}_{\bold{j}_*}\\
& &-Q^{(i+2,i+3)}_{\bold{j}_*}W_{\bold{j}_*}\,R^{Bog}_{\bold{j}_*\,;\,i,i}(z)\,\sum_{l_i=0}^{\infty}\Big[\Gamma^{Bog\,}_{\bold{j}_*\,;\,i,i}R^{Bog}_{\bold{j}_*\,;\,i,i}(z)\Big]^{l_{i}}W^*_{\bold{j}_*}Q^{(i+2,i+3)}_{\bold{j}_*}\,.\label{sandwich-2}
\end{eqnarray}
If we insert the expression found for $Q^{(i+2,i+3)}_{\bold{j}_*}\mathscr{K}^{Bog\,(i)}_{\bold{j}_*}(z)Q^{(i+2,i+3)}_{\bold{j}_*}$ into (\ref{third-term}), the (Neumann) expansion in terms of the resolvent
\begin{equation}
Q^{(i+2,i+3)}_{\bold{j}_*}\frac{1}{Q^{(i+2,i+3)}_{\bold{j}_*}(H^{Bog}_{\bold{j}_*}-z)Q^{(i+2,i+3)}_{\bold{j}_*}}Q^{(i+2,i+3)}_{\bold{j}_*}=:R^{Bog\,}_{\bold{j}_*\,;\,i+2,i+2}(z)
\end{equation}
and of the effective interaction
\begin{eqnarray}
& &-Q^{(i+2,i+3)}_{\bold{j}_*}Q^{(>i+1)}_{\bold{j}_*}W_{\bold{j}_*}R^{Bog}_{\bold{j}_*\,;\,i,i}(z)\sum_{l_{i}=0}^{\infty}\Big[\Gamma^{Bog\,}_{\bold{j}_*\,;\,i,i}R^{Bog}_{\bold{j}_*\,;\,i,i}(z)\Big]^{l_{i}}W^*_{\bold{j}_*}Q^{(>i+1)}_{\bold{j}_*} Q^{(i+2,i+3)}_{\bold{j}_*}\quad\quad\quad\quad\\
& =&-W_{\bold{j}_*\,;\,i+2,i}R^{Bog}_{\bold{j}_*\,;\,i,i}(z)\sum_{l_{i}=0}^{\infty}\Big[\Gamma^{Bog\,}_{\bold{j}_*\,;\,i,i}R^{Bog}_{\bold{j}_*\,;\,i,i}(z)\Big]^{l_{i}}W^*_{\bold{j}_*\,;\,i,i+2}\\
&=:&-\Gamma^{Bog}_{\bold{j}_*\,;\,i+2,i+2}
\end{eqnarray}
yields the desired expression for $\mathscr{K}^{Bog\,(i+2)}_{\bold{j}_*}(z)$.
\\

The formal steps used before become rigorous if for  $2 \leq i \leq N-2$ the quantity
\begin{equation}
\sum_{l_{i}=0}^{\infty}\Big[(R^{Bog}_{\bold{j}_*\,;\,i,i}(z))^{\frac{1}{2}}\Gamma^{Bog}_{\bold{j}_*\,;\,i,i}(z)(R^{Bog}_{\bold{j}_*\,;\,i,i}(z))^{\frac{1}{2}}\Big]^{l_{i}}\quad ,\quad z\leq E^{Bog}_{\bold{j}_*}+ \sqrt{\epsilon_{\bold{j}_*}}\phi_{\bold{j}_*}\sqrt{\epsilon_{\bold{j}_*}^2+2\epsilon_{\bold{j}_*}}\,,
\end{equation}
is seen to be a well defined operator. This is not difficult for $i=2$ because, using the definition in (\ref{GammaBog-2}) and the result in Lemma \ref{main-lemma-Bog}, we can easily estimate
\begin{equation}
\|(R^{Bog}_{\bold{j}_*\,;\,2,2}(z))^{\frac{1}{2}}\Gamma^{Bog}_{\bold{j}_*\,;\,2,2}(z)(R^{Bog}_{\bold{j}_*\,;\,2,2}(z))^{\frac{1}{2}}\|<1\,.
\end{equation}

\noindent 
For $N-2\geq i\geq 4$, starting from the definition 
\begin{equation}\label{def-gamma-ii}
\Gamma^{Bog\,}_{\bold{j}_*\,;\,i,i}(z):=W_{\bold{j}_*\,;\,i,i-2}\,R^{Bog}_{\bold{j}_*\,;\,i-2,i-2}(z) \sum_{l_{i-2}=0}^{\infty}\Big[\Gamma^{Bog}_{\bold{j}_*\,;\,i-2,i-2}(z)R^{Bog}_{\bold{j}_*\,;\,i-2,i-2}(z)\Big]^{l_{i-2}}W^*_{\bold{j}_*\,;\,i-2,i}
\end{equation}
 we can write 
\begin{eqnarray}
& &(R^{Bog}_{\bold{j}_*\,;\,i,i}(z))^{\frac{1}{2}}\Gamma^{Bog\,}_{\bold{j}_*\,;\,i,i}(R^{Bog}_{i,i}(z))^{\frac{1}{2}}\\
&=&(R^{Bog}_{\bold{j}_*\,;\,i,i}(z))^{\frac{1}{2}}W_{\bold{j}_*\,;\,i,i-2}\,R^{Bog}_{\bold{j}_*\,;\,i-2,i-2}(z) \sum_{l_{i-2}=0}^{\infty}\Big[\Gamma^{Bog}_{\bold{j}_*\,;\,i-2,i-2}(z)R^{Bog}_{\bold{j}_*\,;\,i-2,i-2}(z)\Big]^{l_{i-2}}W^*_{\bold{j}_*\,;\,i-2,i}(R^{Bog}_{\bold{j}_*\,;\,i,i}(z))^{\frac{1}{2}}\quad\quad\quad\\
&=&(R^{Bog}_{\bold{j}_*\,;\,i,i}(z))^{\frac{1}{2}}W_{\bold{j}_*\,;\,i,i-2}\,(R^{Bog}_{\bold{j}_*\,;\,i-2,i-2}(z))^{\frac{1}{2}}\times\\
& &\quad\times  \sum_{l_{i-2}=0}^{\infty}\Big[(R^{Bog}_{\bold{j}_*\,;\,i-2,i-2}(z))^{\frac{1}{2}}\Gamma^{Bog}_{\bold{j}_*\,;\,i-2,i-2}(z)(R^{Bog}_{\bold{j}_*\,;\,i-2,i-2}(z))^{\frac{1}{2}}\Big]^{l_{i-2}}\times \\
& &\quad\quad\quad\quad\quad \times (R^{Bog}_{\bold{j}_*\,;\,i-2,i-2}(z))^{\frac{1}{2}}W^*_{\bold{j}_*\,;\,i-2,i}(R^{Bog}_{\bold{j}_*\,;\,i,i}(z))^{\frac{1}{2}}\,,
\end{eqnarray}
and
\begin{eqnarray}
& &\sum_{l_{i}=0}^{\infty}[(R^{Bog}_{\bold{j}_*\,;\,i,i}(z))^{\frac{1}{2}}\Gamma^{Bog\,}_{\bold{j}_*\,;\,i,i}(R^{Bog}_{\bold{j}_*\,;\,i,i}(z))^{\frac{1}{2}}]^{l_i}\label{gammacheck}\\
&=&\sum_{l_{i}=0}^{\infty}\Big[(R^{Bog}_{\bold{j}_*\,;\,i,i}(z))^{\frac{1}{2}}W_{\bold{j}_*\,;\,i,i-2}\,(R^{Bog}_{\bold{j}_*\,;\,i-2,i-2}(z))^{\frac{1}{2}} \times\\
& &\quad\quad\quad\times\sum_{l_{i-2}=0}^{\infty}\Big[(R^{Bog}_{\bold{j}_*\,;\,i-2,i-2}(z))^{\frac{1}{2}}\Gamma^{Bog}_{\bold{j}_*\,;\,i-2,i-2}(z)(R^{Bog}_{\bold{j}_*\,;\,i-2,i-2}(z))^{\frac{1}{2}}\Big]^{l_{i-2}}\times \quad\quad\quad\\
& &\quad\quad\quad\quad\quad\quad\quad\quad\times (R^{Bog}_{\bold{j}_*\,;\,i-2,i-2}(z))^{\frac{1}{2}}W^*_{\bold{j}_*\,;\,i-2,i}(R^{Bog}_{\bold{j}_*\,;\,i,i}(z))^{\frac{1}{2}}\Big]^{l_i}\,.\nonumber
%&=&(R^{Bog}_{i,i}(z))^{\frac{1}{2}}W_{i,i-2}\,(R^{Bog}_{i-2,i-2}(z))^{\frac{1}{2}}\times\\
%& &\quad\times  \sum_{l_{i-2}=0}^{\infty}\Big[(R^{Bog}_{i-2,i-2}(z))^{\frac{1}{2}}\Gamma^{Bog}_{i-2,i-2}(z)R^{Bog}_{i-2,i-2}(z)\Big]^{l_{i-2}}\times \\
%& &\quad \times (R^{Bog}_{i-2,i-2}(z))^{\frac{1}{2}}W^*_{i-2,i}(R^{Bog}_{i,i}(z))^{\frac{1}{2}}\,.
\end{eqnarray}
Hence, it is enough to show that 
\begin{equation}\label{condition-gamma}
\|(R^{Bog}_{\bold{j}_*\,;\,i,i}(z))^{\frac{1}{2}}W_{\bold{j}_*\,;\,i,i-2}(R^{Bog}_{\bold{j}_*\,;\,i-2,i-2}(z))^{\frac{1}{2}}\|^2\,\|\sum_{l_{i-2}=0}^{\infty}\Big[(R^{Bog}_{\bold{j}_*\,;\,i-2,i-2}(z))^{\frac{1}{2}}\Gamma^{Bog}_{\bold{j}_*\,;\,i-2,i-2}(z)(R^{Bog}_{\bold{j}_*\,;\,i-2,i-2}(z))^{\frac{1}{2}}\Big]^{l_{i-2}}\|<1
\end{equation}
so that we can estimate
%we assume $N$ sufficiently large such that .... and, by induction, we prove that for $i\geq 2$
\begin{eqnarray}\label{ineq-gammacheck-1}
& &\|\sum_{l_{i}=0}^{\infty}\Big[(R^{Bog}_{\bold{j}_*\,;\,i,i}(z))^{\frac{1}{2}}\Gamma^{Bog\,}_{\bold{j}_*\,;\,i,i}(R^{Bog}_{\bold{j}_*\,;\,i,i}(z))^{\frac{1}{2}}\Big]^{l_i}\|\\
&\leq & \frac{1}{1-\|(R^{Bog}_{\bold{j}_*\,;\,i,i}(z))^{\frac{1}{2}}W_{\bold{j}_*\,;\,i,i-2}(R^{Bog}_{i-2,i-2}(z))^{\frac{1}{2}}\|^2\|\sum_{l_{i-2}=0}^{\infty}\Big[(R^{Bog}_{\bold{j}_*\,;\,i-2,i-2}(z))^{\frac{1}{2}}\Gamma^{Bog}_{\bold{j}_*\,;\,i-2,i-2}(z)(R^{Bog}_{\bold{j}_*\,;\,i-2,i-2}(z))^{\frac{1}{2}}\Big]^{l_{i-2}}\|}\,.\nonumber
\end{eqnarray}
To this purpose, we define
\begin{equation}\label{gamma-check-a}
\check{\Gamma}^{Bog\,}_{\bold{j}_*\,;\,i,i}:=\sum_{l_{i}=0}^{\infty}[(R^{Bog}_{\bold{j}_*\,;\,i,i}(z))^{\frac{1}{2}}\Gamma^{Bog\,}_{\bold{j}_*\,;\,i,i}(R^{Bog}_{\bold{j}_*\,;\,i,i}(z))^{\frac{1}{2}}]^{l_i}\quad\text{for}\,\,i\geq 2\,,
\end{equation}
and
\begin{equation}
\check{\Gamma}^{Bog\,}_{\bold{j}_*\,;\,0,0}:=\charf\,.\label{gamma-check-b}
\end{equation}
By induction, we shall prove that the R-H-S in (\ref{gamma-check-a}) is a well defined bounded operator.
Notice that, using the definitions in (\ref{def-gamma-ii}) and (\ref{gamma-check-a}), for $i\geq 4$ we have the identity
\begin{eqnarray}
& &\check{\Gamma}^{Bog\,}_{\bold{j}_*\,;\,i,i}(z)\\
%&=&\sum_{l_{i}=0}^{\infty}[(R^{Bog}_{\bold{j}_*\,;\,i,i}(z))^{\frac{1}{2}}W_{\bold{j}_*\,;\,i,i-2}\,R^{Bog}_{\bold{j}_*\,;\,i-2,i-2}(z) \sum_{l_{i-2}=0}^{\infty}\Big[\Gamma^{Bog}_{\bold{j}_*\,;\,i-2,i-2}(z)R^{Bog}_{\bold{j}_*\,;\,i-2,i-2}(z)\Big]^{l_{i-2}}W^*_{\bold{j}_*\,;\,i-2,i}(R^{Bog}_{\bold{j}_*\,;\,i,i}(z))^{\frac{1}{2}}]^{l_i}\nonumber \\
&=&\sum_{l_{i}=0}^{\infty}[(R^{Bog}_{\bold{j}_*\,;\,i,i}(z))^{\frac{1}{2}}W_{\bold{j}_*\,;\,i,i-2}\,(R^{Bog}_{\bold{j}_*\,;\,i-2,i-2}(z))^{\frac{1}{2}} \check{\Gamma}^{Bog}_{\bold{j}_*\,;\,i-2,i-2}(z)(R^{Bog}_{\bold{j}_*\,;\,i-2,i-2}(z))^{\frac{1}{2}}W^*_{\bold{j}_*\,;\,i-2,i}(R^{Bog}_{\bold{j}_*\,;\,i,i}(z))^{\frac{1}{2}}]^{l_i}\,.\nonumber
\end{eqnarray}
Due to the definitions in  (\ref{GammaBog-2}) and (\ref{gamma-check-b}), and taking (\ref{gamma-check-a}) into account, an analogous identity holds for $i=2$:  
\begin{eqnarray}
& &\check{\Gamma}^{Bog\,}_{\bold{j}_*\,;\,2,2}\\
& =&\sum_{l_{2}=0}^{\infty}\Big[(R^{Bog}_{\bold{j}_*\,;\,2,2}(z))^{\frac{1}{2}}W_{\bold{j}_*\,;\,2,0}\,(R^{Bog}_{\bold{j}_*\,;\,0,0}(z))^{\frac{1}{2}} (R^{Bog}_{\bold{j}_*\,;\,0,0}(z))^{\frac{1}{2}}W^*_{\bold{j}_*\,;\,0,2}(R^{Bog}_{\bold{j}_*\,;\,2,2}(z))^{\frac{1}{2}}\Big]^{l_2}\\
&= &\sum_{l_{2}=0}^{\infty}\Big[(R^{Bog}_{\bold{j}_*\,;\,2,2}(z))^{\frac{1}{2}}W_{\bold{j}_*\,;\,2,0}\,(R^{Bog}_{\bold{j}_*\,;\,0,0}(z))^{\frac{1}{2}}\check{\Gamma}^{Bog\,}_{\bold{j}_*\,;\,0,0} (R^{Bog}_{\bold{j}_*\,;\,0,0}(z))^{\frac{1}{2}}W^*_{\bold{j}_*\,;\,0,2}(R^{Bog}_{\bold{j}_*\,;\,2,2}(z))^{\frac{1}{2}}\Big]^{l_2}\,.\quad\quad
\end{eqnarray}
Thus, for $i\geq 2$, the inequality in (\ref{ineq-gammacheck-1}) is equivalent to
\begin{equation}\label{equiv-ineq}
 \frac{1}{\|\check{\Gamma}^{Bog\,}_{\bold{j}_*\,;\,i,i}(z)\|}\geq 1-\|(R^{Bog}_{\bold{j}_*\,;\,i,i}(z))^{\frac{1}{2}}W_{\bold{j}_*\,;\,i,i-2}(R^{Bog}_{\bold{j}_*\,;\,i-2,i-2}(z))^{\frac{1}{2}}\|^2\|\check{\Gamma}^{Bog\,}_{\bold{j}_*\,;\,i-2,i-2}(z)\|\,.
\end{equation}
Furthermore,  an upper bound to  $\|\check{\Gamma}^{Bog\,}_{\bold{j}_*\,;\,i,i}(z)\|$ implies that the Feshbach-Schur Hamiltonian $\mathscr{K}^{Bog\,(i)}_{\bold{j}_*}(z)$ is well defined. 

In order to show inequality (\ref{equiv-ineq}) and the existence of an upper bound to $\|\check{\Gamma}^{Bog\,}_{\bold{j}_*\,;\,i,i}(z)\|$ we consider the sequence, $\{X_i\}$,  defined in Lemma \ref{lemma-sequence} with $\epsilon\equiv \epsilon_{\bold{j}_{*}}$,
%\begin{eqnarray}
%x_{2j+2}&:=&1-\frac{1}{4(1+a_{\epsilon_{\bold{j}_{*}}}-\frac{2b_{\epsilon_{\bold{j}_{*}}}}{N-2j+1}-\frac{1-c_{\epsilon_{\bold{j}_{*}}}}{(N-2j+1)^2})x_{2j}}
%\end{eqnarray}
starting from $X_{0}\equiv 1$. 
% In the sequel, we consider only the sequence with even indeces. The same argument applies to the other sequence.
(We recall that $N$ is assumed to be even.)

\noindent
We must verify  that, for $0\leq i-2\leq N-4$ with $i$ even, and $z\leq E^{Bog}_{\bold{j}_*}+ \sqrt{\epsilon_{\bold{j}_*}}\phi_{\bold{j}_*}\sqrt{\epsilon_{\bold{j}_*}^2+2\epsilon_{\bold{j}_*}}$, if 
\begin{equation} 
\frac{1}{\|\check{\Gamma}^{Bog\,}_{\bold{j}_*\,;\,i-2,i-2}(z)\|}\geq X_{i-2}
\end{equation} then 
\begin{equation}
\frac{1}{\|\check{\Gamma}^{Bog\,}_{\bold{j}_*\,;\,i,i}(z)\|}\geq X_{i}.\label{Gamma-ineq}
\end{equation}
From (\ref{bound-even}) in Lemma \ref{lemma-sequence} we know that 
for $ N\geq N-i+2 \geq 4$ and $\epsilon_{\bold{j}_*}$ small enough
\begin{equation}\label{x-ineq-1}
X_{i-2}\geq\frac{1}{2}\Big[1+\sqrt{\eta a_{\epsilon_{\bold{j}_*}}}-\frac{b_{\epsilon_{\bold{j}_*}}/\sqrt{\eta a_{\epsilon_{\bold{j}_*}}}}{N-i+2-\xi}\Big]\geq \frac{3}{8}+o(1)\,,
\end{equation}
where $\eta=1-\sqrt{\epsilon_{\bold{j}_*}}$ and $\xi=\epsilon_{\bold{j}_*}^{\Theta}$ with $\Theta:=\min\{2(\nu-\frac{11}{8})\,;\,\frac{1}{4}\}$. Hence, for $(0<)\epsilon_{\bold{j}_{*}}$ sufficiently small and $2\leq i \leq N-2$ ($\Rightarrow N-i+1\geq 3$)
\begin{eqnarray}
& &\|(R^{Bog}_{\bold{j}_*\,;\,i,i}(z))^{\frac{1}{2}}W_{\bold{j}_*\,;\,i,i-2}(R^{Bog}_{\bold{j}_*\,;\,i-2,i-2}(z))^{\frac{1}{2}}\|^2\|\check{\Gamma}^{Bog\,}_{\bold{j}_*\,;\,i-2,i-2}(z)\|\label{estimate-series-1}\\
&\leq  &\frac{1}{4(1+a_{\epsilon_{\bold{j}_*}}-\frac{2b_{\epsilon_{\bold{j}_{*}}}}{N-i+1}-\frac{1-c_{\bold{j}_{*}}}{(N-i+1)^2})X_{i-2}}\label{estimate-serie-mezzo}\\
&\leq &\frac{3}{4}+o(1)\,, \label{estimate-series-2}
\end{eqnarray}
where we have used that\footnote{The second inequality in (\ref{doppia}) holds because $-\frac{2b^{(\delta)}_{\epsilon_{\bold{j}_{*}}}}{N-i+1}-\frac{1-c^{(\delta)}_{\bold{j}_{*}}}{(N-i+1)^2}$ is nonincreasing as a function of $\delta$ in the considered  range $\delta\leq 1+\sqrt{\epsilon_{\bold{j}_*}}$ provided  $\epsilon_{\bold{j}_*}$ is sufficiently small.} for $z\leq E^{Bog}_{\bold{j}_*}+ (\delta-1)\phi_{\bold{j}_*}\sqrt{\epsilon_{\bold{j}_*}^2+2\epsilon_{\bold{j}_*}}$,  $\delta\leq 1+\sqrt{\epsilon_{\bold{j}_*}}$, and $\epsilon_{\bold{j}_*}$ small
\begin{equation}\label{doppia}
\|(R^{Bog}_{\bold{j}_*\,;\,i,i}(z))^{\frac{1}{2}}W_{\bold{j}_*\,;\,i,i-2}(R^{Bog}_{\bold{j}_*\,;\,i-2,i-2}(z))^{\frac{1}{2}}\|^2\leq \frac{1}{4(1+a_{\epsilon_{\bold{j}_*}}-\frac{2b^{(\delta)}_{\epsilon_{\bold{j}_{*}}}}{N-i+1}-\frac{1-c^{(\delta)}_{\bold{j}_{*}}}{(N-i+1)^2})}\leq \frac{1}{4(1+a_{\epsilon_{\bold{j}_*}}-\frac{2b_{\epsilon_{\bold{j}_{*}}}}{N-i+1}-\frac{1-c_{\bold{j}_{*}}}{(N-i+1)^2})}\,.
\end{equation} Hence, we can conclude that (\ref{ineq-gammacheck-1}) holds. Next, by means of (\ref{ineq-gammacheck-1}) and Lemma \ref{lemma-sequence} we estimate
\begin{eqnarray}
 \frac{1}{\|\check{\Gamma}^{Bog\,}_{\bold{j}_*\,;\,i,i}(z)\|}&\geq &1-\|(R^{Bog}_{\bold{j}_*\,;\,i,i}(z))^{\frac{1}{2}}W_{\bold{j}_*\,;\,i,i-2}(R^{Bog}_{\bold{j}_*\,;\,i-2,i-2}(z))^{\frac{1}{2}}\|^2\|\check{\Gamma}^{Bog\,}_{\bold{j}_*\,;\,i-2,i-2}(z)\|\\
 &\geq & 1-\frac{1}{4(1+a_{\epsilon_{\bold{j}_{*}}}-\frac{2b_{\epsilon_{\bold{j}_{*}}}}{N-i+1}-\frac{1}{(N-i+1)^2})X_{i-2}} \label{estimate-serie-mezzo-2}\\
 &=&X_{i}\\
 &\geq &\frac{1}{4}+o(1)\,.
\end{eqnarray}
We observe that the property holds at the first step because
\begin{equation}\label{estiGammaN-2}
 \frac{1}{\|\check{\Gamma}^{Bog\,}_{\bold{j}_*\,;\,0,0}(z)\|}=1=X_0\,.
\end{equation}

Thus, in the range considered for $z$, for $\epsilon_{\bold{j}_*}$ sufficiently small and fulfilling the assumptions of  Lemma \ref{lemma-sequence},  the Neumann expansions used on the R-H-S of (\ref{KappaBog-i}) are well defined for $i\leq N-2$. Moreover,  $\|\check{\Gamma}^{Bog\,}_{\bold{j}_*\,;\,i,i}(z)\|$ (with $i\leq N-2$)  does not diverge as $\epsilon_{\bold{j}_{*}}\to 0$. \qed

\noindent
At each step the isospectrality property holds for the  map $\mathscr{F}^{(i)}$, $0\leq i \leq N-2$,  applied to $\mathscr{K}^{Bog\,(i-2)}_{\bold{j}_*}(z)$ because (see \cite{BFS}):
\begin{enumerate}
\item
$\mathscr{P}^{(i)}\mathscr{K}^{Bog\,(i-2)}_{\bold{j}_*}(z)\overline{{\mathscr{P}}^{(i)}}$ and $\overline{{\mathscr{P}}^{(i)}}\mathscr{K}^{Bog\,(i-2)}_{\bold{j}_*}(z)\mathscr{P}^{(i)}$ are bounded operators on $\mathcal{F}^N$; 
\item
the operator  $\overline{{\mathscr{P}}^{(i)}}\mathscr{K}^{Bog\,(i-2)}_{\bold{j}_*}(z)\overline{{\mathscr{P}}^{(i)}}$ is bounded invertible on $\overline{{\mathscr{P}}^{(i)}}\mathcal{F}^N$;
\item  
$\mathscr{P}^{(i)}(H^{Bog}_{\bold{j}_*}-\sum_{\bold{j}\in\mathbb{Z}^d} k^2_{\bold{j}}a_{\bold{j}}^{*}a_{\bold{j}})\mathscr{P}^{(i)}$ is a bounded operator on $\mathcal{F}^N$ and    $\mathscr{P}^{(i)}\sum_{\bold{j}\in\mathbb{Z}^d} k^2_{\bold{j}}a_{\bold{j}}^{*}a_{\bold{j}}\mathscr{P}^{(i)}$ is a closed operator on  $\mathscr{P}^{(i)}\mathcal{F}^N$.
\end{enumerate}
%Conditions in 1. and 2. above turn out to be satisfied as byproduct of Theorem \ref{theorem-Bog} in next section.

%\begin{remark} {\color{red}Probabilmente non necessario. }Using the results of Lemma  {lemma-sequence} for $\xi={\color{red}\frac{1}{2}}$  one gets the bound from below  $\frac{1}{\|\check{\Gamma}^{Bog\,}_{\bold{j}_*\,;\,i,i}(z)\|}>\frac{1}{4}>0$, $0\leq i\leq N-2$, for a lower $z$ but still larger than the Bogoliubov energy. This ensures not only that the expansion holds for arbitrarily small values of $\epsilon_{\bold{j}_{*}}(>0)$ but also that $\|\check{\Gamma}^{Bog\,}_{\bold{j}_*\,;\,i,i}(z)\|$ does not diverge as $\epsilon_{\bold{j}_{*}}\to 0$.
%&\end{remark}
\section{Construction of the ground state of $H^{Bog}_{\bold{j}_*}$}\label{groundstate}
\setcounter{equation}{0}

We remind that, for $i=N-2$,  $Q^{(> i+1)}_{\bold{j}_*} \equiv Q^{(> N-1)}_{\bold{j}_*}$ is the projection onto the subspace where less than $N-i=N-N+1=1$ particles in the modes $\bold{j}_{*}$ and $-\bold{j}_{*}$ are present, i.e., where no particles in the modes $\bold{j}_{*}$ and $-\bold{j}_{*}$ are present. 

\subsection{Last step:  fixed point and ground state energy}\label{last-projection}
For the step from $i=N-2$ to $i=N$ we consider  the projections $\mathscr{P}^{(N)}:=\mathscr{P}_{\eta}:=|\eta \rangle \langle \eta |$ and $\overline{\mathscr{P}^{(N)}}:=\overline{\mathscr{P}_{\eta}}$ such that
\begin{equation}
\mathscr{P}^{(N)}+\overline{\mathscr{P}^{(N)}}=\charf_{Q^{(>N-1)}_{\bold{j}_*}\mathcal{F}^N}\,.
\end{equation}
Formally, we get
\begin{eqnarray}
& &\mathscr{K}^{Bog\,(N)}_{\bold{j}_*}(z)\\
&:=&\mathscr{F}^{(N)}(\mathscr{K}^{Bog\,(N-2)}_{\bold{j}_*}(z))\\
&=&\mathscr{P}_{\eta}(H^{Bog}_{\bold{j}_*}-z)\mathscr{P}_{\eta}\label{K-last-step}\\
& &-\mathscr{P}_{\eta}W_{\bold{j}_*}\,R^{Bog}_{\bold{j}_*\,;\,N-2,N-2}(z)\sum_{l_{N-2}=0}^{\infty}[\Gamma^{Bog}_{\bold{j}_*\,;\,N-2,N-2}(z) R^{Bog}_{\bold{j}_*\,;\,N-2,N-2}(z)]^{l_{N-2}}\,W^*_{\bold{j}_*}\mathscr{P}_{\eta}\quad \nonumber\\
& &-\mathscr{P}_{\eta}W_{\bold{j}_*}\,\overline{\mathscr{P}_{\eta}}\,\frac{1}{\overline{\mathscr{P}_{\eta}}\mathscr{K}^{Bog\,(N-2)}_{\bold{j}_*}(z)\overline{\mathscr{P}_{\eta}}}\overline{\mathscr{P}_{\eta}}W^*_{\bold{j}_*}\mathscr{P}_{\eta}  \label{invers}
\end{eqnarray}
because
\begin{equation}
\mathscr{P}_{\eta}W_{\bold{j}_*}\,R^{Bog}_{\bold{j}_*\,;\,N-2,N-2}(z)\sum_{l_{N-2}=0}^{\infty}[\Gamma^{Bog}_{\bold{j}_*\,;\,N-2,N-2}(z) R^{Bog}_{\bold{j}_*\,;,N-2,N-2}(z)]^{l_{N-2}}\, W^*_{\bold{j}_*}\overline{\mathscr{P}_{\eta}}=0
\end{equation}
due to
\begin{equation}
\Big[ a^*_{\bold{0}}a_{\bold{0}}\,,\,W_{\bold{j}_*}\,R^{Bog}_{\bold{j}_*\,;\,N-2,N-2}(z)\sum_{l_{N-2}=0}^{\infty}[\Gamma^{Bog}_{\bold{j}_*\,;\,N-2,N-2}(z) R^{Bog}_{\bold{j}_*\,;,N-2,N-2}(z)]^{l_{N-2}}\,W^*_{\bold{j}_*}\Big]=0
\end{equation}
combined with $a^*_{\bold{0}}a_{\bold{0}} \mathscr{P}_{\eta}=N\mathscr{P}_{\eta}$ and $\overline{\mathscr{P}_{\eta}}a^*_{\bold{0}}a_{\bold{0}}\overline{\mathscr{P}_{\eta}}\leq (N-1)\overline{\mathscr{P}_{\eta}}$. 

\noindent
The Hamiltonian $\mathscr{K}^{Bog\,(N)}_{\bold{j}_*}(z)$ is well defined if $$\overline{\mathscr{P}_{\eta}}\frac{1}{\overline{\mathscr{P}_{\eta}}\mathscr{K}^{Bog\,(N-2)}_{\bold{j}_*}(z)\overline{\mathscr{P}_{\eta}}}\overline{\mathscr{P}_{\eta}}$$ in (\ref{invers}) is well defined. In this case,  using  $\overline{\mathscr{P}_{\eta}}W^*_{\bold{j}_*}\mathscr{P}_{\eta}=0$  and $\mathscr{P}_{\eta}(H^{Bog}_{\bold{j}_*}-z)\mathscr{P}_{\eta}=-z\mathscr{P}_{\eta}$,  finally we would get
\begin{eqnarray}
& &\mathscr{K}^{Bog\,(N)}_{\bold{j}_*}(z)\\
%&=&\mathscr{P}_{\eta}(H^{Bog}_{\bold{j}_*}-z)\mathscr{P}_{\eta}\\
%& &-\mathscr{P}_{\eta}W_{\bold{j}_*}\,Q^{(N-2)}_{\bold{j}_*}\,R^{Bog}_{\bold{j}_*\,;\,N-2,N-2}(z)\sum_{l_{N-2}=0}^{\infty}[\Gamma^{Bog}_{\bold{j}_*\,;\,N-2,N-2}(z) R^{Bog}_{\bold{j}_*\,;\,N-2,N-2}(z)]^{l_{N-2}}\, Q^{(N-2)}_{\bold{j}_*}W^*_{\bold{j}_*}\mathscr{P}_{\eta}\nonumber\\
&=&-z\mathscr{P}_{\eta}\\
& &-\mathscr{P}_{\eta}W_{\bold{j}_*}\,R^{Bog}_{\bold{j}_*\,;\,N-2,N-2}(z)\sum_{l_{N-2}=0}^{\infty}[\Gamma^{Bog}_{\bold{j}_*\,;\,N-2,N-2}(z) R^{Bog}_{\bold{j}_*\,;\,N-2,N-2}(z)]^{l_{N-2}}\, W^*_{\bold{j}_*}\mathscr{P}_{\eta}\,.\nonumber\,
%& &-Q^{(>N-1)}W\,Q^{(N-2)}\,R^{Bog}_{N-2,N-2}(z)\sum_{l_{N-2}=0}^{\infty}[\Gamma^{Bog}_{N-2,N-2}(z) R^{Bog}_{N-2,N-2}(z)]^{l_{N-2}}\, Q^{(N-2)}W^*Q^{(>N-1)}\quad \quad\quad\quad\label{first-term-last-Bog}\\
%& &-\mathscr{V}^{(N-1)}(z)\,R_{N-1,N-1}(z)\sum_{l_{N-1}=0}^{\infty}[\Gamma_{N-1,N-1}(z) R_{N-1,N-1}(z)]^{l_{N-1}}\,(\mathscr{V}^{(N-1)}(z) )^*\quad\quad \label{last-term-last-Bog}
\end{eqnarray}
Therefore, the operator $\mathscr{K}^{Bog\,(N)}_{\bold{j}_*}(z)$ would be a multiple of the projection $|\eta \rangle \langle \eta |$, i.e.,
\begin{equation}\label{final-H}
\mathscr{K}^{Bog\,(N)}_{\bold{j}_*}(z)=f_{\bold{j}_*}(z)|\eta \rangle \langle \eta |
\end{equation}
where 
\begin{eqnarray}\label{fp-function}
f_{\bold{j}_*}(z)
&:=&-z\\
& &-\langle \eta\,,\,W_{\bold{j}_*}\,R^{Bog}_{\bold{j}_*\,;\,N-2,N-2}(z)\sum_{l_{N-2}=0}^{\infty}[\Gamma^{Bog}_{\bold{j}_*\,;\,N-2,N-2}(z) R^{Bog}_{\bold{j}_*\,;\,N-2,N-2}(z)]^{l_{N-2}}\,W^*_{\bold{j}_*}\eta\rangle.\quad\quad\quad\label{fp-function-rhs}
\end{eqnarray}
 Notice that $f_{\bold{j}_*}(z)>0$ for $|z|$ sufficiently large  (with $z\leq E^{Bog}_{\bold{j}_*}+ \sqrt{\epsilon_{\bold{j}_*}}\phi_{\bold{j}_*}\sqrt{\epsilon_{\bold{j}_*}^2+2\epsilon_{\bold{j}_*}}$)  because 
 \begin{equation}
\lim_{z\to -\infty} \langle \eta\,,\,W_{\bold{j}_*}\,R^{Bog}_{\bold{j}_*\,;\,N-2,N-2}(z)\sum_{l_{N-2}=0}^{\infty}[\Gamma^{Bog}_{\bold{j}_*\,;\,N-2,N-2}(z) R^{Bog}_{\bold{j}_*\,;\,N-2,N-2}(z)]^{l_{N-2}}\,W^*_{\bold{j}_*}\eta\rangle=0\,.
 \end{equation}
%%and $|(\ref{fp-function-rhs})|$ is bounded for 
After determining  the (fixed point) solution, $z_{*}$, to the equation  
% $\overline{\mathscr{P}_{\eta}}\mathscr{K}^{Bog\,(N-2)}(z)\overline{\mathscr{P}_{\eta}}$ This operator is invertible if....
%Assuming for the moment that
\begin{equation}\label{fp-equation}
f_{\bold{j}_*}(z)=0
\end{equation}
we shall show that the last step is implementable for 
\begin{equation}\label{range-z}
z<\min\,\Big\{ z_{*}+\frac{\Delta_0}{2}\,;\,E^{Bog}_{\bold{j}_*}+ \sqrt{\epsilon_{\bold{j}_*}}\phi_{\bold{j}_*}\sqrt{\epsilon_{\bold{j}_*}^2+2\epsilon_{\bold{j}_*}}\Big\}\,,
\end{equation}
 where 
\begin{equation}
\Delta_0:=\min\, \Big\{k_{\bold{j}}^2\,|\,\bold{j}\in \mathbb{Z}^d\setminus\{\bold{0}\}\Big\}\,.
\end{equation}
%Condition 3.2 in Definition \ref{def-pot} ensures $(1-\frac{\phi_{\bold{j}_{*}}}{\Delta_0}\frac{ N^{\mu}}{N})>\frac{1}{2}$.}

\subsubsection{Fixed point}\label{fixed point}

We observe that in the scalar product
\begin{eqnarray}
& &\langle \eta\,,\,W_{\bold{j}_*}\,R^{Bog}_{\bold{j}_*\,;\,N-2,N-2}(z)\sum_{l_{N-2}=0}^{\infty}[\Gamma^{Bog}_{\bold{j}_*\,;\,N-2,N-2}(z) R^{Bog}_{\bold{j}_*\,;\,N-2,N-2}(z)]^{l_{N-2}}\,W^*_{\bold{j}_*}\eta \rangle\quad\nonumber\\
&=&\langle \eta\,,\,W_{\bold{j}_*}\,(R^{Bog}_{\bold{j}_*\,;\,N-2,N-2}(z))^{\frac{1}{2}}\check{\Gamma}^{Bog}_{\bold{j}_*\,;\,N-2,N-2}(z) (R^{Bog}_{\bold{j}_*\,;\,N-2,N-2}(z))^{\frac{1}{2}}\,W^*_{\bold{j}_*}\eta \rangle \label{scalar-prod}
\end{eqnarray}
the operators of the type 
\begin{eqnarray}
& &(R^{Bog}_{\bold{j}_*\,;\,i,i}(z))^{\frac{1}{2}}W_{\bold{j}_*\,;\,i,i-2}\,(R^{Bog}_{\bold{j}_*\,;\,i-2,i-2}(z))^{\frac{1}{2}}\quad,\quad(R^{Bog}_{\bold{j}_*\,;\,i-2,i-2}(z))^{\frac{1}{2}}W^*_{\bold{j}_*\,;\,i-2,i}(R^{Bog}_{\bold{j}_*\,;\,i,i}(z))^{\frac{1}{2}}
%&=&(R^{Bog}_{\bold{j}_*\,;\,i,i}(z))^{\frac{1}{2}}\phi_{\bold{j}}\frac{a^*_{\bold{0}}a^*_{\bold{0}}a_{\bold{j}}a_{-\bold{j}}}{N}\,R^{Bog}_{\bold{j}_*\,;\,i-2,i-2}(z)\,\phi_{\bold{j}}\frac{a_{\bold{0}}a_{\bold{0}}a^*_{\bold{j}}a^*_{-\bold{j}}}{N}(R^{Bog}_{\bold{j}_*\,;\,i,i}(z))^{\frac{1}{2}}\quad
\end{eqnarray}
pop up when we expand $\check{\Gamma}^{Bog}_{\bold{j}_*\,;\,N-2,N-2}(z)$ by iteration of the identity 
\begin{eqnarray}
& &\check{\Gamma}^{Bog\,}_{\bold{j}_*\,;\,i,i}(z)\label{ident}\\
%&=&\sum_{l_{i}=0}^{\infty}[(R^{Bog}_{\bold{j}_*\,;\,i,i}(z))^{\frac{1}{2}}W_{\bold{j}_*\,;\,i,i-2}\,R^{Bog}_{\bold{j}_*\,;\,i-2,i-2}(z) \sum_{l_{i-2}=0}^{\infty}\Big[\Gamma^{Bog}_{\bold{j}_*\,;\,i-2,i-2}(z)R^{Bog}_{\bold{j}_*\,;\,i-2,i-2}(z)\Big]^{l_{i-2}}W^*_{\bold{j}_*\,;\,i-2,i}(R^{Bog}_{\bold{j}_*\,;\,i,i}(z))^{\frac{1}{2}}]^{l_i}\nonumber \\
&=&\sum_{l_{i}=0}^{\infty}[(R^{Bog}_{\bold{j}_*\,;\,i,i}(z))^{\frac{1}{2}}W_{\bold{j}_*\,;\,i,i-2}\,(R^{Bog}_{\bold{j}_*\,;\,i-2,i-2}(z))^{\frac{1}{2}} \check{\Gamma}^{Bog}_{\bold{j}_*\,;\,i-2,i-2}(z)(R^{Bog}_{\bold{j}_*\,;\,i-2,i-2}(z))^{\frac{1}{2}}W^*_{\bold{j}_*\,;\,i-2,i}(R^{Bog}_{\bold{j}_*\,;\,i,i}(z))^{\frac{1}{2}}]^{l_i}\,.\quad\quad\quad\quad\label{gamma-exp}
\end{eqnarray}
Following the same arguments that have been used to re-express (\ref{operator-cnumbers}),  in the scalar product (\ref{scalar-prod}) the operator 
\begin{equation}
W_{\bold{j}_*}\,R^{Bog}_{\bold{j}_*\,;\,N-2,N-2}(z)\sum_{l_{N-2}=0}^{\infty}[\Gamma^{Bog}_{\bold{j}_*\,;\,N-2,N-2}(z) R^{Bog}_{\bold{j}_*\,;\,N-2,N-2}(z)]^{l_{N-2}}\, W^*_{\bold{j}_*}
\end{equation}
 can be replaced with a function of the number operators $a^*_{\bold{j}_*}a_{\bold{j}_*}$, $a^*_{-\bold{j}_*}a_{-\bold{j}_*}$, and $a^*_{\bold{0}}a_{\bold{0}}$ only.  Furthermore,  these (number) operators can be replaced by c-numbers because they act  on vectors with definite number of particles in the modes $\bold{j}_*$, $-\bold{j}_*$ and $\bold{0}$. This is due to the projections contained in the definition of $(R^{Bog}_{\bold{j}_*\,;\,i,i}(z))^{\frac{1}{2}}$ and to the fact that  $\eta$ is a product state with all the particles in the zero mode. It turns out that the couple of companion operators
\begin{equation}
(R^{Bog}_{\bold{j}_*\,;\,i,i}(z))^{\frac{1}{2}}\phi_{\bold{j}_*}\frac{a^*_{\bold{0}}a^*_{\bold{0}}a_{\bold{j}_{*}}a_{-\bold{j}_{*}}}{N}\,(R^{Bog}_{\bold{j}_*\,;\,i-2,i-2}(z))^{\frac{1}{2}}\quad,\quad(R^{Bog}_{\bold{j}_*\,;\,i-2,i-2}(z))^{\frac{1}{2}}\,\phi_{\bold{j}_*}\frac{a_{\bold{0}}a_{\bold{0}}a^*_{\bold{j}_{*}}a^*_{-\bold{j}_{*}}}{N}(R^{Bog}_{\bold{j}_*\,;\,i,i}(z))^{\frac{1}{2}}
\end{equation}
can be replaced with  the c-number
\begin{eqnarray}
& &\mathcal{W}_{\bold{j}_*\,;\,i,i-2}(z)\mathcal{W}^*_{\bold{j}_*\,;\,i-2,i}(z)\\
&:= &\frac{(n_{\bold{j}_0}-1)n_{\bold{j}_0}}{N^2}\,\phi^2_{\bold{j}_{*}}\,\frac{ (n_{\bold{j}_{*}}+1)(n_{-\bold{j}_{*}}+1)}{\Big[(\frac{n_{\bold{j}_0}}{N}\phi_{\bold{j}_{*}}+k_{\bold{j}_{*}}^2)(n_{\bold{j}_{*}}+n_{-\bold{j}_{*}})-z\Big]}\label{def-Wcal-1}\\
& &\quad\quad\quad \times\frac{1}{\Big[(\frac{(n_{\bold{j}_0}-2)}{N}\phi_{\bold{j}_{*}}+k_{\bold{j}_{*}}^2)(n_{\bold{j}_{*}}+n_{-\bold{j}_{*}})+2(\frac{(n_{\bold{j}_0}-2)}{N}\phi_{\bold{j}_{*}}+k_{\bold{j}_{*}}^2)-z\Big]}
%&:=& \frac{(n_{\bold{j}_0}+2)(n_{\bold{j}_0}+1)}{N^2}\,\phi^2_{\bold{j}_{*}}\,\frac{ (n_{\bold{j}_{*}}+1)(n_{-\bold{j}_{*}}+1)}{\Big[(\frac{n_{\bold{j}_0}}{N}\phi_{\bold{j}_{*}}+(k_{\bold{j}_{*}}^2))(n_{\bold{j}_{*}}+n_{-\bold{j}_{*}})-z\Big]}\times\\
%& &\times \frac{1}{\Big[(\frac{(n_{\bold{j}_0}+2)}{N}\phi_{\bold{j}_{*}}+(k_{\bold{j}_{*}}^2))(n_{\bold{j}_{*}}+n_{-\bold{j}_{*}})-2(\frac{(n_{\bold{j}_0}+2)}{N}\phi_{\bold{j}_{*}}+(k_{\bold{j}_{*}}^2))-z\Big]}\quad\quad\quad
\end{eqnarray} 
where 
\begin{equation}
n_{\bold{j}_{*}}+n_{-\bold{j}_{*}}=N-i\quad \text{with}\,\,i\,\,\text{even}\quad;\quad n_{\bold{j}_{*}}=n_{-\bold{j}_{*}}\quad;\quad n_{\bold{j}_0}=i\,.\label{def-Wcal-2}
\end{equation}

\noindent
%\begin{equation}
%(n_{\bold{j}_{*}}+1)(N-i-n_{\bold{j}_{*}}+1)=-(n_{\bold{j}_{*}})^2+(N-i)n_{\bold{j}_{*}}+N-i+1
%\end{equation}
This argument is made rigorous in Proposition \ref{induction-G} where we show the identity
\begin{equation}
(\ref{scalar-prod})=(1-\frac{1}{N})\frac{\phi_{\bold{j}_{*}}}{2\epsilon_{\bold{j}_*}+2-\frac{4}{N}-\frac{z}{\phi_{\bold{j}_{*}}}}\check{\mathcal{G}}_{\bold{j}_*\,;\,N-2,N-2}(z)\,.
\end{equation}
Here (see Proposition \ref{induction-G}), under the assumptions of Theorem \ref{theorem-Bog}, $\check{\mathcal{G}}_{\bold{j}_*\,;\,i,i}(z)$, with $0\leq i\leq N-2$ and even, is defined recursively by the relation
\begin{equation}\label{def-G}
\check{\mathcal{G}}_{\bold{j}_*\,;\,i,i}(z):=\sum_{l_{i}=0}^{\infty}[\mathcal{W}_{\bold{j}_*\,;i,i-2}(z)\mathcal{W}^*_{\bold{j}_*\,;i-2,i}(z)\check{\mathcal{G}}_{\bold{j}_*\,;\,i-2,i-2}(z)]^{l_i}\,
\end{equation}
with initial condition $\check{\mathcal{G}}_{\bold{j}_*\,;\,0,0}(z)\equiv 1$.
%we can also replace the operators
%\begin{equation}
%W_{\bold{j}_*}\,(R^{Bog}_{\bold{j}_*\,;\,N-2,N-2}(z))^{\frac{1}{2}}\quad ,\quad (R^{Bog}_{\bold{j}_*\,;\,N-2,N-2}(z))^{\frac{1}{2}}W^*_{\bold{j}_*}
%\end{equation}}
%with c-numbers, and, in turn, the equation in (\ref{fp-equation}) corresponds to\
Thus, we have derived that the fixed point equation $f_{\bold{j}_*}(z)=0$  corresponds to
\begin{eqnarray}\label{fp-equation-2}
z&=&-(1-\frac{1}{N})\frac{\phi_{\bold{j}_{*}}}{2\epsilon_{\bold{j}_*}+2-\frac{4}{N}-\frac{z}{\phi_{\bold{j}_{*}}}}\check{\mathcal{G}}_{\bold{j}_*\,;\,N-2,N-2}(z)\\
&=&-\frac{\phi_{\bold{j}_{*}}}{2\epsilon_{\bold{j}_*}+2-\frac{z}{\phi_{\bold{j}_{*}}}+\mathcal{O}(\frac{1}{N})}\check{\mathcal{G}}_{\bold{j}_*\,;\,N-2,N-2}(z)\,.\nonumber
\end{eqnarray}

We observe that 
\begin{eqnarray}
\frac{\partial \check{\mathcal{G}}_{\bold{j}_{*}\,;\,i,i}(z)}{\partial z}
&=&
[\check{\mathcal{G}}_{\bold{j}_{*}\,;\,i,i}(z)]^2\times \label{increasing-in}\\
%\Big(\frac{1}{1-[\mathcal{W}_{\bold{j}_{*}\,;i,i-2}(z)\mathcal{W}^*_{\bold{j}_1\,;i-2,i}(z)]_{\Delta n_{\bold{j}_{\bold{0}}}}\,[\check{\mathcal{G}}_{\bold{j}_{*}\,;\,i-2,i-2}(z)]_{\Delta n_{\bold{j}_{\bold{0}}}}}\Big)^2\times \\
& &\quad\quad \times \Big\{\frac{\partial [\mathcal{W}_{\bold{j}_{*}\,;i,i-2}(z)\mathcal{W}^*_{\bold{j}_*\,;i-2,i}(z)]}{\partial z}\, \check{\mathcal{G}}_{\bold{j}_{*}\,;\,i-2,i-2}(z) \\
& &\quad\quad\quad\quad + [\mathcal{W}_{\bold{j}_{*}\,;i,i-2}(z)\mathcal{W}^*_{\bold{j}_*\,;i-2,i}(z)]\,\frac{\partial [\check{\mathcal{G}}_{\bold{j}_{*}\,;\,i-2,i-2}(z)]}{\partial z}\Big\}\,
\end{eqnarray}
with \begin{equation}\frac{\partial [\mathcal{W}_{\bold{j}_{*}\,;i,i-2}(z)\mathcal{W}^*_{\bold{j}_*\,;i-2,i}(z)]}{\partial z}\geq 0\,. \label{increasing-fin}\end{equation}
\begin{remark}\label{increasing}
Starting from (\ref{increasing-in})-(\ref{increasing-fin}), it is easy to show  by induction that   $\frac{d\check{\mathcal{G}}_{\bold{j}_*\,;\,i,i}(z)}{dz}\geq  0$ under the standing assumptions on $\epsilon_{\bold{j}_*}$ and $N$. Consequently, $\check{\mathcal{G}}_{\bold{j}_*\,;\,i,i}(z)$ is nondecreasing  and $f'_{\bold{j}_{*}}(z)\leq -1$ (see (\ref{fp-function})-(\ref{fp-function-rhs}))  in the considered domain.
\end{remark}

\subsection{Lower bound of $\check{\mathcal{G}}_{\bold{j}_{*}\,;\,N-2,N-2}(z)$}\label{section-lower}
We recall that in Lemma \ref{lemma-sequence} we have derived a lower bound to $X_i$. This has been used to show  that $\|\check{\Gamma}^{Bog\,}_{\bold{j}_*\,;\,i,i}(z)\|$ stays bounded (see Theorem \ref{theorem-Bog}) and the Feshbach-Schur flow is well defined. Now,  we must show that $ \check{\mathcal{G}}_{\bold{j}_*\,;\,N-2,N-2}(z)$ is large enough to conclude that there is a solution, $z_*$, to the equation in (\ref{fp-equation-2}).

\noindent
To this purpose, for $0<\gamma<1$ and $z= E^{Bog}_{\bold{j}_*}+ (\delta-1)\phi_{\bold{j}^*}\sqrt{\epsilon_{\bold{j}_*}^2+2\epsilon_{\bold{j}_*}}$ with $$1+(\frac{2\sqrt{2}+3}{6})\sqrt{\epsilon_{\bold{j}_*}}\leq \delta\leq 1+\sqrt{\epsilon_{\bold{j}_*}},$$ we consider the positive quantity
\begin{eqnarray}
\mathcal{W}_{\bold{j}_*\,;\,i,i-2}^{\,(\gamma)}(z)\mathcal{W}_{\bold{j}_*\,;\,i-2,i}^{*\,(\gamma)}(z)
%&= & \frac{(i+2)(i+1)}{N^2}\,\phi^2_{\bold{j}_{*}}\\
%&  &\times\,\frac{(N-i)^2}{4\Big[(\frac{i-2}{N}\phi_{\bold{j}_{*}}+\epsilon_{\bold{j}_*}^2)(N-i)-z\Big]\Big[(\frac{i}{N}\phi_{\bold{j}_{*}}+\epsilon_{\bold{j}_*}^2))(N-i)-2(\frac{i}{N}\phi_{\bold{j}_{*}}+\epsilon_{\bold{j}_*}^2)-z\Big]}\quad\quad\quad\\
%&= &\frac{1}{4\Big[\frac{i-2}{i-1}+\frac{N}{i-1}\epsilon_{\bold{j}_*}-\frac{N}{(i-1)(N-i)}\frac{z}{\phi_{\bold{j}_{*}}}\Big]}\frac{1}{ \Big[1+\frac{N}{i}\epsilon_{\bold{j}_*}-\frac{2}{N-i+2}\Big(1+\frac{N}{i}\epsilon_{\bold{j}_*} \Big)-\frac{1}{N-i+2}\frac{N}{i}\frac{z}{\phi_{\bold{j}_{*}}}\Big]}\\
&:=&\frac{1}{4\Big[1+a^{(\gamma)}_{\epsilon_{\bold{j}_{*}}}-\frac{2b^{(\delta)}_{\epsilon_{\bold{j}_{*}}}}{N-i+2}-\frac{1-c^{(\delta)}_{\epsilon_{\bold{j}_{*}}}}{(N-i+2)^2}\Big]}
% \frac{1}{4 \Big[1+\epsilon_{\bold{j}_*}+\frac{c}{N^{\gamma}}-\frac{1}{(N-i+2)}\Big(\epsilon_{\bold{j}_*}+1+\delta \sqrt{\epsilon_{\bold{j}_*}^2+2\epsilon_{\bold{j}_*}}\,\Big)\Big]}\times\\
%& &\quad\times \frac{1}{\Big[1+\epsilon_{\bold{j}_*}+\frac{c}{N^{\gamma}}+\frac{1}{(N-i+2)}\Big(\epsilon_{\bold{j}_*}+1-\delta\sqrt{\epsilon_{\bold{j}_*}^2+2\epsilon_{\bold{j}_*}}\,\Big)\Big]}
\end{eqnarray}
where $a^{(\gamma)}_{\epsilon_{\bold{j}_{*}}}:=2\epsilon_{\bold{j}_{*}}+c_{\gamma}[\frac{\epsilon_{\bold{j}_*}}{N^{\gamma}}+\frac{1}{N}+\epsilon^2_{\bold{j}_{*}}]$ with $c_{\gamma}> 0$, and the coefficient $b^{(\delta)}_{\epsilon_{\bold{j}_{*}}}$, $c^{(\delta)}_{\epsilon_{\bold{j}_{*}}}$  are given in (\ref{b})-(\ref{c}). For simplicity, we assume that $\gamma$ is such that $N^{1-\gamma}$ is an even number. 
%Starting from (\ref{starting-from})
In Lemma \ref{accessori}, for $z= E^{Bog}_{\bold{j}_*}+ (\delta-1)\phi_{\bold{j}^*}\sqrt{\epsilon_{\bold{j}_*}^2+2\epsilon_{\bold{j}_*}}$ with $1+(\frac{2\sqrt{2}+3}{6})\sqrt{\epsilon_{\bold{j}_*}}\leq \delta\leq 1+\sqrt{\epsilon_{\bold{j}_*}}$,  we prove that the inequality
\begin{equation}\label{ineq-W}
\mathcal{W}_{\bold{j}_*\,;\,i,i-2}^{\,(\gamma)}(z)\mathcal{W}_{\bold{j}_*\,;\,i-2,i}^{*\,(\gamma)}(z)\leq \mathcal{W}_{\bold{j}_*\,;\,i,i-2}(z)\mathcal{W}^*_{\bold{j}_*\,;\,i-2,i}(z)
\end{equation}
holds for $i\geq N-N^{1-\gamma}$ provided $c_{\gamma}$ is sufficiently large. \\

Next, we introduce a sequence of real numbers $\{\tilde{X}^{(\gamma, \delta)}_i\}$ associated with $\mathcal{W}_{\bold{j}_*\,;\,i,i-2}^{\,(\gamma)}(z)\mathcal{W}_{\bold{j}_*\,;\,i-2,i}^{*\,(\gamma)}(z) $ that will be used to estimate  $\check{\mathcal{G}}_{\bold{j}_{*}\,;\,N-2,N-2}(z)$ from below:
%Next, we choose $\gamma$ sufficiently close to $1$ so that 
%\begin{equation}
%\frac{\epsilon_{\bold{j}_*}}{N^{\gamma}}+\frac{1}{N}=o(\epsilon_{\bold{j}_*}\sqrt{\epsilon_{\bold{j}_*}})\,.\label{condition-gamma}
%\end{equation}
%\begin{remark}\label{cond-gamma}
%Notice that the condition in (\ref{condition-gamma}) holds at fixed density $\rho$ if the dimension $d$ is strictly larger than $3$ and $\gamma \geq \frac{1}{3}$. 
%\end{remark}
$\tilde{X}^{(\gamma,\delta)}_i$ -- with $i$ even -- is defined by
\begin{eqnarray}
\tilde{X}_{i+2}^{(\gamma,\delta)}&:=&1-\frac{1}{4(1+a_{\epsilon}^{(\gamma)}-\frac{2b^{(\delta)}_{\epsilon}}{N-i}-\frac{1-c^{(\delta)}_{\epsilon}}{(N-i)^2})\tilde{X}^{(\gamma,\delta)}_{i}}
%x_{2j+3}^{(\gamma)}&:=&1-\frac{1}{4(1+a_{\epsilon}^{(\gamma)}-\frac{2b_{\epsilon}}{N-2j-1}-\frac{1-c_{\epsilon}}{(N-2j-1)^2})x_{2j+1}^{(\gamma)}}
\end{eqnarray}
starting from $\tilde{X}^{(\gamma,\delta)}_{N-N^{1-\gamma}}=1$. Lemma \ref{lemma-sequence-2} below provides an upper bound to $\tilde{X}^{(\gamma,\delta)}_i$.
% that together with (\ref{fixed-p-ineq}) provides the information that is needed.

\begin{lemma}\label{lemma-sequence-2}
 Let $0<\gamma<1$ and, for simplicity, assume that $N ^{1-\gamma}, \frac{N ^{1-\gamma}}{2}$ are both even. Let  $i_0\equiv N-N ^{1-\gamma}$ and consider  for $j\in \mathbb{N}$ and $j\geq \frac{i_0}{2}$ the sequence defined iteratively according to the relation
\begin{eqnarray}\label{sequence}
\tilde{X}^{(\gamma,\delta)}_{2j+2}&:=&1-\frac{1}{4(1+a_{\epsilon}^{(\gamma)}-\frac{2b^{(\delta)}_{\epsilon}}{N-2j}-\frac{1-c^{(\delta)}_{\epsilon}}{(N-2j)^2})\tilde{X}^{(\gamma,\delta)}_{2j}}
%x^{(\gamma)}_{2j+3}&:=&1-\frac{1}{4(1+a_{\epsilon}^{(\gamma)}-\frac{2b_{\epsilon}}{N-2j-1}-\frac{1-c_{\epsilon}}{(N-2j-1)^2})x^{(\gamma)}_{2j+1}}
\end{eqnarray}
with the initial condition $\tilde{X}^{(\gamma,\delta)}_{i_0}=1$ up to  $\tilde{X}^{(\gamma,\delta)}_{2j=N-2}$.  Here, 
\begin{equation}\label{aegamma}
a_{\epsilon}^{(\gamma)}:=2\epsilon+c_{\gamma}[\frac{\epsilon}{N^{\gamma}}+\frac{1}{N}+\epsilon^2]\,,\quad c_{\gamma}>0,
\end{equation}
\begin{equation}
b^{(\delta)}_{\epsilon}:=(1+\epsilon)\delta\sqrt{\epsilon^2+2\epsilon}\,,
\end{equation}
and
\begin{equation}
c^{(\delta)}_{\epsilon}:=-(1-\delta^2)(\epsilon^2+2\epsilon)
\end{equation}
where:
\begin{itemize}
\item
$\epsilon$ is sufficiently small and such that 
\begin{equation} \label{gamma-condition}
\epsilon^2+\frac{\epsilon}{N^{\gamma}}+\frac{1}{N}\leq k_{\gamma}\epsilon \sqrt{\epsilon}\quad\,,\,\quad\frac{1}{N^{1-\gamma}}\leq k_{\gamma}\epsilon\,,
\end{equation} for some constant $k_{\gamma}$ sufficiently small;
\item $1+\frac{2\sqrt{2}+3}{6}\sqrt{\epsilon}\leq \delta\leq 1+\sqrt{\epsilon}$\,.
\end{itemize}
%and $c_{\gamma}$ is a constant sufficiently large to ensure the inequality in (\ref{ineq-W}) for $i\geq N-N^{1-\gamma}$. 
Then, $\tilde{X}^{(\gamma,\delta)}_{2j}>0$ and for $2\leq N-2j\leq \frac{N ^{1-\gamma}}{2}$ the following estimate holds true
\begin{equation}\label{bound-seq-gamma}
(0<)\tilde{X}^{(\gamma,\delta)}_{2j}\leq\frac{1}{2}\Big[1+\sqrt{a_{\epsilon}^{(\gamma)}}-\frac{1}{N-2j+1- b^{(\delta)}_{\epsilon}}\Big]\,.
\end{equation}
%\[
%{\color{red}?x^{(\gamma)}_{2j+1}\leq\frac{1}{2}\Big[1+\sqrt{a_{\epsilon}}-\frac{1}{N-2j-1- b_{\epsilon}}\Big]?\,;}
%\]
%\noindent
%For  $2\left \lfloor{\frac{b_{\epsilon}}{a_{\epsilon}}}\right \rfloor \leq N-2j \leq 2\frac{rb_{\epsilon}}{a_{\epsilon}}$ with $r$ such that $2\frac{rb_{\epsilon}}{a_{\epsilon}}=N$
%\[
%{\color{red}x_{2j}^{(\gamma)}\leq\frac{1}{2}\Big[1+r\frac{b_{\epsilon}}{r-1}-\frac{a_{\epsilon}(N-2j)}{2(r-1)}-\frac{1}{N-2j}\Big]}
%\]
%\[
%{\color{red}x_{2j+1}^{(\gamma)}\leq\frac{1}{2}\Big[1+r\frac{b_{\epsilon}}{r-1}-\frac{a_{\epsilon}(N-2j-1)}{2(r-1)}-\frac{1}{N-2j-1}\Big]}
%\]

\end{lemma}

\noindent
\emph{Proof}

\noindent
See Lemma \ref{lemma-sequence-upper-bound-new} in the Appendix.
\qed
\\

In the next corollary we relate $\check{\mathcal{G}}_{\bold{j}_*\,;\,i,i}(z)$ to the element $\tilde{X}^{(\gamma,\delta)}_i$ of the sequence defined in Lemma \ref{lemma-sequence-2}.
\begin{corollary}
Let $z= E^{Bog}_{\bold{j}_*}+ (\delta-1)\phi_{\bold{j}^*}\sqrt{\epsilon_{\bold{j}_*}^2+2\epsilon_{\bold{j}_*}}$ with $1+(\frac{2\sqrt{2}+3}{6})\sqrt{\epsilon_{\bold{j}_*}}\leq \delta\leq 1+\sqrt{\epsilon_{\bold{j}_*}}$. For some $0<\gamma<1$, consider a constant  $c_{\gamma}$  sufficiently large to ensure the inequality in (\ref{ineq-W}) for $i\geq N-N^{1-\gamma}$. For the chosen $c_{\gamma}$ consider the sequence in (\ref{sequence})
and assume the condition in (\ref{gamma-condition}) with $\epsilon_{\bold{j}_*}$ sufficiently small. 
Then,  ,
the inequality
\begin{equation}\label{fixed-p-ineq}
\check{\mathcal{G}}_{\bold{j}_*\,;\,i,i}(z)\geq \frac{1}{\tilde{X}^{(\gamma,\delta)}_i}\quad,\,\, N-N^{1-\gamma}=:i_0 \leq i\leq N-2 \quad (\text{with}\,\,\epsilon\equiv \epsilon_{\bold{j}_{*}})\,,
\end{equation}
holds true.
\end{corollary}

\noindent
\emph{Proof}

We observe that the condition in (\ref{gamma-condition})  implies that $\frac{1}{N}\leq \epsilon^{\nu}_{\bold{j}_*}$ for some $\nu>\frac{11}{8}$ and, due to Theorem \ref{theorem-Bog}, the functions $\check{\mathcal{G}}_{\bold{j}_*\,;\,i,i}(z)$ are well defined (recall $0\leq \check{\mathcal{G}}_{\bold{j}_*\,;\,i,i}(z)\leq \|\check{\Gamma}_{\bold{j}_*\,;\,i,i}(z)\|$).  From (\ref{def-G}) and $\check{\mathcal{G}}_{\bold{j}_*\,;\,0,0}(z)\equiv 1$, one can deduce that 
\begin{equation}
\check{\mathcal{G}}_{\bold{j}_*\,;\,N-N^{1-\gamma},N-N^{1-\gamma}}(z)\geq 1= \frac{1}{\tilde{X}^{(\gamma,\delta)}_{N-N^{1-\gamma}}}\,.
\end{equation} Then, by using (\ref{def-G}) and (\ref{ineq-W}), the result follows from an inductive argument.
% analogous to Theorem \ref{theorem-Bog}.
\qed

%by the same inductive argument of Theorem \ref{theorem-Bog} using (\ref{def-G}) and (\ref{ineq-W}).

Next, we prove that there is a (unique) fixed point $z_{*}<E^{Bog}_{\bold{j}_*}+ (\frac{2\sqrt{2}+3}{6})\sqrt{\epsilon_{\bold{j}_*}}\phi_{\bold{j}_*}\sqrt{\epsilon_{\bold{j}_*}^2+2\epsilon_{\bold{j}_*}}$. 
\begin{thm}\label{fixed-p-thm}
Let
\begin{equation}\label{range-z}
z\leq E^{Bog}_{\bold{j}_*}+ \sqrt{\epsilon_{\bold{j}_*}}\phi_{\bold{j}_*}\sqrt{\epsilon_{\bold{j}_*}^2+2\epsilon_{\bold{j}_*}}(<0)\,.
\end{equation}For some $0<\gamma<1$, consider a constant  $c_{\gamma}$ sufficiently large to ensure the inequality in (\ref{ineq-W}) for $i\geq N-N^{1-\gamma}$. For the chosen $\gamma$ and $c_{\gamma}$, consider the sequence defined in (\ref{sequence}) and assume  the condition in (\ref{gamma-condition}) with $\epsilon_{\bold{j}_*}$ sufficiently small. Then $f_{\bold{j}_{*}}(z)$ is well defined and there is only one point $z_*$ such that $f_{\bold{j}_{*}}(z_*)=0$, furthermore 
\begin{equation}
z_*<E^{Bog}_{\bold{j}_*}+(\frac{2\sqrt{2}+3}{6})\sqrt{\epsilon_{\bold{j}_*}}\phi_{\bold{j}_*}\sqrt{\epsilon_{\bold{j}_*}^2+2\epsilon_{\bold{j}_*}}. \label{upper-bound-zstar}
\end{equation}
\end{thm}

\noindent
\emph{Proof}

We observe that the condition in (\ref{gamma-condition})  implies that $\frac{1}{N}\leq \epsilon^{\nu}_{\bold{j}_*}$ for some $\nu>\frac{11}{8}$ and, due to Theorem \ref{theorem-Bog}, the functions $\check{\mathcal{G}}_{\bold{j}_*\,;\,i,i}(z)$ are well defined.
In Section \ref{fixed point}  (see (\ref{fp-equation-2})), we have observed that the fixed point equation $f_{\bold{j}_{*}}(z)=0$ (see (\ref{fp-function})-(\ref{fp-function-rhs}))
%\begin{eqnarray}
%f_{\bold{j}_{*}}(z)=0\quad \Longleftrightarrow \quad 0&=&-z-\langle \eta\,,\,W_{\bold{j}_*}\,Q^{(N-2)}_{\bold{j}_*}\,R^{Bog}_{\bold{j}_*\,;\,N-2,N-2}(z)\sum_{l_{N-2}=0}^{\infty}[\Gamma^{Bog}_{\bold{j}_*\,;\,N-2,N-2}(z) R^{Bog}_{\bold{j}_*\,;\,N-2,N-2}(z)]^{l_{N-2}}\, Q^{(N-2)}_{\bold{j}_*}W^*_{\bold{j}_*}\eta\rangle \nonumber\\
%&=&-z-\frac{\phi_{\bold{j}_{*}}^2}{2(k_{\bold{j}_*})^2+2\phi_{\bold{j}_*}-z}\check{\mathcal{G}}_{\bold{j}_*\,;\,N-2,N-2}(z)
%\end{eqnarray}
can be written
\begin{equation}\label{R-H-S-fixed}
0=-\frac{z}{\phi_{\bold{j}_{*}}}-\frac{1}{2\epsilon_{\bold{j}_*}+2-\frac{z}{\phi_{\bold{j}_{*}}}+\mathcal{O}(\frac{1}{N})}\check{\mathcal{G}}_{\bold{j}_*\,;\,N-2,N-2}(z)\,.
\end{equation}
In the mean field limit the solution  $z_*$ to $(\ref{R-H-S-fixed})$ is expected to be located at $E^{Bog}_{\bold{j}_{*}}$ (see \cite{Se1}), i.e., for $N$ large
\begin{equation}
\frac{z_*}{\phi_{\bold{j}_{*}}}\simeq \frac{E^{Bog}_{\bold{j}_{*}}}{\phi_{\bold{j}_{*}}}=-\Big[\epsilon_{\bold{j}_*}+1-\sqrt{\epsilon_{\bold{j}_*}^2+2\epsilon_{\bold{j}_*}}\,\Big]\,.
\end{equation}
Under the assumptions of Lemma \ref{lemma-sequence-2}, for $\epsilon_{\bold{j}_*}^2+\frac{\epsilon_{\bold{j}_*}}{N^{\gamma}}+\frac{1}{N}\leq k_{\gamma}\epsilon_{\bold{j}_*} \sqrt{\epsilon_{\bold{j}_*}}$ with $\epsilon_{\bold{j}_*}$ and $k_{\gamma}$ sufficiently small, we deduce the bound in  (\ref{bound-seq-gamma}) at $2j=N-2$:
\begin{eqnarray}
\tilde{X}^{(\gamma, \delta)}_{N-2}&\leq &\frac{1}{2}\Big(1+\sqrt{d^{(\gamma)}_{\epsilon_{\bold{j}_*}}+2\epsilon_{\bold{j}_*}}-\frac{1}{3-(1+\epsilon_{\bold{j}_*})\delta\sqrt{\epsilon_{\bold{j}_*}^2+2\epsilon_{\bold{j}_*}}}\Big)\\
&\leq & \frac{1+\frac{3}{2}\sqrt{d^{(\gamma)}_{\epsilon_{\bold{j}_*}}+2\epsilon_{\bold{j}_*}}-\frac{\delta}{2}\sqrt{\epsilon_{\bold{j}_*}^2+2\epsilon_{\bold{j}_*}}}{3-(1+\epsilon_{\bold{j}_*})\delta\sqrt{\epsilon_{\bold{j}_*}^2+2\epsilon_{\bold{j}_*}}}\,
\end{eqnarray}
where $d^{(\gamma)}_{\epsilon_{\bold{j}_*}}:=c_{\gamma}[\epsilon_{\bold{j}_*}^2+\frac{\epsilon}{N^{\gamma}}+\frac{1}{N}]$. 
%Notice that because of the assumptions $0\leq d_{\epsilon_{\bold{j}_*},\rho}\leq C \frac{\epsilon_{\bold{j}_*}\sqrt{\epsilon_{\bold{j}_*}}}{\rho^{\gamma}}$ where $C$ is a universal constant.
Hence,   using (\ref{fixed-p-ineq}),  we can estimate 
\begin{equation}\label{low-G}
\check{\mathcal{G}}_{\bold{j}_*\,;\,N-2,N-2}(z)\geq \frac{1}{\tilde{X}^{(\gamma, \delta)}_{N-2}}\geq\frac{3-(1+\epsilon_{\bold{j}_*})\delta\sqrt{\epsilon_{\bold{j}_*}^2+2\epsilon_{\bold{j}_*}}}{1+\frac{3}{2}\sqrt{d^{(\gamma)}_{\epsilon_{\bold{j}_*}}+2\epsilon_{\bold{j}_*}}-\frac{\delta}{2}\sqrt{\epsilon_{\bold{j}_*}^2+2\epsilon_{\bold{j}_*}}}
\end{equation}
for $z= E^{Bog}_{\bold{j}_*}+ (\delta-1)\phi_{\bold{j}_*}\sqrt{\epsilon_{\bold{j}_*}^2+2\epsilon_{\bold{j}_*}}$ where $1+(\frac{2\sqrt{2}+3}{6})\sqrt{\epsilon_{\bold{j}_*}}\leq \delta \leq 1+\sqrt{\epsilon_{\bold{j}_*}}\,.$

Since $E^{Bog}_{\bold{j}}:=-\Big[k^2_{\bold{j}}+\phi_{\bold{j}}-\sqrt{(k^2_{\bold{j}})^2+2\phi_{\bold{j}}k^2_{\bold{j}}}\Big]$, using (\ref{low-G}) for the considered values of $z$ (i.e., $z= E^{Bog}_{\bold{j}_*}+ (\delta-1)\phi_{\bold{j}_*}\sqrt{\epsilon_{\bold{j}_*}^2+2\epsilon_{\bold{j}_*}}$ where $1+(\frac{2\sqrt{2}+3}{6})\sqrt{\epsilon_{\bold{j}_*}}\leq \delta \leq 1+\sqrt{\epsilon_{\bold{j}_*}}$)  we can write
\begin{eqnarray}
\frac{f_{\bold{j}_{*}}(z)}{\phi_{\bold{j}^*}}&\leq &\Big[\epsilon_{\bold{j}_*}+1-\delta\sqrt{\epsilon_{\bold{j}_*}^2+2\epsilon_{\bold{j}_*}}\,\Big]-\frac{1}{3\epsilon_{\bold{j}_*}+3-\delta \sqrt{\epsilon_{\bold{j}_*}^2+2\epsilon_{\bold{j}_*}}} \frac{1}{X^{(\gamma, \delta)}_{N-2}}+\mathcal{O}(\frac{1}{N})\\
&\leq &\Big[\epsilon_{\bold{j}_*}+1-\delta\sqrt{\epsilon_{\bold{j}_*}^2+2\epsilon_{\bold{j}_*}}\,\Big]-\frac{1}{3\epsilon_{\bold{j}_*}+3-\delta\sqrt{\epsilon_{\bold{j}_*}^2+2\epsilon_{\bold{j}_*}}} \Big\{\frac{3-(1+\epsilon_{\bold{j}_*})\delta\sqrt{\epsilon_{\bold{j}_*}^2+2\epsilon_{\bold{j}_*}}}{1+\frac{3}{2}\sqrt{d^{(\gamma)}_{\epsilon_{\bold{j}_*}}+2\epsilon_{\bold{j}_*}}-\frac{\delta}{2}\sqrt{\epsilon_{\bold{j}_*}^2+2\epsilon_{\bold{j}_*}}}\Big\}+\mathcal{O}(\frac{1}{N})\nonumber
\\
&= &\Big[\epsilon_{\bold{j}_*}+1-\delta\sqrt{\epsilon_{\bold{j}_*}^2+2\epsilon_{\bold{j}_*}}\,\Big]-\frac{3-(1+\epsilon_{\bold{j}_*})\delta\sqrt{\epsilon_{\bold{j}_*}^2+2\epsilon_{\bold{j}_*}}+3\epsilon_{\bold{j}_*}-3\epsilon_{\bold{j}_*}}{3\epsilon_{\bold{j}_*}+3-\delta\sqrt{\epsilon_{\bold{j}_*}^2+2\epsilon_{\bold{j}_*}}}\Big\{ \frac{1}{1+\frac{3}{2}\sqrt{d^{(\gamma)}_{\epsilon_{\bold{j}_*}}+2\epsilon_{\bold{j}_*}}-\frac{\delta}{2}\sqrt{\epsilon_{\bold{j}_*}^2+2\epsilon_{\bold{j}_*}}}\Big\}\nonumber\\
& &+\mathcal{O}(\frac{1}{N})\nonumber\\
&=&\Big[\epsilon_{\bold{j}_*}+1-\delta\sqrt{\epsilon_{\bold{j}_*}^2+2\epsilon_{\bold{j}_*}}\,\Big]-\Big\{1-\frac{\epsilon_{\bold{j}_*}\delta\sqrt{\epsilon_{\bold{j}_*}^2+2\epsilon_{\bold{j}_*}}+3\epsilon_{\bold{j}_*}}{3\epsilon_{\bold{j}_*}+3-\delta\sqrt{\epsilon_{\bold{j}_*}^2+2\epsilon_{\bold{j}_*}}}\Big\} \frac{1}{1+\frac{3}{2}\sqrt{d^{(\gamma)}_{\epsilon_{\bold{j}_*}}+2\epsilon_{\bold{j}_*}}-\frac{\delta}{2}\sqrt{\epsilon_{\bold{j}_*}^2+2\epsilon_{\bold{j}_*}}}+\mathcal{O}(\frac{1}{N})\nonumber\\
&\leq &\Big[\epsilon_{\bold{j}_*}+1-\delta\sqrt{2\epsilon_{\bold{j}_*}}\,\Big]-\Big\{1-\epsilon_{\bold{j}_*}+o(\epsilon_{\bold{j}})\Big\} \Big\{1-\frac{3}{2}\sqrt{d^{(\gamma)}_{\epsilon_{\bold{j}_*}}+2\epsilon_{\bold{j}_*}}+\frac{\delta}{2}\sqrt{\epsilon_{\bold{j}_*}^2+2\epsilon_{\bold{j}_*}}\Big\}+\mathcal{O}(\frac{1}{N})\\
&=&2\epsilon_{\bold{j}_*}+\frac{3-3\delta}{2}\sqrt{2\epsilon_{\bold{j}_*}}+\frac{3}{2}
\sqrt{2\epsilon_{\bold{j}}}
\cdot \mathcal{O}(\frac{d^{(\gamma)}_{\epsilon_{\bold{j}_*}}}{\epsilon_{\bold{j}_*}})
+o(\epsilon_{\bold{j}})+\mathcal{O}(\frac{1}{N})\\
&\leq &2\epsilon_{\bold{j}_*}+\Big[\frac{3-3\delta}{2}\sqrt{2\epsilon_{\bold{j}_*}}\Big|_{\delta=1+(\frac{2\sqrt{2}+3}{6})\sqrt{\epsilon_{\bold{j}_*}}}+\frac{3}{2}
\sqrt{2\epsilon_{\bold{j}}}
\cdot \mathcal{O}(\frac{d^{(\gamma)}_{\epsilon_{\bold{j}_*}}}{\epsilon_{\bold{j}_*}})
+o(\epsilon_{\bold{j}})+\mathcal{O}(\frac{1}{N})\\
%&=&\frac{3-3\delta}{2}\sqrt{\epsilon_{\bold{j}_*}^2+2\epsilon_{\bold{j}_*}}+2\epsilon_{\bold{j}_*}-\frac{(3-\delta)^2}{8}2\epsilon_{\bold{j}_*}+o(\epsilon_{\bold{j}_*})\\
&\leq &[2-(\frac{2\sqrt{2}+3}{2\sqrt{2}})+\mathcal{O}(\frac{d^{(\gamma)}_{\epsilon_{\bold{j}_*}}}{\epsilon_{\bold{j}_*}\sqrt{\epsilon_{\bold{j}_*}}})]\epsilon_{\bold{j}_*}+o(\epsilon_{\bold{j}_*})+\mathcal{O}(\frac{1}{N})\\
&=&[\frac{2\sqrt{2}-3}{2\sqrt{2}}+\mathcal{O}(\frac{d^{(\gamma)}_{\epsilon_{\bold{j}_*}}}{\epsilon_{\bold{j}_*}\sqrt{\epsilon_{\bold{j}_*}}})]\epsilon_{\bold{j}_*}+o(\epsilon_{\bold{j}_*})+\mathcal{O}(\frac{1}{N})\label{finale}
%&=&\Big[\epsilon_{\bold{j}_*}+1-\sqrt{\epsilon_{\bold{j}_*}^2+2\epsilon_{\bold{j}_*}}\,\Big]-\frac{1}{3(1+\frac{2}{3}\epsilon_{\bold{j}_*}-\frac{1}{3}\sqrt{\epsilon_{\bold{j}_*}^2+2\epsilon_{\bold{j}_*}})}\frac{3(1+\frac{\zeta \sqrt{\epsilon}}{3})}{1+\frac{3b_{\epsilon}+\zeta \sqrt{\epsilon}}{2}}
\end{eqnarray}
Due to the assumption in (\ref{gamma-condition}), for $k_{\gamma}$ and $\epsilon_{\bold{j}_*}$ sufficiently small the R-H-S in (\ref{finale}) is strictly negative. Since $f_{\bold{j}_{*}}(z)$ is continuous and decreasing (see Remark \ref{increasing}) and $f_{\bold{j}_{*}}(z)>0$ for $|z|$ sufficiently large, we conclude that for $\epsilon_{\bold{j}_*}$ and $k_{\gamma}$ sufficiently small there is a unique (fixed) point $z_*$ in the range (\ref{range-z}) such that $f_{\bold{j}_{*}}(z_*)=0$ with 
\begin{equation}z_*<E^{Bog}_{\bold{j}_*}+ (\frac{2\sqrt{2}+3}{6})\sqrt{\epsilon_{\bold{j}_*}}\phi_{\bold{j}_*}\sqrt{\epsilon_{\bold{j}_*}^2+2\epsilon_{\bold{j}_*}}\,.\label{bound-z*}\end{equation} 
%We notice that $z_*< E^{Bog}_{\bold{j}_*}+ (\delta-1)\phi_{\bold{j}^*}\sqrt{\epsilon_{\bold{j}_*}^2+2\epsilon_{\bold{j}_*}}$ with $\delta=...$
\qed
\\

We can now justify the last step of the iteration for $z$ fulfilling the constraint in (\ref{range}). 
%We also consider $N$ large enough such that $(1-\frac{\phi_{\bold{j}_{*}}}{\Delta_0}\frac{1}{N})\geq \frac{1}{2}$.
% $$z<\min\,\Big\{ z_{*}+\Delta_0(1-\frac{\phi_{\bold{j}_{*}}}{\Delta_0}\frac{1}{N})\,;\,E^{Bog}_{\bold{j}_*}+ \sqrt{\epsilon_{\bold{j}_*}}\phi_{\bold{j}^*}\sqrt{\epsilon_{\bold{j}_*}^2+2\epsilon_{\bold{j}_*}}\Big\}$$ where  $\Delta_0:=\min \{(k_{\bold{j}})^2\,|\,\bold{j}\in \mathbb{Z}^d\setminus \{\bold{0}\}\}$.
\begin{lemma}\label{inversion}
Assume the condition in (\ref{gamma-condition}) (for some $0<\gamma<1$) and Condition 3.2) in Definition \ref{def-pot}:
\begin{equation}\frac{\phi_{\bold{j}_{*}}}{\Delta_0}\frac{N^{\mu}}{N(N-N^{\mu})}<\frac{1}{2}\quad,\quad  \frac{1}{N^{\mu}}\leq \mathcal{O}((\sqrt{\epsilon_{\bold{j}_*}})^{1+\theta})\,,\label{mu-cond}\end{equation}
 for some $1>\mu>0\,,\,\theta>0$. Consider the remainder $(\ref{shift-2-bis})=\mathcal{O}(\frac{1}{\sqrt{\epsilon_{\bold{j}_*}}}(\frac{1}{1+c\sqrt{\epsilon_{\bold{j}_*}}})^{N^{\mu}})$ (where $c>0$) from  Corollary \ref{shift}. For $\epsilon_{\bold{j}_*}$ sufficiently small such that
\begin{equation} 
\frac{\Delta_0}{2}+(\ref{shift-2-bis})=\frac{\Delta_0}{2}+\mathcal{O}(\frac{1}{\sqrt{\epsilon_{\bold{j}_*}}}(\frac{1}{1+c\sqrt{\epsilon_{\bold{j}_*}}})^{N^{\mu}})>0\label{cond-4.59}
\end{equation} 
 and for \begin{equation}\label{interval-of-def}
z<\min\,\Big\{ z_{*}+\frac{\Delta_0}{2}\,;\,E^{Bog}_{\bold{j}_*}+ \sqrt{\epsilon_{\bold{j}_*}}\phi_{\bold{j}_*}\sqrt{\epsilon_{\bold{j}_*}^2+2\epsilon_{\bold{j}_*}}\Big\} \label{range}
\end{equation}
 the Hamiltonian $\mathscr{K}^{Bog\,(N)}_{\bold{j}_*}(z):=\mathscr{F}^{(N)}(\mathscr{K}^{Bog\,(N-2)}_{\bold{j}_*}(z))$ is well defined and corresponds to $f_{\bold{j}_*}(z)|\eta \rangle \langle \eta |$.
The point $z_*$ belongs to the interval (\ref{range}).
\end{lemma}

\noindent
\emph{Proof}

We observe that $z_*$ belongs to the interval (\ref{range}) due to the inequality in (\ref{bound-z*}):
\begin{equation}
z_*<E^{Bog}_{\bold{j}_*}+ (\frac{2\sqrt{2}+3}{6})\sqrt{\epsilon_{\bold{j}_*}}\phi_{\bold{j}_*}\sqrt{\epsilon_{\bold{j}_*}^2+2\epsilon_{\bold{j}_*}}z_*<E^{Bog}_{\bold{j}_*}+ \sqrt{\epsilon_{\bold{j}_*}}\phi_{\bold{j}_*}\sqrt{\epsilon_{\bold{j}_*}^2+2\epsilon_{\bold{j}_*}}\,.
\end{equation}

\noindent
In order to justify the last step\footnote{With the given assumptions the previous steps are well defined due to Theorem \ref{theorem-Bog}.} of the Feshbach-Schur flow, it is enough  to show that $\overline{\mathscr{P}_{\eta}}\mathscr{K}^{Bog\,(N-2)}_{\bold{j}_*}(z)\overline{\mathscr{P}_{\eta}}$ is bounded invertible on $\overline{\mathscr{P}_{\eta}}\mathcal{F}_N$ because as seen in the preliminary discussion (see (\ref{final-H})) this implies $\mathscr{K}^{Bog\,(N)}_{\bold{j}_*}(z)=f_{\bold{j}_*}(z)|\eta\rangle \langle \eta |$. For $\epsilon_{\bold{j}_*}$ sufficiently small such that condition (\ref{cond-4.59}) holds, we derive
\begin{eqnarray}
& &\overline{\mathscr{P}_{\eta}}\mathscr{K}^{Bog\,(N-2)}_{\bold{j}_*}(z)\overline{\mathscr{P}_{\eta}}\\
&=&\overline{\mathscr{P}_{\eta}}(H^{Bog}_{\bold{j}_*}-z)\overline{\mathscr{P}_{\eta}}\\
& &-\overline{\mathscr{P}_{\eta}}W_{\bold{j}_*}\,R^{Bog}_{\bold{j}_*\,;\,N-2,N-2}(z)\sum_{l_{N-2}=0}^{\infty}[\Gamma^{Bog}_{\bold{j}_*\,;\,N-2,N-2}(z) R^{Bog}_{\bold{j}_*\,;\,N-2,N-2}(z)]^{l_{N-2}}\, W^*_{\bold{j}_*}\overline{\mathscr{P}_{\eta}}\quad\quad\quad \\
&=&\overline{\mathscr{P}_{\eta}}(H^{Bog}_{\bold{j}_*}-z)\overline{\mathscr{P}_{\eta}}\\
& &-\overline{\mathscr{P}_{\eta}}W_{\bold{j}_*}\,(R^{Bog}_{\bold{j}_*\,;\,N-2,N-2}(z))^{\frac{1}{2}}\,\check{\Gamma}^{Bog}_{\bold{j}_*\,;\,N-2,N-2}(z)\, (R^{Bog}_{\bold{j}_*\,;\,N-2,N-2}(z))^{\frac{1}{2}}\, W^*_{\bold{j}_*}\overline{\mathscr{P}_{\eta}}\quad\quad\quad \\
%&\geq &-z-\frac{3\phi_{\bold{j}_{*}}^2}{3(k_{\bold{j}_*})^2+3\phi_{\bold{j}_{*}}-z}\\
&\geq &(\Delta_0-z)\overline{\mathscr{P}_{\eta}}\label{last-step}\\
& &-\overline{\mathscr{P}_{\eta}}W_{\bold{j}_*}\,(R^{Bog}_{\bold{j}_*\,;\,N-2,N-2}(z))^{\frac{1}{2}}\,\check{\Gamma}^{Bog}_{\bold{j}_*\,;\,N-2,N-2}(z)\, (R^{Bog}_{\bold{j}_*\,;\,N-2,N-2}(z))^{\frac{1}{2}}\, W^*_{\bold{j}_*}\overline{\mathscr{P}_{\eta}}\quad\quad\quad\label{laststep-2}\\
%&\geq&\phi_{\bold{j}_{*}}\Big[\epsilon_{\bold{j}_*}+1-\delta \sqrt{\epsilon _{\bold{j}_*} ^2+2\epsilon _{\bold{j}_*}}-\frac{1}{2\epsilon_{\bold{j}_{*}}+3-\delta\sqrt{\epsilon _{\bold{j}_*} ^2+2\epsilon _{\bold{j}_*}}}\frac{1}{1+2b_{\epsilon_{\bold{j}_*}}}\Big]\\
&> &(\frac{\Delta_0}{2}-z_{*}+\mathcal{O}(\frac{1}{\sqrt{\epsilon_{\bold{j}_*}}}(\frac{1}{1+c\sqrt{\epsilon_{\bold{j}_*}}})^{N^{\mu}})
)\overline{\mathscr{P}_{\eta}}\label{last-step-bis}\\
& &-\langle \eta|W_{\bold{j}_*}\,(R^{Bog}_{\bold{j}_*\,;\,N-2,N-2}(z_{*}))^{\frac{1}{2}}\,\check{\Gamma}^{Bog}_{\bold{j}_*\,;\,N-2,N-2}(z_{*})\, (R^{Bog}_{\bold{j}_*\,;\,N-2,N-2}(z_{*}))^{\frac{1}{2}}\, W^*_{\bold{j}_*}|\eta\rangle\,\overline{\mathscr{P}_{\eta}}\quad\quad\label{laststep-1}\\
&=&(\frac{\Delta_0}{2}+\mathcal{O}(\frac{1}{\sqrt{\epsilon_{\bold{j}_*}}}(\frac{1}{1+c\sqrt{\epsilon_{\bold{j}_*}}})^{N^{\mu}}))\overline{\mathscr{P}_{\eta}}\label{last}\,\\
&>&0
\end{eqnarray}
for some $c>0$. The step from  (\ref{last-step})-(\ref{laststep-2}) to (\ref{last-step-bis})-(\ref{laststep-1}) is legitimate because
\begin{itemize}
\item  $\Delta_0-z>\frac{\Delta_0}{2}-z_*$ for $z$ in the range given in (\ref{range});
\item  $\overline{\mathscr{P}_{\eta}}$ projects onto a subspace of vectors with no particles in the modes $\pm\bold{j}_{*}$ and orthogonal to $\eta$. Hence, as it is proven in Corollary \ref{shift},
\begin{eqnarray}
& &\|\overline{\mathscr{P}_{\eta}}W_{\bold{j}_*}\,(R^{Bog}_{\bold{j}_*\,;\,N-2,N-2}(z))^{\frac{1}{2}}\,\check{\Gamma}^{Bog}_{\bold{j}_*\,;\,N-2,N-2}(z)\, (R^{Bog}_{\bold{j}_*\,;\,N-2,N-2}(z))^{\frac{1}{2}}\, W^*_{\bold{j}_*}\overline{\mathscr{P}_{\eta}}\|\quad  \label{shift-1}\\
&\leq &\|\mathscr{P}_{\eta}W_{\bold{j}_*}\,(R^{Bog}_{\bold{j}_*\,;\,N-2,N-2}(z-\frac{\Delta_0}{2})^{\frac{1}{2}}\,\check{\Gamma}^{Bog}_{\bold{j}_*\,;\,N-2,N-2}(z-\frac{\Delta_0}{2})\times \quad \quad\quad\\
& &\quad\quad\quad \times (R^{Bog}_{\bold{j}_*\,;\,N-2,N-2}(z-\frac{\Delta_0}{2})^{\frac{1}{2}}\, W^*_{\bold{j}_*}\mathscr{P}_{\eta}\| \nonumber\\
& & +\mathcal{O}(\frac{1}{\sqrt{\epsilon_{\bold{j}_*}}}(\frac{1}{1+c\sqrt{\epsilon_{\bold{j}_*}}})^{N^{\mu}})
\,\quad\quad\quad \label{shift-2}
\end{eqnarray}
holds true if the condition in (\ref{mu-cond}) is satisfied.
The argument in Corollary \ref{shift} makes use of the re-expansion of $\check{\Gamma}^{Bog}_{\bold{j}_*\,;\,N-2,N-2}(z)$ which is the content of  Proposition \ref{lemma-expansion-proof-0};
\item $z-\frac{\Delta_0}{2}<z_*$ for $z$ in the range given in (\ref{range}) and
\begin{eqnarray}
& &\|\mathscr{P}_{\eta}W_{\bold{j}_*}\,(R^{Bog}_{\bold{j}_*\,;\,N-2,N-2}(w))^{\frac{1}{2}}\,\check{\Gamma}^{Bog}_{\bold{j}_*\,;\,N-2,N-2}(w)\, (R^{Bog}_{\bold{j}_*\,;\,N-2,N-2}(w))^{\frac{1}{2}}\,W^*_{\bold{j}_*}\mathscr{P}_{\eta}\|\quad\\
&=&\langle \eta \,,\, W_{\bold{j}_*}\,(R^{Bog}_{\bold{j}_*\,;\,N-2,N-2}(w))^{\frac{1}{2}}\,\check{\Gamma}^{Bog}_{\bold{j}_*\,;\,N-2,N-2}(w)\, (R^{Bog}_{\bold{j}_*\,;\,N-2,N-2}(w))^{\frac{1}{2}}\,W^*_{\bold{j}_*} \eta \rangle
\end{eqnarray}
is nondecreasing for $w\leq E^{Bog}_{\bold{j}_*}+ \sqrt{\epsilon_{\bold{j}_*}}\phi_{\bold{j}_*}\sqrt{\epsilon_{\bold{j}_*}^2+2\epsilon_{\bold{j}_*}}$ (see Remark \ref{increasing}), therefore (recall that $z_*$ belongs to the interval (\ref{range}))
\begin{eqnarray}
& &-\|\mathscr{P}_{\eta}W_{\bold{j}_*}\,(R^{Bog}_{\bold{j}_*\,;\,N-2,N-2}(z-\frac{\Delta_0}{2}))^{\frac{1}{2}}\,\check{\Gamma}^{Bog}_{\bold{j}_*\,;\,N-2,N-2}(z-\frac{\Delta_0}{2})\, (R^{Bog}_{\bold{j}_*\,;\,N-2,N-2}(z-\frac{\Delta_0}{2}))^{\frac{1}{2}}\,W^*_{\bold{j}_*}\mathscr{P}_{\eta}\|\quad\quad\quad\\
&\geq &-\langle \eta|W_{\bold{j}_*}\,(R^{Bog}_{\bold{j}_*\,;\,N-2,N-2}(z_{*}))^{\frac{1}{2}}\,\check{\Gamma}^{Bog}_{\bold{j}_*\,;\,N-2,N-2}(z_{*})\, (R^{Bog}_{\bold{j}_*\,;\,N-2,N-2}(z_{*}))^{\frac{1}{2}}\, W^*_{\bold{j}_*}|\eta\rangle\,.
\end{eqnarray}
\end{itemize}
In the step from (\ref{last-step-bis})-(\ref{laststep-1}) to (\ref{last})
we use the identity $f_{\bold{j}_*}(z_*)=0$, 
\begin{equation}
0=-z_*-\langle \eta|W_{\bold{j}_*}\,(R^{Bog}_{\bold{j}_*\,;\,N-2,N-2}(z_{*}))^{\frac{1}{2}}\,\check{\Gamma}^{Bog}_{\bold{j}_*\,;\,N-2,N-2}(z_{*})\, (R^{Bog}_{\bold{j}_*\,;\,N-2,N-2}(z_{*}))^{\frac{1}{2}}\, W^*_{\bold{j}_*}|\eta\rangle\,\,.
\end{equation}
\qed

\begin{remark} \label{cond-gamma}
For any dimension $d\geq 1$, condition (\ref{gamma-condition}) (for any $0<\gamma<1$) and condition (\ref{mu-cond}) can be fulfilled  in the mean field limiting regime.  

\noindent
At fixed particle density $\rho$ and  for $d\geq 2$, condition (\ref{mu-cond}) can be fulfilled  if $L$ is sufficiently large. 

\noindent
Concerning the condition in (\ref{gamma-condition}) when the size of the box tends to infinity, it yields a relation between the particle density and the size of the box. For example, choosing $\gamma=\frac{1}{3}$ the density $\rho$ must scale like $L^{3-d}$.  

\noindent
To conclude, we point out that by choosing $\gamma=\frac{1}{3}$, $\mu=\frac{2}{3}$ and $\rho$ sufficiently large but independent of $L(>1)$, the conditions in (\ref{gamma-condition}) and (\ref{mu-cond}) are satisfied for $d\geq 3$ and arbitrarily  large $1<L<\infty$, i.e, in the thermodynamic limit. 
\end{remark}
\noindent
%By the expansion of 
%\begin{equation}
%R^{Bog}_{N-2,N-2}(z)\sum_{l_{N-2}=0}^{\infty}[\Gamma^{Bog}_{N-2,N-2}(z) R^{Bog}_{N-2,N-2}(z)]^{l_{N-2}}\
%\end{equation}
%it will be proven (--to be done, but see also my comment in Remark \ref{expansions}--) that for $z<\xi^{Bog} E^{Bog}$ there exists a unique $z_*$ such that
%\begin{eqnarray}
%z_*&=&-\langle \eta\,,\,W\,Q^{(N-2)}\,R^{Bog}_{N-2,N-2}(z_*)\sum_{l_{N-2}=0}^{\infty}[\Gamma^{Bog}_{N-2,N-2}(z_*)R^{Bog}_{N-2,N-2}(z_*)]^{l_{N-2}}\, Q^{(N-2)}W^*\eta\rangle \quad \label{fixed-point}
%&&-\langle \eta\,,\,\mathscr{V}^{(N-1)}(z_*)\,R_{N-1,N-1}(z_*)\sum_{l_{N-1}=0}^{\infty}[\Gamma_{N-1,N-1}(z_*) R_{N-1,N-1}(z_*)]^{l_{N-1}}\,(\mathscr{V}^{(N-1)}(z_*) )^*\eta\rangle\quad\quad
%\end{eqnarray}
%and that (as we already know from Seiringer's paper)  $z_*=E^{Bog}+o(1)$.
%Then, the condition in (\ref{condition}) is fulfilled for $\xi$ sufficiently close to $1$ and $\mathscr{F}_{\eta}(\mathscr{K}^{Bog\,(N-1)}(z))(z_*)$ is well defined.
%Moreover, because of its uniqueness we can conclude that $z_*$  is the ground state energy of $\mathscr{F}_{\eta}(\mathscr{K}^{Bog\,(N-1)}(z))(z_*)$.

\subsection{Isospectrality and construction of the ground state vector}
The isospectrality property  (see the comment after Theorem \ref{theorem-Bog}) holds up to the last step. Hence, if $\mathscr{K}^{Bog\,(N)}_{\bold{j}_*}(z_*)\eta=0$ then also the Hamiltonian $\mathscr{K}^{Bog\,(N-2)}_{\bold{j}_*}(z_*)$ has eigenvalue zero and the corresponding eigenvector is
\begin{equation}
\Big[\mathscr{P}_{\eta}-\frac{1}{\overline{\mathscr{P}_{\eta}}\mathscr{K}^{Bog\,(N-2)}_{\bold{j}_*}(z_*)\overline{\mathscr{P}_{\eta}}}\overline{\mathscr{P}_{\eta}}\mathscr{K}^{Bog\,(N-2)}_{\bold{j}_*}(z_*)\mathscr{P}_{\eta}\Big]\eta\,\equiv\eta\,.
\end{equation}%As from the previous section we know that zero is the ground state energy of $\mathscr{F}_{\eta}(\mathscr{K}^{Bog\,(N-1)}_{\bold{j}_*}(z_*))$, we can conclude that zero is also the ground state energy of $\mathscr{K}^{Bog\,(N-2)}_{\bold{j}_*}(z_*)$. 
Furthermore, since $\mathscr{K}^{Bog\,(N)}_{\bold{j}_*}(z)$ is bounded invertible for $z<z_*$ so $\mathscr{K}^{Bog\,(N-2)}_{\bold{j}_*}(z)$ is.

Iterating this isospectrality argument, we get that
%Furthermore, if zero is the ground state energy of $\mathscr{K}^{Bog\,(N-2)}_{\bold{j}_*}(z_*)$ then zero is also the ground state energy of $\mathscr{K}^{Bog\,(N-3)}_{\bold{j}_*}(z_*)$ and the corresponding ground state eigenvector is
%\begin{eqnarray}
%& &\Big[Q^{(>N-2)}_{\bold{j}_*}-\frac{1}{Q^{(N-2)}_{\bold{j}_*}\mathscr{K}^{Bog\,(N-3)}_{\bold{j}_*}(z_*)Q^{(N-2)}_{\bold{j}_*}}Q^{(N-2)}_{\bold{j}_*}\mathscr{K}^{Bog\,(N-3)}_{\bold{j}_*}(z_*)Q^{(>N-2)}_{\bold{j}_*}\Big]\times \quad \quad \\
%& &\quad\quad\quad\times \Big[Q^{(>N-1)}_{\bold{j}_*}-\frac{1}{Q^{(N-1)}_{\bold{j}_*}\mathscr{K}^{Bog\,(N-2)}_{\bold{j}_*}(z_*)Q^{(N-1)}_{\bold{j}_*}}Q^{(N-1)}_{\bold{j}_*}\mathscr{K}^{Bog\,(N-2)}_{\bold{j}_*}(z_*)Q^{(>N-1)}_{\bold{j}_*}\Big]\eta\,.
%\end{eqnarray}
$H^{Bog}_{\bold{j}_*}-z_*$ has ground state energy zero, i.e., $H^{Bog}_{\bold{j}_*}$ has ground state energy $z_*$, and the corresponding eigenvector is
\begin{eqnarray}
& &\psi^{Bog}_{\bold{j}_*}\\
&:=&\Big[Q^{(>1)}_{\bold{j}_*}-\frac{1}{Q^{(0,1)}_{\bold{j}_*}(H^{Bog}_{\bold{j}_*}-z_*)Q^{(0,1)}_{\bold{j}_*}}Q^{(0,1)}_{\bold{j}_*}(H^{Bog}_{\bold{j}_*}-z_*)Q^{(>1)}_{\bold{j}_*}\Big]\times \label{gs-vector}\\
& &\quad\quad\quad\times \Big\{ \prod_{i=0\,,\, i\,\,\text{even}}^{N-4}\Big[Q^{(>i+3)}_{\bold{j}_*}-\frac{1}{Q^{(i+2,i+3)}_{\bold{j}_*}\mathscr{K}^{Bog\,(i)}_{\bold{j}_*}(z_*)Q^{(i+2,i+3)}_{\bold{j}_*}}Q^{(i+2,i+3)}_{\bold{j}_*}\mathscr{K}^{Bog\,(i)}_{\bold{j}_*}(z_*)Q^{(>i+3)}_{\bold{j}_*}\Big]\Big\}\eta \,.\nonumber
\end{eqnarray}

In the next corollary we collect the results that hold for $N$ and $\epsilon_{\bold{j}_*}>0$ fulfilling the assumptions of Theorem \ref{fixed-p-thm} and Lemma \ref{inversion}.  In addition, we include the result proven in Lemma  \ref{eigenvalue}.
\begin{corollary}\label{col-hbog}
%Assume the conditions in (\ref{gamma-condition}), (\ref{mu-cond}),  and 
Let $\epsilon_{\bold{j}_*}$ be sufficiently small. Then, the following properties hold true:

\noindent
a) In the mean field limiting regime for any space  dimension $d\geq 1$,  the (nondegenerate) ground state energy (of  $H^{Bog}_{\bold{j}_*}$) is $z_*$  and approaches $E^{Bog}_{\bold{j}_*}$ as $N=\rho L^d\to \infty$. In this limit,  the spectral gap above $z_*$ is not smaller than
\begin{equation}
\frac{\Delta_0}{2}\,.\label{spectral-gap-1}
\end{equation}
b) In dimension $d\geq 3$, at fixed (but large) $\rho$ and arbitrarily large $1<L<\infty$, the nondegenerate ground state energy (of $H^{Bog}_{\bold{j}_*}$) is $z_*$  and the spectral gap above it can be estimated not smaller than
\begin{equation}
\min\,\Big\{\frac{\Delta_0}{2}\,;\,  (\frac{-2\sqrt{2}+3}{6})\sqrt{\epsilon_{\bold{j}_*}}\phi_{\bold{j}_*}\sqrt{\epsilon_{\bold{j}_*}^2+2\epsilon_{\bold{j}_*}}\Big\}\,.\label{spectral-gap-2}
\end{equation}
Lemma  \ref{eigenvalue} implies that for $d\geq 4$ and at fixed $\rho$  the ground state energy $z_*$ tends to $E^{Bog}_{\bold{j}_*}$ as $N=\rho L^d \to \infty$. In the case $d=3$, at fixed (large) $\rho$ it follows from  the argument outlined in Remark \ref{EBoglimit}. 
%and $(1-\frac{\phi_{\bold{j}_{*}}N^{\mu}}{\Delta_0}\frac{1}{N})> \frac{1}{2}$.
%Then, for \begin{equation}
%z\leq E^{Bog}_{\bold{j}_*}+ (\delta-1)\phi_{\bold{j}^*}\sqrt{\epsilon_{\bold{j}_*}^2+2\epsilon_{\bold{j}_*}}\,,
%\end{equation} with $\delta=1+\sqrt{\epsilon_{\bold{j}_*}}$, there is only one point $z_*$ such that $f_{\bold{j}_{*}}(z_*)=0$.  

\noindent
c) Whenever the ground state energy $z_*$ exists,  the corresponding eigenvector is
\begin{eqnarray}
& &\psi^{Bog}_{\bold{j}_*}\\
&=&\eta \label{gs-1}\\
& &-\frac{1}{Q^{(N-2,N-1)}_{\bold{j}_*}\mathscr{K}^{Bog\,(N-4)}_{\bold{j}_*}(z_*)Q^{(N-2,N-1)}_{\bold{j}_*}}Q^{(N-2,N-1)}_{\bold{j}_*}W^*_{\bold{j}_*}\eta \label{gs-1-bis}\\
& &-\sum_{j=2}^{N/2}\Big\{\prod^{2}_{r=j}\Big[-\frac{1}{Q^{(N-2r,N-2r+1)}_{\bold{j}_*}\mathscr{K}^{Bog\,(N-2r-2)}_{\bold{j}_*}(z_*)Q^{(N-2r,N-2r+1)}_{\bold{j}_*}}W^*_{\bold{j}_*\,;\,N-2r,N-2r+2}\Big]\Big\}\times \quad\quad\quad \label{gs-2}\\
& &\quad\quad\quad\quad\quad \quad\quad\times \frac{1}{Q^{(N-2,N-1)}_{\bold{j}_*}\mathscr{K}^{Bog\,(N-4)}_{\bold{j}_*}(z_*)Q^{(N-2,N-1)}_{\bold{j}_*}}Q^{(N-2, N-1)}_{\bold{j}_*}W^*_{\bold{j}_*}\eta
\nonumber
\end{eqnarray}
where $\mathscr{K}^{Bog\,(-2)}_{\bold{j}_*}(z_*):=H^{Bog}_{\bold{j}_*}-z_*$.
\end{corollary}

\noindent
\emph{Proof}

As pointed out in Remark \ref{cond-gamma}, in both cases a) and b) the assumptions of Theorem \ref{fixed-p-thm} and  Lemma \ref{inversion} can be satisfied. 
%for $\epsilon_{\bold{j}_*}$ sufficiently small.. 
The existence and uniqueness of the fixed point $z_*$ has been established in Theorem \ref{fixed-p-thm}. Lemma \ref{inversion} implies that $\mathscr{K}^{Bog\,(N)}_{\bold{j}_*}(z)$ is well defined for $z$ in the interval (\ref{interval-of-def}) and $\mathscr{K}^{Bog\,(N)}_{\bold{j}_*}(z)=f_{\bold{j}_*}(z)|\eta\rangle \langle \eta |$.  From the isospectrality  property of the Feshbach-Schur map and from $f_{\bold{j}_*}(z) \neq 0$ for $z<z_*$ we derive that the Hamiltonian $H^{Bog}_{\bold{j}_*}$ has nondegenerate ground state energy $z_*$ with the corresponding eigenvector given by the formula in (\ref{gs-1})-(\ref{gs-2}). In Lemma \ref{eigenvalue} we prove that $z_*\to E^{Bog}_{\bold{j}_*}$ as $N\to \infty$  in the mean field limit. The same result holds  at fixed  $\rho$ if $d\geq 4$. For the convergence at fixed (large) $\rho$ and $d=3$ see Remark \ref{EBoglimit}.
%{\color{red} Estimates (\ref{spectral-gap-1}) and (\ref{spectral-gap-2}) of the spectral gap above the ground state energy follows from: 1) the uniqueness of the fixed point $z_*$ in the given interval of $z$ (see (\ref{range})) where the final Feshbach Hamiltonian is defined; 2)  the bound in (\ref{bound-z*}) and Lemma \ref{eigenvalue}; 3) the isospectrality of the Feshbach map. }

Concerning the estimate of the spectral gap above the ground state energy we recall that  the fixed point $z_*$ is the only fixed point in the interval given in (\ref{range}). Indeed, $z_*$ belongs to  (\ref{range}) and $f_{\bold{j}_*}(z)$ is decreasing in this interval (see Remark \ref{increasing}). Hence, due to the isospectrality of the Feshbach-Schur map we can estimate the spectral gap larger than
\begin{equation}\label{diff-gap}
\min\,\Big\{ z_{*}+\frac{\Delta_0}{2}\,;\,E^{Bog}_{\bold{j}_*}+ \sqrt{\epsilon_{\bold{j}_*}}\phi_{\bold{j}_*}\sqrt{\epsilon_{\bold{j}_*}^2+2\epsilon_{\bold{j}_*}}\Big\}-z_*\,.
\end{equation}
We recall that in the mean field limit  the ground state energy $z_*$ tends to $E^{Bog}_{\bold{j}_*}$ as $N=\rho L^d \to \infty$. Therefore, in this limit the difference in (\ref{diff-gap}) tends to $\frac{\Delta_0}{2}$ because $\frac{\Delta_0}{2}< \sqrt{2}k_{\bold{j}_*}^2<\sqrt{\epsilon_{\bold{j}_*}}\phi_{\bold{j}_*}\sqrt{\epsilon_{\bold{j}_*}^2+2\epsilon_{\bold{j}_*}}$. 

In dimension $d\geq 3$ and at fixed (large) $\rho$ we just exploit the upper bound in (\ref{bound-z*}) and estimate
\begin{eqnarray}
& &\min\,\Big\{ z_{*}+\frac{\Delta_0}{2}\,;\,E^{Bog}_{\bold{j}_*}+ \sqrt{\epsilon_{\bold{j}_*}}\phi_{\bold{j}_*}\sqrt{\epsilon_{\bold{j}_*}^2+2\epsilon_{\bold{j}_*}}\Big\}-z_*\\
&\geq & \min\,\Big\{\frac{\Delta_0}{2}\,;\, \sqrt{\epsilon_{\bold{j}_*}}\phi_{\bold{j}^*}\sqrt{\epsilon_{\bold{j}_*}^2+2\epsilon_{\bold{j}_*}}- (\frac{2\sqrt{2}+3}{6})\sqrt{\epsilon_{\bold{j}_*}}\phi_{\bold{j}_*}\sqrt{\epsilon_{\bold{j}_*}^2+2\epsilon_{\bold{j}_*}}\Big\}\\
&=&\min\,\Big\{\frac{\Delta_0}{2}\,;\,  (\frac{-2\sqrt{2}+3}{6})\sqrt{\epsilon_{\bold{j}_*}}\phi_{\bold{j}_*}\sqrt{\epsilon_{\bold{j}_*}^2+2\epsilon_{\bold{j}_*}}\Big\}\,.
\end{eqnarray}

Using the selection rules of $W_{\bold{j}_*}$ and $W^*_{\bold{j}_*}$,  it is straightforward to check that the expression in (\ref{gs-vector}) corresponds to the sum in (\ref{gs-1})-(\ref{gs-2}). Now, we show how to control the expansion of the ground state.  It is not difficult to see that for any $N$
\begin{eqnarray}
%& &\Big\|-\frac{1}{Q^{(N-2)}\mathscr{K}^{Bog\,(N-3)}(z_*)Q^{(N-2)}}Q^{(N-2)}W^*\eta\Big\| \\
& &\sum_{j=2}^{N/2}\Big\|\,\Big\{\prod^{2}_{r=j}\Big[-\frac{1}{Q^{(N-2r, N-2r+1)}_{\bold{j}_*}\mathscr{K}^{Bog\,(N-2r-2)}_{\bold{j}_*}(z_*)Q^{(N-2r,N-2r+1)}_{\bold{j}_*}}W^*_{N-2r,N-2r+2}\,\Big]\Big\}\times \quad\quad \\
& &\quad\quad\quad\quad\quad\quad\times \frac{1}{Q^{(N-2,N-1)}_{\bold{j}_*}\mathscr{K}^{Bog\,(N-4)}_{\bold{j}_*}(z_*)Q^{(N-2,N-1)}_{\bold{j}_*}}Q^{(N-2,N-1)}_{\bold{j}_*}W^*_{\bold{j}_*}\eta \Big\| \nonumber
\end{eqnarray}
is bounded by a  series which is convergent for   $\epsilon_{\bold{j}_{*}}>0$ sufficiently small. Indeed,  using the identity in (\ref{sandwich-1})-(\ref{sandwich-2}) we have 
\begin{eqnarray}
& &\frac{1}{Q^{(N-2r,N-2r+1)}_{\bold{j}_*}\mathscr{K}^{Bog\,(N-2r-2)}_{\bold{j}_*}(z_*)Q^{(N-2r,N-2r+1)}_{\bold{j}_*}}\label{sandwich-1bis}\\
&=&\sum_{l_{N-2r}=0}^{\infty}R^{Bog}_{\bold{j}_*\,;\,N-2r,N-2r}(z_*)\Big[\Gamma^{Bog\,}_{\bold{j}_*\,;\,N-2r,N-2r}(z_*)\,R^{Bog}_{\bold{j}_*\,;\,N-2r,N-2r}(z_*)\Big]^{l_{N-2r}}\\
&=&[R^{Bog}_{\bold{j}_*\,;\,N-2r,N-2r}(z_*)]^{\frac{1}{2}}\sum_{l_{N-2r}=0}^{\infty}\Big[[R^{Bog}_{N-2r,N-2r}(z_*)]^{\frac{1}{2}}\Gamma^{Bog\,}_{\bold{j}_*\,;\,N-2r,N-2r}(z_*)[R^{Bog}_{\bold{j}_*\,;\,N-2r,N-2r}(z_*)]^{\frac{1}{2}}\Big]^{l_{N-2r}}\times\quad\quad\quad \label{sandwich-3}\\
& &\quad\quad\quad\quad \times [R^{Bog}_{\bold{j}_*\,;\,N-2r,N-2r}(z_*)]^{\frac{1}{2}} \nonumber\\
&=&[R^{Bog}_{\bold{j}_*\,;\,N-2r,N-2r}(z_*)]^{\frac{1}{2}}\check{\Gamma}^{Bog\,}_{\bold{j}_*\,;\,N-2r,N-2r}(z_*)[R^{Bog}_{\bold{j}_*\,;\,N-2r,N-2r}(z_*)]^{\frac{1}{2}}\label{sandwich-4}
\end{eqnarray}
where in the step from (\ref{sandwich-3}) to (\ref{sandwich-4}) we have used the definition in (\ref{gamma-check-a}).
Therefore, we can write
\begin{eqnarray}
& &\Big\{\prod^{2}_{l=j}\Big[-\frac{1}{Q^{(N-2l,N-2l+1)}_{\bold{j}_*}\mathscr{K}^{Bog\,(N-2l-2)}_{\bold{j}_*}(z_*)Q^{(N-2l,N-2l+1)}_{\bold{j}_*}}W^*_{\bold{j}_*\,;\,N-2l,N-2l+2}\Big]\Big\}\times\\
& &\quad\quad\quad\quad\quad \times \frac{1}{Q^{(N-2, N-1)}_{\bold{j}_*}\mathscr{K}^{Bog\,(N-4)}(z_*)Q^{(N-2, N-1)}_{\bold{j}_*}}Q^{(N-2, N-1)}_{\bold{j}_*}W_{\bold{j}_*}^*\eta  \nonumber\\
&=&\Big\{\prod^{2}_{l=j}\Big[-[R^{Bog}_{\bold{j}_*\,;\,N-2l,N-2l}(z_*)]^{\frac{1}{2}}\check{\Gamma}^{Bog\,}_{N-2l,N-2l}(z_*)[R^{Bog}_{\bold{j}_*\,;\,N-2l,N-2l}(z_*)]^{\frac{1}{2}}W^*_{\bold{j}_*\,;\,N-2l,N-2l+2}\Big]\Big\}\times \quad\quad\quad\quad\quad \\
& &\quad\quad\quad \times [R^{Bog}_{\bold{j}_*\,;\,N-2,N-2}(z_*)]^{\frac{1}{2}}\check{\Gamma}^{Bog\,}_{N-2,N-2}(z_*)[R^{Bog}_{\bold{j}_*\,;\,N-2,N-2}(z_*)]^{\frac{1}{2}}W_{\bold{j}_*}^*\eta\,. \nonumber
\end{eqnarray}
Hence, we estimate
\begin{eqnarray}
%& &\Big\|\Big\{\prod^{2}_{i=j}\Big[-\frac{1}{Q^{(N-2i)}_{\bold{j}_*}\mathscr{K}^{Bog\,(N-2i+1)}_{\bold{j}_*}(z_*)Q^{(N-2i)}}W^*_{\bold%{j}_*\,;\,N-2i,N-2i+2}\Big]\Big\}\frac{1}{Q^{(N-2)}_{\bold{j}_*}\mathscr{K}^{Bog\,(N-3)}_{\bold{j}_*}(z_*)Q^{(N-2)}_{\bold{j}_*}}Q^%{(N-2)}_{\bold{j}_*}W_{\bold{j}_*}^*\eta\Big\|  \nonumber\\
& &\Big\|\Big\{\prod^{2}_{l=j}\Big[-\frac{1}{Q^{(N-2l, N-2l+1)}_{\bold{j}_*}\mathscr{K}^{Bog\,(N-2l-2)}_{\bold{j}_*}(z_*)Q^{(N-2l,N-2l+1)}_{\bold{j}_*}}W^*_{\bold{j}_*\,;\,N-2l,N-2l+2}\Big]\Big\}\times \\
& &\quad\quad\quad\quad\quad\quad\times \frac{1}{Q^{(N-2,N-1)}_{\bold{j}_*}\mathscr{K}^{Bog\,(N-4)}_{\bold{j}_*}(z_*)Q^{(N-2,N-1)}_{\bold{j}_*}}Q^{(N-2, N-1)}_{\bold{j}_*}W^*_{\bold{j}_*}\eta \Big\| \nonumber\\
&\leq&\Big\|[R^{Bog}_{\bold{j}_*\,;\,N-2j,N-2j}(z_*)]^{\frac{1}{2}}\Big\|\times \nonumber \\
& &\quad\quad\times\Big\{\prod^{2}_{l=j}\,\Big\|\check{\Gamma}^{Bog\,}_{\bold{j}_*\,;\,N-2l,N-2l}(z_*)\Big\|\,\Big\|[R^{Bog}_{\bold{j}_*\,;\,N-2l,N-2l}(z_*)]^{\frac{1}{2}}W^*_{\bold{j}_*\,;\,N-2l,N-2l+2}[R^{Bog}_{\bold{j}_*\,;\,N-2l+2,N-2l+2}(z_*)]^{\frac{1}{2}}\Big\|\Big\}\times \quad\quad\quad \nonumber\\
& &\quad\quad \times\,\Big\|\check{\Gamma}^{Bog\,}_{\bold{j}_*\,;\,N-2,N-2}(z_*)\Big\|\,\Big\|[R^{Bog}_{\bold{j}_*\,;\,N-2,N-2}(z_*)]^{\frac{1}{2}}Q^{(N-2,N-1)}_{\bold{j}_*}W^*_{\bold{j}_*}\eta \Big\|\,. \nonumber
\end{eqnarray}
Next, we invoke  the estimate
\begin{eqnarray}\label{key-estimate}
&  &\Big\|[R^{Bog}_{\bold{j}_*\,;\,N-2l,N-2l}(z_*)]^{\frac{1}{2}}W^*_{\bold{j}_*\,;\,N-2l,N-2l+2}[R^{Bog}_{\bold{j}_*\,;\,N-2l+2,N-2l+2}(z_*)]^{\frac{1}{2}}\Big\|\\
%&=&\Big\|[R^{Bog}_{\bold{j}_*\,;\,N-2l,N-2l}(z_*)]^{\frac{1}{2}}\Big\|\,\Big\|W_{\bold{j}_*\,;\,N-2l+2,N-2l}[R^{Bog}_{\bold{j}_*\,;\,N-2l,N-2l}(z_*)]W^*_{\bold{j}_*\,;\,N-2l,N-2l+2}\Big\|^{\frac{1}{2}}\\
&\leq&\frac{1}{2\Big[1+a_{\epsilon_{\bold{j}_*}}-\frac{2b_{\epsilon_{\bold{j}_*}}}{2l-1}-\frac{1-c_{\epsilon_{\bold{j}_*}}}{(2l-1)^2}]^{\frac{1}{2}}}
%\sqrt{(\ref{def-deltabog})}
\end{eqnarray}
from Lemma \ref{main-lemma-Bog}; recall (\ref{doppia}) and the definition of $b_{\epsilon_{\bold{j}_*}}$ and $c_{\epsilon_{\bold{j}_*}}$ in (\ref{bdelta}) and (\ref{cdelta}), respectively. In addition, from (\ref{Gamma-ineq}) and (\ref{x-ineq-1}) we know that 
%for $z\leq \xi E^{Bog}$
%\[
%???x_{2j}\geq\frac{1}{2}\Big[1+b_{\epsilon}-\frac{1}{N-2j+1-\xi}\Big]\,;???
%\]
\begin{equation}
\|\check{\Gamma}^{Bog\,}_{\bold{j}_*\,;\,N-2l,N-2l}(z_*)\|\leq  \frac{2}{\Big[1+\sqrt{\eta a_{\epsilon_{\bold{j}_*}}}-\frac{b_{\epsilon_{\bold{j}_*}}/ \sqrt{\eta a_{\epsilon_{\bold{j}_*}}}}{2l-\epsilon^{\Theta}_{\bold{j}_*}}\Big]}\,,
\end{equation}
and, using arguments like in  Lemma \ref{main-lemma-Bog},
\begin{equation}
\Big\|[R^{Bog}_{\bold{j}_*\,;\,N-2j,N-2j}(z_*)]^{\frac{1}{2}}\Big\|\Big\|[R^{Bog}_{\bold{j}_*\,;\,N-2,N-2}(z_*)]^{\frac{1}{2}}Q^{(N-2,N-1)}_{\bold{j}_*}W^*_{\bold{j}_*}\eta \Big\|\leq \mathcal{O}(1)\,.
\end{equation}
With the same ingredients we get $\|(\ref{gs-1-bis})\|\leq \mathcal{O}(1)$.

Combining these estimates, we conclude that the sum
\begin{eqnarray}
& &\sum_{j=2}^{N/2}\Big\|\Big\{\prod^{2}_{l=j}\Big[-\frac{1}{Q^{(N-2l, N-2l+1)}_{\bold{j}_*}\mathscr{K}^{Bog\,(N-2l-2)}_{\bold{j}_*}(z_*)Q^{(N-2l,N-2l+1)}_{\bold{j}_*}}W^*_{\bold{j}_*\,;\,N-2l,N-2l+2}\Big]\Big\}\times \quad\quad \\
& &\quad\quad\quad\quad\quad\quad\times \frac{1}{Q^{(N-2,N-1)}_{\bold{j}_*}\mathscr{K}^{Bog\,(N-4)}_{\bold{j}_*}(z_*)Q^{(N-2,N-1)}_{\bold{j}_*}}Q^{(N-2, N-1)}_{\bold{j}_*}W^*_{\bold{j}_*}\eta \Big\|
%\Big\{\prod^{2}_{l=j}\Big\|[R^{Bog}_{\bold{j}_*\,;\,N-2l,N-2l}(z_*)]^{\frac{1}{2}}\Big\|\,\Big\|\check{\Gamma}^{Bog\,}_{\bold{j}_*\,;\,N-2l,N-2l}(z_*)\Big\|\,\Big\|[R^{Bog}_{\bold{j}_*\,;\,N-2l,N-2l}(z_*)]^{\frac{1}{2}}W^*_{\bold{j}_*\,;\,N-2l,N-2l+2}\Big\|\Big\}\quad\quad\quad
\end{eqnarray}
is bounded by a universal constant times the series
\begin{eqnarray}\label{convergent-series}
& &\sum_{j=2}^{\infty}c_j:=\sum_{j=2}^{\infty}\Big\{\prod^{2}_{l=j}\frac{1}{\Big[1+\sqrt{\eta a_{\epsilon_{\bold{j}_*}}}-\frac{b_{\epsilon_{\bold{j}_*}}/ \sqrt{\eta a_{\epsilon_{\bold{j}_*}}}}{2l-\epsilon^{\Theta}_{\bold{j}_*}}\Big]\Big[1+a_{\epsilon_{\bold{j}_*}}-\frac{2b_{\epsilon_{\bold{j}_*}}}{2l-1}-\frac{1-c_{\epsilon_{\bold{j}_*}}}{(2l-1)^2}\Big]^{\frac{1}{2}}}\Big\}
\end{eqnarray}
which is convergent because 
\begin{equation}\label{ratioc}
\frac{c_{j}}{c_{j-1}}=\frac{1}{\Big[1+\sqrt{\eta a_{\epsilon_{\bold{j}_*}}}-\frac{b_{\epsilon_{\bold{j}_*}}/ \sqrt{\eta a_{\epsilon_{\bold{j}_*}}}}{2j-\epsilon^{\Theta}_{\bold{j}_*}}\Big]\Big[1+a_{\epsilon_{\bold{j}_*}}-\frac{2b_{\epsilon_{\bold{j}_*}}}{2j-1}-\frac{1-c_{\epsilon_{\bold{j}_*}}}{(2j-1)^2}\Big]^{\frac{1}{2}}}<1
\end{equation}
for $j$ sufficiently large.\qed
\begin{remark}\label{expansions}
The sum of the series in (\ref{convergent-series})  is clearly divergent in the limit $\epsilon_{\bold{j}_*} \to 0$. Nevertheless, for any $\epsilon_{\bold{j}_*}>0$ the expansion (\ref{gs-1})-(\ref{gs-2}) of $\psi^{Bog}_{\bold{j}_*}$ is well defined and  controlled in terms of the parameter $\Sigma_{\epsilon_{\bold{j}_*}}:=\frac{1}{1+\sqrt{\epsilon_{\bold{j}_*}}+o(\sqrt{\epsilon_{\bold{j}_*}})}$. On the contrary,  the R-H-S of (\ref{R-H-S-fixed}) is not divergent as $\epsilon_{\bold{j}_*}$ tends to zero (and $N\to \infty$).
\end{remark}
\begin{remark}
In the mean field limit (see \cite{Se1}, \cite{LNSS}), and in the diagonal limit considered in \cite{DN}, the control of the excitation spectrum that is derived in the quoted papers provides a much more accurate estimate of the spectral gap.
\end{remark}
%\begin{corollary}\label{same-gs}
%Under the assumptions of Corollary \ref{col-hbog}, $\psi^{Bog}_{{\bold{j}_*}}$ and $z_*$ are also ground state and ground state energy  of $\hat{H}^{Bog}_{{\bold{j}_*}}$, respectively.
%\end{corollary}

%\noindent
%\emph{Proof}
\subsection{Re-expansion of the  operators  $\Gamma^{Bog}_{\bold{j}_*\,;\,i,i}(z)$ and convergent expansion of the ground state}\label{conv-exp}
This section is mostly devoted  to the \emph{re-expansion} of the  operators  $\Gamma^{Bog}_{\bold{j}_*\,;\,i,i}(z)$,  for $z\leq E^{Bog}_{\bold{j}_*}+ (\delta -1)\phi_{\bold{j}_*}\sqrt{\epsilon_{\bold{j}_*}^2+2\epsilon_{\bold{j}_*}}$ with $\delta\leq 1+\sqrt{\epsilon_{\bold{j}_*}}$, which is stated in Proposition \ref{lemma-expansion-proof-0}. From Proposition \ref{lemma-expansion-proof-0} we derive  Corollary \ref{shift} that is a crucial ingredient for  Lemma \ref{inversion} and Lemma \ref{eigenvalue}. In this respect, we stress that Proposition \ref{lemma-expansion-proof-0} holds whenever the assumptions of Theorem \ref{theorem-Bog} are satisfied.  

In the sequel, at first we informally explain how to re-expand $\Gamma^{Bog}_{\bold{j}_*\,;\,6,6}(z)$. Next, in Proposition \ref{lemma-expansion-proof-0} we show how to do it for any $\Gamma^{Bog}_{\bold{j}_*\,;\,i,i}(z)$.  For sake of brevity, in Sections \ref{informal} and \ref{informal-1} we drop the label $\bold{j}_*$ in the notation used for $\Gamma^{Bog}_{\bold{j}_*\,;\,i,i}(z)$, $W_{\bold{j}_*\,;\,i,i-2}$, $W^*_{\bold{j}_*\,;\,i-2,i}$, and $R^{Bog}_{\bold{j}_*\,;\,i,i}(z)$. We re-introduce the complete notation in Corollary \ref{shift}.

In Section \ref{expansion-gs} we show how to expand the ground state vector in terms of the bare operators and, in general, how to provide a more explicit expression for it.

\subsubsection{Informal description}\label{informal}
Suppose that we want to approximate
\begin{eqnarray}
\Gamma^{Bog}_{6,6}(z)&=&W_{6,4}\,(R^{Bog}_{4,4}(z))^{\frac{1}{2}}\times \label{example}\\
& &\quad\times \sum_{l_4=0}^{\infty}\Big[(R^{Bog}_{4,4}(z))^{\frac{1}{2}}W_{4,2}\,(R^{Bog}_{2,2}(z))^{\frac{1}{2}}\sum_{l_2=0}^{\infty}\Big[(R^{Bog}_{2,2}(z))^{\frac{1}{2}}W_{2,0}\,R_{0,0}^{Bog}(z)W_{0,2}^*(R^{Bog}_{2,2}(z))^{\frac{1}{2}}
\Big]^{l_2}\times \nonumber \\
& &\quad\quad\quad\quad\times(R^{Bog}_{2,2}(z))^{\frac{1}{2}}W^*_{2,4}(R^{Bog}_{4,4}(z))^{\frac{1}{2}}\Big]^{l_4}(R^{Bog}_{4,4}(z))^{\frac{1}{2}}W^*_{4,6}\nonumber
\end{eqnarray}
up to a remainder the norm of which we estimate of order $c^h$ where $0<c<1$, $h\in \mathbb{N}$ and $h\geq 2$. 
We start observing that if $l_4=0$ then there is no summation in $l_2$. Then, we proceed by implementing the following steps:
\begin{itemize}
%\item Using Lemma \ref{main-lemma-Bog}, we can easily derive the estimate
%\begin{equation}
%\Big\|\sum_{l_2=r}^{\infty}\Big[(R^{Bog}_{2,2}(z))^{\frac{1}{2}}W_{2,0}\,R_{0,0}^{Bog}(z)W_{0,2}^*(R^{Bog}_{2,2}(z))^{\frac{1}{2}}\Big]^{l_2}\Big\|\leq (\frac{1}{3})^r\sum_{j=0}^{\infty}(\frac{1}{3})^j=\frac{3}{2}(\frac{1}{3})^r
%\end{equation}
\item
We isolate a first remainder
\begin{eqnarray}
& &[\Gamma^{Bog\,}_{6,6}(z)]_{(4,h_+)}\\
&:= &W_{6,4}\,(R^{Bog}_{4,4}(z))^{\frac{1}{2}}\times \label{remainder-1}\\
& &\quad\times \sum_{l_4=h}^{\infty} \Big[(R^{Bog}_{4,4}(z))^{\frac{1}{2}}W_{4,2}\,(R^{Bog}_{2,2}(z))^{\frac{1}{2}}\sum_{l_2=0}^{\infty}\Big[(R^{Bog}_{2,2}(z))^{\frac{1}{2}}W_{2,0}\,R_{0,0}^{Bog}(z)W_{0,2}^*(R^{Bog}_{2,2}(z))^{\frac{1}{2}}
\Big]^{l_2}\times \nonumber \\
& &\quad\quad\quad\quad\quad\quad \times (R^{Bog}_{2,2}(z))^{\frac{1}{2}}W^*_{2,4}(R^{Bog}_{4,4}(z))^{\frac{1}{2}}\Big]^{l_4}(R^{Bog}_{4,4}(z))^{\frac{1}{2}}W^*_{4,6}\nonumber
\end{eqnarray}
and define $[\Gamma^{Bog\,}_{6,6}(z)]^{(0)}_{(4,h_-)}:=W_{6,4}\,R^{Bog}_{4,4}(z)W^*_{4,6}$.
%where $ \check{\sum}_{l_4=1}^{\infty}$ means that at least one of the factors 
%$$(R^{Bog}_{4,4}(z))^{\frac{1}{2}}W_{4,2}\,(R^{Bog}_{2,2}(z))^{\frac{1}{2}}\sum_{l_2=1}^{\infty}\Big[(R^{Bog}_{2,2}(z))^{\frac{1}{2}}W_{2,0}\,R_{0,0}^{Bog}(z)W_{0,2}^*(R^{Bog}_{2,2}(z))^{\frac{1}{2}}]^{l_2}(R^{Bog}_{2,2}(z))^{\frac{1}{2}}W^*_{2,4}(R^{Bog}_{4,4}(z))^{\frac{1}{2}}$$
%is replaced by
%$$(R^{Bog}_{4,4}(z))^{\frac{1}{2}}W_{4,2}\,(R^{Bog}_{2,2}(z))^{\frac{1}{2}}\sum_{l_2=h}^{\infty}\Big[(R^{Bog}_{2,2}(z))^{\frac{1}{2}}W_{2,0}\,R_{0,0}^{Bog}(z)W_{0,2}^*(R^{Bog}_{2,2}(z))^{\frac{1}{2}}]^{l_2}(R^{Bog}_{2,2}(z))^{\frac{1}{2}}W^*_{2,4}(R^{Bog}_{4,4}(z))^{\frac{1}{2}}$$
%We call this remainder $[\Gamma^{Bog\,}_{8,8}(z)]_{(2,h_+)}$.
\item
In the quantity that remains, i.e.,
\begin{eqnarray}
& &[\Gamma^{Bog\,}_{6,6}(z)]^{(0)}_{(4,h_-)} \label{zero-leading}\\
& &+W_{6,4}\,(R^{Bog}_{4,4}(z))^{\frac{1}{2}}\times \label{uno-leading}\\
& &\quad\times \sum_{l_4=1}^{h-1}\Big[(R^{Bog}_{4,4}(z))^{\frac{1}{2}}W_{4,2}\,(R^{Bog}_{2,2}(z))^{\frac{1}{2}}\sum_{l_2=0}^{\infty}\Big[(R^{Bog}_{2,2}(z))^{\frac{1}{2}}W_{2,0}\,R_{0,0}^{Bog}(z)W_{0,2}^*(R^{Bog}_{2,2}(z))^{\frac{1}{2}}
\Big]^{l_2}\times \nonumber \\
& &\quad\quad\quad\quad\quad\quad \times(R^{Bog}_{2,2}(z))^{\frac{1}{2}}W^*_{2,4}(R^{Bog}_{4,4}(z))^{\frac{1}{2}}\Big]^{l_4}  (R^{Bog}_{4,4}(z))^{\frac{1}{2}}W^*_{4,6}\,,\nonumber
\end{eqnarray}
for each of the $l_4$ factors in the product 
\begin{eqnarray}
%&&\Big[(R^{Bog}_{6,6}(z))^{\frac{1}{2}}W_{6,4}\,(R^{Bog}_{4,4}(z))^{\frac{1}{2}}\times\\
& &\Big[(R^{Bog}_{4,4}(z))^{\frac{1}{2}}W_{4,2}\,(R^{Bog}_{2,2}(z))^{\frac{1}{2}}\sum_{l_2=0}^{\infty}\Big[(R^{Bog}_{2,2}(z))^{\frac{1}{2}}W_{2,0}\,R_{0,0}^{Bog}(z)W_{0,2}^*(R^{Bog}_{2,2}(z))^{\frac{1}{2}}
\Big]^{l_2}\times \quad\quad \quad\quad \label{product} \\
& &\quad\quad\quad\quad\quad\quad \times (R^{Bog}_{2,2}(z))^{\frac{1}{2}}W^*_{2,4}(R^{Bog}_{4,4}(z))^{\frac{1}{2}}\Big]^{l_4}\nonumber
\end{eqnarray}
we split the summation $\sum_{l_2=0}^{\infty}$ into $\sum_{l_2=0}^{h-1}+\sum_{l_2=h}^{\infty}$.
\item
We isolate a second remainder
\begin{eqnarray}
& &[\Gamma^{Bog\,}_{6,6}(z)]_{(2,h_+;4,h_-)}\label{remainder-1-bis}\\
&:= &W_{6,4}\,(R^{Bog}_{4,4}(z))^{\frac{1}{2}}\times \\
& &\quad\times \hat{\sum}_{l_4=1}^{h-1}\Big[(R^{Bog}_{4,4}(z))^{\frac{1}{2}}W_{4,2}\,(R^{Bog}_{2,2}(z))^{\frac{1}{2}}\sum_{l_2=0}^{\infty}\Big[(R^{Bog}_{2,2}(z))^{\frac{1}{2}}W_{2,0}\,R_{0,0}^{Bog}(z)W_{0,2}^*(R^{Bog}_{2,2}(z))^{\frac{1}{2}}
\Big]^{l_2}\times \nonumber \\
& &\quad\quad\quad\quad\quad\quad \times (R^{Bog}_{2,2}(z))^{\frac{1}{2}}W^*_{2,4}(R^{Bog}_{4,4}(z))^{\frac{1}{2}}\Big]^{l_4}(R^{Bog}_{4,4}(z))^{\frac{1}{2}}W^*_{4,6}\nonumber
\end{eqnarray}
where the symbol $\hat{\sum}_{l_4=1}^{h-1}$ stands for the sum of all the summands in $\sum_{l_4=1}^{h-1}$  where at least in one of the $l_4$ factors of the  product in (\ref{product}) the sum over $l_2$ is replaced with the sum starting from $l_2=h$.
\item
In the remaining quantity
\begin{eqnarray}\label{remaining-quantity}
& &(\ref{uno-leading})-[\Gamma^{Bog\,}_{6,6}(z)]_{(2,h_+;4,h_-)}
\end{eqnarray}
we isolate the term
\begin{eqnarray}
& &[\Gamma^{Bog\,}_{6,6}(z)]_{(2,h_-;4,h_-)}\\
&:= &W_{6,4}\,(R^{Bog}_{4,4}(z))^{\frac{1}{2}}\times \label{remaing-leading}\\
& &\quad\times \check{\sum}_{l_4=1}^{h-1}\Big[(R^{Bog}_{4,4}(z))^{\frac{1}{2}}W_{4,2}\,(R^{Bog}_{2,2}(z))^{\frac{1}{2}}\sum_{l_2=0}^{h-1}\Big[(R^{Bog}_{2,2}(z))^{\frac{1}{2}}W_{2,0}\,R_{0,0}^{Bog}(z)W_{0,2}^*(R^{Bog}_{2,2}(z))^{\frac{1}{2}}
\Big]^{l_2}\times \nonumber \\
& &\quad\quad\quad\quad\quad\quad \times(R^{Bog}_{2,2}(z))^{\frac{1}{2}}W^*_{2,4}(R^{Bog}_{4,4}(z))^{\frac{1}{2}}\Big]^{l_4}  (R^{Bog}_{4,4}(z))^{\frac{1}{2}}W^*_{4,6}\nonumber
\end{eqnarray}
where $\check{\sum}_{l_4=1}^{h-1}$ denotes the sum of all summands where at least in one factor of the product 
\begin{eqnarray}
%&&\Big[(R^{Bog}_{6,6}(z))^{\frac{1}{2}}W_{6,4}\,(R^{Bog}_{4,4}(z))^{\frac{1}{2}}\times\\
& &\Big[(R^{Bog}_{4,4}(z))^{\frac{1}{2}}W_{4,2}\,(R^{Bog}_{2,2}(z))^{\frac{1}{2}}\sum_{l_2=0}^{h-1}\Big[(R^{Bog}_{2,2}(z))^{\frac{1}{2}}W_{2,0}\,R_{0,0}^{Bog}(z)W_{0,2}^*(R^{Bog}_{2,2}(z))^{\frac{1}{2}}
\Big]^{l_2}\times \quad\quad\quad\quad  \label{remainder-2} \\
& &\quad\quad\quad\quad\quad\quad \times (R^{Bog}_{2,2}(z))^{\frac{1}{2}}W^*_{2,4}(R^{Bog}_{4,4}(z))^{\frac{1}{2}}\Big]^{l_4}\nonumber
\end{eqnarray}
the sum over $l_2$  is replaced with $\sum_{l_2=1}^{h-1}$, so that we can write
\begin{eqnarray}
& &(\ref{remaining-quantity})\\
&=&W_{6,4}\,(R^{Bog}_{4,4}(z))^{\frac{1}{2}}\sum_{l_4=1}^{h-1}\Big[(R^{Bog}_{4,4}(z))^{\frac{1}{2}}W_{4,2}\,R^{Bog}_{2,2}(z)W^*_{2,4}(R^{Bog}_{4,4}(z))^{\frac{1}{2}}\Big]^{l_4}(R^{Bog}_{4,4}(z))^{\frac{1}{2}}W^*_{4,6}\label{first/+/leading}\\
& &+W_{6,4}\,(R^{Bog}_{4,4}(z))^{\frac{1}{2}}\times \label{inductive-first}\\
& &\quad\quad\quad\times \check{\sum}_{l_4=1}^{h-1}\Big[(R^{Bog}_{4,4}(z))^{\frac{1}{2}}W_{4,2}\,(R^{Bog}_{2,2}(z))^{\frac{1}{2}}\sum_{l_2=0}^{h-1}\Big[(R^{Bog}_{2,2}(z))^{\frac{1}{2}}W_{2,0}\,R_{0,0}^{Bog}(z)W_{0,2}^*(R^{Bog}_{2,2}(z))^{\frac{1}{2}}
\Big]^{l_2}\times \nonumber \\
& &\quad\quad\quad\quad\quad\quad \times(R^{Bog}_{2,2}(z))^{\frac{1}{2}}W^*_{2,4}(R^{Bog}_{4,4}(z))^{\frac{1}{2}}\Big]^{l_4}  (R^{Bog}_{4,4}(z))^{\frac{1}{2}}W^*_{4,6}\,.\nonumber
\end{eqnarray}
We define
\begin{equation}
[\Gamma^{Bog\,}_{6,6}(z)]^{(>0)}_{(4,h_-)}:= (\ref{first/+/leading})
\end{equation}
and  
\begin{equation}
[\Gamma^{Bog\,}_{6,6}(z)]_{(4,h_-)}:=[\Gamma^{Bog\,}_{6,6}(z)]^{(0)}_{(4,h_-)}+[\Gamma^{Bog\,}_{6,6}(z)]^{(>0)}_{(4,h_-)}=(\ref{zero-leading})+(\ref{first/+/leading})\,.
\end{equation}
\end{itemize}
Thus, we have split the original expression in (\ref{example}) into the sum of two (leading) contributions each of them containing only finite sums
\begin{eqnarray}
& &W_{6,4}\,(R^{Bog}_{4,4}(z))^{\frac{1}{2}}\sum_{l_4=0}^{h-1}\Big[(R^{Bog}_{4,4}(z))^{\frac{1}{2}}W_{4,2}\,R^{Bog}_{2,2}(z)W^*_{2,4}(R^{Bog}_{4,4}(z))^{\frac{1}{2}}\Big]^{l_4}(R^{Bog}_{4,4}(z))^{\frac{1}{2}}W^*_{4,6}\\
%& &W_{8,6}\,(R^{Bog}_{6,6}(z))^{\frac{1}{2}}\sum_{l_6=0}^{h-1}\Big[(R^{Bog}_{6,6}(z))^{\frac{1}{2}}W_{6,4}\,R^{Bog}_{4,4}(z)W^*_{4,6}(R^{Bog}_{6,6}(z))^{\frac{1}{2}}\Big]^{l_6}(R^{Bog}_{6,6}(z))^{\frac{1}{2}}W^*_{6,8}\\
%& &+W_{8,6}\,(R^{Bog}_{6,6}(z))^{\frac{1}{2}}\check{\sum}_{l_6=1}^{h-1}\Big[(R^{Bog}_{6,6}(z))^{\frac{1}{2}}W_{6,4}\,(R^{Bog}_{4,4}(z))^{\frac{1}{2}}\times \\
& &+W_{6,4}\,(R^{Bog}_{4,4}(z))^{\frac{1}{2}}\times \label{inductive-first}\\
& &\quad\quad\quad\times \check{\sum}_{l_4=1}^{h-1}\Big[(R^{Bog}_{4,4}(z))^{\frac{1}{2}}W_{4,2}\,(R^{Bog}_{2,2}(z))^{\frac{1}{2}}\sum_{l_2=0}^{h-1}\Big[(R^{Bog}_{2,2}(z))^{\frac{1}{2}}W_{2,0}\,R_{0,0}^{Bog}(z)W_{0,2}^*(R^{Bog}_{2,2}(z))^{\frac{1}{2}}
\Big]^{l_2}\times \nonumber \\
& &\quad\quad\quad\quad\quad\quad \times(R^{Bog}_{2,2}(z))^{\frac{1}{2}}W^*_{2,4}(R^{Bog}_{4,4}(z))^{\frac{1}{2}}\Big]^{l_4}  (R^{Bog}_{4,4}(z))^{\frac{1}{2}}W^*_{4,6}\nonumber
\end{eqnarray}
plus the two remainders (\ref{remainder-1-bis}) and (\ref{remainder-1}). Making use of the definitions, we have derived the identity
\begin{eqnarray}
\Gamma^{Bog\,}_{6,6}(z)
&=&[\Gamma^{Bog\,}_{6,6}(z)]_{(4,h_-)}+[\Gamma^{Bog\,}_{6,6}(z)]_{(4,h_+)}\label{identity-1}\\
& &+[\Gamma^{Bog\,}_{6,6}(z)]_{(2,h_-;4,h_-)}+[\Gamma^{Bog\,}_{6,6}(z)]_{(2,h_+;4,h_-)}\,.\label{identity-2}
%& &+[\Gamma^{Bog\,}_{8,8}(z)]_{(2,h_-;4,h_-;6,h_-)}+[\Gamma^{Bog\,}_{8,8}(z)]_{(2,h_+;4,h_-;6,h_-)}\label{identity-3}
\end{eqnarray}
%\begin{remark}
%Notice the ordering  (from the left to the right) 
%\begin{eqnarray}
%& &[\Gamma^{Bog\,}_{6,6}(z)]_{(4,h_+)}\rightarrow  [\Gamma^{Bog\,}_{8,8}(z)]_{(4,h_+;6,h_-)} \rightarrow [\Gamma^{Bog\,}_{8,8}(z)]_{(6,h_-)} \label{ordering-1}\\
%& &\rightarrow [\Gamma^{Bog\,}_{8,8}(z)]_{(2,h_+;4,h_-;6,h_-)} \rightarrow [\Gamma^{Bog\,}_{8,8}(z)]_{(4,h_-;6,h_-)} \rightarrow [\Gamma^{Bog\,}_{8,8}(z)]_{(2,h_-;4,h_-;6,h_-)}\label{ordering-2}
%\end{eqnarray}
%that has been followed for the definition of the various contributions in (\ref{identity-1})-(\ref{identity-3}).
%\end{remark}

In the last part of this discussion we establish relations between the quantities in (\ref{identity-1})-(\ref{identity-2}) and the analogous quantities associated with $\Gamma^{Bog\,}_{4,4}(z)$. 

\noindent
We observe that
\begin{eqnarray}\label{decomp-4}
\Gamma^{Bog\,}_{4,4}(z)&=&[\Gamma^{Bog\,}_{4,4}(z)]_{(2,h_-)}+[\Gamma^{Bog\,}_{4,4}(z)]_{(2,h_+)}\\
\quad\quad &=&[\Gamma^{Bog\,}_{4,4}(z)]^{(0)}_{(2,h_-)}+[\Gamma^{Bog\,}_{4,4}(z)]^{(>0)}_{(2,h_-)}+[\Gamma^{Bog\,}_{4,4}(z)]_{(2,h_+)} \nonumber 
\end{eqnarray}
where
\begin{equation}
[\Gamma^{Bog\,}_{4,4}(z)]^{(0)}_{(2,h_-)}:=W_{4,2}R^{Bog}_{2,2}(z)W^*_{2,4}\,,
\end{equation}
\begin{equation}
[\Gamma^{Bog\,}_{4,4}(z)]^{(>0)}_{(2,h_-)}:=W_{4,2}\,(R^{Bog}_{2,2}(z))^{\frac{1}{2}}\sum_{l_2=1}^{h-1}\Big[(R^{Bog}_{2,2}(z))^{\frac{1}{2}}W_{2,0}\,R^{Bog}_{0,0}(z)W^*_{0,2}(R^{Bog}_{2,2}(z))^{\frac{1}{2}}\Big]^{l_2}(R^{Bog}_{2,2}(z))^{\frac{1}{2}}W^*_{2,4}\,,
\end{equation}
\begin{equation}
[\Gamma^{Bog\,}_{4,4}(z)]_{(2,h_+)}:=W_{4,2}\,(R^{Bog}_{2,2}(z))^{\frac{1}{2}}\sum_{l_2=h}^{\infty}\Big[(R^{Bog}_{2,2}(z))^{\frac{1}{2}}W_{2,0}\,R^{Bog}_{0,0}(z)W^*_{0,2}(R^{Bog}_{2,2}(z))^{\frac{1}{2}}\Big]^{l_2}(R^{Bog}_{2,2}(z))^{\frac{1}{2}}W^*_{2,4}\,.
\end{equation}
Consequently,
\begin{eqnarray}
& &\sum_{l_4=0}^{h-1}\Big[(R^{Bog}_{4,4}(z))^{\frac{1}{2}}\Gamma^{Bog\,}_{4,4}(z)(R^{Bog}_{4,4}(z))^{\frac{1}{2}}\Big]^{l_4}\\
&=&\sum_{l_4=0}^{h-1}\Big[(R^{Bog}_{4,4}(z))^{\frac{1}{2}}\{[\Gamma^{Bog\,}_{4,4}(z)]^{(0)}_{(2,h_-)}+[\Gamma^{Bog\,}_{4,4}(z)]^{(>0)}_{(2,h_-)}+[\Gamma^{Bog\,}_{4,4}(z)]_{(2,h_+)}\}(R^{Bog}_{4,4}(z))^{\frac{1}{2}}\Big]^{l_4}\,.\quad\quad \label{sum-l4}
\end{eqnarray}
Furthermore, we can write
\begin{eqnarray}
%&&\sum_{l_4=0}^{h-1}\Big[(R^{Bog}_{4,4}(z))^{\frac{1}{2}}\{[\Gamma^{Bog\,}_{4,4}(z)]^{(0)}_{(2,h_-)}+[\Gamma^{Bog\,}_{4,4}(z)]^{(>0)}_{(2,h_-)}+[\Gamma^{Bog\,}_{4,4}(z)]_{(2,h_+)}\}(R^{Bog}_{4,4}(z))^{\frac{1}{2}}\Big]^{l_4}\quad \\
& &(\ref{sum-l4})\\
&=&\sum_{l_4=0}^{h-1}\Big[(R^{Bog}_{4,4}(z))^{\frac{1}{2}}[\Gamma^{Bog\,}_{4,4}(z)]_{(2,h_-)}^{(0)}(R^{Bog}_{4,4}(z))^{\frac{1}{2}}\Big]^{l_4}\\
& &+\check{\sum}_{l_4=1}^{h-1}\Big[(R^{Bog}_{4,4}(z))^{\frac{1}{2}}[\Gamma^{Bog\,}_{4,4}(z)]_{(2,h_-)}(R^{Bog}_{4,4}(z))^{\frac{1}{2}}\Big]^{l_4}\label{gamma44}\\
& &+\hat{\sum}_{l_4=1}^{h-1}\Big[(R^{Bog}_{4,4}(z))^{\frac{1}{2}}[\Gamma^{Bog\,}_{4,4}(z)]_{(2,h_+)}(R^{Bog}_{4,4}(z))^{\frac{1}{2}}\Big]^{l_4}\,\label{gamma44-bis}
\end{eqnarray}
provided:
\begin{itemize}
\item
The symbol $\check{\sum}_{l_4=1}^{h-1}$ in (\ref{gamma44}) means summing from $l_4=1$ up to $h-1$ all the products
\begin{equation}
\Big[(R^{Bog}_{4,4}(z))^{\frac{1}{2}}\mathcal{X}(R^{Bog}_{4,4}(z))^{\frac{1}{2}} \Big]^{l_4}
\end{equation}
that are obtained by replacing  $\mathcal{X}$ (for each factor) with the operators  of the type $[\Gamma^{Bog\,}_{4,4}(z)]^{(0)}_{(2,h_-)}$ and $[\Gamma^{Bog\,}_{4,4}(z)]^{(>0)}_{(2,h_-)}$, with the constraint that $\mathcal{X}$ is replaced with $[\Gamma^{Bog\,}_{4,4}(z)]^{(>0)}_{(2,h_-)}$ in at least one factor.
\item
The symbol $\hat{\sum}_{l_4=1}^{h-1}$ in (\ref{gamma44-bis}) means summing  from $l_4=1$ up to $h-1$  all the products
\begin{equation}
\Big[(R^{Bog}_{4,4}(z))^{\frac{1}{2}}\mathcal{X}(R^{Bog}_{4,4}(z))^{\frac{1}{2}} \Big]^{l_4}
\end{equation}
that are obtained by replacing  $\mathcal{X}$ (for each factor) with the operators  of the type $[\Gamma^{Bog\,}_{4,4}(z)]_{(2,h_-)}$ and $[\Gamma^{Bog\,}_{4,4}(z)]_{(2,h_+)}$, with the constraint that $\mathcal{X}$ is replaced with $[\Gamma^{Bog\,}_{4,4}(z)]_{(2,h_+)}$ in one factor at least.
\end{itemize}
Thereby, we have derived the identities
\begin{eqnarray}
[\Gamma^{Bog\,}_{6,6}(z)]_{(4,h_-)}&= &W_{6,4}\,(R^{Bog}_{4,4}(z))^{\frac{1}{2}}\sum_{l_4=0}^{h-1}\Big[(R^{Bog}_{4,4}(z))^{\frac{1}{2}}[\Gamma^{Bog\,}_{4,4}(z)]_{(2,h_-)}^{(0)}(R^{Bog}_{4,4}(z))^{\frac{1}{2}}\Big]^{l_4} (R^{Bog}_{4,4}(z))^{\frac{1}{2}}W^*_{4,6}\nonumber
\end{eqnarray}
\begin{eqnarray}
[\Gamma^{Bog\,}_{6,6}(z)]_{(2,h_-;4,h_-)}&= &W_{6,4}\,(R^{Bog}_{4,4}(z))^{\frac{1}{2}}\check{\sum}_{l_4=1}^{h-1}\Big[(R^{Bog}_{4,4}(z))^{\frac{1}{2}}[\Gamma^{Bog\,}_{4,4}(z)]_{(2,h_-)}(R^{Bog}_{4,4}(z))^{\frac{1}{2}}\Big]^{l_4} (R^{Bog}_{4,4}(z))^{\frac{1}{2}}W^*_{4,6}\nonumber
\end{eqnarray}
\begin{eqnarray}
[\Gamma^{Bog\,}_{6,6}(z)]_{(2,h_+;4,h_-)}&= &W_{6,4}\,(R^{Bog}_{4,4}(z))^{\frac{1}{2}}\hat{\sum}_{l_4=1}^{h-1}\Big[(R^{Bog}_{4,4}(z))^{\frac{1}{2}}[\Gamma^{Bog\,}_{4,4}(z)]_{(2,h_+)}(R^{Bog}_{4,4}(z))^{\frac{1}{2}}\Big]^{l_4}(R^{Bog}_{4,4}(z))^{\frac{1}{2}}W^*_{4,6}\,.\nonumber
\end{eqnarray}

%the norm of which we can estimate less than
%\begin{eqnarray}
%& &\Big\|\sum_{l_2=1}^{h-1}\Big[(R^{Bog}_{2,2}(z))^{\frac{1}{2}}W_{2,0}\,R_{0,0}^{Bog}(z)W_{0,2}^*(R^{Bog}_{2,2}(z))^{\frac{1}{2}}
%\Big]^{l_2}\Big\|\times \\
%& &\times \sum_{l_6=1}^{\infty}\Big[\Big\|(R^{Bog}_{6,6}(z))^{\frac{1}{2}}W_{6,4}\,(R^{Bog}_{4,4}(z))^{\frac{1}{2}}\Big\|^2  \sum_{l_4=m}^{\infty}\Big\|(R^{Bog}_{4,4}(z))^{\frac{1}{2}}W_{4,2}\,(R^{Bog}_{2,2}(z))^{\frac{1}{2}}(R^{Bog}_{2,2}(z))^{\frac{1}{2}}W^*_{2,4}(R^{Bog}_{4,4}(z))^{\frac{1}{2}}\Big\|^{l_4}\Big]^{l_6}\nonumber\\
%&\leq& (\frac{1}{3}\times \frac{3}{2})  (\frac{1}{3}\times \frac{3}{2})(\frac{1}{3})^m
%\end{eqnarray}

\subsubsection{Re-expansion of the  operators  $\Gamma^{Bog}_{\bold{j}_*\,;\,i,i}(z)$ in the general case}\label{informal-1}

We adapt the strategy used to re-expand $\Gamma^{Bog\,}_{6,6}(z)$ to the general case in Proposition \ref{lemma-expansion-proof-0}, and provide estimates both for the leading and for the remainder terms. To this purpose, first we need some definitions.
\begin{definition} \label{def-sums}

\noindent
Let $h\in \mathbb{N}$, $h\geq 2$, and $z\leq E^{Bog}_{\bold{j}_*}+ (\delta -1)\phi_{\bold{j}_*}\sqrt{\epsilon_{\bold{j}_*}^2+2\epsilon_{\bold{j}_*}}$ with $\delta\leq 1+\sqrt{\epsilon_{\bold{j}_*}}$. Let  $\frac{1}{N}\leq \epsilon^{\nu}_{\bold{j}_*}$ for some  $\nu >\frac{11}{8}$  and $\epsilon_{\bold{j}_*}\equiv \epsilon $ be sufficiently small. We define:
\begin{enumerate}
\item For $N-2\geq  j \geq 4$ with $j$ even
\begin{equation}
[\Gamma^{Bog\,}_{j,j}(z)]_{(j-2, h_-)}:=[\Gamma^{Bog\,}_{j,j}(z)]^{(0)}_{(j-2, h_-)}+[\Gamma^{Bog\,}_{j,j}(z)]^{(>0)}_{(j-2, h_-)}
\end{equation}
where
\begin{eqnarray}
& &[\Gamma^{Bog\,}_{j,j}(z)]^{(0)}_{(j-2, h_-)}:=W_{j,j-2}R^{Bog}_{j-2,j-2}(z) W^*_{j-2,j}\quad \text{for}\,\, j\geq 2\,
\end{eqnarray}
and
\begin{eqnarray}
& &[\Gamma^{Bog\,}_{j,j}(z)]^{(>0)}_{(j-2, h_-)}\\
&:=&W_{j,j-2}\,(R^{Bog}_{j-2,j-2}(z))^{\frac{1}{2}}\times \\
& &\quad \times  \sum_{l_{j-2}=1}^{h-1}\Big[(R^{Bog}_{j-2,j-2}(z))^{\frac{1}{2}}W_{j-2,j-4}\,R_{j-4,j-4}^{Bog}(z)W_{j-4,j-2}^*(R^{Bog}_{j-2,j-2}(z))^{\frac{1}{2}} \Big]^{l_{j-2}}(R^{Bog}_{j-2,j-2}(z))^{\frac{1}{2}} W^*_{j-2,j}\nonumber\\
&=&W_{j,j-2}\,(R^{Bog}_{j-2,j-2}(z))^{\frac{1}{2}}\times \\
& &\quad \times  \sum_{l_{j-2}=1}^{h-1}\Big[(R^{Bog}_{j-2,j-2}(z))^{\frac{1}{2}}[\Gamma^{Bog\,}_{j-2,j-2}(z)]^{(0)}_{(j-4, h_-)}(R^{Bog}_{j-2,j-2}(z))^{\frac{1}{2}} \Big]^{l_{j-2}}(R^{Bog}_{j-2,j-2}(z))^{\frac{1}{2}} W^*_{j-2,j}\,. \nonumber
\end{eqnarray}
For $N-2\geq  j \geq 4$ with $j$ even
\begin{eqnarray}
& &[\Gamma^{Bog\,}_{j,j}(z)]_{(j-2, h_+)}\\
&:=&W_{j,j-2}\,(R^{Bog}_{j-2,j-2}(z))^{\frac{1}{2}}\sum_{l_{j-2}=h}^{\infty}\Big[(R^{Bog}_{j-2,j-2}(z))^{\frac{1}{2}}\Gamma^{Bog\,}_{j-2,j-2}(z)(R^{Bog}_{j-2,j-2}(z))^{\frac{1}{2}} \Big]^{l_{j-2}}\times \nonumber \\
& &\quad\quad\quad \times (R^{Bog}_{j-2,j-2}(z))^{\frac{1}{2}} W^*_{j-2,j}\,.\nonumber
\end{eqnarray}
\item
For $N-2\geq  j \geq 6$  and $2\leq r \leq j-4$ with $r$ and $j$ even
\begin{eqnarray}
& &[\Gamma^{Bog\,}_{j,j}(z)]_{(r,h_-; r+2,h_-;\dots;j-4,h_-;j-2,h_-)}\\
&:= &W_{j,j-2}\,(R^{Bog}_{j-2,j-2}(z))^{\frac{1}{2}} \check{\sum}_{l_{j-2}=1}^{h-1}\Big[(R^{Bog}_{j-2,j-2}(z))^{\frac{1}{2}}[\Gamma^{Bog\,}_{j-2,j-2}(z)]_{(r,h_-;r+2,h_-;\dots;j-4,h_-)}(R^{Bog}_{j-2,j-2}(z))^{\frac{1}{2}} \Big]^{l_{j-2}}\times \nonumber\\
& &\quad\quad\quad\quad \times(R^{Bog}_{j-2,j-2}(z))^{\frac{1}{2}} W^*_{j-2,j}\,. \label{collection}
\end{eqnarray}
Here, the  symbol $\check{\sum}^{h-1}_{l_{j-2}=1}$ stands for a sum of terms resulting from operations $\mathcal{A}1$ and $\mathcal{A}2$ below:
\begin{itemize}
\item[$\mathcal{A}1)$]
At fixed $1\leq l_{j-2} \leq h-1$ summing all the products
\begin{equation}
\Big[(R^{Bog}_{j-2,j-2}(z))^{\frac{1}{2}}\mathcal{X}(R^{Bog}_{j-2,j-2}(z))^{\frac{1}{2}} \Big]^{l_{j-2}}
\end{equation}
that are obtained by replacing  $\mathcal{X}$ for each factor with the operators (iteratively defined) of the type $[\Gamma^{Bog\,}_{j-2,j-2}(z)]_{(m,h_-;m+2,h_-;\dots;j-4,h_-)}$ with $r\leq m\leq j-4$ where $m$ is even, and with the constraint that if $r\leq j-6$ then $\mathcal{X}$ is replaced with $[\Gamma^{Bog\,}_{j-2,j-2}(z)]_{(r,h_-;r+2,h_-;\dots;j-4,h_-)}$ in one factor at least, whereas if $r= j-4$ then  $\mathcal{X}$ is replaced with $[\Gamma^{Bog\,}_{j-2,j-2}(z)]^{(>0)}_{(j-4,h_-)}$ in one factor at least;
\item[$\mathcal{A}2)$]
Summing from $l_{j-2}=1$ up to $l_{j-2}=h-1$.
\end{itemize}
\item
For $N-2\geq  j \geq 6$  and $2\leq r \leq j-4$ with $r$ and $j$ even
\begin{eqnarray}
& &[\Gamma^{Bog\,}_{j,j}(z)]_{(r,h_+; r+2,h_-;\dots;j-4,h_-;j-2,h_-)}\\
&:= &W_{j,j-2}\,(R^{Bog}_{j-2,j-2}(z))^{\frac{1}{2}} \check{\sum}_{l_{j-2}=1}^{+\infty}\Big[(R^{Bog}_{j-2,j-2}(z))^{\frac{1}{2}}[\Gamma^{Bog\,}_{j-2,j-2}(z)]_{(r,h_+;r+2,h_-;\dots;j-4,h_-)}\times \nonumber\\
& &\quad\quad \quad\quad\quad\quad\quad\quad\quad\quad\quad\quad\times (R^{Bog}_{j-2,j-2}(z))^{\frac{1}{2}} \Big]^{l_{j-2}}(R^{Bog}_{j-2,j-2}(z))^{\frac{1}{2}} W^*_{j-2,j}\,.\label{collection-bis}
\end{eqnarray}
Here, the  symbol $\check{\sum}^{h-1}_{l_{j-2}=1}$ stands for a sum of terms resulting from operations $\mathcal{B}1$ and $\mathcal{B}2$ below:
\begin{itemize}
\item[$\mathcal{B}1)$]
At fixed $1\leq l_{j-2} \leq h-1$, summing all the products
\begin{equation}
\Big[(R^{Bog}_{j-2,j-2}(z))^{\frac{1}{2}}\mathcal{X}(R^{Bog}_{j-2,j-2}(z))^{\frac{1}{2}} \Big]^{l_{j-2}}
\end{equation}
that are obtained by replacing  $\mathcal{X}$ for each factor with the operators (iteratively defined) of the type $[\Gamma^{Bog\,}_{j-2,j-2}(z)]_{(m,h_+;m+2,h_-;\dots;j-4,h_-)}$ and $[\Gamma^{Bog\,}_{j-2,j-2}(z)]_{(m',h_-;m'+2,h_-;\dots;j-4,h_-)}$  with $r\leq m\leq j-4$ and $2\leq m'\leq j-4$ where $m$ and $m'$ are even, and with the constraint   that $\mathcal{X}$ is replaced with $[\Gamma^{Bog\,}_{j-2,j-2}(z)]_{(r,h_+;r+2,h_-;\dots;j-4,h_-)}$ in one factor at least.
\item[$\mathcal{B}2)$]
Summing from $l_{j-2}=1$ up to $h-1$.
\end{itemize}
\end{enumerate}
The definitions of above can be adapted in an obvious manner to the case $h=\infty$, in particular the terms $[\Gamma^{Bog\,}_{j,j}(w)]_{(l,h_+; l+2,h_-;\dots;j-4,h_-;j-2,h_-)}$ are absent.
%The definitions of above can be adapted in an obvious manner to the case $h=\infty$.
\end{definition}
\begin{proposition}\label{lemma-expansion-proof-0} Let $\frac{1}{N}\leq \epsilon^{\nu}_{\bold{j}_*}$ for some  $\nu >\frac{11}{8}$  and $\epsilon_{\bold{j}_*}\equiv \epsilon $ be sufficiently small.
For any fixed $2\leq h \in \mathbb{N}$ and for $N-2\geq i\geq 4$ and even,  the splitting
\begin{eqnarray}
\Gamma^{Bog\,}_{i,i}(z)&=&\sum_{r=2,\,r\, even}^{i-2}[\Gamma^{Bog\,}_{i,i}(z)]_{(r,h_-; r+2,h_-;\dots ; i-2,h_-)}+\sum_{r=2\,,\,r\, even}^{i-2}[\Gamma^{Bog\,}_{i,i}(z)]_{(r,h_+; r+2,h_-;\dots; i-2,h_-)}\quad\quad\quad \label{decomposition-0}
\end{eqnarray}
holds true for $z\leq E^{Bog}_{\bold{j}_*}+ (\delta -1)\phi_{\bold{j}_m}\sqrt{\epsilon_{\bold{j}_*}^2+2\epsilon_{\bold{j}_*}}$ with $\delta\leq 1+\sqrt{\epsilon_{\bold{j}_*}}$. Moreover, for $2\leq r \leq i-2$ and even,  the estimates
\begin{eqnarray}\label{gamma-exp-1-0}
& &\Big\|(R^{Bog}_{i,i}(z))^{\frac{1}{2}}[\Gamma^{Bog\,}_{i,i}(z)]_{(r,h_-; r+2, h_-;\, \dots \,; i-2,h_-)}(R^{Bog}_{i,i}(z))^{\frac{1}{2}}\Big\|\\
& &\leq  \prod_{f=r+2\,,\, f-r\,\text{even}}^{i}\frac{K_{f,\epsilon}}{(1-Z_{f-2,\epsilon})^2}\nonumber
%\Big(\frac{2}{3}+\mathcal{O}(\sqrt{\epsilon})\Big)^{\frac{i-l}{2}}\prod_{f=l+2\,,\, f-l\,\text{even}}^{i-2}(1+a_{\epsilon}-\frac{2b_{\epsilon}}{N-f-1}-\frac{1-c_{\epsilon}}{(N-f-1)^2})^{-1}\nonumber
\end{eqnarray}
and
%\begin{equation}
%\Big\|(R^{Bog}_{i,i}(z))^{\frac{1}{2}}[\Gamma^{Bog\,}_{i,i}(z)]_{(l,h_-;\dots ; j,h_-;j-4,h_+)}
%(R^{Bog}_{i,i}(z))^{\frac{1}{2}}\Big\|\leq {\color{red}...}
%\end{equation}
\begin{eqnarray}\label{gamma-exp-2-0}
& &\|(R^{Bog}_{i,i}(z))^{\frac{1}{2}}[\Gamma^{Bog\,}_{i,i}(z)]_{(r,h_+; r+2,h_-;\dots;i-2,h_-)}(R^{Bog}_{i,i}(z))^{\frac{1}{2}}\|\\
&\leq& (Z_{r,\epsilon})^h\,\prod_{f=r+2\,,\, f-r\,\text{even}}^{i}\frac{K_{f,\epsilon}}{(1-Z_{f-2,\epsilon})^2} \nonumber
%& &\quad\times \prod_{f=l+2\,,\, f-l\,\text{even}}^{i-2}(1+a_{\epsilon}-\frac{2b_{\epsilon}}{N-f-1}-\frac{1-c_{\epsilon}}{(N-f-1)^2})^{-1}\nonumber
 \end{eqnarray}
hold true, where 
\begin{equation}\label{KZ-0}
K_{i,\epsilon}:=\frac{1}{4(1+a_{\epsilon}-\frac{2b_{\epsilon}}{N-i+1}-\frac{1-c_{\epsilon}}{(N-i+1)^2})}\quad,\quad Z_{i-2,\epsilon}:=\frac{1}{4(1+a_{\epsilon}-\frac{2b_{\epsilon}}{N-i+3}-\frac{1-c_{\epsilon}}{(N-i+3)^2})}\frac{2}{\Big[1+\sqrt{\eta a_{\epsilon}}-\frac{b_{\epsilon}/\sqrt{\eta a_{\epsilon}}}{N-i+4-\epsilon^{\Theta}}\Big]}
\end{equation}
where $a_{\epsilon},b_{\epsilon}$, and $c_{\epsilon}$ are defined in (\ref{adelta})-(\ref{bdelta})-(\ref{cdelta}) and $\Theta:=\min\{2(\nu-\frac{11}{8})\,;\,\frac{1}{4}\}$.
\end{proposition}

\noindent
\emph{Proof}

\noindent
See Proposition \ref{lemma-expansion-proof-bis} in the Appendix.
\qed

\begin{remark}\label{prod-control}
From the definitions in (\ref{KZ-0}) and the $\epsilon-$dependence of $a_{\epsilon}$, $b_{\epsilon}$, and $c_{\epsilon}$ (see (\ref{adelta})-(\ref{bdelta})-(\ref{cdelta})),  it is evident that there exist constants $C,c>0$ such that (assuming $\epsilon$ sufficiently small)
\begin{equation}\label{KZ}
\frac{K_{f,\epsilon}}{(1-Z_{f-2,\epsilon})^2}\leq \frac{1}{1+c\sqrt{\epsilon}}
\end{equation}
for $N-f>\frac{C}{\sqrt{\epsilon}}$\,.  With a similar computation, one can check that for $N-2\geq i> N-\frac{C}{\sqrt{\epsilon}}$ and some $c'>0$ 
\begin{equation}
\frac{K_{f,\epsilon}}{(1-Z_{f-2,\epsilon})^2}\leq (1+c'\frac{\sqrt{\epsilon}}{N-f}+\mathcal{O}(\frac{1}{(N-f)^2}))
\end{equation}
In consequence, for $N-2\geq i> N-\frac{C}{\sqrt{\epsilon}}$ (and assuming for simplicity that $N-\frac{C}{\sqrt{\epsilon}}$ is an even number) the inequality 
\begin{eqnarray}
& &\prod_{f=N-\frac{C}{\sqrt{\epsilon}}\,,\, f\,\text{even}}^{i}\frac{K_{f,\epsilon}}{(1-Z_{f-2,\epsilon})^2}\\
&\leq & \prod_{f=N-\frac{C}{\sqrt{\epsilon}}\,,\, f\,\text{even}}^{i}(1+c'\frac{\sqrt{\epsilon}}{N-f}+\mathcal{O}(\frac{1}{(N-f)^2}))\\
&=&\prod_{f=N-\frac{C}{\sqrt{\epsilon}}\,,\, f\,\text{even}}^{i}\exp[\ln\Big(1+c'\frac{\sqrt{\epsilon}}{N-f}+\mathcal{O}(\frac{1}{(N-f)^2})\Big)]\leq \mathcal{O}(1)\,
\end{eqnarray}
holds true. Therefore, we can conclude that:

\noindent
1) If $N-\frac{C}{\sqrt{\epsilon}}\geq i\geq r+2$
\begin{equation}
\prod_{f=r+2\,,\, f\,\text{even}}^{i}\frac{K_{f,\epsilon}}{(1-Z_{f-2,\epsilon})^2}\leq\mathcal{O}((\frac{1}{1+c\sqrt{\epsilon}})^{i-r-2})\quad ;  
\end{equation}
2) If $N-2\geq i>N-\frac{C}{\sqrt{\epsilon}}\geq r+2$, then
\begin{eqnarray}
& &\prod_{f=r+2\,,\, f\,\text{even}}^{i}\frac{K_{f,\epsilon}}{(1-Z_{f-2,\epsilon})^2}\\
& =&\Big[\prod_{f=r+2\,,\, f\,\text{even}}^{N-\frac{C}{\sqrt{\epsilon}}}\frac{K_{f,\epsilon}}{(1-Z_{f-2,\epsilon})^2}\Big]\,\Big[\prod_{f=N-\frac{C}{\sqrt{\epsilon}}+2\,,\, f\,\text{even}}^{i}\frac{K_{f,\epsilon}}{(1-Z_{f-2,\epsilon})^2} \Big]\\
&\leq & \mathcal{O}((\frac{1}{1+c\sqrt{\epsilon}})^{N-\frac{C}{\sqrt{\epsilon}}-r-2})\,.
\quad\end{eqnarray}
\end{remark}
\begin{remark}\label{estimation-proc}
In this remark we explain how to provide an estimate  of
\begin{equation}\label{estimate-E}
\|(R^{Bog}_{i,i}(z))^{\frac{1}{2}}\,\sum_{r=2,\,r\, even}^{i-2}[\Gamma^{Bog\,}_{i,i}(w)]_{(r,h_-; r+2,h_-;\dots ; i-2,h_-)}\,(R^{Bog}_{i,i}(z))^{\frac{1}{2}}\|
\end{equation}
without using  (\ref{gamma-exp-1-0}) and the computations in Remark \ref{prod-control}. Indeed, this would make the estimate worse.

\noindent
%We observe that the bound in (\ref{def-deltabog}) of Lemma \ref{main-lemma-Bog} can be employed to provide an upper bound to (\ref{estimate-E}) since 
The operator in (\ref{estimate-E}) can be expressed as a sum of products of operators of the type in (\ref{estimate-main-lemma-Bog}). We call ``blocks" the operators of the type in (\ref{estimate-main-lemma-Bog}) and define
\begin{equation}\label{E-est}
\mathcal{E}(\|(R^{Bog}_{i,i}(z))^{\frac{1}{2}}\,\sum_{r=2,\, r\, even}^{i-2}[\Gamma^{Bog\,}_{i,i}(w)]_{(r,h_-; r+2,h_-;\dots ; i-2,h_-)}\,(R^{Bog}_{i,i}(z))^{\frac{1}{2}}\|)\,
\end{equation}
the upper bound obtained estimating the norm of the sum (of the operators) with the sum of the norms of the summands, and the norm of each operator product with the product of the norms of the blocks. The estimate of the norm of each block is provided by Lemma \ref{main-lemma-Bog}.

%summing up the estimate of the norm of the operator products that follows from Lemma \ref{main-lemma-Bog}.

\noindent
Next, we point out that
\begin{itemize}
  \item by using the decomposition in (\ref{decomposition-0})  of Proposition \ref{lemma-expansion-proof-0}  for $h'\equiv \infty$, we get 
  % up to a remainder of arbitrarily small norm we can approximate 
\begin{eqnarray}
& &(R^{Bog}_{i,i}(w))^{\frac{1}{2}}\Gamma^{Bog\,}_{i,i}(w)(R^{Bog}_{i,i}(w))^{\frac{1}{2}}\label{complete-zero} \\
& =&(R^{Bog}_{i,i}(w))^{\frac{1}{2}}\sum_{r=2,\,r\, even}^{i-2}[\Gamma^{Bog\,}_{i,i}(w)]_{(r,h'_-; r+2,h'_-;\dots ; i-2,h'_-)}(R^{Bog}_{i,i}(w))^{\frac{1}{2}}\,,\label{partial-0-0}
 \end{eqnarray}
% by choosing an $h'$ sufficiently large. 
and
\begin{equation} (R^{Bog}_{i,i}(w))^{\frac{1}{2}}\sum_{r=2,\,r\, even}^{i-2}[\Gamma^{Bog\,}_{i,i}(w)]_{(r,h_-; r+2,h_-;\dots ; i-2,h_-)}(R^{Bog}_{i,i}(w))^{\frac{1}{2}}\,,\quad h<\infty , \, \label{partial-0}
\end{equation} is by construction a partial sum of the terms in (\ref{partial-0-0});
 \item
both for the estimate of  the norm of (\ref{complete-zero}) provided in Theorem  \ref{theorem-Bog} and for $\mathcal{E}(\|(\ref{partial-0})\|)$ we use the same procedure: by Lemma \ref{main-lemma-Bog} we estimate the operator norm of the  blocks and of the products of blocks;  then for each sum of products of blocks we sum up the (estimates of the) operator norms of the products. 
\end{itemize}
Hence, we can conclude that
\begin{eqnarray}\label{estimate-E-bis}
& &\mathcal{E}(\|(R^{Bog}_{i,i}(z))^{\frac{1}{2}}\,\sum_{r=2,\,r\, even}^{i-2}[\Gamma^{Bog\,}_{i,i}(w)]_{(r,h_-; r+2,h_-;\dots ; i-2,h_-)}\,(R^{Bog}_{i,i}(z))^{\frac{1}{2}}\|)\\
&\leq &\mathcal{E}(\|(R^{Bog}_{i,i}(z))^{\frac{1}{2}}\,\Gamma^{Bog\,}_{i,i}(w)\,(R^{Bog}_{i,i}(z))^{\frac{1}{2}}\|)\\
&\leq &\frac{4}{5}
\end{eqnarray}
where the last step follows  for $\epsilon $ sufficiently small from the identity
\begin{eqnarray}
& &(R^{Bog}_{i,i}(w))^{\frac{1}{2}}\Gamma^{Bog\,}_{i,i}(R^{Bog}_{i,i}(w))^{\frac{1}{2}}\label{gammacheck}\\
&=&(R^{Bog}_{i,i}(w))^{\frac{1}{2}}W_{\bold{j}_m\,;\,i,i-2}\,(R^{Bog}_{i-2,i-2}(w))^{\frac{1}{2}} \times\\
& &\quad\quad\quad\times\sum_{l_{i-2}=0}^{\infty}\Big[(R^{Bog}_{i-2,i-2}(w))^{\frac{1}{2}}\Gamma^{Bog}_{i-2,i-2}(w)(R^{Bog}_{i-2,i-2}(w))^{\frac{1}{2}}\Big]^{l_{i-2}}\times \quad\quad\quad\\
& &\quad\quad\quad\quad\quad\quad\quad\quad\times (R^{Bog}_{i-2,i-2}(w))^{\frac{1}{2}}W^*_{i-2,i}(R^{Bog}_{i,i}(w))^{\frac{1}{2}}\,.\nonumber
%&=&(R^{Bog}_{i,i}(z))^{\frac{1}{2}}W_{i,i-2}\,(R^{Bog}_{i-2,i-2}(z))^{\frac{1}{2}}\times\\
%& &\quad\times  \sum_{l_{i-2}=0}^{\infty}\Big[(R^{Bog}_{i-2,i-2}(z))^{\frac{1}{2}}\Gamma^{Bog}_{i-2,i-2}(z)R^{Bog}_{i-2,i-2}(z)\Big]^{l_{i-2}}\times \\
%& &\quad \times (R^{Bog}_{i-2,i-2}(z))^{\frac{1}{2}}W^*_{i-2,i}(R^{Bog}_{i,i}(z))^{\frac{1}{2}}\,.
\end{eqnarray}
and from estimates (\ref{Gamma-ineq}), (\ref{def-deltabog}), and (\ref{doppia}).
\end{remark}
As a byproduct of the control of the decomposition in (\ref{decomposition-0}) and of the estimates in (\ref{gamma-exp-1-0})-(\ref{gamma-exp-2-0}), in the sequel we prove the estimate in (\ref{shift-1})-(\ref{shift-2}) used in Lemma \ref{inversion} for the invertibility of $\overline{\mathscr{P}_{\eta}}\mathscr{K}^{Bog\,(N-2)}_{\bold{j}_*}(z)\overline{\mathscr{P}_{\eta}}$ on $\overline{\mathscr{P}_{\eta}}\mathcal{F}^N$.
\begin{corollary}\label{shift}
Let $\epsilon_{\bold{j}_*}$ be sufficiently small and $z$ in the range defined in (\ref{range}).  Assume  $\frac{1}{N}\leq \epsilon^{\nu}_{\bold{j}_*}$ for some $\nu >\frac{11}{8}$  and $\mu$ as in Condition 3.2) of Definition \ref{def-pot}. Then
\begin{eqnarray}
& &\|\overline{\mathscr{P}_{\eta}}W_{\bold{j}_*}\,(R^{Bog}_{\bold{j}_*\,;\,N-2,N-2}(z))^{\frac{1}{2}}\,\check{\Gamma}^{Bog}_{\bold{j}_*\,;\,N-2,N-2}(z)\, (R^{Bog}_{\bold{j}_*\,;\,N-2,N-2}(z))^{\frac{1}{2}}\, W^*_{\bold{j}_*}\overline{\mathscr{P}_{\eta}}\|\quad  \label{shift-1-bis}\\
&\leq &\|\mathscr{P}_{\eta}W_{\bold{j}_*}\,(R^{Bog}_{\bold{j}_*\,;\,N-2,N-2}(z-\frac{\Delta_0}{2})))^{\frac{1}{2}}\,\check{\Gamma}^{Bog}_{\bold{j}_*\,;\,N-2,N-2}(z-\frac{\Delta_0}{2}))\times \quad\quad\quad \quad \label{ineq-shift}\\
& &\quad\quad\quad \times (R^{Bog}_{\bold{j}_*\,;\,N-2,N-2}(z-\frac{\Delta_0}{2}))^{\frac{1}{2}}\, W^*_{\bold{j}_*}\mathscr{P}_{\eta}\| \nonumber\\
& &+\mathcal{O}(\frac{1}{\sqrt{\epsilon_{\bold{j}_*}}}(\frac{1}{1+c\sqrt{\epsilon_{\bold{j}_*}}})^{N^{\mu}})\, \,\quad\quad\quad \label{shift-2-bis}
\end{eqnarray}
where $c>0$.
\end{corollary}

\noindent
\emph{Proof}

\noindent
For any normalized vector $\varphi \in \overline{\mathscr{P}_{\eta}}\mathcal{F}^N$ with definite number of particles in the modes $\bold{j}\in \mathbb{Z}^d$ we consider the scalar product 
\begin{equation}
\langle \varphi\,,\,W_{\bold{j}_*}(R^{Bog}_{\bold{j}_*\,;\,N-2,N-2}(z))^{\frac{1}{2}}\,\check{\Gamma}^{Bog}_{\bold{j}_*\,;\,N-2,N-2}(z)\, (R^{Bog}_{\bold{j}_*\,;\,N-2,N-2}(z))^{\frac{1}{2}}\, W^*_{\bold{j}_*}\varphi \rangle\,.
\end{equation} We make use of the decomposition in (\ref{decomposition-0}) along with the estimates in (\ref{gamma-exp-1-0}), (\ref{gamma-exp-2-0}),  and take Remark \ref{prod-control} and Remark \ref{estimation-proc} into account.
% and the observation in Remark \ref{estimation-proc}. 
Next, we invoke the result  in Corollary \ref{truncated} for $N^{1-\beta} \equiv N^{\mu}$, and we get 
\begin{eqnarray}
& &\langle \varphi\,,\,W_{\bold{j}_*}(R^{Bog}_{\bold{j}_*\,;\,N-2,N-2}(z))^{\frac{1}{2}}\,\check{\Gamma}^{Bog}_{\bold{j}_*\,;\,N-2,N-2}(z)\, (R^{Bog}_{\bold{j}_*\,;\,N-2,N-2}(z))^{\frac{1}{2}}\, W^*_{\bold{j}_*}\varphi \rangle\label{vector-phi}\\
&=&\sum_{l_{N-2}=0}^{\infty}\langle \varphi\,,\,W_{\bold{j}_*}R^{Bog}_{\bold{j}_*\,;\,N-2,N-2}(z)\,\Big\{\Gamma^{Bog\,}_{\bold{j}_*\,;\,N-2,N-2}(z)\, R^{Bog}_{\bold{j}_*\,;\,N-2,N-2}(z)\,\Big\}^{l_{N-2}}\, W^*_{\bold{j}_*}\varphi \rangle\\
&=&\sum_{l_{N-2}=0}^{\infty}\langle \varphi\,,\,W_{\bold{j}_*}R^{Bog}_{\bold{j}_*\,;\,N-2,N-2}(z)\,\Big\{\sum_{r=N-N^{\mu},\, r\, even}^{N-4}\,[\Gamma^{Bog\,}_{\bold{j}_*\,;\,N-2,N-2}(z)]_{(r, h_-; r+2,h_-;\dots ; N-4,h_-)}|_{h\equiv \infty}\, R^{Bog}_{\bold{j}_*\,;\,N-2,N-2}(z)\,\Big\}^{l_{N-2}}\, W^*_{\bold{j}_*}\varphi \rangle\,\,\nonumber \\
& &+\mathcal{O}(\frac{1}{\sqrt{\epsilon_{\bold{j}_*}}}(\frac{1}{1+c\sqrt{\epsilon_{\bold{j}_*}}})^{N^{\mu}})\label{first-scalar}
\end{eqnarray}
where: 1) The symbol $|_{h\equiv \infty}$ means that in the decomposition of $\Gamma^{Bog\,}_{\bold{j}_*\,;\,N-2,N-2}(z)$ (see Proposition \ref{lemma-expansion-proof-0}) we have chosen $h\equiv \infty$. We omit this symbol in the rest of the proof; 2) $\frac{1}{N^{\mu}}= o(\sqrt{\epsilon_{\bold{j}*}})$ (see Condition 3.2) in Definition \ref{def-pot}), and for simplicity we have assumed that $N-N^{\mu}$ is even.
Similarly to the procedure used for the expression in (\ref{scalar-prod}),  we observe that the scalar product
\begin{eqnarray}
& &\langle \varphi\,,\,W_{\bold{j}_*}(R^{Bog}_{\bold{j}_*\,;\,N-2,N-2}(z))^{\frac{1}{2}}\times \nonumber\\
& &\quad \times \,\Big\{(R^{Bog}_{\bold{j}_*\,;\,N-2,N-2}(z))^{\frac{1}{2}}\sum_{r=N-N^{\mu},\,r\, even}^{N-2}\,[\Gamma^{Bog\,}_{\bold{j}_*\,;\,N-2,N-2}(z)]_{(r,h_-; r+2,h_-;\dots ; N-2,h_-)}\, (R^{Bog}_{\bold{j}_*\,;\,N-2,N-2}(z))^{\frac{1}{2}}\,\Big\}^{l_{N-2}}\, \times \nonumber\\
& &\quad\quad\quad \times (R^{Bog}_{\bold{j}_*\,;\,N-2,N-2}(z))^{\frac{1}{2}}W^*_{\bold{j}_*}\varphi \rangle
\end{eqnarray}
corresponds to the same expression where:
\begin{enumerate}
\item
The vector $\varphi$ is replaced with $\eta$.
\item Each couple of companion operators
\begin{equation}
(R^{Bog}_{\bold{j}_*\,;\,i,i}(z))^{\frac{1}{2}}\phi_{\bold{j}}\frac{a^*_{\bold{0}}a^*_{\bold{0}}a_{\bold{j}_{*}}a_{-\bold{j}_{*}}}{N}\,(R^{Bog}_{\bold{j}_*\,;\,i-2,i-2}(z))^{\frac{1}{2}}\quad,\quad(R^{Bog}_{\bold{j}_*\,;\,i-2,i-2}(z))^{\frac{1}{2}}\,\phi_{\bold{j}}\frac{a_{\bold{0}}a_{\bold{0}}a^*_{\bold{j}_{*}}a^*_{-\bold{j}_{*}}}{N}(R^{Bog}_{\bold{j}_*\,;\,i,i}(z))^{\frac{1}{2}}
\end{equation}
that pop up from the re-expansion of $$(R^{Bog}_{\bold{j}_*\,;\,N-2,N-2}(z))^{\frac{1}{2}}[\Gamma^{Bog\,}_{\bold{j}_*\,;\,N-2,N-2}(z)]_{(r,h_-; r+2,h_-;\dots ; N-2,h_-)}(R^{Bog}_{\bold{j}_*\,;\,N-2,N-2}(z))^{\frac{1}{2}}$$ is replaced with the c-number
\begin{eqnarray}
& &[\mathcal{W}_{\bold{j}_*\,;\,i,i-2}(z)\mathcal{W}^*_{\bold{j}_*\,;\,i-2,i}(z)]_{\varphi} \label{Wvar}\\
&:= &\frac{(n_{\bold{j}_0}-1)n_{\bold{j}_0}}{N^2}\,\phi^2_{\bold{j}_{*}}\,\frac{ (n_{\bold{j}_{*}}+1)(n_{-\bold{j}_{*}}+1)}{\Big[E_{\varphi}+(\frac{n_{\bold{j}_0}}{N}\phi_{\bold{j}_{*}}+k_{\bold{j}_{*}}^2)(n_{\bold{j}_{*}}+n_{-\bold{j}_{*}})-z\Big]}\times\\
& &\quad\quad\quad \times\frac{1}{\Big[E_{\varphi}+(\frac{(n_{\bold{j}_0}-2)}{N}\phi_{\bold{j}_{*}}+k_{\bold{j}_{*}}^2)(n_{\bold{j}_{*}}+n_{-\bold{j}_{*}})+2(\frac{(n_{\bold{j}_0}-2)}{N}\phi_{\bold{j}_{*}}+k_{\bold{j}_{*}}^2)-z\Big]}\\
&= &\frac{1}{N^2}\,\phi^2_{\bold{j}_{*}}\,\frac{ (n_{\bold{j}_{*}}+1)(n_{-\bold{j}_{*}}+1)}{\Big[\frac{E_{\varphi}}{n_{\bold{j}_0}}+(\frac{1}{N}\phi_{\bold{j}_{*}}+\frac{k_{\bold{j}_{*}}^2}{n_{\bold{j}_0}})(n_{\bold{j}_{*}}+n_{-\bold{j}_{*}})-\frac{z}{n_{\bold{j}_0}}\Big]}\times\\
& &\quad\quad\quad \times\frac{1}{\Big[\frac{E_{\varphi}}{n_{\bold{j}_0}-1}+(\frac{1}{N}\phi_{\bold{j}_{*}}+\frac{k_{\bold{j}_{*}}^2}{n_{\bold{j}_0}-1}-\frac{\phi_{\bold{j}_{*}}}{N(n_{\bold{j}_0}-1)})(n_{\bold{j}_{*}}+n_{-\bold{j}_{*}}+2)-\frac{z}{n_{\bold{j}_0}-1}\Big]}
%&:=& \frac{(n_{\bold{j}_0}+2)(n_{\bold{j}_0}+1)}{N^2}\,\phi^2_{\bold{j}_{*}}\,\frac{ (n_{\bold{j}_{*}}+1)(n_{-\bold{j}_{*}}+1)}{\Big[(\frac{n_{\bold{j}_0}}{N}\phi_{\bold{j}_{*}}+(k_{\bold{j}_{*}}^2))(n_{\bold{j}_{*}}+n_{-\bold{j}_{*}})-z\Big]}\times\\
%& &\times \frac{1}{\Big[(\frac{(n_{\bold{j}_0}+2)}{N}\phi_{\bold{j}_{*}}+(k_{\bold{j}_{*}}^2))(n_{\bold{j}_{*}}+n_{-\bold{j}_{*}})-2(\frac{(n_{\bold{j}_0}+2)}{N}\phi_{\bold{j}_{*}}+(k_{\bold{j}_{*}}^2))-z\Big]}\quad\quad\quad
\end{eqnarray} 
where $n_{\bold{j}_0}>1$ (otherwise the expression vanishes) and
\begin{itemize}
\item
\begin{equation}
n_{\bold{j}_{*}}+n_{-\bold{j}_{*}}=N-i\quad \text{and}\,\quad n_{\bold{j}_{*}}=n_{-\bold{j}_{*}}\,,
\end{equation}
\item
$1<n_{\bold{j}_0}<i$ equals $i-s$ where $s$, $i-1>s\geq 1$,  is the number of particles in the modes $\bold{j}\notin \{\bold{0},\pm \bold{j}_*\}$ contained in the vector $\varphi$,
\item $E_{\varphi}\geq s\Delta_0$ is the kinetic energy of the state $\varphi$ which is by assumption an eigenvector of the kinetic energy operator.
\end{itemize}
We now show that (see (\ref{Wvar})) 
\begin{equation}
[\mathcal{W}_{\bold{j}_*\,;i,i-2}(z)\mathcal{W}^*_{\bold{j}_*\,;i-2,i}(z)]_{\varphi} \leq \mathcal{W}_{\bold{j}_*\,;i,i-2}\Big(z-\frac{\Delta_0}{2}\Big)\mathcal{W}^*_{\bold{j}_*\,;i-2,i}\Big(z-\frac{\Delta_0}{2}\Big)\,
\end{equation}
for $z$ in the range defined in (\ref{range}), where $\mathcal{W}_{\bold{j}_*\,;i,i-2}(z)\mathcal{W}^*_{\bold{j}_*\,;i,i-2}(z)$  is defined in (\ref{def-Wcal-1})-(\ref{def-Wcal-2}).
The only nontrivial inequality  to be proven is
\begin{eqnarray}
& &\frac{E_{\varphi}}{n_{\bold{j}_0}-1}+(\frac{1}{N}\phi_{\bold{j}_{*}}+\frac{k_{\bold{j}_{*}}^2}{n_{\bold{j}_0}-1}-\frac{\phi_{\bold{j}_{*}}}{N(n_{\bold{j}_0}-1)})(n_{\bold{j}_{*}}+n_{-\bold{j}_{*}}+2)-\frac{z}{n_{\bold{j}_0}-1}\\
&\geq &\frac{\Delta_0}{2(i-1)}+(\frac{1}{N}\phi_{\bold{j}_{*}}+\frac{k_{\bold{j}_{*}}^2}{i-1}-\frac{\phi_{\bold{j}_{*}}}{N(i-1)})(n_{\bold{j}_{*}}+n_{-\bold{j}_{*}}+2)-\frac{z}{i-1}\,.
\end{eqnarray}
Since $E_{\varphi}\geq s\Delta_0$ and $n_{\bold{j}_0}=i-s<i$ it is enough to show that
\begin{equation}\label{ineq-den}
\frac{(s-\frac{1}{2})\Delta_0}{(i-s-1)}-\frac{\phi_{\bold{j}_{*}}}{N(i-s-1)}(n_{\bold{j}_{*}}+n_{-\bold{j}_{*}}+2)\geq -\frac{\phi_{\bold{j}_{*}}}{N(i-1)}(n_{\bold{j}_{*}}+n_{-\bold{j}_{*}}+2)
\end{equation} 
which is equivalent to
\begin{equation}\label{equivalent}
\Delta_0\geq \frac{s\phi_{\bold{j}_{*}}}{N(s-\frac{1}{2})(i-1)}(n_{\bold{j}_{*}}+n_{-\bold{j}_{*}}+2)\,.
\end{equation}
 Since $s\geq 1$, $n_{\bold{j}_{*}}+n_{-\bold{j}_{*}}=N-i$,  and (recall $i-2\geq r\geq N-N^{\mu}$)
 $$i\geq N-N^{\mu}+2\,\,\Rightarrow\,\, n_{\bold{j}_{*}}+n_{-\bold{j}_{*}}+2\leq N^{\mu}$$ the inequality in (\ref{equivalent}) holds because we assume $\frac{\phi_{\bold{j}_*}N^{\mu}}{\Delta_0 N(N-N^{\mu})}<\frac{1}{2}$ (see Condition 3.2 in Definition \ref{def-pot}). Hence, the inequality in (\ref{ineq-den}) holds too.
\item
Due to an analogous argument, the operator
\begin{equation}
(R^{Bog}_{\bold{j}_*\,;\,N-2,N-2}(z))^{\frac{1}{2}}W^*_{\bold{j}_*}
\end{equation}
next to the vector $\varphi$ is replaced with
\begin{equation}
\frac{\sqrt{(N-s)(N-s-1)}}{N}\frac{\phi_{\bold{j}_{*}}}{[E_{\varphi}+(k^2_{\bold{j}_*}+\frac{N-2-s}{N}\phi_{\bold{j}_*})2-z]^{\frac{1}{2}}}\
\end{equation}
which is less than
\begin{equation}
\sqrt{1-\frac{1}{N}}\frac{\phi_{\bold{j}_{*}}}{[(k^2_{\bold{j}_*}+\frac{N-2}{N}\phi_{\bold{j}_*})2-z+\frac{\Delta_0}{2}]^{\frac{1}{2}}}\
\end{equation}
due to the inequalities in (\ref{ineq-den}) and (\ref{equivalent}) with $n_{\bold{j}_*}+n_{-\bold{j}_*}=0$ and $i=N$.
\end{enumerate}
Consequently, we can estimate
\begin{eqnarray}
& &\langle \varphi\,,\,W_{\bold{j}_*}R^{Bog}_{\bold{j}_*\,;\,N-2,N-2}(z)\,\Big\{\sum_{r=N-N^{\mu},\,r\, even}^{N-2}\,[\Gamma^{Bog\,}_{\bold{j}_*\,;\,N-2,N-2}(z)]_{(r,h_-; r+2,h_-;\dots ; N-2,h_-)}\, R^{Bog}_{\bold{j}_*\,;\,N-2,N-2}(z)\,\Big\}^{l_{N-2}}\, W^*_{\bold{j}_*}\varphi \rangle\quad\quad\quad\quad\\
&\leq &\langle \eta\,,\,W_{\bold{j}_*}R^{Bog}_{\bold{j}_*\,;\,N-2,N-2}(w)\,\Big\{\sum_{r=N-N^{\mu},\,r\, even}^{N-2}\,[\Gamma^{Bog\,}_{\bold{j}_*\,;\,N-2,N-2}(w)]_{(r,h_-; r+2,h_-;\dots ; N-2,h_-)}\, R^{Bog}_{\bold{j}_*\,;\,N-2,N-2}(w)\,\Big\}^{l_{N-2}}\, W^*_{\bold{j}_*}\eta \rangle \label{first-scalar-bis}
\end{eqnarray}
where $w\equiv z-\frac{\Delta_0}{2} $. Next, we add  the positive quantity -- indeed we know that after the re-expansion both quantities (\ref{complete}) and (\ref{partial})  below can be written as a sum of positive summands and the quantity in (\ref{partial}) is a partial sum of the terms in (\ref{complete}) -- :
\begin{eqnarray}
& &\sum_{l_{N-2}=0}^{\infty}\langle \eta\,,\,W_{\bold{j}_*}R^{Bog}_{\bold{j}_*\,;\,N-2,N-2}(w)\,\Big\{\sum_{r=2,\,r\, even}^{N-2}\,[\Gamma^{Bog\,}_{\bold{j}_*\,;\,N-2,N-2}(w)]_{(r,h_-; r+2,h_-;\dots ; N-2,h_-)}\, R^{Bog}_{\bold{j}_*\,;\,N-2,N-2}(w)\,\Big\}^{l_{N-2}}\, W^*_{\bold{j}_*}\eta \rangle\quad\quad \label{complete}\\
& &-\sum_{l_{N-2}=0}^{\infty}\langle \eta\,,\,W_{\bold{j}_*}R^{Bog}_{\bold{j}_*\,;\,N-2,N-2}(w)\,\Big\{\sum_{r=N-N^{\mu},\,r\, even}^{N-2}\,[\Gamma^{Bog\,}_{\bold{j}_*\,;\,N-2,N-2}(w)]_{(r,h_-; r+2,h_-;\dots ; N-2,h_-)}\, R^{Bog}_{\bold{j}_*\,;\,N-2,N-2}(w)\,\Big\}^{l_{N-2}}\, W^*_{\bold{j}_*}\eta\rangle\quad\quad \quad\label{partial}
\end{eqnarray}
to (\ref{first-scalar-bis}). 
Hence, we have shown that
\begin{eqnarray}
& &\langle \varphi\,,\,W_{\bold{j}_*}(R^{Bog}_{\bold{j}_*\,;\,N-2,N-2}(z))^{\frac{1}{2}}\,\check{\Gamma}^{Bog}_{\bold{j}_*\,;\,N-2,N-2}(z)\, (R^{Bog}_{\bold{j}_*\,;\,N-2,N-2}(z))^{\frac{1}{2}}\, W^*_{\bold{j}_*}\varphi \rangle \label{ineq-fin-0}\\
&\leq&\langle \eta\,,\,W_{\bold{j}_*}(R^{Bog}_{\bold{j}_*\,;\,N-2,N-2}(w))^{\frac{1}{2}}\,\check{\Gamma}^{Bog}_{\bold{j}_*\,;\,N-2,N-2}(w)\, (R^{Bog}_{\bold{j}_*\,;\,N-2,N-2}(w))^{\frac{1}{2}}\, W^*_{\bold{j}_*}\eta \rangle\\
& &+\mathcal{O}(\frac{1}{\sqrt{\epsilon_{\bold{j}_*}}}(\frac{1}{1+c\sqrt{\epsilon_{\bold{j}_*}}})^{N^{\mu}})\label{ineq-fin-2}
\end{eqnarray}
with $w\equiv z-\frac{\Delta_0}{2} $. 

From  (\ref{ineq-fin-0})-(\ref{ineq-fin-2}), the inequality in (\ref{ineq-shift}) follows straightforwardly because the positive  operator $$W_{\bold{j}_*}\,(R^{Bog}_{\bold{j}_*\,;\,N-2,N-2}(z))^{\frac{1}{2}}\,\check{\Gamma}^{Bog}_{\bold{j}_*\,;\,N-2,N-2}(z)\, (R^{Bog}_{\bold{j}_*\,;\,N-2,N-2}(z))^{\frac{1}{2}}\, W^*_{\bold{j}_*}$$ commutes with all number operators $a^*_{\bold{j}}a_{\bold{j}}$.
\qed

\subsubsection{Convergent expansion of the ground state}\label{expansion-gs}

 Starting from expression (\ref{gs-1})-(\ref{gs-2}), for a given $\zeta>0$ we want to define a vector, $(\psi^{Bog}_{\bold{j}_*})_{\zeta}$,  in terms of the vector $\eta$ and of a finite sum of products of the interaction terms $W^*_{\bold{j}_*}\,,\,W_{\bold{j}_*}$, and of the resolvent $\frac{1}{\hat{H}^0_{\bold{j}_*}-z_*}$ (see (\ref{H0j})), that approximates $\psi^{Bog}_{{\bold{j}_*}}$ up to a quantity in norm less than $\mathcal{O}(\zeta)$ provided $N$ is sufficiently large. Since we want to consider arbitrarily small $\zeta$, we restrict the analysis  to the mean field limiting regime. The operations to be implemented are described below for $N$ sufficiently large:
\begin{itemize}
\item
Using the convergence of the series in  (\ref{convergent-series}) we truncate the sum in (\ref{gs-2}) at some $\zeta-$dependent $\bar{j}$;
\item
In each summand obtained from the expression (\ref{gs-2}) after the truncation,    for some $\zeta-$dependent $h$ and $j_{\#}$, we replace the  operators  $\Gamma^{Bog}_{\bold{j}_*\,;\,i,i}(z_*)$ (see (\ref{sandwich-1bis})-(\ref{sandwich-4})) with 
$$ \sum_{r=j_{\#},\,r\, even}^{i-2}[\Gamma^{Bog\,}_{\bold{j}_*\,;\,i,i}(z_*)]_{(r,h_-; r+2,h_-;\dots ; i-2,h_-)}$$
up to a remainder of sufficiently small operator norm depending on $\zeta$. (Here, we make use of Proposition \ref{lemma-expansion-proof-0} and Remark \ref{prod-control} in Section \ref{informal-1}.) Thereby, we replace $\Gamma^{Bog}_{\bold{j}_*\,;\,i,i}(z_*)$ with a  finite ($\zeta-$dependent) sum of products of the operators $W_{\bold{j}_*\,;\,j,j-2}$, $W^*_{\bold{j}_*\,;\,j-2,j}$  (where $j_{\#}+2\leq j<i$)  and $R^{Bog}_{\bold{j}_*\,;\,j,j}(z_*)$  (where $j_{\#}\leq j<i$).
%\item 
%With some further work one can derive bounds from above and below for $\check{\mathcal{G}}_{\bold{j}_*\,;\,N-2,N-2}(E^{Bog}_{\bold{j}_*})$. Then, the property mentioned in Remark \ref{increasing} implies a straightforward control of the difference between $E^{Bog}_{\bold{j}_*}$ and $z_*$.
%\item
%$\bar{j}$ and the expansion of the operators $\Gamma^{Bog}_{\bold{j}_*\,;\,i,i}(z)$ are chosen so that the sum of the remainder terms is an operator in norm less than $\delta$.
\end{itemize}
Furthermore, for the ($\zeta$-dependent) finite sum of terms that we have isolated we can invoke the result of Lemma \ref{eigenvalue} to approximate $z_*$ with $E^{Bog}_{\bold{j}_*}$  up to an arbitrarily small error for $N$ sufficiently large. Therefore,  in the mean field limit the approximation of $\psi^{Bog}_{\bold{j}_*}$ in terms of the vector $\eta$ and of a finite sum of products of the interaction terms $W^*_{\bold{j}_*}\,,\,W_{\bold{j}_*}$, and of the resolvent $\frac{1}{\hat{H}^0_{\bold{j}_*}-E^{Bog}_{\bold{j}*}}$ is up to any desired precision. 

In order to get a more compact expression, after observing that the operators
\begin{equation}
Q^{(N-2l, N-2l+1)}_{\bold{j}_*}\frac{1}{Q^{(N-2l, N-2l+1)}_{\bold{j}_*}\mathscr{K}^{Bog\,(N-2l-2)}_{\bold{j}_*}(z_*)Q^{(N-2l,N-2l+1)}_{\bold{j}_*}}Q^{(N-2l, N-2l+1)}_{\bold{j}_*} \label{4.11}
\end{equation}
commute with each  number operator $a^*_{\bold{j}}a_{\bold{j}}$, we point out that the operators of the type in (\ref{4.11}) and contained  in expression (\ref{gs-1})-(\ref{gs-2})   can be replaced with  $c-$ numbers. Similarly to the treatment of  the quantity $\check{\mathcal{G}}_{\bold{j}_*\,;\,N-2,N-2}(z_*)$ in Lemma  \ref{eigenvalue}, for $2l$ much smaller than $N$, these  $c-$ numbers  can be computed up to an arbitrarily small $\zeta$-dependent remainder  provided $N$ is sufficiently large. Indeed, Lemma  \ref{eigenvalue} could be generalized to compute these quantities.
%Regarding the approximation of $z_*$ with $E^{Bog}_{\bold{j}_*}$ up to $\frac{\zeta}{2}$, this is possible using either the result in \cite{Se1} or the result in Lemma \ref{eigenvalue} which follows from the construction of the Feshbach flow and Proposition \ref{lemma-expansion-proof-0}. At fixed particle density and dimension $d\geq 4$  the result in Lemma \ref{eigenvalue} also we can approximate $z_*$ with $E^{Bog}_{\bold{j}_*}$ for a sufficiently large $L$.
\\

%%%%%%%%%%%%%%%%%%%%%%%%%%%%%%%%%%%%%%%%%%%
%%%%%%%%%%%%A P P E N D I X%%%%%%%%%%%%%%%%%%%%%%%
%%%%%%%%%%%%%%%%%%%%%%%%%%%%%%%%%%%%%%%%%%%%
\section{Appendix}\label{appendix}
\setcounter{equation}{0}
In the first lemma we provide some lengthy computations needed in Lemma \ref{main-lemma-Bog} and in the proof of inequality (\ref{ineq-W})  in Section \ref{section-lower}.
\begin{lemma}\label{accessori}
The  step from (\ref{meno-1}) to (\ref{meno-0}) is justified under the assumptions of Lemma  \ref{main-lemma-Bog}. Furthermore, the inequality in (\ref{ineq-W}) is verified for $i\geq N-N^{1-\gamma}$ with $0<\gamma<1$  provided the (positive) constant $c_{\gamma}$ is sufficiently large. 
\end{lemma}

\noindent
\emph{Proof}

\noindent
The step from (\ref{meno-1}) to (\ref{meno-0})  is completed by the identities below where we assume $0\leq \delta<2$ and $\frac{1}{N}\leq \epsilon^{\nu}$,
\begin{eqnarray}
& &\Big[1+\epsilon_{\bold{j}_{*}}-\frac{\Big[\epsilon_{\bold{j}_{*}}+1+\delta\sqrt{\epsilon_{\bold{j}_{*}}^2+2\epsilon_{\bold{j}_{*}}}\,\Big]}{N-i+1}\Big]\Big[1+\epsilon_{\bold{j}_{*}}-\frac{1}{N}+\frac{\Big[\epsilon_{\bold{j}_{*}}+1-\delta\sqrt{\epsilon_{\bold{j}_{*}}^2+2\epsilon_{\bold{j}_{*}}}\,\Big]}{N-i+1}\Big]\quad\quad\quad\label{comp-in}\\
& =&(1+\epsilon_{\bold{j}_{*}})^2+\mathcal{O}(\epsilon^{\nu}_{\bold{j}_{*}})+\frac{\Big[\epsilon_{\bold{j}_{*}}+1-\delta\sqrt{\epsilon_{\bold{j}_{*}}^2+2\epsilon_{\bold{j}_{*}}}\,\Big]}{N-i+1}(1+\epsilon_{\bold{j}_{*}})-\frac{\Big[\epsilon_{\bold{j}_{*}}+1+\delta\sqrt{\epsilon_{\bold{j}_{*}}^2+2\epsilon_{\bold{j}_{*}}}\,\Big]}{N-i+1}(1+\epsilon_{\bold{j}_{*}})\quad\quad\\
& &-\frac{\Big[\epsilon_{\bold{j}_{*}}+1-\delta\sqrt{\epsilon_{\bold{j}_{*}}^2+2\epsilon_{\bold{j}_{*}}}\,\Big]}{N-i+1}\frac{\Big[\epsilon_{\bold{j}_{*}}+1+\delta\sqrt{\epsilon_{\bold{j}_{*}}^2+2\epsilon_{\bold{j}_{*}}}\,\Big]}{N-i+1}\\
& =&(1+\epsilon_{\bold{j}_{*}})^2+\mathcal{O}(\epsilon^{\nu}_{\bold{j}_{*}})-\frac{\Big[2(\epsilon_{\bold{j}_{*}}+1)\delta\sqrt{\epsilon_{\bold{j}_{*}}^2+2\epsilon_{\bold{j}_{*}}}\,\Big]}{N-i+1}-\frac{\Big[(\epsilon_{\bold{j}_{*}}+1)^2-\delta^2(\epsilon_{\bold{j}_{*}}^2+2\epsilon_{\bold{j}_{*}})\,\Big]}{(N-i+1)^2}\\
&=&1+a_{\epsilon_{\bold{j}_{*}}}-\frac{2b^{(\delta)}_{\epsilon_{\bold{j}_{*}}}}{N-i+1}-\frac{1-c^{(\delta)}_{\epsilon_{\bold{j}_{*}}}}{(N-i+1)^2}\label{comp-fin}
\end{eqnarray}
using the definitions in (\ref{a}),(\ref{b}), and (\ref{c}).
\\

With regard to the inequality in (\ref{ineq-W}), by picking
\begin{equation}
z= E^{Bog}_{\bold{j}_*}+ (\delta-1)\phi_{\bold{j}_*}\sqrt{\epsilon_{\bold{j}_*}^2+2\epsilon_{\bold{j}_*}}
\end{equation} with $1+\frac{2\sqrt{2}+3}{6}\sqrt{\epsilon}\leq \delta\leq 1+\sqrt{\epsilon_{\bold{j}_*}}$, we can write $$-\frac{z}{\phi_{\bold{j}_*}}=\Big[\epsilon _{\bold{j}_*} +1-\delta \sqrt{\epsilon _{\bold{j}_*} ^2+2\epsilon _{\bold{j}_*}}\,\Big]\,$$ and (for $2\leq i\leq N-2$)
\begin{eqnarray}
& &\mathcal{W}_{\bold{j}_*\,;\,i,i-2}(z)\mathcal{W}^*_{\bold{j}_*\,;\,i-2,i}(z)|_{z=E^{Bog}_{\bold{j}_*}+ (\delta-1)\phi_{\bold{j}_*}\sqrt{\epsilon_{\bold{j}_*}^2+2\epsilon_{\bold{j}_*}}}\label{5.7}\\
&= & \frac{(i-1)i}{N^2}\,\phi^2_{\bold{j}_{*}}\times \\
&  &\times\,\frac{(N-i+2)^2}{4\Big[(\frac{i}{N}\phi_{\bold{j}_{*}}+(k_{\bold{j}_*})^2)(N-i)-z\Big]\Big[(\frac{i-2}{N}\phi_{\bold{j}_{*}}+(k_{\bold{j}_*})^2)(N-i)+2(\frac{i-2}{N}\phi_{\bold{j}_{*}}+(k_{\bold{j}_*})^2)-z\Big]}|_{z=E^{Bog}_{\bold{j}_*}+ (\delta-1)\phi_{\bold{j}^*}\sqrt{\epsilon_{\bold{j}_*}^2+2\epsilon_{\bold{j}_*}}}\quad\quad\quad\\
&=&\frac{1}{4\Big[(1+\frac{N}{i}\epsilon_{\bold{j}_*})(1-\frac{2}{N-i+2})-\frac{N}{i(N-i+2)}\frac{z}{\phi_{\bold{j}_{*}}}\Big]}\frac{1}{ \Big[1+\frac{N}{i-1}\epsilon_{\bold{j}_*}-\frac{1}{i-1}-\frac{1}{N-i+2}\frac{N}{i-1}\frac{z}{\phi_{\bold{j}_{*}}}\Big]}|_{z=E^{Bog}_{\bold{j}_*}+ (\delta-1)\phi_{\bold{j}^*}\sqrt{\epsilon_{\bold{j}_*}^2+2\epsilon_{\bold{j}_*}}}\\
&=&\frac{1}{4\Big\{(1+\frac{N}{i}\epsilon_{\bold{j}_*})(1-\frac{2}{N-i+2})+\frac{N}{i(N-i+2)}\Big[\epsilon_{\bold{j}_*}+1-\delta \sqrt{\epsilon_{\bold{j}_*}^2+2\epsilon_{\bold{j}_*}}\,\Big]\Big\}}\times\\
& &\quad\quad \times  \frac{1}{ \Big\{1+\frac{N}{i-1}\epsilon_{\bold{j}_*}-\frac{1}{i-1}+\frac{N}{(i-1)(N-i+2)}\Big[\epsilon_{\bold{j}_*}+1-\delta \sqrt{\epsilon_{\bold{j}_*}^2+2\epsilon_{\bold{j}_*}}\,\Big]\Big\}}\nonumber\\
&=&\frac{1}{4\Big\{1+\frac{N}{i}\epsilon_{\bold{j}_*}-\frac{2}{N-i+2}+\frac{N}{i(N-i+2)}\Big[-\epsilon_{\bold{j}_*}+1-\delta\sqrt{\epsilon_{\bold{j}_*}^2+2\epsilon_{\bold{j}_*}}\,\Big]\Big\}}\times \label{starting-from}\\
& &\quad\quad \times \frac{1}{ \Big\{1+\frac{N}{i-1}\epsilon_{\bold{j}_*}-\frac{1}{i-1}+\frac{N}{(i-1)(N-i+2)}\Big[\epsilon_{\bold{j}_*}+1-\delta\sqrt{\epsilon_{\bold{j}_*}^2+2\epsilon_{\bold{j}_*}}\,\Big]\Big\}}\,. \nonumber
\end{eqnarray}
Then, it is enough to show that for $a^{(\gamma)}_{\epsilon_{\bold{j}_{*}}}:=2\epsilon_{\bold{j}_*}+c_{\gamma}[\epsilon^2_{\bold{j}_{*}}+\frac{\epsilon_{\bold{j}_*}}{N^{\gamma}}+\frac{1}{N}]$
\begin{eqnarray}
&&\Big\{1+\frac{N}{i}\epsilon_{\bold{j}_*}-\frac{2}{N-i+2}+\frac{N}{i(N-i+2)}\Big[-\epsilon_{\bold{j}_*}+1-\delta\sqrt{\epsilon_{\bold{j}_*}^2+2\epsilon_{\bold{j}_*}}\,\Big]\Big\}\times \label{starting-from}\\
& &\quad\quad \times  \Big\{1+\frac{N}{i-1}\epsilon_{\bold{j}_*}-\frac{1}{i-1}+\frac{N}{(i-1)(N-i+2)}\Big[\epsilon_{\bold{j}_*}+1-\delta\sqrt{\epsilon_{\bold{j}_*}^2+2\epsilon_{\bold{j}_*}}\,\Big]\Big\}\,\\
&\leq &1+a^{(\gamma)}_{\epsilon_{\bold{j}_{*}}}-\frac{2b^{(\delta)}_{\epsilon_{\bold{j}_{*}}}}{N-i+2}-\frac{1-c^{(\delta)}_{\epsilon_{\bold{j}_{*}}}}{(N-i+2)^2}
\end{eqnarray}
provided $i\geq N-N^{1-\gamma}$ and the (positive) constant $c_{\gamma}$ is sufficiently large. We observe that for $i\geq N-N^{1-\gamma}$
\begin{eqnarray}
&&\Big\{1+\frac{N}{i}\epsilon_{\bold{j}_*}-\frac{2}{N-i+2}+\frac{N}{i(N-i+2)}\Big[-\epsilon_{\bold{j}_*}+1-\delta\sqrt{\epsilon_{\bold{j}_*}^2+2\epsilon_{\bold{j}_*}}\,\Big]\Big\}\times \label{pre-step}\\
& &\quad\quad \times  \Big\{1+\frac{N}{i-1}\epsilon_{\bold{j}_*}-\frac{1}{i-1}+\frac{N}{(i-1)(N-i+2)}\Big[\epsilon_{\bold{j}_*}+1-\delta\sqrt{\epsilon_{\bold{j}_*}^2+2\epsilon_{\bold{j}_*}}\,\Big]\Big\}\,\nonumber\\
&=&\Big\{1+\frac{N}{i}\epsilon_{\bold{j}_*}+\frac{2N}{i(N-i+2)}-\frac{2}{N-i+2}-\frac{2N}{i(N-i+2)}+\frac{N}{i(N-i+2)}\Big[-\epsilon_{\bold{j}_*}+1-\delta\sqrt{\epsilon_{\bold{j}_*}^2+2\epsilon_{\bold{j}_*}}\,\Big]\Big\}\times \quad\quad\quad\quad \\
& &\quad\quad \times  \Big\{1+\frac{N}{i-1}\epsilon_{\bold{j}_*}-\frac{1}{i-1}+\frac{N}{(i-1)(N-i+2)}\Big[\epsilon_{\bold{j}_*}+1-\delta\sqrt{\epsilon_{\bold{j}_*}^2+2\epsilon_{\bold{j}_*}}\,\Big]\Big\}\,\quad\quad\nonumber\\
&= &\Big\{1+\epsilon_{\bold{j}_*}+\mathcal{O}(\frac{\epsilon_{\bold{j}_*}}{N^{\gamma}})+\mathcal{O}(\frac{1}{N})-\frac{N}{i(N-i+2)}\Big[\epsilon_{\bold{j}_*}+1+\delta\sqrt{\epsilon_{\bold{j}_*}^2+2\epsilon_{\bold{j}_*}}\,\Big]\Big\}\times \label{inter-step}\\
& &\quad\quad \times  \Big\{1+\epsilon_{\bold{j}_*}+\mathcal{O}(\frac{\epsilon_{\bold{j}_*}}{N^{\gamma}})+\mathcal{O}(\frac{1}{N})+\frac{N}{(i-1)(N-i+2)}\Big[\epsilon_{\bold{j}_*}+1-\delta\sqrt{\epsilon_{\bold{j}_*}^2+2\epsilon_{\bold{j}_*}}\,\Big]\Big\}\,\nonumber
\end{eqnarray}
where in the step from (\ref{pre-step}) to (\ref{inter-step}) we have exploited $\frac{N}{i}\epsilon_{\bold{j}_*}=\epsilon_{\bold{j}_*}+\mathcal{O}(\frac{\epsilon_{\bold{j}_*}}{N^{\gamma}})$ and $\frac{2N}{i(N-i+2)}-\frac{2}{N-i+2}=\mathcal{O}(\frac{1}{N})$. Next, making use of $\frac{N}{(i-1)(N-i+2)}-\frac{N}{i(N-i+2)}=\mathcal{O}(\frac{1}{N})$, we estimate
\begin{eqnarray}
& &(\ref{inter-step})\\
&= &(1+\epsilon_{\bold{j}_*})^2+\mathcal{O}(\frac{\epsilon_{\bold{j}_*}}{N^{\gamma}})+\mathcal{O}(\frac{1}{N})\\
& &+(1+\epsilon_{\bold{j}_*})\frac{N}{(i-1)(N-i+2)}\Big[\epsilon_{\bold{j}_*}+1-\delta\sqrt{\epsilon_{\bold{j}_*}^2+2\epsilon_{\bold{j}_*}}\,\Big]\Big\}\\
& &-(1+\epsilon_{\bold{j}_*})\frac{N}{i(N-i+2)}\Big[\epsilon_{\bold{j}_*}+1+\delta\sqrt{\epsilon_{\bold{j}_*}^2+2\epsilon_{\bold{j}_*}}\,\Big]\,\\
& &-\frac{N}{(i-1)(N-i+2)}\frac{N}{i(N-i+2)}\Big[\epsilon_{\bold{j}_*}+1-\delta\sqrt{\epsilon_{\bold{j}_*}^2+2\epsilon_{\bold{j}_*}}\,\Big]\Big[\epsilon_{\bold{j}_*}+1+\delta\sqrt{\epsilon_{\bold{j}_*}^2+2\epsilon_{\bold{j}_*}}\,\Big]\quad\quad\quad\\
&=&(1+\epsilon_{\bold{j}_*})^2+\mathcal{O}(\frac{\epsilon_{\bold{j}_*}}{N^{\gamma}})+\mathcal{O}(\frac{1}{N})\label{W-a}\\
& &-(1+\epsilon_{\bold{j}_*})\frac{N}{(i-1)(N-i+2)}\Big[2\delta\sqrt{\epsilon_{\bold{j}_*}^2+2\epsilon_{\bold{j}_*}}\,\Big]\,\label{W-b}\\
& &-\frac{N}{(i-1)(N-i+2)}\frac{N}{i(N-i+2)}\Big[\epsilon_{\bold{j}_*}+1-\delta\sqrt{\epsilon_{\bold{j}_*}^2+2\epsilon_{\bold{j}_*}}\,\Big]\Big[\epsilon_{\bold{j}_*}+1+\delta\sqrt{\epsilon_{\bold{j}_*}^2+2\epsilon_{\bold{j}_*}}\,\Big]\quad\quad\label{W-c}\\
&<&1+2\epsilon_{\bold{j}_*}+\epsilon_{\bold{j}_*}^2+\mathcal{O}(\frac{\epsilon_{\bold{j}_*}}{N^{\gamma}})+\mathcal{O}(\frac{1}{N})\\
& &-\frac{1}{(N-i+2)}\Big[2\delta(1+\epsilon_{\bold{j}_*})\sqrt{\epsilon_{\bold{j}_*}^2+2\epsilon_{\bold{j}_*}}\,\Big]\,\\
& &-\frac{1}{(N-i+2)}\frac{1}{(N-i+2)}\Big[\epsilon_{\bold{j}_*}+1-\delta\sqrt{\epsilon_{\bold{j}_*}^2+2\epsilon_{\bold{j}_*}}\,\Big]\Big[\epsilon_{\bold{j}_*}+1+\delta\sqrt{\epsilon_{\bold{j}_*}^2+2\epsilon_{\bold{j}_*}}\,\Big]\quad\quad\\
&=&1+2\epsilon_{\bold{j}_*}+\epsilon_{\bold{j}_*}^2+\mathcal{O}(\frac{\epsilon_{\bold{j}_*}}{N^{\gamma}})+\mathcal{O}(\frac{1}{N})-\frac{2b^{(\delta)}_{\epsilon_{\bold{j}_*}}}{(N-i+2)}-\frac{1-c^{(\delta)}_{\epsilon_{\bold{j}_*}}}{(N-i+2)^2}\,.
\end{eqnarray}
Hence, the inequality in (\ref{ineq-W}) holds for a sufficiently large constant $c_{\gamma}$.
\qed

\begin{lemma}\label{lemma-sequence-lower-bound}
Assume $\epsilon>0$ sufficiently small. Consider  for $j\in \mathbb{N}_{0}$ the sequence defined iteratively according to the relation
\begin{eqnarray}\label{sequence-1}
X_{2j+2}&:=&1-\frac{1}{4(1+a_{\epsilon}-\frac{2b_{\epsilon}}{N-2j-1}-\frac{1-c_{\epsilon}}{(N-2j-1)^2})X_{2j}}
%x_{2j+3}&:=&1-\frac{1}{4(1+a_{\epsilon}-\frac{2b_{\epsilon}}{N-2j-1}-\frac{1-c_{\epsilon}}{(N-2j-1)^2})x_{2j+1}}
\end{eqnarray}
with the initial condition $X_{0}=1$  up to $X_{N-2}$. (We recall that $N$ is assumed to be even.)  Here, 
\begin{equation}
a_{\epsilon}:=2\epsilon+\mathcal{O}(\epsilon^{\nu})\,,\quad  \nu >\frac{11}{8}\,,\label{a-final}
\end{equation}
\begin{equation}
b_{\epsilon}:=(1+\epsilon)\delta \, \chi_{[0,2)}(\delta)\sqrt{\epsilon^2+2\epsilon}\,\Big|_{\delta=1+\sqrt{\epsilon}}\label{b-final}
\end{equation}
and
\begin{equation}
c_{\epsilon}:=-(1-\delta^2 \, \chi_{[0,2)}(\delta))(\epsilon^2+2\epsilon)\,\Big|_{\delta=1+\sqrt{\epsilon}}\,\label{c-final}
\end{equation}
with $\chi_{[0,2)}$ the characteristic function of the interval $[0,2)$.
Then, the following estimate holds true for $2\leq N-2j\leq  N $,
\begin{equation}
X_{2j}\geq\frac{1}{2}\Big[1+\sqrt{\eta a_{\epsilon}}-\frac{b_{\epsilon}/ \sqrt{\eta a_{\epsilon}}}{N-2j-\xi} \Big]\,.\label{inf-bound}
\end{equation}
with $\eta=1-\sqrt{\epsilon}$, $\xi=\epsilon^{\Theta}$ where  $\Theta:=\min\{2(\nu-\frac{11}{8})\,;\,\frac{1}{4}\}$.
\end{lemma}
%\noindent
%For  $2\left \lfloor{\frac{b_{\epsilon}}{a_{\epsilon}}}\right \rfloor \leq N-2j \leq 2\frac{rb_{\epsilon}}{a_{\epsilon}}$ with $r$ such that $2\frac{rb_{\epsilon}}{a_{\epsilon}}=N$
%\[
%x_{2j}\geq\frac{1}{2}\Big[1+r\frac{b_{\epsilon}}{r-1}-\frac{a_{\epsilon}(N-2j)}{2(r-1)}-\frac{1}{N-2j+1-\xi}\Big]
%\]
%\end{lemma}

\noindent
\emph{Proof}

\noindent
%We show the proof for the sequence $x_{2j}$ and in the case $N$ even. The other cases can be treated in the same way.
%\\

By setting $2l:=N-2j$ and $Y_{2l}:=X_{2j}$, the statement of the lemma can be re-phrased in terms of the sequence defined by the relation
\begin{eqnarray}\label{formula-y}
Y_{2l-2}&:=&1-\frac{1}{4(1+a_{\epsilon}-\frac{2b_{\epsilon}}{2l-1}-\frac{1-c_{\epsilon}}{(2l-1)^2})Y_{2l}}\,
\end{eqnarray}
and starting from $Y_{N}\equiv 1$ down to $Y_{2}$.

\begin{remark}
We observe that for $\epsilon=0$ and initial value $Y_{N}=\frac{1}{2}(1-\frac{1}{N})$,  the sequence $Y_{2l}$ can be explicitly computed. Indeed, for $Y_{2l}=\frac{1}{2}(1-\frac{1}{2l})$ we have
\begin{eqnarray}\label{exact-sol}
Y_{2l-2}=\frac{1}{2}(1-\frac{1}{2l-2})&=&1-\frac{1}{4(1-\frac{1}{(2l-1)^2})\frac{1}{2}(1-\frac{1}{2l})}=1-\frac{1}{4(1-\frac{1}{(2l-1)^2})Y_{2l}}\,.
\end{eqnarray}
\end{remark}
%We distinguish two ranges, $R1$ and $R2$, of the index $l$, which are associated with two different inductive hypotheses:
%\begin{itemize}
%\item the range $1\leq l\leq \left \lfloor{\frac{b_{\epsilon}}{a_{\epsilon}}} \right \rfloor $ where $ \left \lfloor{q}\right \rfloor $ is the integer part of $q\in \mathbb{R}$
%\item the range $\left \lfloor{\frac{b_{\epsilon}}{a_{\epsilon}}}\right \rfloor \leq l \leq \frac{rb_{\epsilon}}{a_{\epsilon}}$ with $r$ such that $\frac{2rb_{\epsilon}}{a_{\epsilon}}=N$
%\end{itemize}

%\noindent
%\emph{\underline{Range $R1$}}

For $\epsilon$ small enough we consider the following inductive
hypothesis 
\begin{equation}\label{inductive-hp}
Y_{2l}\geq\frac{1}{2}\Big[1+\sqrt{\eta a_{\epsilon}}-\frac{b_{\epsilon}/ \sqrt{\eta a_{\epsilon}}}{2l-\xi }\Big]\,,
\end{equation}
with $\eta=1-\sqrt{\epsilon}$ and  $0< \xi < 1$. % where $ \left \lfloor{q}\right \rfloor $ is the integer part of $q\in \mathbb{R}$.
 % and $h=2-l\left \lfloor{\frac{a_{\epsilon}}{b_{\epsilon}}} \right \rfloor$ is a number between $1$ and $2$ to be determined.

We observe that (\ref{inductive-hp}) is fulfilled for $2l=N$ and $\epsilon$ sufficiently small by the initial condition $Y_N\equiv 1$. The inductive proof amounts to check the inequality
$$
1-\frac{1}{4(1+a_{\epsilon}-\frac{2b_{\epsilon}}{2l-1}-\frac{1-c_{\epsilon}}{(2l-1)^2})\frac{1}{2}\Big[1+\sqrt{\eta a_{\epsilon}}-\frac{b_{\epsilon}/ \sqrt{\eta a_{\epsilon}}}{2l-\xi }\Big]}\geq \frac{1}{2}\Big[1+\sqrt{\eta a_{\epsilon}}-\frac{b_{\epsilon}/ \sqrt{\eta a_{\epsilon}}}{2l-2-\xi }\Big]
$$
%\begin{eqnarray}
% &  & (1-b_{\epsilon}+\frac{1}{2l-2-\xi })(1+b_{\epsilon}-\frac{1}{2l-\xi })(1+a_{\epsilon}-\frac{2b_{\epsilon}}{2l-1}-\frac{1-c_{\epsilon}}{(2l-1)^{2}})
%  \geq  1
%\end{eqnarray}
that is equivalent to
\begin{eqnarray}\label{inductive-rel}
 &  &f(l):= (1-\sqrt{\eta a_{\epsilon}}+\frac{b_{\epsilon}/\sqrt{\eta a_{\epsilon}}}{2l-2-\xi })(1+\sqrt{\eta a_{\epsilon}}-\frac{b_{\epsilon}/\sqrt{\eta a_{\epsilon}}}{2l-\xi })(1+a_{\epsilon}-\frac{2b_{\epsilon}}{2l-1}-\frac{1-c_{\epsilon}}{(2l-1)^{2}})
  \geq  1 \quad\quad\quad
\end{eqnarray}
for any $2\leq l\leq \frac{N}{2} $.
A lenghty calculation shows that
\begin{eqnarray}
f(l) & = & (1-\sqrt{\eta a_{\epsilon}}+\frac{b_{\epsilon}}{2l-2-\xi }+\sqrt{\eta a_{\epsilon}}-\eta a_{\epsilon}+\frac{b_{\epsilon}/\sqrt{\eta a_{\epsilon}}}{2l-2-\xi }-\frac{b_{\epsilon}/\sqrt{\eta a_{\epsilon}}}{2l-\xi }+\frac{b_{\epsilon}}{2l-\xi }-\frac{b^2_{\epsilon}/(\eta a_{\epsilon})}{(2l-2-\xi)(2l-\xi)})\quad\quad\quad\\
 &  &\quad\quad \times(1+a_{\epsilon}-\frac{2b_{\epsilon}}{2l-1}-\frac{1-c_{\epsilon}}{(2l-1)^{2}})\\
& = & 1+a_{\epsilon}(1-\eta)-\eta a_{\epsilon}^{2}+\frac{2\eta b_{\epsilon}a_{\epsilon}}{2l-1}+\frac{\eta a_{\epsilon}(1-c_{\epsilon})}{(2l-1)^{2}}\\
 &  & +\frac{b_{\epsilon}}{2l-\xi }+\frac{a_{\epsilon}b_{\epsilon}}{2l-\xi }-\frac{2b_{\epsilon}^{2}}{(2l-\xi )(2l-1)}-\frac{b_{\epsilon}(1-c_{\epsilon})}{(2l-\xi )(2l-1)^{2}}+\frac{b_{\epsilon}}{2l-2-\xi }-\frac{2b_{\epsilon}}{2l-1}+\frac{a_{\epsilon}b_{\epsilon}}{2l-2-\xi }\\
 &  & -\frac{2b_{\epsilon}^{2}}{(2l-2-\xi )(2l-1)}-\frac{b_{\epsilon}(1-c_{\epsilon})}{(2l-2-\xi )(2l-1)^{2}}+\frac{2(b_{\epsilon}/\sqrt{\eta a_{\epsilon}})-(b_{\epsilon}/\sqrt{\eta  a_{\epsilon}})^2}{(2l-2-\xi)(2l-\xi)}-\frac{1-c_{\epsilon}}{(2l-1)^{2}}\label{third-line}\\
 &  & +\frac{a_{\epsilon}[2(b_{\epsilon}/\sqrt{\eta  a_{\epsilon}})-(b_{\epsilon}/\sqrt{\eta  a_{\epsilon}})^2]}{(2l-2-\xi)(2l-\xi)}-\frac{2b_{\epsilon}[2(b_{\epsilon}/\sqrt{\eta  a_{\epsilon}})-(b_{\epsilon}/\sqrt{\eta  a_{\epsilon}})^2]}{(2l-1)(2l-2-\xi)(2l-\xi)}\label{fourth-line}\\
 & &-\frac{(1-c_{\epsilon})[2(b_{\epsilon}/\sqrt{\eta  a_{\epsilon}})-(b_{\epsilon}/\sqrt{\eta  a_{\epsilon}})^2]}{(2l-1)^{2}(2l-2-\xi)(2l-\xi)}\label{fifth-line}\\
 & = & 1\\
 & &+a_{\epsilon}(1-\eta)-\eta a^2_{\epsilon}\\
 & &+\frac{2(b_{\epsilon}/\sqrt{\eta a_{\epsilon}})-(b_{\epsilon}/\sqrt{\eta  a_{\epsilon}})^2}{(2l-2-\xi)(2l-\xi)}-\frac{1}{(2l-1)^{2}}\label{epsilon-indep}\\
 & &+\frac{b_{\epsilon}}{2l-\xi }+\frac{b_{\epsilon}}{2l-2-\xi }-\frac{2b_{\epsilon}}{2l-1} \label{control-2}\\
 &  &+\frac{2\eta b_{\epsilon}a_{\epsilon}}{2l-1}+\frac{\eta a_{\epsilon}(1-c_{\epsilon})}{(2l-1)^{2}}+\frac{a_{\epsilon}b_{\epsilon}}{2l-\xi }+\frac{a_{\epsilon}b_{\epsilon}}{2l-2-\xi } +\frac{a_{\epsilon}[2(b_{\epsilon}/\sqrt{\eta a_{\epsilon}})-(b_{\epsilon}/\sqrt{\eta a_{\epsilon}})^2]}{(2l-2-\xi)(2l-\xi)} +\frac{c_{\epsilon}}{(2l-1)^{2}}\label{all-pos}\\
 && -\frac{2(b_{\epsilon}/\sqrt{\eta a_{\epsilon}})-(b_{\epsilon}/\sqrt{\eta  a_{\epsilon}})^2}{(2l-2-\xi)(2l-\xi)}\frac{(1-c_{\epsilon})}{(2l-1)^{2}}\label{all-pos-2}\\
 &  &-\frac{2b_{\epsilon}^{2}}{(2l-\xi )(2l-1)}-\frac{2b_{\epsilon}^{2}}{(2l-2-\xi )(2l-1)}\label{fifth-line-bis}\\
 & &-\frac{b_{\epsilon}(1-c_{\epsilon})}{(2l-\xi )(2l-1)^{2}}-\frac{b_{\epsilon}(1-c_{\epsilon})}{(2l-2-\xi )(2l-1)^{2}} -\frac{2b_{\epsilon}[2(b_{\epsilon}/\sqrt{\eta a_{\epsilon}})-(b_{\epsilon}/\sqrt{\eta a_{\epsilon}})^2]}{(2l-1)(2l-2-\xi)(2l-\xi)}\label{fifth-line-bis-bis}
\end{eqnarray}
%\begin{equation}
%b_{\epsilon}:=(1+\epsilon)\delta\sqrt{\epsilon^2+2\epsilon}
%\end{equation}
%where the term in (\ref{epsilon-indep}) comes from the last two terms in (\ref{third-line}) and the last term in (\ref{fourth-line}).
Using the definitions of $a_{\epsilon}$, $b_{\epsilon}$, and $c_{\epsilon}$, we observe that:
\begin{itemize}
\item For $\eta=1-\sqrt{\epsilon}$ and $\epsilon$ small enough
\begin{equation}
a_{\epsilon}(1-\eta)-\eta a^2_{\epsilon}>c_1\epsilon^{\frac{3}{2}};
\end{equation}
for some $c_1>0$ independent of $\epsilon$;
\item In the considered ranges for $\xi$ and $l$, and for $\epsilon$ small,
\begin{equation}
2(b_{\epsilon}/\sqrt{\eta a_{\epsilon}})-(b_{\epsilon}/\sqrt{\eta a_{\epsilon}})^2=1+\mathcal{O}(\epsilon^{2(\nu-1)})+\mathcal{O}(\epsilon)
\end{equation}
so that 
\begin{eqnarray}
(\ref{epsilon-indep})&= &\frac{4l\xi -2\xi -\xi^2+1}{(2l-2-\xi )(2l-1)^2(2l-\xi)}+\mathcal{O}(\frac{1}{l^2}\epsilon^{2(\nu-1)})+\mathcal{O}(\frac{1}{l^2}\epsilon)\\
&>&c_2\frac{\xi}{l^3}+\frac{1}{(2l-2-\xi )(2l-1)^2(2l-\xi)}+\mathcal{O}(\frac{1}{l^2}\epsilon^{2(\nu-1)})+\mathcal{O}(\frac{1}{l^2}\epsilon)
\end{eqnarray}
for some $c_2>0$ independent of $\epsilon$, $\xi$, and $l$;
%\item $\frac{(h-1)2l^2-(1+c_{\epsilon})[(hl)h(l-1)+h-1]}{hl(2l)^{2}h(l-1)}>0$ if $\frac{2(h-1)}{h^2}>(1+c_{\epsilon})=1+(1-\delta^2)(\epsilon^2+2\epsilon)$
%\item  $a_{\epsilon}-b_{\epsilon}^{2}=\epsilon^2+2\epsilon-\mathcal{O}(\epsilon^2)-(1+\epsilon)^2\delta^2(\epsilon^2+2\epsilon)=(1-\delta^2)(\epsilon^2+2\epsilon)+\mathcal{O}(\epsilon^2)$  
\item In the considered ranges for $l$ and $\xi$ 
\begin{equation}
(\ref{control-2})=\frac{b_{\epsilon}}{2l-\xi }+\frac{b_{\epsilon}}{2l-2-\xi }-\frac{2b_{\epsilon}}{2l-1}=\frac{b_{\epsilon}(2+4l\xi-2\xi-2\xi^2)}{(2l-\xi)(2l-2-\xi)(2l-1) }>0\,; \label{control-b}
\end{equation}
%for some $c_3>0$ independent of $\epsilon $, $\xi$, and $l$;
\item In the considered ranges for $\xi$ and $l$, and for $\epsilon$ small, the terms in (\ref{all-pos}) are all positive.
\item  In the considered ranges for $\xi$ and $l$, and for $\epsilon$ small,
\begin{equation}
(\ref{all-pos-2})=-\frac{1}{(2l-2-\xi )(2l-1)^2(2l-\xi)} +\mathcal{O}(\frac{1}{l^4}\epsilon^{2(\nu-1)})+\mathcal{O}(\frac{1}{l^4}\epsilon)
\end{equation}
\item The terms in (\ref{fifth-line-bis}) and (\ref{fifth-line-bis-bis}) are $\mathcal{O}(\frac{1}{l^2}\epsilon)$ and $\mathcal{O}(\frac{1}{l^3}\sqrt{\epsilon})$, respectively.
\end{itemize}
In conclusion, we have to require that 
\begin{equation}\label{ineq-teta}
c_1 \epsilon^{\frac{3}{2}}+c_2\frac{\xi}{l^3}+\mathcal{O}(\frac{1}{l^2}\epsilon^{2(\nu-1)})+\mathcal{O}(\frac{1}{l^3}\sqrt{\epsilon})+\mathcal{O}(\frac{1}{l^2}\epsilon)>0\,.
\end{equation}
This is verified for 
\begin{equation}
2(\nu-1)-\frac{3}{4}=\Theta'>0\quad \text{and}\quad  
\xi=\epsilon^{\Theta}
\end{equation}
with $\Theta:=\min\{\Theta'\,;\,\frac{1}{4}\}$ and $\epsilon$ sufficiently small because:
\begin{itemize}
\item 
$c_2\frac{\xi}{l^3}-\mathcal{O}(\frac{1}{l^3}\sqrt{\epsilon})\geq c_3\frac{\xi}{l^3}$ for some $c_3>0$;
\item
$$\frac{1}{l^2}\epsilon=\epsilon^{\frac{1}{8}}\frac{1}{l^2}\epsilon^{\frac{3}{4}}\epsilon^{\frac{1}{8}}\leq \epsilon^{\frac{1}{8}}[\frac{1}{2}\epsilon^{\frac{3}{2}}+\frac{\epsilon^{\frac{1}{4}}}{2l^4}];$$
%hence 
%$$\mathcal{O}(\frac{1}{l^2}\epsilon)\leq  \mathcal{O}(\epsilon^{\frac{1}{8}}\frac{1}{2}\epsilon^{\frac{3}{2}}\,+\, \epsilon^{\frac{1}{8}}\frac{\epsilon^{\frac{1}{4}}}{2l^4})\,;$$
\item
$$\frac{1}{l^2}\epsilon^{2(\nu-1)}=\frac{1}{l^2}\epsilon^{2(\nu-1)-\frac{3}4}\epsilon^{\frac{3}{4}}=\frac{1}{l^2}\epsilon^{\Theta'}\epsilon^{\frac{3}{4}}=\frac{1}{l^2}\epsilon^{\frac{\Theta'}{2}}\epsilon^{\frac{\Theta'}{2}}\epsilon^{\frac{3}{4}}\leq \epsilon^{\frac{\Theta'}{2}}[\frac{1}{2l^4}\epsilon^{\Theta'}+\frac{\epsilon^{\frac{3}{2}}}{2}]\,.$$
%hence
%$$\mathcal{O}(\frac{1}{l^2}\epsilon^{2(\nu-1)})\leq \mathcal{O}( \epsilon^{\frac{\Theta'}{2}}[\frac{1}{2l^4}\epsilon^{\Theta'}+\frac{\epsilon^{\frac{3}{2}}}{2}])\leq  \mathcal{O}( \epsilon^{\frac{\Theta}{2}}[\frac{1}{2l^4}\epsilon^{\Theta}+\frac{\epsilon^{\frac{3}{2}}}{2}])\,.$$
\end{itemize}
Therefore, for  $\epsilon$ small enough (\ref{ineq-teta}) is fulfilled if
\begin{equation}\label{ineq-teta-bis}
c_1 \epsilon^{\frac{3}{2}}+c_3\frac{\xi}{l^3}+ \mathcal{O}( \epsilon^{\frac{\Theta}{2}}[\frac{1}{2l^4}\epsilon^{\Theta}+\frac{\epsilon^{\frac{3}{2}}}{2}])+ \mathcal{O}(\epsilon^{\frac{1}{8}}\frac{1}{2}\epsilon^{\frac{3}{2}}\,+\, \epsilon^{\frac{1}{8}}\frac{\epsilon^{\frac{1}{4}}}{2l^4})>0\,.
\end{equation}
In fact,  for  $\epsilon$ small enough and $\xi=\epsilon^{\Theta}$,  the sum $c_1 \epsilon^{\frac{3}{2}}+c_2\frac{\xi}{l^3}$ dominates the remaining terms and the inequality in (\ref{ineq-teta}) holds true.
\\

For an analogous sequence $X^{(\delta)}_{2j+2}$ defined by the initial condition $X^{(\delta)}_{i_0}\equiv 1$ for some (even) $0\leq i_0<N-2$ and by the relation (\ref{sequence-1}) but with $b_{\epsilon}^{(\delta)}, c_{\epsilon}^{(\delta)}$ (see (\ref{b})-(\ref{c})) replacing $b_{\epsilon}$ and $c_{\epsilon}$, respectively, we  can show that if $\delta\leq 1+\sqrt{\epsilon}$  then $X_{2j}^{(\delta)}\geq X_{2j}$ for $i_0\leq 2j\leq N-2$. This holds because $1=X^{(\delta)}_{i_0}\geq X_{i_0}(>0)$ and 
%We observe that $b_{\epsilon,\delta}\geq b_{\epsilon,\delta'}$ and $c_{\epsilon,\delta}\geq c_{\epsilon,\delta'}$. Furthermore, the property is true for $2j=0$. Hence,
assuming that the property holds for $2j$ we derive
\begin{eqnarray}
X^{(\delta)}_{2j+2}&=&1-\frac{1}{4(1+a_{\epsilon}-\frac{2b_{\epsilon}^{(\delta)}}{N-2j-1}-\frac{1-c_{\epsilon}^{(\delta)}}{(N-2j-1)^2})X_{2j}^{(\delta)}}\label{2.54}\\
& \geq& 1-\frac{1}{4(1+a_{\epsilon}-\frac{2b_{\epsilon}}{N-2j-1}-\frac{1-c_{\epsilon}}{(N-2j-1)^2})X_{2j}^{(\delta)}}\label{2.55}\\
& \geq &1-\frac{1}{4(1+a_{\epsilon}-\frac{2b_{\epsilon}}{N-2j-1}-\frac{1-c_{\epsilon}}{(N-2j-1)^2})X_{2j}}\nonumber\\
&=&X_{2j+2}\\
&\geq&\frac{1}{2}\Big[1+\sqrt{\eta a_{\epsilon}}-\frac{b_{\epsilon}/ \sqrt{\eta a_{\epsilon}}}{N-2j-\xi} \Big]\
\end{eqnarray}
where in the step from (\ref{2.54}) to (\ref{2.55}) we have use that $-\frac{2b_{\epsilon}^{(\delta)}}{N-2j-1}-\frac{1-c_{\epsilon}^{(\delta)}}{(N-2j-1)^2}$ is nonincreasing in $\delta$, for $\epsilon$ and $\delta$ in the assumed ranges.
%the sum controls all the negative terms in (\ref{fifth-line-bis}), (\ref{fifth-line-bis-bis}) and in $(\ref{epsilon-indep})+(\ref{all-pos-2})$ if  $\epsilon$ is small enough and $C_{\{x\}}>0$ is large enough with the constraint $C_{\{x\}}\epsilon^{\frac{1}{2}}<1$.

%Therefore, $h(y)+q(y)$ is always increasing in the given interval and the minimum is attained at $\left \lfloor{\frac{b_{\epsilon}}{a_{\epsilon}}}\right \rfloor$ where $g(\left \lfloor{\frac{b_{\epsilon}}{a_{\epsilon}}}\right \rfloor)\equiv f(\left \lfloor{\frac{b_{\epsilon}}{a_{\epsilon}}}\right \rfloor)\geq1$ where $f$ is defined in the next paragraph \emph{\underline{Range $R2$}} and the last inequality is proven provided $\epsilon$ is sufficiently small. 
\qed

\begin{lemma}\label{lemma-sequence-upper-bound-new}
 Let $0<\gamma<1$ and, for notational simplicity, assume that $N ^{1-\gamma}, \frac{N ^{1-\gamma}}{2}$ are both even. Let  $i_0\equiv N-N ^{1-\gamma}$ and consider  for $j\in \mathbb{N}$ and $j\geq \frac{i_0}{2}$ the sequence defined iteratively according to the relation
\begin{eqnarray}
\tilde{X}^{(\gamma,\delta)}_{2j+2}&:=&1-\frac{1}{4(1+a_{\epsilon}^{(\gamma)}-\frac{2b^{(\delta)}_{\epsilon}}{N-2j}-\frac{1-c^{(\delta)}_{\epsilon}}{(N-2j)^2})\tilde{X}^{(\gamma, \delta)}_{2j}}
%x^{(\gamma)}_{2j+3}&:=&1-\frac{1}{4(1+a_{\epsilon}^{(\gamma)}-\frac{2b_{\epsilon}}{N-2j-1}-\frac{1-c_{\epsilon}}{(N-2j-1)^2})x^{(\gamma)}_{2j+1}}
\end{eqnarray}
with the initial condition $\tilde{X}^{(\gamma,\delta)}_{i_0}=1$ up to  $\tilde{X}^{(\gamma,\delta)}_{2j=N-2}$.  Here, 
\begin{equation}
a_{\epsilon}^{(\gamma)}:=2\epsilon+c_{\gamma}[\frac{\epsilon}{N^{\gamma}}+\frac{1}{N}+\epsilon^2]\,,\quad c_{\gamma}>0,
\end{equation}
\begin{equation}\label{bdelta-0}
b^{(\delta)}_{\epsilon}:=(1+\epsilon)\delta\sqrt{\epsilon^2+2\epsilon}\,,
\end{equation}
and
\begin{equation}\label{cdelta-0}
c^{(\delta)}_{\epsilon}:=-(1-\delta^2)(\epsilon^2+2\epsilon)
\end{equation}
where:
\begin{itemize}
\item
$\epsilon$ is sufficiently small and such that 
\begin{equation} \label{gamma-condition-bis}
\epsilon^2+\frac{\epsilon}{N^{\gamma}}+\frac{1}{N}\leq k_{\gamma}\epsilon \sqrt{\epsilon}\quad\,,\,\quad\frac{1}{N^{1-\gamma}}\leq k_{\gamma}\epsilon\,,
\end{equation} for some constant $k_{\gamma}$ sufficiently small;
\item $1+\frac{2\sqrt{2}+3}{6}\sqrt{\epsilon}\leq \delta\leq 1+\sqrt{\epsilon}$\,.
\end{itemize}
%and $c_{\gamma}$ is a constant sufficiently large to ensure the inequality in (\ref{ineq-W}) for $i\geq N-N^{1-\gamma}$. 
Then, $\tilde{X}^{(\gamma,\delta)}_{2j}>0$ and for $2\leq N-2j\leq \frac{N ^{1-\gamma}}{2}$ the following estimate holds true
\[
(0<)\tilde{X}^{(\gamma,\delta)}_{2j}\leq\frac{1}{2}\Big[1+\sqrt{a_{\epsilon}^{(\gamma)}}-\frac{1}{N-2j+1- b^{(\delta)}_{\epsilon}}\Big]\,.
\]
%\[
%{\color{red}?x^{(\gamma)}_{2j+1}\leq\frac{1}{2}\Big[1+\sqrt{a_{\epsilon}}-\frac{1}{N-2j-1- b_{\epsilon}}\Big]?\,;}
%\]
%\noindent
%For  $2\left \lfloor{\frac{b_{\epsilon}}{a_{\epsilon}}}\right \rfloor \leq N-2j \leq 2\frac{rb_{\epsilon}}{a_{\epsilon}}$ with $r$ such that $2\frac{rb_{\epsilon}}{a_{\epsilon}}=N$
%\[
%{\color{red}x_{2j}^{(\gamma)}\leq\frac{1}{2}\Big[1+r\frac{b_{\epsilon}}{r-1}-\frac{a_{\epsilon}(N-2j)}{2(r-1)}-\frac{1}{N-2j}\Big]}
%\]
%\[
%{\color{red}x_{2j+1}^{(\gamma)}\leq\frac{1}{2}\Big[1+r\frac{b_{\epsilon}}{r-1}-\frac{a_{\epsilon}(N-2j-1)}{2(r-1)}-\frac{1}{N-2j-1}\Big]}
%\]
\end{lemma}

\noindent
\emph{Proof}

\noindent
By setting $2l:=N-2j$ and $Y^{(\gamma, \delta)}_{2l}:=\tilde{X}^{(\gamma, \delta)}_{2j}$, the statement of the lemma can be re-phrased in terms of the sequence defined by the relation
\begin{eqnarray}
Y^{(\gamma,\delta)}_{2l-2}&:=&1-\frac{1}{4(1+a_{\epsilon}^{(\gamma)}-\frac{2b^{(\delta)}_{\epsilon}}{2l}-\frac{1-c^{(\delta)}_{\epsilon}}{4l^2})Y^{(\gamma,\delta)}_{2l}}\,,
\end{eqnarray}
starting from $Y^{(\gamma,\delta)}_{2l=N ^{1-\gamma}}=1$ down to $Y^{(\gamma,\delta)}_{2l-2\equiv 2}$. Since $1+\frac{2\sqrt{2}+3}{6}\sqrt{\epsilon}\leq \delta\leq 1+\sqrt{\epsilon}$, the same arguments of Lemma \ref{lemma-sequence-lower-bound} ensure that $1\geq Y^{(\gamma,\delta)}_{2l}>0$ if $\epsilon$ is small enough, so that the sequence is well defined in the considered range for $2l$.

Provided $\epsilon$ is small enough,  and provided the inequalities in (\ref{gamma-condition-bis}) are satisfied,  we shall prove that
\[
Y^{(\gamma,\delta)}_{2l}\leq\frac{1}{2}\Big[1+\sqrt{a_{\epsilon}^{(\gamma)}}-\frac{1}{2l+1- b^{(\delta)}_{\epsilon}}\Big]\,
\]
for $2\leq 2l<\frac{N ^{1-\gamma}}{2}$ assuming that it is true for $2l=\frac{N ^{1-\gamma}}{2}$. The latter assumption will be shown to be satisfied in the final part of the lemma.
 % where $ \left \lfloor{q}\right \rfloor $ is the integer part of $q\in \mathbb{R}$.
 % and $h=2-l\left \lfloor{\frac{a_{\epsilon}}{b_{\epsilon}}} \right \rfloor$ is a number between $1$ and $2$ to be determined.
 \\
 
Similarly to (\ref{inductive-rel}) in Lemma \ref{lemma-sequence-lower-bound}, it is enough to check that, for $4\leq 2l\leq  \frac{N ^{1-\gamma}}{2}$,  the maximum of 
\begin{eqnarray}
 &  & f(l): =  (1-\sqrt{a_{\epsilon}^{(\gamma)}}+\frac{1}{2l-1-b^{(\delta)}_{\epsilon} })(1+\sqrt{a_{\epsilon}^{(\gamma)}}-\frac{1}{2l+1- b^{(\delta)}_{\epsilon} })(1+a^{(\gamma)}_{\epsilon}-\frac{b^{(\delta)}_{\epsilon}}{l}-\frac{1-c^{(\delta)}_{\epsilon}}{(2l)^{2}})\quad\quad\quad
\end{eqnarray}
is smaller than or equal to $1$. In the computation below is helpful to recall that  $a^{(\gamma)}_{\epsilon}=\mathcal{O}(\epsilon), b^{(\delta)}_{\epsilon}=\mathcal{O}(\epsilon^{\frac{1}{2}})$, and  $c^{(\delta)}_{\epsilon}=\mathcal{O}(\epsilon^{\frac{1}{2}}\epsilon)$ in the considered range of $\delta$. We get
\begin{eqnarray}
f(l) & = & 1+\frac{(4lb^{(\delta)}_{\epsilon}-b^2_{\epsilon})(1-c^{(\delta)}_{\epsilon})+4l^2c^{(\delta)}_{\epsilon}}{(2l)^{2}(4l^2-1-4lb^{(\delta)}_{\epsilon} +(b^{(\delta)}_{\epsilon})^2 )}\\
 & &+\frac{\sqrt{a_{\epsilon}^{(\gamma)}}}{2l+1-b^{(\delta)}_{\epsilon}}+\frac{\sqrt{a_{\epsilon}^{(\gamma)}}}{2l-1-b^{(\delta)}_{\epsilon} }-\frac{b^{(\delta)}_{\epsilon}}{l}\\
 &  & -(a_{\epsilon}^{(\gamma)})^{2}+\frac{a_{\epsilon}^{(\gamma)}b^{(\delta)}_{\epsilon}}{l}+\frac{a_{\epsilon}^{(\gamma)}(1-c^{(\delta)}_{\epsilon})}{(2l)^{2}}+\frac{a_{\epsilon}^{(\gamma)}\sqrt{a_{\epsilon}^{(\gamma)}}}{2l+1-b^{(\delta)}_{\epsilon} }-\frac{b^{(\delta)}_{\epsilon}\sqrt{a_{\epsilon}^{(\gamma)}}}{(2l-1-b^{(\delta)}_{\epsilon} )l}-\frac{\sqrt{a_{\epsilon}^{(\gamma)}}(1-c^{(\delta)}_{\epsilon})}{(2l-1-b^{(\delta)}_{\epsilon})(2l)^{2}}+\frac{a_{\epsilon}^{(\gamma)}\sqrt{a_{\epsilon}^{(\gamma)}}}{2l-1-b^{(\delta)}_{\epsilon}}\nonumber \\
 &  & -\frac{b^{(\delta)}_{\epsilon}\sqrt{a_{\epsilon}^{(\gamma)}}}{(2l+1-b^{(\delta)}_{\epsilon} )l}-\frac{\sqrt{a_{\epsilon}^{(\gamma)}}(1-c^{(\delta)}_{\epsilon})}{(2l+1-b^{(\delta)}_{\epsilon} )(2l)^{2}}+\frac{a_{\epsilon}^{(\gamma)}}{4l^2-1-4lb^{(\delta)}_{\epsilon}+(b^{(\delta)}_{\epsilon})^2 }-\frac{b^{(\delta)}_{\epsilon}}{l(4l^2-1-4lb^{(\delta)}_{\epsilon} +(b^{(\delta)}_{\epsilon})^2 )}\nonumber \\
 & = & 1-\frac{(b^{(\delta)}_{\epsilon})^2}{(2l)^{2}(4l^2-1)}-(a_{\epsilon}^{(\gamma)})^{2}\\
& & +\frac{a_{\epsilon}^{(\gamma)}}{(2l)^{2}}+\frac{a_{\epsilon}^{(\gamma)}}{4l^2-1 }-\frac{b^{(\delta)}_{\epsilon}\sqrt{a_{\epsilon}^{(\gamma)}}}{(2l-1)l} -\frac{b^{(\delta)}_{\epsilon}\sqrt{a_{\epsilon}^{(\gamma)}}}{(2l+1)l}\label{gamma-sequence-1}\\
 & &+\frac{\sqrt{a_{\epsilon}^{(\gamma)}}}{2l+1-b^{(\delta)}_{\epsilon}}+\frac{\sqrt{a_{\epsilon}^{(\gamma)}}}{2l-1-b^{(\delta)}_{\epsilon} }-\frac{b^{(\delta)}_{\epsilon}}{l}\label{gamma-sequence-2}\\
 &  &-\frac{\sqrt{a_{\epsilon}^{(\gamma)}}}{(2l+1-b^{(\delta)}_{\epsilon} )(2l)^{2}}-\frac{\sqrt{a_{\epsilon}^{(\gamma)}}}{(2l-1-b^{(\delta)}_{\epsilon})(2l)^{2}}\label{gamma-sequence-3}\\
% &  & +\frac{a_{\epsilon}^{(\gamma)}}{(2l)^{2}}+\frac{a_{\epsilon}^{(\gamma)}}{4l^2-1} \label{gamma-sequence-4}\\
 & &-\frac{b^{(\delta)}_{\epsilon}}{l(4l^2-1-4lb^{(\delta)}_{\epsilon})}+\frac{4lb^{(\delta)}_{\epsilon}}{(2l)^{2}(4l^2-1-4lb^{(\delta)}_{\epsilon})}\label{gamma-sequence-5}\\
 & &+\frac{1}{l}o(\epsilon)
\end{eqnarray}
First we observe that due to the assumption in (\ref{gamma-condition-bis}) we can write  
\begin{equation}\label{bound-epsilon}
b^{(\delta)}_{\epsilon}\geq \sqrt{a_{\epsilon}^{(\gamma)}}+[\frac{2\sqrt{2}+3}{6}+k'_{\gamma}]\epsilon
\end{equation}
 where $|k'_{\gamma}|>0$ can be made arbitrarily small provided $k_{\gamma}>0$ is sufficiently small, in particular we consider $|k'_{\gamma}|<\frac{2\sqrt{2}+3}{6}$. We point out that:

\begin{itemize}
\item
Because of (\ref{bound-epsilon}) the sum of the terms in (\ref{gamma-sequence-1}) is negative;
\item
The term in (\ref{gamma-sequence-5}) is identically zero;
\item 
As far as (\ref{gamma-sequence-2}) and (\ref{gamma-sequence-3}) are concerned, due to (\ref{bound-epsilon}) we can write \begin{eqnarray}
(\ref{gamma-sequence-2})&\leq &\frac{\sqrt{a_{\epsilon}^{(\gamma)}}}{2l+1-b_{\epsilon}}+\frac{\sqrt{a_{\epsilon}^{(\gamma)}}}{2l-1-b_{\epsilon} }-\frac{\sqrt{a_{\epsilon}^{(\gamma)}}}{l}-\frac{[(\frac{2\sqrt{2}+3}{6})+k'_{\gamma}]\epsilon}{l}\\
%&= &\frac{\sqrt{a_{\epsilon}^{(\gamma)}}}{2l+1-b_{\epsilon}}+\frac{\sqrt{a_{\epsilon}^{(\gamma)}}}{2l-1-b_{\epsilon} }-\frac{\sqrt{a_{\epsilon}^{(\gamma)}}}{l}-\frac{\sqrt{2}\epsilon}{l}-\mathcal{O}(\frac{\epsilon\sqrt{\epsilon}}{l})\\
&=&\frac{\sqrt{a_{\epsilon}^{(\gamma)}}(2lb^{(\delta)}_{\epsilon}-(b^{(\delta)}_{\epsilon})^2+1)}{(2l+1-b^{(\delta)}_{\epsilon})(2l-1-b^{(\delta)}_{\epsilon})l}-\frac{[(\frac{2\sqrt{2}+3}{6})+k'_{\gamma}]\epsilon}{l}\end{eqnarray}
and
\begin{eqnarray}
(\ref{gamma-sequence-3})&=&-\frac{\sqrt{a_{\epsilon}^{(\gamma)}}}{(2l+1-b^{(\delta)}_{\epsilon} )(2l)^{2}}-\frac{\sqrt{a_{\epsilon}^{(\gamma)}}}{(2l-1-b^{(\delta)}_{\epsilon})(2l)^{2}}\\
& =&-\frac{\sqrt{a_{\epsilon}^{(\gamma)}}(4l-2b^{(\delta)}_{\epsilon})}{(2l+1-b^{(\delta)}_{\epsilon} )(2l-1-b^{(\delta)}_{\epsilon})(2l)^{2}}\,,
\end{eqnarray}
hence
\begin{eqnarray}
(\ref{gamma-sequence-2})+(\ref{gamma-sequence-3})&\leq &\frac{\sqrt{a_{\epsilon}^{(\gamma)}}(4l^2b^{(\delta)}_{\epsilon}-2l(b^{(\delta)}_{\epsilon})^2+b^{(\delta)}_{\epsilon})}{(2l+1-b^{(\delta)}_{\epsilon})(2l-1-b^{(\delta)}_{\epsilon})2l^2}-\frac{[\frac{2\sqrt{2}+3}{6}+k'_{\gamma}]\epsilon}{l}\,; \label{sum}
\end{eqnarray}
\item
Concerning (\ref{sum}),  we notice that
\begin{eqnarray}
& &\frac{\sqrt{a_{\epsilon}^{(\gamma)}}(4l^2b^{(\delta)}_{\epsilon}+b^{(\delta)}_{\epsilon})}{(2l+1-b^{(\delta)}_{\epsilon})(2l-1-b^{(\delta)}_{\epsilon})2l^2}\\
&=&\frac{4\epsilon}{(2l+1-b^{(\delta)}_{\epsilon})(2l-1-b^{(\delta)}_{\epsilon})}+\frac{\epsilon}{(2l+1-b^{(\delta)}_{\epsilon})(2l-1-b^{(\delta)}_{\epsilon})l^2}\\
& &+\frac{1}{l^2}o(\epsilon)\,.
\end{eqnarray}
Furthermore, since $l\geq 2$, for $\epsilon$ and $|k'_{\gamma}|$ sufficiently small
\begin{equation}
\frac{4\epsilon}{(2l+1-b^{(\delta)}_{\epsilon})(2l-1-b^{(\delta)}_{\epsilon})}+\frac{\epsilon}{(2l+1-b^{(\delta)}_{\epsilon})(2l-1-b^{(\delta)}_{\epsilon})l^2}-\frac{[\frac{2\sqrt{2}+3}{6}+k'_{\gamma}]\epsilon}{l}< -c\frac{\epsilon}{l}
\end{equation}
for some $c>0$.
\end{itemize}

These observations show that $f(l)<1$ for $\epsilon$ sufficiently small.
\\

Now we prove that in fact $Y^{(\gamma,\delta)}_{2l}\leq \frac{1}{2}\Big[1+\sqrt{a^{(\gamma)}_{\epsilon}}-\frac{1}{2l+1- b^{(\delta)}_{\epsilon}}\Big]$ for $2l=\frac{N ^{1-\gamma}}{2}$ and $\epsilon$ sufficiently small. Starting from the definition
\begin{eqnarray}
Y^{(\gamma,\delta)}_{2l-2}&:=&1-\frac{1}{4(1+a^{(\gamma)}_{\epsilon}-\frac{2b^{(\delta)}_{\epsilon}}{2l}-\frac{1-c^{(\delta)}_{\epsilon}}{4l^2})Y^{(\gamma,\delta)}_{2l}}\,,\quad Y^{(\gamma,\delta)}_{2l=N ^{1-\gamma}}=1\,,
\end{eqnarray}
we observe that for $\frac{N ^{1-\gamma}}{2}\leq 2l\leq N ^{1-\gamma}$ the inequality $Y^{(\gamma,\delta)}_{2l}\leq \check{Y}_{2l}$ holds where $\check{Y}_{2l}$ is defined by
\begin{eqnarray}\label{check-y-sequence}
\check{Y}_{2l-2}&:=&1-\frac{1}{4(1+a^{(\gamma)}_{\epsilon})\check{Y}_{2l}}\,
\end{eqnarray}
with $\check{Y}_{N ^{1-\gamma}} \equiv 1$.
Furthermore, the bound
\begin{eqnarray}\label{bound-fixed}
\check{Y}_{2l}&\geq &\check{Y}:=\frac{1}{2}+\frac{1}{2}\sqrt{\frac{a^{(\gamma)}_{\epsilon}}{1+a^{(\gamma)}_{\epsilon}}}\,,
\end{eqnarray}
holds true, where $\check{Y}$ solves the equation
\begin{eqnarray}\label{check-y}
y&=&1-\frac{1}{4(1+a^{(\gamma)}_{\epsilon})y}\,.
\end{eqnarray}
Hence, using (\ref{check-y-sequence}), (\ref{check-y}) and the bound in (\ref{bound-fixed}) we can estimate
\begin{equation}\label{fixed-point-arg}
|\check{Y}-\check{Y}_{2l-2}|= \frac{1}{4(1+a^{(\gamma)}_{\epsilon})}\frac{|\check{Y}-\check{Y}_{2l}|}{\check{Y}\cdot\check{Y}_{2l}}\leq \frac{1}{(1+c\epsilon^{\frac{1}{2}})}|\check{Y}-\check{Y}_{2l}|\leq  [\frac{1}{1+c\epsilon^{\frac{1}{2}}}]^{(N^{1-\gamma}-2l+2)/2}\,|\check{Y}-\check{Y}_{N^{1-\gamma}}|
\end{equation}
for some $c>0$.
Finally, due to the second inequality in (\ref{gamma-condition-bis}), we can conclude that if $\kappa_{\gamma}$ is sufficiently small then
$$
Y^{(\gamma,\delta)}_{\frac{N ^{1-\gamma}}{2}}\leq \check{Y}_{\frac{N ^{1-\gamma}}{2}}=\check{Y}_{\frac{N ^{1-\gamma}}{2}}-\check{Y}+\check{Y}\leq \mathcal{O}( [\frac{1}{1+c\epsilon^{\frac{1}{2}}}]^{\frac{N^{1-\gamma}}{4}})+\frac{1}{2}+\frac{1}{2}\sqrt{\frac{a^{(\gamma)}_{\epsilon}}{1+a^{(\gamma)}_{\epsilon}}}\leq \frac{1}{2}\Big[1+\sqrt{a^{(\gamma)}_{\epsilon}}-\frac{1}{\frac{N ^{1-\gamma}}{2}+1- b^{(\delta)}_{\epsilon}}\Big]\,
$$
for $\epsilon$ sufficiently small.
\qed

In the next lemma we estimate the difference between the ground state energy, $z_*$, of $H^{Bog}_{\bold{j}_*}$ and $E^{Bog}_{\bold{j}_*}$. 
\begin{lemma}\label{eigenvalue}
Let $\epsilon_{\bold{j}_*}$ be sufficiently small  and $N$ be sufficiently large to fulfill the assumptions of Theorem \ref{fixed-p-thm} (therefore we implicitly assume Proposition \ref{lemma-expansion-proof-0}, Lemma \ref{inversion}, and Lemma \ref{lemma-sequence-lower-bound}).   Then, for some $c>0$ the estimate
\begin{equation}
|z_*-E^{Bog}_{\bold{j}_*}|\leq \mathcal{O}(\frac{1}{\epsilon_{\bold{j}_*}N^{\beta}})+\mathcal{O}(\frac{1}{\sqrt{\epsilon_{\bold{j}_*}}}[\frac{1}{1+c\sqrt{\epsilon_{\bold{j}_*}}}]^{N^{1-\beta}})+\mathcal{O}(\frac{1}{N})\,,\quad 0<\beta <1\,,\label{diff}
\end{equation}
holds true  provided $\frac{1}{N^{\beta}}=o(\epsilon_{\bold{j}_*})\,,\,\frac{1}{N^{1-\beta}}=o(\sqrt{\epsilon_{\bold{j}_*}})$. 
%and  $$\frac{2b_{\epsilon_{\bold{j}_*}}}{N^{1-\beta}/2}+\frac{1-c_{\epsilon_{\bold{j}_*}}}{(N^{1-\beta}/2)^2}\leq \frac{a'_{\epsilon_{\bold{j}_*}}}{2}\,,\quad a'_{\epsilon_{\bold{j}_*}}:=\epsilon_{\bold{j}_*}^2+2\epsilon_{\bold{j}_*}\,,$$ where $b_{\epsilon_{\bold{j}_*}}$ and $c_{\epsilon_{\bold{j}_*}}$ are calculated at $\delta=1$.
\end{lemma}

\noindent
\emph{Proof}

\noindent
From Remark \ref{increasing} we know that $$f_{\bold{j}_*}(z):=-z-(1-\frac{1}{N})\frac{\phi_{\bold{j}_*}^2}{\phi_{\bold{j}_*}(2\epsilon_{\bold{j}_*}+2-\frac{4}{N})-z}\check{\mathcal{G}}_{\bold{j}_*\,;\,N-2,N-2}(z)$$ has derivative not larger than $-1$ for $z$ in the interval (\ref{interval-of-def}). Hence, for $z$ in this interval we can estimate
\begin{eqnarray}\label{deriv}
& &|z-E^{Bog}_{\bold{j}_*}|\\
&\leq &|f_{\bold{j}_*}(z)-f_{\bold{j}_*}(E^{Bog}_{\bold{j}_*}) |\nonumber\\
&= &|z+(1-\frac{1}{N})\frac{\phi_{\bold{j}_*}^2}{\phi_{\bold{j}_*}(2\epsilon_{\bold{j}_*}+2-\frac{4}{N})-z}\check{\mathcal{G}}_{\bold{j}_*\,;\,N-2,N-2}(z)\\
& &\quad\quad\quad\quad -[E^{Bog}_{\bold{j}_*}+(1-\frac{1}{N})\frac{\phi_{\bold{j}_*}^2}{\phi_{\bold{j}_*}(2\epsilon_{\bold{j}_*}+2-\frac{4}{N})-E^{Bog}_{\bold{j}_*}}\check{\mathcal{G}}_{\bold{j}_*\,;\,N-2,N-2}(E^{Bog}_{\bold{j}_*})]|\nonumber \,.
\end{eqnarray}
We recall that $f_{\bold{j}_*}(z_*)=0$ by definition of $z_*$. Thus, we deduce that
%using (\ref{EBOG-1})-(\ref{EBOG-2}) and (\ref{deriv}), we can bound
\begin{eqnarray}
|z_*-E^{Bog}_{\bold{j}_*}|&\leq & |-E^{Bog}_{\bold{j}_*}-(1-\frac{1}{N})\frac{\phi_{\bold{j}_*}}{(2\epsilon_{\bold{j}_*}+2-\frac{4}{N})-\frac{E^{Bog}_{\bold{j}_*}}{\phi_{\bold{j}_*}}}\check{\mathcal{G}}_{\bold{j}_*\,;\,N-2,N-2}(E^{Bog}_{\bold{j}_*})|\,.\label{diff-0}
%&\leq &\mathcal{O}(\frac{1}{\epsilon_{\bold{j}_*}N^{\beta}})+\mathcal{O}(\frac{1}{\epsilon_{\bold{j}_*}}[\frac{1}{1+c\sqrt{\epsilon}_{\bold{j}_*}}]^{(N^{1-\beta}-2l_{\epsilon_{\bold{j}_*}})})\,.\label{diff}
\end{eqnarray}
In the remaining part of the proof, we provide an approximated expression, $\frac{1}{[Y_{2l}]_B}$, to $\check{\mathcal{G}}_{\bold{j}_*\,;\,N-2,N-2}(E^{Bog}_{\bold{j}_*})$ through successive steps.
Finally, using this approximation we show that 
\begin{equation}
|(\ref{diff-0})|\leq \mathcal{O}(\frac{1}{\epsilon_{\bold{j}_*}N^{\beta}})+\mathcal{O}(\frac{1}{\sqrt{\epsilon_{\bold{j}_*}}}[\frac{1}{1+c\sqrt{\epsilon}_{\bold{j}_*}}]^{N^{1-\beta}})+\mathcal{O}(\frac{1}{N})\,.
\end{equation}
The approximation of $\check{\mathcal{G}}_{\bold{j}_*\,;\,N-2,N-2}(E^{Bog}_{\bold{j}_*})$ consists of three steps:
\begin{eqnarray}
& &\check{\mathcal{G}}_{\bold{j}_*\,;\,N-2,N-2}(z)\quad \rightarrow \quad [\check{\mathcal{G}}_{\bold{j}_*\,;\,N-2,N-2}]_T(z)\\
& &[\check{\mathcal{G}}_{\bold{j}_*\,;\,N-2,N-2}]_T(z)|_{z=E^{Bog}_{\bold{j}_*}} \quad  \rightarrow \quad \frac{1}{[Y_{2}]_*}\quad\quad\\
& &\frac{1}{[Y_{2}]_*}\quad  \rightarrow \quad \frac{1}{[Y_{2}]_B}
\end{eqnarray}
where the quantities $[\check{\mathcal{G}}_{\bold{j}_*\,;\,N-2,N-2}]_T(z)$, $\frac{1}{[Y_{2}]_*}$, and $\frac{1}{[Y_{2}]_B}$ are defined at points 1), 2), and 3) below where we outline  each step.

\noindent
\emph{For expository convenience, in the following we  assume that $N^{1-\beta}$ is an even number and avoid to introduce $\lfloor N^{1-\beta} \rfloor$ or $\lfloor N^{1-\beta} \rfloor-1$.}
\begin{itemize}
\item [1)] For $z\leq E^{Bog}_{\bold{j}_*}+ \sqrt{\epsilon_{\bold{j}_*}}\phi_{\bold{j}_*}\sqrt{\epsilon_{\bold{j}_*}^2+2\epsilon_{\bold{j}_*}}$, we estimate the difference between $\check{\mathcal{G}}_{\bold{j}_*\,;\,N-2,N-2}(z)$ and $[\check{\mathcal{G}}_{\bold{j}_*\,;\,N-2,N-2}]_T(z)$ where $\check{\mathcal{G}}_{\bold{j}_*\,;\,N-2,N-2}(z)$ is the element corresponding to $i=N-2$ of the sequence defined by
\begin{equation}\label{relation-G}
\check{\mathcal{G}}_{\bold{j}_*\,;\,i,i}(z):=\sum_{l_{i}=0}^{\infty}[\mathcal{W}_{\bold{j}_*\,;i,i-2}(z)\mathcal{W}^*_{\bold{j}_*\,;i-2,i}(z)\check{\mathcal{G}}_{\bold{j}_*\,;\,i-2,i-2}(z)]^{l_i}
\end{equation}
for $i\geq 2$ with $\check{\mathcal{G}}_{\bold{j}_*\,;\,0,0}(z)\equiv 1$, whereas $[\check{\mathcal{G}}_{\bold{j}_*\,;\,N-2,N-2}]_T(z)$ is obtained from the same sequence defined in (\ref{relation-G}) but starting from $i=N-N^{1-\beta}$ and with initial condition $\check{\mathcal{G}}_{\bold{j}_*\,;\,N-N^{1-\beta}-2,N-N^{1-\beta}-2}(z)\equiv 1$.
\item [2)] We consider the sequence $[Y_{2l}]_*$ that is defined by the relation  \begin{eqnarray}\label{reference}
Y_{2l-2}&:=&1-\frac{1}{4(1+a'_{\epsilon}-\frac{2b^{(\delta)}_{\epsilon}}{2l}-\frac{1-c^{(\delta)}_{\epsilon}}{4l^2})Y_{2l}}\,,\quad a'_{\epsilon}:=\epsilon^2+2\epsilon\,,
\end{eqnarray}
where $b^{(\delta)}_{\epsilon},c^{(\delta)}_{\epsilon}$ are calculated at $\delta\equiv 1$ and $\epsilon\equiv \epsilon_{\bold{j}_*}$, and with initial condition  $[Y_{N^{1-\beta}+2}]_*\equiv 1$.

\noindent
{\bf{Warning}}: We recall the reader that the index $i$ is increasing, the index $l$ is decreasing, and their relation is $2l=N-i$.

\noindent
In this step, from the inequality $$\Big|(\mathcal{W}_{\bold{j}_*\,;i,i-2}(z)\mathcal{W}^*_{\bold{j}_*\,;i-2,i}(z)|_{z\equiv E^{Bog}_{\bold{j}_*}}-(\frac{1}{4(1+a'_{\epsilon}-\frac{2b^{(\delta)}_{\epsilon}}{N-i}-\frac{1-c^{(\delta)}_{\epsilon}}{(N-i)^2})}|_{\epsilon\equiv \epsilon_{\bold{j}_*}, \delta=1}\Big|\leq\mathcal{O}(\frac{1}{N^{\beta}}),$$
that holds for $i\geq N-N^{1-\beta}$ we infer $$\Big|\frac{1}{[\check{\mathcal{G}}_{\bold{j}_*\,;\,N-2,N-2}]_T(z)|_{z=E^{Bog}_{\bold{j}_*}}}-[Y_{2}]_*\Big|\leq \mathcal{O}(\frac{1}{\epsilon_{\bold{j}_*} N^{\beta}})\,.$$
% are very close at the step $2l=\frac{N^{\gamma}}{2}$ and then remain close up to $2l=2$
\item [3)] We construct an explicit solution, $[Y_{2l}]_B$, of 
\begin{eqnarray}\label{reference}
Y_{2l-2}&:=&1-\frac{1}{4(1+a'_{\epsilon_{\bold{j}_*}}-\frac{2b^{(\delta)}_{\epsilon_{\bold{j}_*}}}{2l}-\frac{1-c^{(\delta)}_{\epsilon_{\bold{j}_*}}}{4l^2})Y_{2l}}\,
\end{eqnarray}
where $b^{(\delta)}_{\epsilon_{\bold{j}_*}},c^{(\delta)}_{\epsilon_{\bold{j}_*}}$ are calculated at $\delta\equiv 1$. The solution $[Y_{2l}]_B$ of (\ref{reference}) has initial condition $$[Y_{N^{1-\beta}+2}]_B\equiv \frac{1}{2}(1+\frac{\sqrt{a'_{\epsilon}}}{\sqrt{1+a'_{\epsilon}}}-\frac{1}{(N^{1-\beta}+3)(1+a'_{\epsilon})-\sqrt{a'_{\epsilon}}\sqrt{1+a'_{\epsilon}}})|_{\epsilon\equiv \epsilon_{\bold{j}_*}}$$ whereas the solution $[Y_{2l}]_*$ at point 2) has initial condition  $[Y_{N^{1-\beta}+2}]_*\equiv1$. Firstly, we compare them at the step $2l\equiv2l_{\epsilon_{\bold{j}_*}}$ (defined later), secondly we estimate $|[Y_{2l}]_B-[Y_{2l}]_*|$ at $2l\equiv 2$.
\end{itemize}
Now, we implement the three steps outlined above.
\begin{itemize}
\item[1)] Assume $z\leq E^{Bog}_{\bold{j}_*}+ \sqrt{\epsilon_{\bold{j}_*}}\phi_{\bold{j}_*}\sqrt{\epsilon_{\bold{j}_*}^2+2\epsilon_{\bold{j}_*}}$. We define $\psi_{N-2; 1,1}$ the normalized vector with $N-2$ particles in the mode $\bold{0}$ and one particle in each mode $\bold{j}_*$ and $-\bold{j}_*$. We recall: {\bf{a)}} the proof of Proposition \ref{induction-G}  used in Section \ref{fixed point} to replace operators with c-numbers; {\bf{b)}} Definition \ref{def-sums} and the identity  in (\ref{decomposition-0}). Hence, we observe (and explain later) that, by construction,
\begin{equation}\label{G}
\check{\mathcal{G}}_{\bold{j}_*\,;\,N-2,N-2}(z)=\langle \psi_{N-2; 1,1}\,,\,\sum_{l_{N-2}=0}^{\infty}[(R^{Bog}_{\bold{j}_*\,;\,N-2,N-2}(z))^{\frac{1}{2}}\Gamma^{Bog\,}_{\bold{j}_*\,;\,N-2,N-2}(R^{Bog}_{\bold{j}_*\,;\,N-2,N-2}(z))^{\frac{1}{2}}]^{l_{N-2}}\,\psi_{N-2; 1,1} \rangle
\end{equation}
and
\begin{eqnarray}
& &[\check{\mathcal{G}}_{\bold{j}_*\,;\,N-2,N-2}]_T(z)\\
&=&\langle \psi_{N-2; 1,1}\,,\,\sum_{l_{N-2}=0}^{\infty}\Big\{(R^{Bog}_{\bold{j}_*\,;\,N-2,N-2}(z))^{\frac{1}{2}}\,\times\label{Gt}\\
& &\quad \times \sum_{r=N-N^{1-\beta},\,r\, even}^{N-4}[\Gamma^{Bog\,}_{\bold{j}_*\,;\,N-2,N-2}(z)]_{(r,h_-; r+2,h_-;\dots ; N-4,h_-)}|_{h\equiv \infty}\,(R^{Bog}_{\bold{j}_*\,;\,N-2,N-2}(z))^{\frac{1}{2}}\Big\}^{l_{N-2}}\,\psi_{N-2; 1,1} \rangle\,,\quad\quad\quad \nonumber
\end{eqnarray}
where the symbol $|_{h \equiv \infty}$ means that $h\equiv \infty $ in the re-expansion of $\Gamma^{Bog\,}_{\bold{j}_*\,;\,N-2,N-2}(z)$ (see Definition \ref{def-sums}).

Regarding the R-H-S of (\ref{G}), this is evident using  {\bf{a)}}. Concerning (\ref{Gt}), notice that, by construction (see Definition \ref{def-sums}),
$$ \sum_{r=N-N^{1-\beta},\,r\, even}^{N-4}[\Gamma^{Bog\,}_{\bold{j}_*\,;\,N-2,N-2}(z)]_{(r,h_-; r+2,h_-;\dots ; N-4,h_-)}|_{h \equiv \infty}$$
corresponds to the subset of terms obtained from the re-expansion of $\Gamma^{Bog\,}_{\bold{j}_*\,;\,N-2,N-2}(z)$ that do not contain the operators $W_{\bold{j}_*\,;\,i,i-2}$ and $W^*_{\bold{j}_*\,;\,i-2,i}$ with $i< N-N^{1-\beta}$. After the replacement with c-numbers in the R-H-S of (\ref{Gt}), those terms must coincide with the R-H-S of (\ref{relation-G}) with initial condition $\check{\mathcal{G}}_{\bold{j}_*\,;\,N-N^{1-\beta}-2,N-N^{1-\beta}-2}(z)\equiv 1$.
%By means of the result in Proposition \ref{lemma-expansion-proof-0} we can write 
%\begin{eqnarray}
%& &\check{\mathcal{G}}_{\bold{j}_*\,;\,N-2,N-2}(z)-[\check{\mathcal{G}}_{\bold{j}_*\,;\,N-2,N-2}]_T(z)\\
%&=&\langle \eta\,,\,\sum_{l_{N-2}=0}^{\infty}\Big\{(R^{Bog}_{\bold{j}_*\,;\,N-2,N-2}(z))^{\frac{1}{2}}\,\times\\
%& &\quad \times \sum_{l=2,\,l\, even}^{N-N^{1-\beta}}[\Gamma^{Bog\,}_{\bold{j}_*\,;\,N-2,N-2}(z)]_{(l,h_-;l+2,h_-;\dots ; N-4,h_-)}(R^{Bog}_{\bold{j}_*\,;\,N-2,N-2}(z))^{\frac{1}{2}}\Big\}^%{l_{N-2}}\,\eta \rangle|_{h_- \equiv \infty}\quad\quad\quad  \label{diff-G}
%&= &\langle \eta\,,\,\Big\{\Gamma^{Bog\,}_{N-2,N-2}(z)-\sum_{l=N-N^{1-\beta},\,l\, even}^{N-4}[\Gamma^{Bog\,}_{\bold{j}_*\,;\,N-2,N-2}(z)]_{(l,h_-;l+2,h_-;\dots ; N-4,h_-)}\Big\}\eta \rangle|_{h_- \equiv \infty}\quad\quad
%\\
%&=&\langle \eta\,,\,\Big\{\sum_{l=2,\,l\, even}^{N-N^{1-\beta}}[\Gamma^{Bog\,}_{\bold{j}_*\,;\,N-2,N-2}(z)]_{(l,h_-;l+2,h_-;\dots ; N-4,h_-)}\Big\}\eta \rangle|_{h_- \equiv \infty}\,.
%\end{eqnarray}

Finally, Corollary \ref{truncated} implies 
\begin{equation}
|\check{\mathcal{G}}_{\bold{j}_*\,;\,N-2,N-2}(z)-[\check{\mathcal{G}}_{\bold{j}_*\,;\,N-2,N-2}]_T(z)|\leq \mathcal{O}(\frac{1}{\sqrt{\epsilon_{\bold{j}_*}}}(\frac{1}{1+c\sqrt{\epsilon_{\bold{j}_*}}})^{N^{1-\beta}})\,.
\end{equation}
\item[2)] For $i\geq N-N^{1-\beta}$,  and  for $\delta\equiv 1$,  we consider the computation in Lemma  \ref{accessori} (see (\ref{5.7})) with $\gamma\equiv \beta$ and deduce
\begin{eqnarray}
\mathcal{W}_{\bold{j}_*\,;i,i-2}(z)\mathcal{W}^*_{\bold{j}_*\,;i-2,i}(z)|_{z\equiv E^{Bog}_{\bold{j}_*}}&=&\frac{1}{4[(\ref{W-a})+(\ref{W-b})+(\ref{W-c})]}\\
&=&\frac{1}{4(1+a'_{\epsilon_{\bold{j}_*}}+\mathcal{O}(\frac{1}{N^{\beta}})-\frac{2b^{(1)}_{\epsilon_{\bold{j}_*}}}{2l}-\frac{1-c^{(1)}_{\epsilon_{\bold{j}_*}}}{4l^2})}
\end{eqnarray}
where $2l=N-i$ and $a'_{\epsilon}$ is defined in (\ref{reference}). 
%Next, we set $l_{\epsilon_{\bold{j}_*}}=\mathcal{O}(\frac{1}{\sqrt{\epsilon_{\bold{j}_*}}})$ the smallest natural number such that
%$$a'_{\epsilon_{\bold{j}_*}}-\frac{2b^{(1)}_{\epsilon_{\bold{j}_*}}}{2l_{\epsilon_{\bold{j}_*}}}-\frac{1-c^{(1)}_{\epsilon_{\bold{j}_*}}}{4l_{\epsilon_{\bold{j}_*}}^2}\geq \frac{2a'_{\epsilon_{\bold{j}_*}}}{3}\,.$$
For $2l$ even and decreasing from $N^{1-\beta}+2$ down to $2$, we make use of (\ref{relation-G}) and (\ref{reference}) to study the difference
\begin{eqnarray}
& &\Big|[\check{\mathcal{G}}_{\bold{j}_*\,;\,N-2l+2,N-2l+2}]_T(E^{Bog}_{\bold{j}_*})]^{-1}-[Y_{2l-2}]_*\Big|\\
&= &\Big|1-\mathcal{W}_{\bold{j}_*\,;N-2l+2,N-2l}(z)\mathcal{W}^*_{\bold{j}_*\,;N-2l+2,N-2l}(z)|_{z\equiv E^{Bog}_{\bold{j}_*}}[\check{\mathcal{G}}_{\bold{j}_*\,;\,N-2l,N-2l}]_T(E^{Bog}_{\bold{j}_*})]\\
& &\quad\quad -1+ \frac{1}{4(1+a'_{\epsilon_{\bold{j}_*}}-\frac{2b^{(1)}_{\epsilon_{\bold{j}_*}}}{2l}-\frac{1-c^{(1)}_{\epsilon_{\bold{j}_*}}}{(2l)^2})[Y_{2l}]_*}\Big|\\
&=&\Big|\Big\{\frac{1}{4(1+a'_{\epsilon_{\bold{j}_*}}-\frac{2b^{(1)}_{\epsilon_{\bold{j}_*}}}{2l}-\frac{1-c^{(1)}_{\epsilon_{\bold{j}_*}}}{(2l)^2})}-\mathcal{W}_{\bold{j}_*\,;N-2l+2,N-2l}(z)\mathcal{W}^*_{\bold{j}_*\,;N-2l+2,N-2l}(z)|_{z\equiv E^{Bog}_{\bold{j}_*}}\Big\}[\check{\mathcal{G}}_{\bold{j}_*\,;\,N-2l,N-2l}]_T(E^{Bog}_{\bold{j}_*})]\nonumber\\
& &\quad\quad - \frac{1}{4(1+a'_{\epsilon_{\bold{j}_*}}-\frac{2b^{(1)}_{\epsilon_{\bold{j}_*}}}{2l}-\frac{1-c^{(1)}_{\epsilon_{\bold{j}_*}}}{(2l)^2})}\Big\{\frac{1}{[\check{\mathcal{G}}_{\bold{j}_*\,;\,N-2l,N-2l}]_T(E^{Bog}_{\bold{j}_*})]^{-1}}-\frac{1}{[Y_{2l}]_*}\Big\}\Big|\\
&=&\Big|\Big\{\frac{1}{4(1+a'_{\epsilon_{\bold{j}_*}}-\frac{2b^{(1)}_{\epsilon_{\bold{j}_*}}}{2l}-\frac{1-c^{(1)}_{\epsilon_{\bold{j}_*}}}{(2l)^2})}-\frac{1}{4(1+a'_{\epsilon_{\bold{j}_*}}+\mathcal{O}(\frac{1}{N^{\beta}})-\frac{2b^{(1)}_{\epsilon_{\bold{j}_*}}}{2l}-\frac{1-c^{(1)}_{\epsilon_{\bold{j}_*}}}{4l^2})}\Big\}[\check{\mathcal{G}}_{\bold{j}_*\,;\,N-2l,N-2l}]_T(E^{Bog}_{\bold{j}_*})] \nonumber\\
& &\quad\quad - \frac{1}{4(1+a'_{\epsilon_{\bold{j}_*}}-\frac{2b^{(1)}_{\epsilon_{\bold{j}_*}}}{2l}-\frac{1-c^{(1)}_{\epsilon_{\bold{j}_*}}}{(2l)^2})}\Big\{\frac{[Y_{2l}]_*-[\check{\mathcal{G}}_{\bold{j}_*\,;\,N-2l,N-2l}]_T(E^{Bog}_{\bold{j}_*})]^{-1}}{[\check{\mathcal{G}}_{\bold{j}_*\,;\,N-2l,N-2l}]_T(E^{Bog}_{\bold{j}_*})]^{-1}\,[Y_{2l}]_*}\Big\}\Big|\,.\label{diff-doppia}
\end{eqnarray} 

Notice that
\begin{equation}
\frac{1}{4(1+a'_{\epsilon_{\bold{j}_*}}-\frac{2b^{(1)}_{\epsilon_{\bold{j}_*}}}{2l}-\frac{1-c^{(1)}_{\epsilon_{\bold{j}_*}}}{(2l)^2})}-\frac{1}{4(1+a'_{\epsilon_{\bold{j}_*}}+\mathcal{O}(\frac{1}{N^{\beta}})-\frac{2b^{(1)}_{\epsilon_{\bold{j}_*}}}{2l}-\frac{1-c^{(1)}_{\epsilon_{\bold{j}_*}}}{4l^2})}=\mathcal{O}(\frac{1}{N^{\beta}})\,.
\end{equation}

\noindent
Next, we split the range $2\leq 2l \leq N^{1-\beta}+2$ into two intervals:  

\begin{equation}
2l_{\epsilon_{\bold{j}_*}}< 2l \leq N^{1-\beta}+2\quad \text{and}\quad 2\leq 2l \leq 2l_{\epsilon_{\bold{j}_*}}\,,
\end{equation}
where $l_{\epsilon_{\bold{j}_*}}=\mathcal{O}(\frac{1}{\sqrt{\epsilon_{\bold{j}_*}}})$ is chosen so that, in the range  $2l_{\epsilon_{\bold{j}_*}}<2l \leq N^{1-\beta}+2$, both
$$a'_{\epsilon_{\bold{j}_*}}-\frac{2b^{(1)}_{\epsilon_{\bold{j}_*}}}{2l_{\epsilon_{\bold{j}_*}}}-\frac{1-c^{(1)}_{\epsilon_{\bold{j}_*}}}{4l_{\epsilon_{\bold{j}_*}}^2}\geq \frac{a'_{\epsilon_{\bold{j}_*}}}{2}$$ and
$$a'_{\epsilon_{\bold{j}_*}}-\frac{2b^{(1)}_{\epsilon_{\bold{j}_*}}}{2l_{\epsilon_{\bold{j}_*}}}+\mathcal{O}(\frac{1}{N^{\beta}})-\frac{1-c^{(1)}_{\epsilon_{\bold{j}_*}}}{4l_{\epsilon_{\bold{j}_*}}^2}\geq \frac{a'_{\epsilon_{\bold{j}_*}}}{2}\,.$$ This is possible because by assumption $\frac{1}{N^{\beta}}=o(\epsilon_{\bold{j}_*})$. Then, in the range $$2l_{\epsilon_{\bold{j}_*}}< 2l \leq N^{1-\beta}+2$$ we can make use of the lower bound of the type in (\ref{bound-fixed}) both for $[\check{\mathcal{G}}_{\bold{j}_*\,;\,N-2l,N-2l}]_T(E^{Bog}_{\bold{j}_*})]^{-1}$ and $[Y_{2l}]_*$ and estimate
\begin{equation}
[\check{\mathcal{G}}_{\bold{j}_*\,;\,N-2l,N-2l}]_T(E^{Bog}_{\bold{j}_*})]^{-1}\,[Y_{2l}]_*\geq \frac{1}{4}(1+c\sqrt{\epsilon_{\bold{j}_*}})
\end{equation}
for some $c>0$. Using this input in (\ref{diff-doppia}) one can check by induction that the following bound holds as long as $2l_{\epsilon_{\bold{j}_*}}<2l \leq N^{1-\beta}+2$
\begin{equation}\Big|\frac{1}{[\check{\mathcal{G}}_{\bold{j}_*\,;\,N-2l,N-2l}]_T(z)|_{z=E^{Bog}_{\bold{j}_*}}}-[Y_{2l}]_*\Big|\leq \frac{C}{N^{\beta}}\sum_{i= 0\,,\, i\,\text{even}}^{N^{1-\beta}+2-2l}(\frac{1}{1+c\sqrt{\epsilon_{\bold{j}_*}}})^{\frac{i}{2}} \leq \frac{C'}{\sqrt{\epsilon_{\bold{j}_*}}N^{\beta}}\label{partial-est}
\end{equation} 
for some positive constants $c, C,C'$. 

In the range  $2 \leq 2l \leq  2l_{\epsilon_{\bold{j}_*}}$, we invoke\footnote{See the observation in the last paragraph of  Lemma \ref{lemma-sequence-lower-bound} and consider that, by construction,  $[Y_{2l}]_* \geq \check{X}^{(1)}_{N-2l}$ where $\check{X}^{(\delta)}_{N-2l}$ is defined like $X^{(\delta)}_{N-2l}$  except for the coefficient $a_{\epsilon_{\bold{j}_*}}$ replaced with $a'_{\epsilon_{\bold{j}_*}}$, and with initial condition $\check{X}^{(\delta)}_{N^{1-\beta}+2}\equiv 1$.  $\check{X}^{(1)}_{N-2l}$ has the same type of lower bound of $X^{(1)}_{N-2l}$, that means $\check{X}^{(1)}_{N-2l}\geq \check{X}^{(1+\sqrt{\epsilon_{\bold{j}_*}})}_{N-2l}\geq \frac{1}{2}\Big[1+\sqrt{\eta a'_{\epsilon_{\bold{j}_*}}}-\frac{b_{\epsilon_{\bold{j}_*}}/ \sqrt{\eta a'_{\epsilon_{\bold{j}_*}}}}{N-2l-\xi} \Big]$ with $\xi=\epsilon^{\Theta}_{\bold{j}_*}$ and $\eta=1-\sqrt{\epsilon_{\bold{j}_*}}$. Concerning $[\check{\mathcal{G}}_{\bold{j}_*\,;\,N-2l,N-2l}]_T(E^{Bog}_{\bold{j}_*})]^{-1}$ 
consider the inequalities $[\check{\mathcal{G}}_{\bold{j}_*\,;\,N-2l,N-2l}]_T(E^{Bog}_{\bold{j}_*})]^{-1}
\geq [\check{\mathcal{G}}_{\bold{j}_*\,;\,N-2l,N-2l}(E^{Bog}_{\bold{j}_*})]^{-1}\geq \frac{1}{\|\check{\Gamma}_{\bold{j}_*\,;\,N-2l,N-2l}(E^{Bog}_{\bold{j}_*})\|} \geq X_{N-2l}$ where in the last step we have exploited (\ref{Gamma-ineq}).} 
%Lemma \ref{lemma-sequence-lower-bound} 
and use the (worse) lower bound of the type in (\ref{inf-bound}) and (\ref{inductive-hp}) to estimate
\begin{equation}\label{arg-pont-2-in}
[\check{\mathcal{G}}_{\bold{j}_*\,;\,N-2l,N-2l}]_T(E^{Bog}_{\bold{j}_*})]^{-1}\,[Y_{2l}]_*\geq \frac{1}{4}(1-\frac{2+\mathcal{O}(\epsilon_{\bold{j}_*}^{\Theta})}{2l})\,,\quad 2l\geq 4\, ,
\end{equation}
in expression (\ref{diff-doppia}). This also implies
\begin{equation}
4(1+a'_{\epsilon_{\bold{j}_*}}-\frac{2b^{(1)}_{\epsilon_{\bold{j}_*}}}{2l}-\frac{1-c^{(1)}_{\epsilon_{\bold{j}_*}}}{(2l)^2})[\check{\mathcal{G}}_{\bold{j}_*\,;\,N-2l,N-2l}]_T(E^{Bog}_{\bold{j}_*})]^{-1}\,[Y_{2l}]_*\geq 1-\frac{2+\mathcal{O}(\epsilon_{\bold{j}_*}^{\Theta})}{2l}+\mathcal{O}(\frac{1}{(2l)^2})>0\,,\quad 2l\geq 4\,,
\end{equation}
due to the assumptions on $\epsilon_{\bold{j}_*}$ and $2l$.
Starting from the result in (\ref{partial-est}),  one can check by induction that the following bound holds for some $C'''>0$ (with $2 \leq 2l \leq  2l_{\epsilon_{\bold{j}_*}}$)
\begin{eqnarray}
& &\Big|[\check{\mathcal{G}}_{\bold{j}_*\,;\,N-2l,N-2l}]_T(E^{Bog}_{\bold{j}_*})]^{-1}-[Y_{2l}]_*\Big| \\
&\leq&\frac{C'''}{N^{\beta}}\sum_{j=N^{1-\beta}-2l_{\epsilon_{\bold{j}_*}}-2}^{N^{1-\beta}-2l-2}\,\,\prod_{r=N^{1-\beta}-2l_{\epsilon_{\bold{j}_*}}-2}^{j}\Big(\frac{1}{1-\frac{2+\mathcal{O}(\epsilon_{\bold{j}_*}^{\Theta})}{N^{1-\beta}-r}+\mathcal{O}(\frac{1}{(N^{1-\beta}-r)^2})}\Big)\quad\quad\quad \\
& &+\frac{C'''}{\sqrt{\epsilon_{\bold{j}_*}}N^{\beta}}\prod_{r=N^{1-\beta}-2l_{\epsilon_{\bold{j}_*}}-2}^{N^{1-\beta}-2l-2}\Big(\frac{1}{1-\frac{2+\mathcal{O}(\epsilon_{\bold{j}_*}^{\Theta})}{N^{1-\beta}-r}+\mathcal{O}(\frac{1}{(N^{1-\beta}-r)^2})}\Big)\quad\quad 
\end{eqnarray} 
where both $r$ and $j$ are even numbers.
Since $2\leq 2l \leq  2l_{\epsilon_{\bold{j}_*}}$ and $l_{\epsilon_{\bold{j}_*}}=\mathcal{O}(\frac{1}{\sqrt{\epsilon_{\bold{j}_*}}})$,  we derive that for $N^{1-\beta}-2l_{\epsilon_{\bold{j}_*}}-2\leq j \leq N^{1-\beta}-2l-2$ 
\begin{eqnarray}\label{arg-pont-2-fin}
& &\prod_{r=N^{1-\beta}-2l_{\epsilon_{\bold{j}_*}}-2\,,\, r\,\text{even}}^{j}(\frac{1}{1-\frac{2+\mathcal{O}(\epsilon_{\bold{j}_*}^{\Theta})}{N^{1-\beta}-r}+\mathcal{O}(\frac{1}{(N^{1-\beta}-r)^2})})\\
& &=\prod_{r=N^{1-\beta}-2l_{\epsilon_{\bold{j}_*}}-2\,,\, r\,\text{even}}^{j}\exp[\ln\Big(\frac{1}{1-\frac{2+\mathcal{O}(\epsilon_{\bold{j}_*}^{\Theta})}{N^{1-\beta}-r}+\mathcal{O}(\frac{1}{(N^{1-\beta}-r)^2})}\Big)]\\
&= &\exp[\sum_{r=N^{1-\beta}-2l_{\epsilon_{\bold{j}_*}}-2\,,\, r\,\text{even}}^{j}\ln\Big(\frac{1}{1-\frac{2+\mathcal{O}(\epsilon_{\bold{j}_*}^{\Theta})}{N^{1-\beta}-r}+\mathcal{O}(\frac{1}{(N^{1-\beta}-r)^2})}\Big)]\leq \mathcal{O}(\frac{1}{\sqrt{\epsilon_{\bold{j}_*}}})
\end{eqnarray}
where we have used that $r$ is even and $2l+2\leq N^{1-\beta}-r\leq 2l_{\epsilon_{\bold{j}_*}}+2\leq \mathcal{O}(\frac{1}{\sqrt{\epsilon_{\bold{j}_*}}})$ so that $\sum_{r=N^{1-\beta}-2l_{\epsilon_{\bold{j}_*}}-2\,,\, r\,\text{even}}^{j}\frac{2}{N^{1-\beta}-r}\leq \ln(c\frac{1}{\sqrt{\epsilon_{\bold{j}_*}}})$ for some $c>0$.  
 Finally, we can estimate
$$\Big|\frac{1}{[\check{\mathcal{G}}_{\bold{j}_*\,;\,N-2,N-2}]_T(z)|_{z=E^{Bog}_{\bold{j}_*}}}-[Y_{2}]_*\Big| \leq \frac{C''''}{\epsilon_{\bold{j}_*}N^{\beta}}$$
for some $C''''>0$.

\item [3)] 
A direct computation shows that (see the supporting file "note-about-exact-solution-relation (5.108).pdf")
\begin{eqnarray}
[Y_{2l}]_B&=&\frac{1}{2}(1+\frac{\sqrt{a'_{\epsilon}}}{\sqrt{1+a'_{\epsilon}}}-\frac{1}{(2l+1)(1+a'_{\epsilon})-\sqrt{a'_{\epsilon}}\sqrt{1+a'_{\epsilon}}})|_{\epsilon\equiv \epsilon_{\bold{j}_*}}
\end{eqnarray}
fulfills the relation in (\ref{reference}).

Using the same argumentation exploited in (\ref{fixed-point-arg}) and a lower bound of the type in (\ref{bound-fixed}), for some $c>0$  we can estimate
\begin{eqnarray}
& &|[Y_{2l_{\epsilon_{\bold{j}_*}}-2}]_*-[Y_{2l_{\epsilon_{\bold{j}_*}}-2}]_B|\label{medium-difference}\\
&=& \frac{1}{4(1+a'_{\epsilon_{\bold{j}_*}}-\frac{2b^{(1)}_{\epsilon_{\bold{j}_*}}}{2l_{\epsilon_{\bold{j}_*}}}-\frac{1-c^{(1)}_{\epsilon_{\bold{j}_*}}}{(2l_{\epsilon_{\bold{j}_*}})^2})[Y_{2l_{\epsilon_{\bold{j}_*}}}]_*[Y_{2l_{\epsilon_{\bold{j}_*}}}]_B}|[Y_{2l_{\epsilon_{\bold{j}_*}}}]_*-[Y_{2l_{\epsilon_{\bold{j}_*}}}]_B|\\
&\leq  & \frac{1}{1+c\sqrt{\epsilon_{\bold{j}_*}}}|[Y_{2l_{\epsilon_{\bold{j}_*}}}]_*-[Y_{2l_{\epsilon_{\bold{j}_*}}}]_B|\\
&\leq &[ \frac{1}{1+c\sqrt{\epsilon_{\bold{j}_*}}}]^{\frac{(N^{1-\beta}+4-2l_{\epsilon_{\bold{j}_*}})}{2}}|[Y_{N^{1-\beta}+2}]_*-[Y_{N^{1-\beta}+2}]_B|\,.
\end{eqnarray}
Next, we implement an argument analogous to point 2) (see (\ref{arg-pont-2-in})-(\ref{arg-pont-2-fin})) for the range $2\leq 2l <2l_{\epsilon_{\bold{j}_*}}-2 $.  Finally, we can estimate $$|[Y_{2}]_*-[Y_{2}]_B|\leq \mathcal{O}(\frac{1}{\sqrt{\epsilon_{\bold{j}_*}}}[\frac{1}{1+c\sqrt{\epsilon_{\bold{j}_*}}}]^{\frac{(N^{1-\beta}+4-2l_{\epsilon_{\bold{j}_*}})}{2}})\,.$$.
\end{itemize}

\noindent
\emph{Conclusion}

The  quantity $\frac{1}{[Y_2]_B}$ that we use to  approximate $\check{\mathcal{G}}_{\bold{j}_*\,;\,N-2,N-2}(E^{Bog}_{\bold{j}_*})$ in (\ref{diff-0}) has the following property 
\begin{equation}\label{exact}
-\frac{E^{Bog}_{\bold{j}_*}}{\phi_{\bold{j}_*}}-\frac{1}{2\epsilon_{\bold{j}_*}+2-\frac{E^{Bog}_{\bold{j}_*}}{\phi_{\bold{j}_*}}}\frac{1}{[Y_2]_B}=0\,,
\end{equation}
that can be verified using 
$$\frac{E^{Bog}_{\bold{j}_*}}{\phi_{\bold{j}_*}}=-\Big[\epsilon_{\bold{j}_*}+1-\sqrt{\epsilon_{\bold{j}_*}^2+2\epsilon_{\bold{j}_*}}\,\Big]$$
and
\begin{eqnarray}
[Y_{2}]_B&=&\frac{1}{2}\Big(\frac{[\sqrt{a'_{\epsilon}}+\sqrt{1+a'_{\epsilon}}]}{\sqrt{1+a'_{\epsilon}}}-\frac{1}{[3\sqrt{1+a'_{\epsilon}}-\sqrt{a'_{\epsilon}}]\sqrt{1+a'_{\epsilon}}}\Big)|_{\epsilon\equiv \epsilon_{\bold{j}_*}}\\
%&=&\frac{1}{2}(1+\frac{\sqrt{a'_{\epsilon}}}{\sqrt{1+a'_{\epsilon}}}-\frac{1}{[3\sqrt{1+a'_{\epsilon}}-\alpha]\sqrt{1+a'_{\epsilon}}})\\
%&=&\frac{1}{2}(\frac{\{3\sqrt{1+a'_{\epsilon}}-\alpha]\sqrt{1+a'_{\epsilon}}+\sqrt{a'_{\epsilon}}[3\sqrt{1+a'_{\epsilon}}-\alpha]-1\}}{[3\sqrt{1+a'_{\epsilon}}-\alpha]\sqrt{1+a'_{\epsilon}}})\\
&=&(\frac{\sqrt{a'_{\epsilon}}+\sqrt{1+a'_{\epsilon}}}{3\sqrt{1+a'_{\epsilon}}-\sqrt{a'_{\epsilon}}})|_{\epsilon\equiv \epsilon_{\bold{j}_*}}\\
&=&\Big(\frac{1}{3(1+\epsilon)-\sqrt{\epsilon^2+2\epsilon}}\Big)\Big(\frac{1}{\epsilon+1-\sqrt{\epsilon^2+2\epsilon}}\Big)|_{\epsilon\equiv \epsilon_{\bold{j}_*}}\,.
\end{eqnarray}
Thus, using the successive approximations of $\check{\mathcal{G}}_{\bold{j}_*\,;\,N-2,N-2}(E^{Bog}_{\bold{j}_*})$ obtained at points 1), 2) and 3), we have proven that
\begin{eqnarray}
(\ref{diff-0})& =&|-E^{Bog}_{\bold{j}_*}-(1-\frac{1}{N})\frac{\phi_{\bold{j}_*}}{(2\epsilon_{\bold{j}_*}+2-\frac{4}{N})-\frac{E^{Bog}_{\bold{j}_*}}{\phi_{\bold{j}_*}}}\check{\mathcal{G}}_{\bold{j}_*\,;\,N-2,N-2}(E^{Bog}_{\bold{j}_*})|\label{EBOG-1}\\
&=&|-E^{Bog}_{\bold{j}_*}-\frac{\phi_{\bold{j}_*}}{(2\epsilon_{\bold{j}_*}+2)-\frac{E^{Bog}_{\bold{j}_*}}{\phi_{\bold{j}_*}}}[\check{\mathcal{G}}_{\bold{j}_*\,;\,N-2,N-2}]_T(E^{Bog}_{\bold{j}_*})|+\mathcal{O}(\frac{1}{\sqrt{\epsilon_{\bold{j}_*}}}(\frac{1}{1+c\sqrt{\epsilon_{\bold{j}_*}}})^{N^{1-\beta}})+\mathcal{O}(\frac{1}{N})\quad\quad\quad \\
&=&|-E^{Bog}_{\bold{j}_*}-\frac{\phi_{\bold{j}_*}}{(2\epsilon_{\bold{j}_*}+2)-\frac{E^{Bog}_{\bold{j}_*}}{\phi_{\bold{j}_*}}}\frac{1}{[Y_2]_*}|+\mathcal{O}(\frac{1}{\epsilon_{\bold{j}_*}N^{\beta}})+\mathcal{O}(\frac{1}{\sqrt{\epsilon_{\bold{j}_*}}}(\frac{1}{1+c\sqrt{\epsilon_{\bold{j}_*}}})^{N^{1-\beta}})+\mathcal{O}(\frac{1}{N})\\
&=&|-E^{Bog}_{\bold{j}_*}-\frac{\phi_{\bold{j}_*}}{(2\epsilon_{\bold{j}_*}+2)-\frac{E^{Bog}_{\bold{j}_*}}{\phi_{\bold{j}_*}}}\frac{1}{[Y_2]_B}|+\mathcal{O}(\frac{1}{\epsilon_{\bold{j}_*}N^{\beta}})+\mathcal{O}(\frac{1}{\sqrt{\epsilon_{\bold{j}_*}}}[\frac{1}{1+c\sqrt{\epsilon}_{\bold{j}_*}}]^{\frac{(N^{1-\beta}+4-2l_{\epsilon_{\bold{j}_*}})}{2}})+\mathcal{O}(\frac{1}{N})\quad\quad\quad\quad \label{EBOG-2-0}\\
&=&\mathcal{O}(\frac{1}{\epsilon_{\bold{j}_*}N^{\beta}})+\mathcal{O}(\frac{1}{\sqrt{\epsilon_{\bold{j}_*}}}[\frac{1}{1+c'\sqrt{\epsilon}_{\bold{j}_*}}]^{N^{1-\beta})})+\mathcal{O}(\frac{1}{N})\,.\label{EBOG-2}
\end{eqnarray}
where in the last step from (\ref{EBOG-2-0}) to (\ref{EBOG-2}) we have used (\ref{exact}), $l_{\epsilon_{\bold{j}_*}}\leq \mathcal{O}(\frac{1}{\sqrt{\epsilon_{\bold{j}_*}}})$, and $\frac{1}{N^{1-\beta}}= o(\sqrt{\epsilon_{\bold{j}_*}})$. 
%Since $f(z):=-z-\frac{\phi_{\bold{j}_*}^2}{\phi_{\bold{j}_*}(2\epsilon_{\bold{j}_*}+2)-z}\check{\mathcal{G}}_{\bold{j}_*\,;\,N-2,N-2}(z)$ has derivative not larger than $-1$ 
%\begin{equation}\label{deriv}
%|z-E^{Bog}_{\bold{j}_*}|\leq |-z-\frac{\phi_{\bold{j}_*}^2}{\phi_{\bold{j}_*}(2\epsilon_{\bold{j}_*}+2)-z}\check{\mathcal{G}}_{\bold{j}_*\,;\,N-2,N-2}(z)-[-E^{Bog}_{\bold{j}_*}-\frac{\phi_{\bold{j}_*}^2}{\phi_{\bold{j}_*}(2\epsilon_{\bold{j}_*}+2)-E^{Bog}_{\bold{j}_*}}\check{\mathcal{G}}_{\bold{j}_*\,;\,N-2,N-2}(E^{Bog}_{\bold{j}_*})]|\,.
%\end{equation}
%We recall that $f(z_*)=0$ by definition of $z_*$. Thus, using (\ref{EBOG-1})-(\ref{EBOG-2}) and (\ref{deriv}), we can bound
%\begin{eqnarray}
%|z_*-E^{Bog}_{\bold{j}_*}|&\leq & |-E^{Bog}_{\bold{j}_*}-\frac{\phi_{\bold{j}_*}}{(2\epsilon_{\bold{j}_*}+2)-\frac{E^{Bog}_{\bold{j}_*}}{\phi_{\bold{j}_*}}}\check{\mathcal{G}}_{\bold{j}_*\,;\,N-2,N-2}(E^{Bog}_{\bold{j}_*})|\\
%&\leq &\mathcal{O}(\frac{1}{\epsilon_{\bold{j}_*}N^{\beta}})+\mathcal{O}(\frac{1}{\epsilon_{\bold{j}_*}}[\frac{1}{1+c\sqrt{\epsilon}_{\bold{j}_*}}]^{(N^{1-\beta}-2l_{\epsilon_{\bold{j}_*}})})\,.\label{diff}
%\end{eqnarray}
\qed
\begin{remark}\label{EBoglimit}
Lemma \ref{eigenvalue} shows that, for any dimension $d\geq 1$, in the mean field limiting regime the difference between the ground state energy, $z_*$, of $H^{Bog}_{\bold{j}_*}$ and $E^{Bog}_{\bold{j}_*}$ is bounded by $\mathcal{O}(\frac{1}{N^{\beta}})$ for any $0<\beta<1$. Notice that, by setting $\beta=\frac{2}{3}$, at fixed $\rho$,    the R-H-S in  (\ref{diff}) goes to zero as $L\to \infty$ in space dimension $d\geq 4$. In space dimension $d=3$, by setting $\beta=\frac{2}{3}$, Lemma  \ref{eigenvalue} ensures the same result provided $\rho=\rho_0 (\frac{L}{L_0})^{s}$ with $s>0$ where $\rho_0>0$ and $L_0=1$.

However, starting from the definition and the monotonicity of $\check{\mathcal{G}}_{\bold{j}_*\,;\,i,i}(z)(\geq 1)$ (see (\ref{def-G}) and Remark \ref{increasing}, respectively), making use of the inequality\footnote{This inequality can be derived from the definition of $\mathcal{W}_{\bold{j}_*\,;i,i-2}(z)\mathcal{W}^*_{\bold{j}_*\,;i-2,i}(z)$ (see (\ref{def-Wcal-1})) and computations like in Lemma \ref{accessori}.}
$$\mathcal{W}_{\bold{j}_*\,;i,i-2}(z)\mathcal{W}^*_{\bold{j}_*\,;i-2,i}(z)\leq \frac{1}{4(1+a_{\epsilon}-\frac{2b_{\epsilon}}{N-i+2}-\frac{1-c_{\epsilon}}{(N-i+2)^2})}\,,$$ and exploting the same type of procedure of Lemma \ref{lemma-sequence-lower-bound},  it is possible to show the inequality (for $0\leq i\leq N-2$ and even) $$\frac{1}{\check{\mathcal{G}}_{\bold{j}_*\,;\,i,i}(z)}\geq \frac{1}{\check{\mathcal{G}}_{\bold{j}_*\,;\,i,i}(z_{\delta=1+\sqrt{\epsilon}})}\geq\frac{1}{2}\Big[1-\frac{1}{N-i+1}\Big] + o_{L\to \infty}(1)\,
$$
for $z\leq z_{\delta=1+\sqrt{\epsilon}}\equiv  E^{Bog}_{\bold{j}_*}+ \sqrt{\epsilon_{\bold{j}_*}}\phi_{\bold{j}_*}\sqrt{\epsilon_{\bold{j}_*}^2+2\epsilon_{\bold{j}_*}}$.  This inequality and the fixed point equation (see (\ref{fp-equation-2})) imply
$$-\frac{z_*}{\phi_{\bold{j}_{*}}}[2\epsilon_{\bold{j}_*}+2-\frac{z_*}{\phi_{\bold{j}_{*}}}+\mathcal{O}(\frac{1}{\rho L^3})]=\check{\mathcal{G}}_{\bold{j}_*\,;\,N-2,N-2}(z_*)\leq 3+ o_{L\to \infty}(1)$$
and, consequently, $\frac{z_*}{\phi_{\bold{j}_*}}\geq -1+o_{L\to \infty}(1)$.  Combining the latter inequality with the bound (\ref{upper-bound-zstar}) in Theorem \ref{fixed-p-thm} that holds if $\rho$ is sufficiently large (but fixed), we derive that  $z_*=-\phi_{\bold{j}_*}+o_{L\to \infty}(1)$. Hence, as expected (see \cite{LSSY}), also in space dimension $d=3$ the ground state energy tends to the Bogoliubov energy in the thermodynamic limit if the assumptions of Corollary \ref{col-hbog} are satisfied.
%In $d=3$ the R-H-S in can be shown to be less than $\mathcal{O}(\frac{1}{\rho})$ uniformly in $L$ by choosing $\beta=\frac{2}{3}$.
%general, if $\rho \geq ...$ The same result holds in dimension $d\geq 4$ for a  density $\rho$ sufficiently large that can be chosen independently of the size of the box. {\color{red}... condition 3.2)...}

\end{remark}
\begin{proposition}\label{induction-G}
Under the assumptions of Theorem \ref{theorem-Bog}, the following identity holds true
\begin{equation}
(\ref{scalar-prod})=(1-\frac{1}{N})\frac{\phi_{\bold{j}_{*}}}{2\epsilon_{\bold{j}_*}+2-\frac{4}{N}-\frac{z}{\phi_{\bold{j}_{*}}}}\check{\mathcal{G}}_{\bold{j}_*\,;\,N-2,N-2}(z)
\end{equation}
where $\check{\mathcal{G}}_{\bold{j}_*\,;\,i,i}(z)$, with $0\leq i\leq N-2$ and even, is defined recursively starting from $\check{\mathcal{G}}_{\bold{j}_*\,;\,0,0}(z)\equiv 1$ and
\begin{equation}\label{def-G-bis}
\check{\mathcal{G}}_{\bold{j}_*\,;\,i,i}(z):=\sum_{l_{i}=0}^{\infty}[\mathcal{W}_{\bold{j}_*\,;i,i-2}(z)\mathcal{W}^*_{\bold{j}_*\,;i-2,i}(z)\check{\mathcal{G}}_{\bold{j}_*\,;\,i-2,i-2}(z)]^{l_i}\,,
\end{equation}
with $\mathcal{W}_{\bold{j}_*\,;i,i-2}(z)\mathcal{W}^*_{\bold{j}_*\,;i-2,i}(z)$ defined in (\ref{def-Wcal-1}).
\end{proposition}

\noindent
\emph{Proof}

\noindent
First, we use the identity in (\ref{ident}) and show by induction that 
\begin{equation}
\check{\Gamma}^{Bog\,}_{\bold{j}_*\,;\,i,i}(z)\tilde{Q}^{(i)}_{\bold{j}_*}Q^{(N-i)}_{\bold{0}}=\check{\mathcal{G}}_{\bold{j}_*\,;\,i,i}(z)\tilde{Q}^{(i)}_{\bold{j}_*}Q^{(N-i)}_{\bold{0}}\,
\end{equation} 
where $\tilde{Q}^{(i)}_{\bold{j}_*}$  projects onto the subspace of vectors with exactly $\frac{N-i}{2}$ particles in the modes $\bold{j}_{*}$ and $-\bold{j_*}$, respectively, and $Q^{(N-i)}_{\bold{0}}$ onto the subspace of vectors with $i$ particles in the mode $\bold{0}$. 
Due to (\ref{gamma-check-b}), the property holds for $i=0$.  Recall the form of $W_{\bold{j}_*\,;\,i-2,i}$, $W^*_{\bold{j}_*\,;\,i-2,i}$ and the fact that $(R^{Bog}_{\bold{j}_*\,;\,i,i}(z))^{\frac{1}{2}}$ commutes with each number operator $a^*_{\bold{j}}a_{\bold{j}}$. Then, assuming that the property is true for $i-2\geq 0$,   we derive
\begin{eqnarray}
& &\check{\Gamma}^{Bog\,}_{\bold{j}_*\,;\,i,i}(z)\tilde{Q}^{(i)}_{\bold{j}_*}Q^{(N-i)}_{\bold{0}}\label{ident-bis}\\
%&=&\sum_{l_{i}=0}^{\infty}[(R^{Bog}_{\bold{j}_*\,;\,i,i}(z))^{\frac{1}{2}}W_{\bold{j}_*\,;\,i,i-2}\,R^{Bog}_{\bold{j}_*\,;\,i-2,i-2}(z) \sum_{l_{i-2}=0}^{\infty}\Big[\Gamma^{Bog}_{\bold{j}_*\,;\,i-2,i-2}(z)R^{Bog}_{\bold{j}_*\,;\,i-2,i-2}(z)\Big]^{l_{i-2}}W^*_{\bold{j}_*\,;\,i-2,i}(R^{Bog}_{\bold{j}_*\,;\,i,i}(z))^{\frac{1}{2}}]^{l_i}\nonumber \\
%&= &Q^{(i)}_{\bold{j}_*}\sum_{l_{i}=0}^{\infty}[(R^{Bog}_{\bold{j}_*\,;\,i,i}(z))^{\frac{1}{2}}W_{\bold{j}_*\,;\,i,i-2}\,(R^{Bog}_{\bold{j}_*\,;\,i-2,i-2}(z))^{\frac{1}{2}} \check{\Gamma}^{Bog}_{\bold{j}_*\,;\,i-2,i-2}(z)(R^{Bog}_{\bold{j}_*\,;\,i-2,i-2}(z))^{\frac{1}{2}}W^*_{\bold{j}_*\,;\,i-2,i}(R^{Bog}_{\bold{j}_*\,;\,i,i}(z))^{\frac{1}{2}}]^{l_i}Q^{(i)}_{\bold{j}_*}\\
%&=&\sum_{l_{i}=0}^{\infty}[(R^{Bog}_{\bold{j}_*\,;\,i,i}(z))^{\frac{1}{2}}W_{\bold{j}_*\,;\,i,i-2}\,(R^{Bog}_{\bold{j}_*\,;\,i-2,i-2}(z))^{\frac{1}{2}}Q^{(i-2)}_{\bold{j}_*} \check{\Gamma}^{Bog}_{\bold{j}_*\,;\,i-2,i-2}(z)Q^{(i-2)}_{\bold{j}_*}(R^{Bog}_{\bold{j}_*\,;\,i-2,i-2}(z))^{\frac{1}{2}}W^*_{\bold{j}_*\,;\,i-2,i}(R^{Bog}_{\bold{j}_*\,;\,i,i}(z))^{\frac{1}{2}}]^{l_i}\,.\quad\quad\quad\label{gamma-exp-bis-bis}\\
%&=&\sum_{l_{i}=0}^{\infty}[(R^{Bog}_{\bold{j}_*\,;\,i,i}(z))^{\frac{1}{2}}W_{\bold{j}_*\,;\,i,i-2}\,(R^{Bog}_{\bold{j}_*\,;\,i-2,i-2}(z))^{\frac{1}{2}}\check{\mathcal{G}}_{\bold{j}_*\,;\,i-2,i-2}(z)Q^{(i-2)}_{\bold{j}_*}(R^{Bog}_{\bold{j}_*\,;\,i-2,i-2}(z))^{\frac{1}{2}}W^*_{\bold{j}_*\,;\,i-2,i}(R^{Bog}_{\bold{j}_*\,;\,i,i}(z))^{\frac{1}{2}}]^{l_i}\\
&=&\sum_{l_{i}=0}^{\infty}[(R^{Bog}_{\bold{j}_*\,;\,i,i}(z))^{\frac{1}{2}}W_{\bold{j}_*\,;\,i,i-2}\,(R^{Bog}_{\bold{j}_*\,;\,i-2,i-2}(z))^{\frac{1}{2}}\times \nonumber\\
& &\quad\quad\quad\times\check{\Gamma}^{Bog\,}_{\bold{j}_*\,;\,i-2,i-2}(z)\tilde{Q}^{(i-2)}_{\bold{j}_*}Q^{(N-i+2)}_{\bold{0}}(R^{Bog}_{\bold{j}_*\,;\,i-2,i-2}(z))^{\frac{1}{2}}W^*_{\bold{j}_*\,;\,i-2,i}(R^{Bog}_{\bold{j}_*\,;\,i,i}(z))^{\frac{1}{2}}\tilde{Q}^{(i)}_{\bold{j}_*}Q^{(N-i)}_{\bold{0}}]^{l_i}\nonumber\\
&=&\sum_{l_{i}=0}^{\infty}[(R^{Bog}_{\bold{j}_*\,;\,i,i}(z))^{\frac{1}{2}}W_{\bold{j}_*\,;\,i,i-2}\,(R^{Bog}_{\bold{j}_*\,;\,i-2,i-2}(z))^{\frac{1}{2}}\check{\mathcal{G}}_{\bold{j}_*\,;\,i-2,i-2}(z)\tilde{Q}^{(i-2)}_{\bold{j}_*}(R^{Bog}_{\bold{j}_*\,;\,i-2,i-2}(z))^{\frac{1}{2}}W^*_{\bold{j}_*\,;\,i-2,i}(R^{Bog}_{\bold{j}_*\,;\,i,i}(z))^{\frac{1}{2}}\tilde{Q}^{(i)}_{\bold{j}_*}Q^{(N-i)}_{\bold{0}}]^{l_i}\nonumber\\
&=&\sum_{l_{i}=0}^{\infty}[\mathcal{W}_{\bold{j}_*\,;\,i,i-2}(z)\mathcal{W}^*_{\bold{j}_*\,;\,i-2,i}(z)\check{\mathcal{G}}_{\bold{j}_*\,;\,i-2,i-2}(z)\tilde{Q}^{(i)}_{\bold{j}_*}Q^{(N-i)}_{\bold{0}}]^{l_i}\\
&=&\sum_{l_{i}=0}^{\infty}[\mathcal{W}_{\bold{j}_*\,;\,i,i-2}(z)\mathcal{W}^*_{\bold{j}_*\,;\,i-2,i}(z)\check{\mathcal{G}}_{\bold{j}_*\,;\,i-2,i-2}(z)]^{l_i}\tilde{Q}^{(i)}_{\bold{j}_*}Q^{(N-i)}_{\bold{0}}\label{def-G-bis}
\end{eqnarray}
where we have made use of the identity (see the definition of $\mathcal{W}_{\bold{j}_*\,;\,i,i-2}(z)\mathcal{W}^*_{\bold{j}_*\,;\,i-2,i}(z)$ in (\ref{def-Wcal-1})) 
\begin{eqnarray}
& &(R^{Bog}_{\bold{j}_*\,;\,i,i}(z))^{\frac{1}{2}}W_{\bold{j}_*\,;\,i,i-2}\,(R^{Bog}_{\bold{j}_*\,;\,i-2,i-2}(z))^{\frac{1}{2}}(R^{Bog}_{\bold{j}_*\,;\,i-2,i-2}(z))^{\frac{1}{2}}W^*_{\bold{j}_*\,;\,i-2,i}(R^{Bog}_{\bold{j}_*\,;\,i,i}(z))^{\frac{1}{2}}\tilde{Q}^{(i)}_{\bold{j}_*}Q^{(N-i)}_{\bold{0}}\quad\quad\quad\quad\\
&=&\mathcal{W}_{\bold{j}_*\,;\,i,i-2}(z)\mathcal{W}^*_{\bold{j}_*\,;\,i-2,i}(z)\tilde{Q}^{(i)}_{\bold{j}_*}Q^{(N-i)}_{\bold{0}}\,.
\end{eqnarray}
Since $$\check{\mathcal{G}}_{\bold{j}_*\,;\,i,i}(z)\leq \|\check{\Gamma}^{Bog\,}_{\bold{j}_*\,;\,i,i}(z)\|\quad \text{and}\quad \mathcal{W}_{\bold{j}_*\,;i,i-2}(z)\mathcal{W}^*_{\bold{j}_*\,;i-2,i}(z)\leq \|(R^{Bog}_{\bold{j}_*\,;\,i,i}(z))^{\frac{1}{2}}W_{\bold{j}_*\,;\,i,i-2}(R^{Bog}_{\bold{j}_*\,;\,i-2,i-2}(z))^{\frac{1}{2}}\|^2\,,$$
the series on the R-H-S of (\ref{def-G-bis}) is convergent under the assumptions of Theorem \ref{theorem-Bog}, and we can readily deduce
\begin{equation}
(\ref{def-G-bis})=\frac{1}{1-\mathcal{W}_{\bold{j}_*\,;i,i-2}(z)\mathcal{W}^*_{\bold{j}_*\,;i-2,i}(z)\check{\mathcal{G}}_{\bold{j}_*\,;\,i-2,i-2}(z)}\tilde{Q}^{(i)}_{\bold{j}_*}Q^{(N-i)}_{\bold{0}}=\check{\mathcal{G}}_{\bold{j}_*\,;\,i,i}(z)\tilde{Q}^{(i)}_{\bold{j}_*}Q^{(N-i)}_{\bold{0}}\,.
\end{equation}
Then, a straightforward computation yields
\begin{eqnarray}
& &\langle \eta\,,\,W_{\bold{j}_*}\,(R^{Bog}_{\bold{j}_*\,;\,N-2,N-2}(z))^{\frac{1}{2}}\check{\Gamma}^{Bog}_{\bold{j}_*\,;\,N-2,N-2}(z) (R^{Bog}_{\bold{j}_*\,;\,N-2,N-2}(z))^{\frac{1}{2}}\,W^*_{\bold{j}_*}\eta \rangle \\
%&=&\langle \eta\,,\,W_{\bold{j}_*}\,(R^{Bog}_{\bold{j}_*\,;\,N-2,N-2}(z))^{\frac{1}{2}}Q^{(i)}_{\bold{j}_*}\check{\Gamma}^{Bog}_{\bold{j}_*\,;\,N-2,N-2}(z) Q^{(i)}_{\bold{j}_*}(R^{Bog}_{\bold{j}_*\,;\,N-2,N-2}(z))^{\frac{1}{2}}\,W^*_{\bold{j}_*}\eta \rangle\\
&=&\langle \eta\,,\,W_{\bold{j}_*}\,(R^{Bog}_{\bold{j}_*\,;\,N-2,N-2}(z))^{\frac{1}{2}} \check{\mathcal{G}}_{\bold{j}_*\,;\,N-2,N-2}(z)\tilde{Q}^{(N-2)}_{\bold{j}_*}(R^{Bog}_{\bold{j}_*\,;\,N-2,N-2}(z))^{\frac{1}{2}}\,W^*_{\bold{j}_*}\eta \rangle \\
&=&(1-\frac{1}{N})\frac{\phi_{\bold{j}_{*}}}{2\epsilon_{\bold{j}_*}+2-\frac{4}{N}-\frac{z}{\phi_{\bold{j}_{*}}}}\check{\mathcal{G}}_{\bold{j}_*\,;\,N-2,N-2}(z)
\end{eqnarray}
that concludes the proof.
\qed

We recall that in Sections \ref{informal} and \ref{informal-1} we have dropped the index $\bold{j}_*$ in the notation used for $\Gamma^{Bog}_{\bold{j}_*\,;\,i,i}(z)$, $W_{\bold{j}_*\,;\,i+2,i}$, and $R^{Bog}_{\bold{j}_*\,;\,i,i}(z)$. 
The notation in the next proposition is consistent with this choice.
%\subsection{Informal presentation}
\begin{proposition}\label{lemma-expansion-proof-bis} Let  $\frac{1}{N}\leq \epsilon^{\nu}_{\bold{j}_*}$ for some $\nu >\frac{11}{8}$ and $\epsilon_{\bold{j}_*}\equiv \epsilon$ be sufficiently small.
For any $2\leq h \in \mathbb{N}$ and for $N-2\geq i\geq 4$ and even,  the splitting
\begin{eqnarray}
\Gamma^{Bog\,}_{i,i}(z)&=&\sum_{r=2,\,l\, even}^{i-2}[\Gamma^{Bog\,}_{i,i}(z)]_{(r,h_-; r+2,h_-;\dots ; i-2,h_-)}+\sum_{r=2\,,\,l\, even}^{i-2}[\Gamma^{Bog\,}_{i,i}(z)]_{(r,h_+; r+2,h_-;\dots; i-2,h_-)}\quad \quad\quad\label{decomposition}
\end{eqnarray}
holds true for $z\leq E^{Bog}_{\bold{j}_*}+ (\delta -1)\phi_{\bold{j}_*}\sqrt{\epsilon_{\bold{j}_*}^2+2\epsilon_{\bold{j}_*}}$ with $\delta\leq 1+\sqrt{\epsilon_{\bold{j}_*}}$. Moreover, for $2\leq r \leq i-2$ and even,  the estimates
\begin{eqnarray}\label{gamma-exp-1}
& &\Big\|(R^{Bog}_{i,i}(z))^{\frac{1}{2}}[\Gamma^{Bog\,}_{i,i}(z)]_{( r,h_-; r+2, h_-;\, \dots \,; i-2,h_-)}(R^{Bog}_{i,i}(z))^{\frac{1}{2}}\Big\|\\
& &\leq  \prod_{f=r+2\,,\, f-r\,\text{even}}^{i}\frac{K_{f,\epsilon}}{(1-Z_{f-2,\epsilon})^2}
%\Big(\frac{2}{3}+\mathcal{O}(\sqrt{\epsilon})\Big)^{\frac{i-l}{2}}\prod_{f=l+2\,,\, f-l\,\text{even}}^{i-2}(1+a_{\epsilon}-\frac{2b_{\epsilon}}{N-f-1}-\frac{1-c_{\epsilon}}{(N-f-1)^2})^{-1}\nonumber
\end{eqnarray}
and
%\begin{equation}
%\Big\|(R^{Bog}_{i,i}(z))^{\frac{1}{2}}[\Gamma^{Bog\,}_{i,i}(z)]_{(l,h_-;\dots ; j,h_-;j-4,h_+)}
%(R^{Bog}_{i,i}(z))^{\frac{1}{2}}\Big\|\leq {\color{red}...}
%\end{equation}
\begin{eqnarray}\label{gamma-exp-2}
& &\|(R^{Bog}_{i,i}(z))^{\frac{1}{2}}[\Gamma^{Bog\,}_{i,i}(z)]_{(r,h_+;r+2,h_-;\dots;i-2,h_-)}(R^{Bog}_{i,i}(z))^{\frac{1}{2}}\|\\
&\leq& (Z_{r,\epsilon})^h\,\prod_{f=r+2\,,\, f-r\,\text{even}}^{i}\frac{K_{f,\epsilon}}{(1-Z_{f-2,\epsilon})^2}\times \nonumber
%& &\quad\times \prod_{f=l+2\,,\, f-l\,\text{even}}^{i-2}(1+a_{\epsilon}-\frac{2b_{\epsilon}}{N-f-1}-\frac{1-c_{\epsilon}}{(N-f-1)^2})^{-1}\nonumber
 \end{eqnarray}
hold true, where 
\begin{equation}\label{def-K-Z}
K_{i,\epsilon}:=\frac{1}{4(1+a_{\epsilon}-\frac{2b_{\epsilon}}{N-i+1}-\frac{1-c_{\epsilon}}{(N-i+1)^2})}\quad,\quad Z_{i-2,\epsilon}:=\frac{1}{4(1+a_{\epsilon}-\frac{2b_{\epsilon}}{N-i+3}-\frac{1-c_{\epsilon}}{(N-i+3)^2})}\frac{2}{\Big[1+\sqrt{\eta a_{\epsilon}}-\frac{b_{\epsilon}/\sqrt{\eta a_{\epsilon}}}{N-i+4-\epsilon^{\Theta}}\Big]}\,.
\end{equation}
where $a_{\epsilon},b_{\epsilon}$, and $c_{\epsilon}$ are defined in (\ref{adelta})-(\ref{bdelta})-(\ref{cdelta}) and $\Theta:=\min\{2(\nu-\frac{11}{8})\,;\,\frac{1}{4}\}$.
\end{proposition}

\noindent
\emph{Proof}

\noindent
In Section \ref{informal} we have proven that the decomposition in (\ref{decomposition}) holds for $i=4$ (see (\ref{decomp-4})). We assume that it holds for all the even numbers $k$ with  $4\leq k\leq i-2\leq N-4$ and we show that it is verified for $i$. Starting from the identity
\begin{eqnarray}
& &\Gamma^{Bog\,}_{i,i}(z)\\
&=&W_{i,i-2}\,(R^{Bog}_{i-2,i-2}(z))^{\frac{1}{2}} \sum_{l_{i-2}=0}^{\infty}\Big[(R^{Bog}_{i-2,i-2}(z))^{\frac{1}{2}} \Gamma^{Bog}_{i-2,i-2}(z)(R^{Bog}_{i-2,i-2}(z))^{\frac{1}{2}} \Big]^{l_{i-2}}(R^{Bog}_{i-2,i-2}(z))^{\frac{1}{2}} W^*_{i-2,i}\quad\quad\quad
\end{eqnarray}
we repeat some steps of the informal discussion of Section \ref{informal}. First, we isolate
\begin{eqnarray}
& &[\Gamma^{Bog\,}_{i,i}(z)]_{(i-2,h_+)}\label{first4identity}\\
&:=&W_{i,i-2}\,(R^{Bog}_{i-2,i-2}(z))^{\frac{1}{2}} \sum_{l_{i-2}=h}^{\infty}\Big[(R^{Bog}_{i-2,i-2}(z))^{\frac{1}{2}} \Gamma^{Bog}_{i-2,i-2}(z)(R^{Bog}_{i-2,i-2}(z))^{\frac{1}{2}} \Big]^{l_{i-2}}(R^{Bog}_{i-2,i-2}(z))^{\frac{1}{2}} W^*_{i-2,i}\nonumber \quad\quad\quad
\end{eqnarray}
and
\begin{equation}\label{second4identity}
[\Gamma^{Bog\,}_{i,i}(z)]^{(0)}_{(i-2,h_-)}:=W_{i,i-2}\,R^{Bog}_{i-2,i-2}(z)W^*_{i-2,i}\,.
\end{equation}
Concerning the remaining quantity
\begin{equation}
%& &W_{i,i-2}\,R^{Bog}_{i-2,i-2}(z)W^*_{i-2,i}\\
W_{i,i-2}\,(R^{Bog}_{i-2,i-2}(z))^{\frac{1}{2}} \sum_{l_{i-2}=1}^{h-1}\Big[(R^{Bog}_{i-2,i-2}(z))^{\frac{1}{2}} \Gamma^{Bog}_{i-2,i-2}(z)(R^{Bog}_{i-2,i-2}(z))^{\frac{1}{2}} \Big]^{l_{i-2}}(R^{Bog}_{i-2,i-2}(z))^{\frac{1}{2}} W^*_{i-2,i}\label{plug}
\end{equation}
we invoke the inductive hypothesis for $\Gamma^{Bog}_{i-2,i-2}(z)$, i.e.,
\begin{equation}
\Gamma^{Bog\,}_{i-2,i-2}(z):=\sum_{r=2,\, r\, even}^{i-4}[\Gamma^{Bog\,}_{i-2,i-2}(z)]_{(r,h_-; r+2,h_-;\dots ; i-4,h_-)}+\sum_{r=2\,,\,r\, even}^{i-4}[\Gamma^{Bog\,}_{i-2,i-2}(z)]_{(r,h_+; r+2,h_-;\dots; i-4,h_-)}\,.
\end{equation}
Making use of of the symbols $\hat{\sum}$ and $\check{\sum}$ introduced in Definition \ref{def-sums}, we can write
\begin{eqnarray}
&&\sum_{l_{i-2}=1}^{h-1}\Big[(R^{Bog}_{i-2,i-2}(z))^{\frac{1}{2}}\Big\{\sum_{r=2,\,r\, even}^{i-4}\Big[[\Gamma^{Bog\,}_{i-2,i-2}(z)]_{(r,h_-;r+2,h_-;\dots ; i-4,h_-)}+[\Gamma^{Bog\,}_{i-2,i-2}(z)]_{(r,h_+; r+2,h_-;\dots; i-4,h_-)}\Big]\Big\}(R^{Bog}_{i-2,i-2}(z))^{\frac{1}{2}}\Big]^{l_{i-2}}\nonumber \label{int-in}\\
&=&\sum_{l_{i-2}=1}^{h-1}\Big[(R^{Bog}_{i-2,i-2}(z))^{\frac{1}{2}}[\Gamma^{Bog\,}_{i-2,i-2}(z)]_{(i-4,h_-)}^{(0)}(R^{Bog}_{i-2,i-2}(z))^{\frac{1}{2}}\Big]^{l_{i-2}}\\
& &+\check{\sum}_{l_{i-2}=1}^{h-1}\Big[(R^{Bog}_{i-2,i-2}(z))^{\frac{1}{2}}[\Gamma^{Bog\,}_{i-2,i-2}(z)]_{(i-4,h_-)}(R^{Bog}_{2,2}(z))^{\frac{1}{2}}\Big]^{l_{i-2}}\\
& &+\hat{\sum}_{l_{i-2}=1}^{h-1}\Big[(R^{Bog}_{i-2,i-2}(z))^{\frac{1}{2}}[\Gamma^{Bog\,}_{i-2,i-2}(z)]_{(i-4,h_+)}(R^{Bog}_{i-2,i-2}(z))^{\frac{1}{2}}\Big]^{l_{i-2}}\\
& &+\check{\sum}_{l_{i-2}=1}^{h-1}\Big[(R^{Bog}_{i-2,i-2}(z))^{\frac{1}{2}}[\Gamma^{Bog\,}_{i-2,i-2}(z)]_{(i-6,h_-;i-4,h_-)}(R^{Bog}_{i-2,i-2}(z))^{\frac{1}{2}}\Big]^{l_{i-2}}\\
& &+\hat{\sum}_{l_{i-2}=1}^{h-1}\Big[(R^{Bog}_{i-2,i-2}(z))^{\frac{1}{2}}[\Gamma^{Bog\,}_{i-2,i-2}(z)]_{(i-6,h_+;i-4,h_-)}(R^{Bog}_{i-2,i-2}(z))^{\frac{1}{2}}\Big]^{l_{i-2}}\\
& &+\dots\\
& &+\check{\sum}_{l_{i-2}=1}^{h-1}\Big[(R^{Bog}_{i-2,i-2}(z))^{\frac{1}{2}}[\Gamma^{Bog\,}_{i-2,i-2}(z)]_{(2,h_-;\dots;i-4,h_-)}(R^{Bog}_{i-2,i-2}(z))^{\frac{1}{2}}\Big]^{l_{i-2}}\\
& &+\hat{\sum}_{l_{i-2}=1}^{h-1}\Big[(R^{Bog}_{i-2,i-2}(z))^{\frac{1}{2}}[\Gamma^{Bog\,}_{i-2,i-2}(z)]_{(2,h_+;\dots ;i-4,h_-)}(R^{Bog}_{i-2,i-2}(z))^{\frac{1}{2}}\Big]^{l_{i-2}}\,.\label{int-fin}
\end{eqnarray}
Next, we  plug (\ref{int-in})-(\ref{int-fin}) into  (\ref{plug}) and due to Definition \ref{def-sums} we derive that 
\begin{eqnarray}
& &W_{i,i-2}\,(R^{Bog}_{i-2,i-2}(z))^{\frac{1}{2}}\sum_{l_{i-2}=1}^{h-1}\Big[(R^{Bog}_{i-2,i-2}(z))^{\frac{1}{2}}[\Gamma^{Bog\,}_{i-2,i-2}(z)]_{(i-4,h_-)}^{(0)}(R^{Bog}_{i-2,i-2}(z))^{\frac{1}{2}}\Big]^{l_{i-2}}(R^{Bog}_{i-2,i-2}(z))^{\frac{1}{2}}W^*_{i-2,i}\nonumber\\
&= &[\Gamma^{Bog\,}_{i,i}(z)]_{(i-2,h_-)}^{(>0)}=[\Gamma^{Bog\,}_{i,i}(z)]_{(i-2,h_-)}-[\Gamma^{Bog\,}_{i,i}(z)]^{(0)}_{(i-2,h_-)}\,,
\end{eqnarray}
\begin{eqnarray}
& &W_{i,i-2}\,(R^{Bog}_{i-2,i-2}(z))^{\frac{1}{2}}\check{\sum}_{l_{i-2}=1}^{h-1}\Big[(R^{Bog}_{i-2,i-2}(z))^{\frac{1}{2}}[\Gamma^{Bog\,}_{i-2,i-2}(z)]_{(i-4,h_-)}(R^{Bog}_{i-2,i-2}(z))^{\frac{1}{2}}\Big]^{l_{i-2}}(R^{Bog}_{i-2,i-2}(z))^{\frac{1}{2}}W^*_{i-2,i}\nonumber\\
& = &[\Gamma^{Bog\,}_{i,i}(z)]_{(i-4, h_-; i-2,h_-)}\,,
\end{eqnarray}
\begin{eqnarray}
& &W_{i,i-2}\,(R^{Bog}_{i-2,i-2}(z))^{\frac{1}{2}}\hat{\sum}_{l_{i-2}=1}^{h-1}\Big[(R^{Bog}_{i-2,i-2}(z))^{\frac{1}{2}}[\Gamma^{Bog\,}_{i-2,i-2}(z)]_{(i-4,h_+)}(R^{Bog}_{i-2,i-2}(z))^{\frac{1}{2}}\Big]^{l_{i-2}}(R^{Bog}_{i-2,i-2}(z))^{\frac{1}{2}}W^*_{i-2,i}\nonumber \\
&= &[\Gamma^{Bog\,}_{i,i}(z)]_{(i-4,h_+;i-2,h_-)}\,.
\end{eqnarray}
In general, we get 
\begin{eqnarray}
& &W_{i,i-2}\,(R^{Bog}_{i-2,i-2}(z))^{\frac{1}{2}}\check{\sum}_{l_{i-2}=1}^{h-1}\Big[(R^{Bog}_{i-2,i-2}(z))^{\frac{1}{2}}[\Gamma^{Bog\,}_{i-2,i-2}(z)]_{(f,h_-;\dots;i-4,h_-)}(R^{Bog}_{i-2,i-2}(z))^{\frac{1}{2}}\Big]^{l_2}(R^{Bog}_{i-2,i-2}(z))^{\frac{1}{2}}W^*_{i-2,i}\nonumber\\
&=  &[\Gamma^{Bog\,}_{i,i}(z)]_{(f,h_-;f-2,h_-;\dots; i-2,h_-)}\,
\end{eqnarray}
for $2\leq f\leq i-4$ and even,
and 
\begin{eqnarray}
& &W_{i,i-2}\,(R^{Bog}_{i-2,i-2}(z))^{\frac{1}{2}}\hat{\sum}_{l_{i-2}=1}^{h-1}\Big[(R^{Bog}_{i-2,i-2}(z))^{\frac{1}{2}}[\Gamma^{Bog\,}_{i-2,i-2}(z)]_{(r,h_+;\dots;i-4,h_-)}(R^{Bog}_{i-2,i-2}(z))^{\frac{1}{2}}\Big]^{l_2}(R^{Bog}_{i-2,i-2}(z))^{\frac{1}{2}}W^*_{i-2,i}\nonumber\\
& = &[\Gamma^{Bog\,}_{i,i}(z)]_{(r,h_+;r-2,h_+;\dots; i-2,h_-)}
\end{eqnarray}
for $2\leq r\leq i-4$ and even.
We conclude that
\begin{equation}
(\ref{plug})
= -[\Gamma^{Bog\,}_{i,i}(z)]^{(0)}_{(i-2,h_-)}+\sum_{r=2,\,r\, even}^{i-2}[\Gamma^{Bog\,}_{i,i}(z)]_{(r,h_-; r+2,h_-;\dots ; i-2,h_-)}+\sum_{r=2\,,\, r\, even}^{i-4}[\Gamma^{Bog\,}_{i,i}(z)]_{(r,h_+; r+2,h_-;\dots; i-2,h_-)}
\end{equation}
By adding the terms in (\ref{first4identity}), (\ref{second4identity}) that have been previously isolated the identity in (\ref{decomposition}) is proven.
\\

Now, we prove the operator norm estimates in (\ref{gamma-exp-1}) and (\ref{gamma-exp-2}). To this purpose, we recall Lemma \ref{main-lemma-Bog}  and (\ref{doppia}), and observe that for $\epsilon$  sufficiently small and $i\leq N-2$
\begin{eqnarray}
& &\|(R^{Bog}_{i,i}(z))^{\frac{1}{2}}[\Gamma^{Bog\,}_{i,i}(z)]^{(>0)}_{(i-2, h_-)}(R^{Bog}_{i,i}(z))^{\frac{1}{2}}\|\label{esti-y-2}\\
%&\leq&\|(R^{Bog}_{i,i}(z))^{\frac{1}{2}}[\Gamma^{Bog\,}_{i,i}(z)]_{(i-2, h_-)}(R^{Bog}_{i,i}(z))^{\frac{1}{2}}\|\\
&=&\|(R^{Bog}_{i,i}(z))^{\frac{1}{2}}W_{i,i-2}\,(R^{Bog}_{i-2,i-2}(z))^{\frac{1}{2}} \sum_{l_{i-2}=1}^{h-1}\Big[(R^{Bog}_{i-2,i-2}(z))^{\frac{1}{2}}W_{i-2,i-4}\,R_{i-4,i-4}^{Bog}(z)W_{i-4,i-2}^*(R^{Bog}_{i-2,i-2}(z))^{\frac{1}{2}} \Big]^{l_{i-2}}\times\nonumber \\
& &\quad\quad \times (R^{Bog}_{i-2,i-2}(z))^{\frac{1}{2}} W^*_{i-2,i}(R^{Bog}_{i,i}(z))^{\frac{1}{2}}\|\nonumber \\
&\leq &\|(R^{Bog}_{i,i}(z))^{\frac{1}{2}}W_{i,i-2}\,(R^{Bog}_{i-2,i-2}(z))^{\frac{1}{2}}\| \sum_{l_{i-2}=0}^{\infty}\Big[\|(R^{Bog}_{i-2,i-2}(z))^{\frac{1}{2}}W_{i-2,i-4}\,R_{i-4,i-4}^{Bog}(z)W_{i-4,i-2}^*(R^{Bog}_{i-2,i-2}(z))^{\frac{1}{2}} \|\Big]^{l_{i-2}}\times\nonumber \\
& &\quad\quad \times \|(R^{Bog}_{i-2,i-2}(z))^{\frac{1}{2}} W^*_{i-2,i}(R^{Bog}_{i,i}(z))^{\frac{1}{2}}\|\nonumber \\
&\leq &\frac{1}{4(1+a_{\epsilon}-\frac{2b_{\epsilon}}{N-i+1}-\frac{1-c_{\epsilon}}{(N-i+1)^2})} \sum_{l_{i-2}=0}^{\infty}\Big[\|(R^{Bog}_{i-2,i-2}(z))^{\frac{1}{2}}W_{i-2,i-4}\,R_{i-4,i-4}^{Bog}(z)W_{i-4,i-2}^*(R^{Bog}_{i-2,i-2}(z))^{\frac{1}{2}} \|\Big]^{l_{i-2}}\nonumber \\
&\leq &\frac{1}{4(1+a_{\epsilon}-\frac{2b_{\epsilon}}{N-i+1}-\frac{1-c_{\epsilon}}{(N-i+1)^2})}\Big[\frac{1}{1-\frac{1}{4(1+a_{\epsilon}-\frac{2b_{\epsilon}}{N-i+3}-\frac{1-c_{\epsilon}}{(N-i+3)^2})}}\Big] \nonumber \\
&<&\frac{K_{i,\epsilon}}{(1-Z_{i-2,\epsilon})^2}\,.\label{esti-y-2-bis}
%& &\quad\quad \times \|(R^{Bog}_{j-2,j-2}(z))^{\frac{1}{2}} W^*_{j-2,j}(R^{Bog}_{j,j}(z))^{\frac{1}{2}}\|\nonumber
\end{eqnarray}
%Then we define
%\begin{equation}
%y^{(l)}_{i-2}:= \|(R^{Bog}_{i-2,i-2}(z))^{\frac{1}{2}}[\Gamma^{Bog\,}_{i-2,i-2}(z)]_{(l,h_-;4,h_-;\dots;i-4,h_-)}(R^{Bog}_{i-2,i-2}(z))^{\frac{1}{2}}\|
%\end{equation}
%{\color{red} and we know that $\|(R^{Bog}_{j-2,j-2}(z))^{\frac{1}{2}}[\Gamma^{Bog\,}_{j-2,j-2}(z)]_{(l,h_-;4,h_-;\dots;j-4,h_-)}(R^{Bog}_{j-2,j-2}(z))^{\frac{1}{2}}\|\geq \frac{1}{4}$

For $r< i-4$, using (\ref{collection})  we estimate
\begin{eqnarray}
& &\|(R^{Bog}_{i,i}(z))^{\frac{1}{2}}[\Gamma^{Bog\,}_{i,i}(z)]_{(r,h_-;r+2,h_-;\dots;i-4,h_-;i-2,h_-)}(R^{Bog}_{i,i}(z))^{\frac{1}{2}}\| \\
&= &\|(R^{Bog}_{i,i}(z))^{\frac{1}{2}}W_{i,i-2}\,(R^{Bog}_{i-2,i-2}(z))^{\frac{1}{2}} \times \label{esti-gamma-1}\\
& &\quad\quad \times\check{\sum}_{l_{i-2}=1}^{h-1}\Big[(R^{Bog}_{i-2,i-2}(z))^{\frac{1}{2}}[\Gamma^{Bog\,}_{i-2,i-2}(z)]_{(r,h_-; r+2,h_-;\dots;i-4,h_-)}(R^{Bog}_{i-2,i-2}(z))^{\frac{1}{2}} \Big]^{l_{i-2}} (R^{Bog}_{i-2,i-2}(z))^{\frac{1}{2}} W^*_{i-2,i}(R^{Bog}_{i,i}(z))^{\frac{1}{2}}\|\nonumber\\
%&\leq &\|(R^{Bog}_{j,j}(z))^{\frac{1}{2}}W_{j,j-2}\,(R^{Bog}_{j-2,j-2}(z))^{\frac{1}{2}}\| \sum_{l_{j-2}=1}^{h-1}\Big[\|(R^{Bog}_{j-2,j-2}(z))^{\frac{1}{2}}[\Gamma^{Bog\,}_{j-2,j-2}(z)]_{(l,h_-;4,h_-;\dots;j-4,h_-)}(R^{Bog}_{j-2,j-2}(z))^{\frac{1}{2}}\| \Big]^{l_{2}}\times \nonumber \\
%& &\quad\quad\quad \times \|(R^{Bog}_{j-2,j-2}(z))^{\frac{1}{2}} W^*_{j-2,j}(R^{Bog}_{j,j}(z))^{\frac{1}{2}}\|\nonumber\\
%&\leq &\frac{1}{4(1+a_{\epsilon}-\frac{2b_{\epsilon}}{N-i+1}-\frac{1-c_{\epsilon}}{(N-i+1)^2})}\Big[ \frac{\|(R^{Bog}_{i-2,i-2}(z))^{\frac{1}{2}}[\Gamma^{Bog\,}_{i-2,i-2}(z)]_{(l,h_-;4,h_-;\dots;i-4,h_-)}(R^{Bog}_{i-2,i-2}(z))^{\frac{1}{2}}\|}{1- \|(R^{Bog}_{i-2,i-2}(z))^{\frac{1}{2}}[\Gamma^{Bog\,}_{i-2,i-2}(z)](R^{Bog}_{i-2,i-2}(z))^{\frac{1}{2}}\|}\Big]  \label{ineq-RGammaR-1}\\
&\leq &\|(R^{Bog}_{i,i}(z))^{\frac{1}{2}}W_{i,i-2}\,(R^{Bog}_{i-2,i-2}(z))^{\frac{1}{2}}\|^2\,\frac{1}{(1-Z_{i-2,\epsilon})^2}\,\|(R^{Bog}_{i-2,i-2}(z))^{\frac{1}{2}}[\Gamma^{Bog\,}_{i-2,i-2}(z)]_{(r,h_-; r+2,h_-;\dots;i-4,h_-)}(R^{Bog}_{i-2,i-2}(z))^{\frac{1}{2}}\|\quad\quad\quad \label{esti-gamma-2}\\
&=&\frac{K_{i,\epsilon}}{(1-Z_{i-2,\epsilon})^2}\,\|(R^{Bog}_{i-2,i-2}(z))^{\frac{1}{2}}[\Gamma^{Bog\,}_{i-2,i-2}(z)]_{(r,h_-; r+2,h_-;\dots;i-4,h_-)}(R^{Bog}_{i-2,i-2}(z))^{\frac{1}{2}}\|
\end{eqnarray}
%If we define
%\begin{equation
%y^{(l)}_{j-2}:=\|(R^{Bog}_{j-2,j-2}(z))^{\frac{1}{2}}[\Gamma^{Bog\,}_{j-2,j-2}(z)]_{(l,h_-;4,h_-;\dots;j-4,h_-)}(R^{Bog}_{j-2,j-2}(z))^{\frac{1}{2}}\|
%\end{equation}
where the step from (\ref{esti-gamma-1}) to (\ref{esti-gamma-2}) follows from the definitions in (\ref{def-K-Z}) and the two observations below:
\begin{itemize}
\item
 By definition of $\check{\sum}_{l_{i-2}=1}^{h-1}$,  
 \begin{equation}
\check{\sum}_{l_{i-2}=1}^{h-1} \Big[(R^{Bog}_{i-2,i-2}(z))^{\frac{1}{2}}[\Gamma^{Bog\,}_{i-2,i-2}(z)]_{( r,h_-; r+2,h_-;\dots;i-4,h_-)}(R^{Bog}_{i-2,i-2}(z))^{\frac{1}{2}} \Big]^{l_{i-2}}\label{esti-gamma-1-bis}
 \end{equation}
 stands for a sum of products where at least one of the factors
 must contain $[\Gamma^{Bog\,}_{i-2,i-2}(z)]_{(r,h_-; r+2,h_-;\dots;i-4,h_-)}$ that, consequently, can be factorized; 
 \item
After the factorization, using the argument of Remark \ref{estimation-proc},  the norm of the sum in (\ref{esti-gamma-1-bis}) can be bounded with
\begin{eqnarray}
& &\mathcal{E}\Big(\sum_{l_{i-2}=0}^{h-2}(l_{i-2}+1)\|(R^{Bog}_{i-2,i-2}(z))^{\frac{1}{2}}[\Gamma^{Bog\,}_{i-2,i-2}(z)](R^{Bog}_{i-2,i-2}(z))^{\frac{1}{2}}\|^{l_{i-2}}\Big)\\
&=&\sum_{l_{i-2}=0}^{h-2}(l_{i-2}+1)\mathcal{E}\Big(\|(R^{Bog}_{i-2,i-2}(z))^{\frac{1}{2}}[\Gamma^{Bog\,}_{i-2,i-2}(z)](R^{Bog}_{i-2,i-2}(z))^{\frac{1}{2}}\|\Big)^{l_{i-2}}\label{doppiageom}
\end{eqnarray}
where the symbol $\mathcal{E}(\dots )$ has been defined in Remark \ref{estimation-proc}. 
%Since $i-2\leq N-4$,  the norm of (\ref{esti-gamma-1-bis}) is bounded by $\frac{8}{3}+\mathcal{O}(\sqrt{\epsilon})$;
%\begin{equation}
%\frac{1}{1-(\frac{1}{2}+\mathcal{O}(\sqrt{\epsilon}))};
%\end{equation}
We know that (see (\ref{def-gamma-ii}), (\ref{Gamma-ineq}), and (\ref{doppia}))
\begin{eqnarray}
& &\mathcal{E}\Big(\|(R^{Bog}_{i-2,i-2}(z))^{\frac{1}{2}}[\Gamma^{Bog\,}_{i-2,i-2}(z)](R^{Bog}_{i-2,i-2}(z))^{\frac{1}{2}}\|\Big)\\
&=&\mathcal{E}\Big(\|(R^{Bog}_{i-2,i-2}(z))^{\frac{1}{2}}W_{\bold{j}_*\,;\,i-2,i-4}(R^{Bog}_{i-4,i-4}(z))^{\frac{1}{2}}\|^2\Big)\mathcal{E}\Big(\|\check{\Gamma}^{Bog\,}_{i-4,i-4}(z)\|\Big)\\
&\leq &\frac{1}{4(1+a_{\epsilon}-\frac{2b_{\epsilon}}{N-i+3}-\frac{1-c_{\epsilon}}{(N-i+3)^2})}\frac{2}{\Big[1+\sqrt{\eta a_{\epsilon}}-\frac{b_{\epsilon}/\sqrt{\eta a_{\epsilon}}}{N-i+4-\epsilon^{\Theta}}\Big]}\\
&=:&Z_{i-2,\epsilon}\,.
\end{eqnarray}
The R-H-S of (\ref{doppiageom}) is therefore bounded by
\begin{equation}
\frac{1}{(1-Z_{i-2,\epsilon})^2}\,.
\end{equation}
\end{itemize}
For $r=i-4$ the R-H-S in (\ref{esti-gamma-2}) is replaced with
\begin{equation}
\frac{K_{i,\epsilon}}{(1-Z_{i-2,\epsilon})^2}\,\|(R^{Bog}_{i-2,i-2}(z))^{\frac{1}{2}}[\Gamma^{Bog}_{i-2,i-2}(z)]^{(>0)}_{(i-4,h_-)}(R^{Bog}_{i-2,i-2}(z))^{\frac{1}{2}}\|\,.
\end{equation}
By iteration we get
\begin{eqnarray}
& &\|(R^{Bog}_{i,i}(z))^{\frac{1}{2}}[\Gamma^{Bog\,}_{i,i}(z)]_{(r,h_-; r+2,h_-;\dots;i-4,h_-;i-2,h_-)}(R^{Bog}_{i,i}(z))^{\frac{1}{2}}\| \\
%&\leq &\|(R^{Bog}_{i,i}(z))^{\frac{1}{2}}W_{i,i-2}\,(R^{Bog}_{i-2,i-2}(z))^{\frac{1}{2}}\|^2\,\frac{1}{(1-Z_{i-2,\epsilon})^2}\,\|(R^{Bog}_{i-2,i-2}(z))^{\frac{1}{2}}[\Gamma^{Bog\,}_{i-2,i-2}(z)]_{(l,h_-;4,h_-;\dots;i-4,h_-)}(R^{Bog}_{i-2,i-2}(z))^{\frac{1}{2}}\|\quad\quad\quad \label{esti-gamma-2}\\
&\leq&\frac{K_{i,\epsilon}}{(1-Z_{i-2,\epsilon})^2}\,\|(R^{Bog}_{i-2,i-2}(z))^{\frac{1}{2}}[\Gamma^{Bog\,}_{i-2,i-2}(z)]_{(r,h_-; r+2,h_-;\dots;i-4,h_-)}(R^{Bog}_{i-2,i-2}(z))^{\frac{1}{2}}\|\\
&\leq&\frac{K_{i,\epsilon}}{(1-Z_{i-2,\epsilon})^2} \,\frac{K_{i-2,\epsilon}}{(1-Z_{i-4,\epsilon})^2}\,\|(R^{Bog}_{i-4,i-4}(z))^{\frac{1}{2}}[\Gamma^{Bog\,}_{i-4,i-4}(z)]_{(r,h_-; r+2,h_-;\dots;i-6,h_-)}(R^{Bog}_{i-4,i-4}(z))^{\frac{1}{2}}\|\quad\quad\quad\quad \\
&\leq&\frac{K_{i,\epsilon}}{(1-Z_{i-2,\epsilon})^2}\,\dots \,\frac{K_{r+4,\epsilon}}{(1-Z_{r+2,\epsilon})^2}\,\|(R^{Bog}_{r+2, r+2}(z))^{\frac{1}{2}}[\Gamma^{Bog}_{r+2, r+2}(z)]^{(>0)}_{(r,h_-)}(R^{Bog}_{r+2, r+2}(z))^{\frac{1}{2}}\|\quad\quad \\
&\leq &\prod_{f=r+2\,,\, f-r\,\text{even}}^{i}\frac{K_{f,\epsilon}}{(1-Z_{f-2,\epsilon})^2}\,
\end{eqnarray}
where in the last step we have  used the estimate in (\ref{esti-y-2-bis}).

 As for the estimate in (\ref{gamma-exp-2}), the argument is very similar. First, we observe that for $\epsilon$  sufficiently small 
 \begin{eqnarray}
& &\|(R^{Bog}_{i,i}(z))^{\frac{1}{2}}[\Gamma^{Bog\,}_{i,i}(z)]_{(i-2, h_+)}(R^{Bog}_{i,i}(z))^{\frac{1}{2}}\|\label{esti-y-2-bis-bis}\\
&=&\|(R^{Bog}_{i,i}(z))^{\frac{1}{2}}W_{i,i-2}\,(R^{Bog}_{i-2,i-2}(z))^{\frac{1}{2}} \sum_{l_{i-2}=h}^{\infty}\Big[(R^{Bog}_{i-2,i-2}(z))^{\frac{1}{2}}[\Gamma^{Bog\,}_{i-2,i-2}(z)](R^{Bog}_{i-2,i-2}(z))^{\frac{1}{2}} \Big]^{l_{i-2}}\times\nonumber \\
& &\quad\quad \times (R^{Bog}_{i-2,i-2}(z))^{\frac{1}{2}} W^*_{i-2,i}(R^{Bog}_{i,i}(z))^{\frac{1}{2}}\|\nonumber \\
&\leq &\|(R^{Bog}_{i,i}(z))^{\frac{1}{2}}W_{i,i-2}\,(R^{Bog}_{i-2,i-2}(z))^{\frac{1}{2}}\|^2 \sum_{l_{i-2}=h}^{\infty}\Big\|(R^{Bog}_{i-2,i-2}(z))^{\frac{1}{2}}[\Gamma^{Bog\,}_{i-2,i-2}(z)](R^{Bog}_{i-2,i-2}(z))^{\frac{1}{2}}\Big\|^{l_{i-2}}\nonumber \\
&\leq &\frac{1}{4(1+a_{\epsilon}-\frac{2b_{\epsilon}}{N-i+1}-\frac{1-c_{\epsilon}}{(N-i+1)^2})} \sum_{l_{i-2}=h}^{\infty}\Big[\|(R^{Bog}_{i-2,i-2}(z))^{\frac{1}{2}}W_{i-2,i-4}\,R_{i-4,i-4}^{Bog}(z)W_{i-4,i-2}^*(R^{Bog}_{i-2,i-2}(z))^{\frac{1}{2}} \|\,\|\check{\Gamma}^{Bog\,}_{i-4,i-4}(z)\|\Big]^{l_{i-2}}\nonumber \\
&\leq &(Z_{i-2,\epsilon})^h\frac{K_{i,\epsilon}}{1-Z_{i-2,\epsilon}}\,
%\Big[\frac{1}{1-\frac{1}{4(1+a_{\epsilon}-\frac{2b_{\epsilon}}{N-i+3}-\frac{1-c_{\epsilon}}{(N-i+3)^2})}}\Big]\,. \nonumber
\end{eqnarray}
Then, we estimate 
\begin{eqnarray}
& &\|(R^{Bog}_{i,i}(z))^{\frac{1}{2}}[\Gamma^{Bog\,}_{i,i}(z)]_{(r,h_+; r+2,h_-;\dots;i-4,h_-;i-2,h_-)}(R^{Bog}_{i,i}(z))^{\frac{1}{2}}\| \\
&= &\|(R^{Bog}_{i,i}(z))^{\frac{1}{2}}W_{i,i-2}\,(R^{Bog}_{i-2,i-2}(z))^{\frac{1}{2}} \hat{\sum}_{l_{i-2}=1}^{h-1}\Big[(R^{Bog}_{i-2,i-2}(z))^{\frac{1}{2}}[\Gamma^{Bog\,}_{i-2,i-2}(z)]_{(r,h_+; r+2,h_-;\dots;i-4,h_-)}(R^{Bog}_{i-2,i-2}(z))^{\frac{1}{2}} \Big]^{l_{i-2}}\times \nonumber \\
& &\quad\quad\quad \times (R^{Bog}_{i-2,i-2}(z))^{\frac{1}{2}} W^*_{i-2,i}(R^{Bog}_{i,i}(z))^{\frac{1}{2}}\|\nonumber\\
&\leq &\|(R^{Bog}_{i,i}(z))^{\frac{1}{2}}W_{i,i-2}\,(R^{Bog}_{i-2,i-2}(z))^{\frac{1}{2}}\|\Big\| \hat{\sum}_{l_{i-2}=1}^{h-1}\Big[(R^{Bog}_{i-2,i-2}(z))^{\frac{1}{2}}[\Gamma^{Bog\,}_{i-2,i-2}(z)]_{(r,h_+; r+2,h_-;\dots;i-4,h_-)}(R^{Bog}_{i-2,i-2}(z))^{\frac{1}{2}}\Big]^{l_{i-2}}\Big\| \times \nonumber \\
& &\quad\quad\quad \times \|(R^{Bog}_{i-2,i-2}(z))^{\frac{1}{2}} W^*_{i-2,i}(R^{Bog}_{i,i}(z))^{\frac{1}{2}}\|\nonumber
%&\leq &\frac{2+\mathcal{O}(\sqrt{\epsilon})}{3(1+a_{\epsilon}-\frac{2b_{\epsilon}}{N-i+2}-\frac{1-c_{\epsilon}}{(N-i+2)^2})}\, \|(R^{Bog}_{i-2,i-2}(z))^{\frac{1}{2}}[\Gamma^{Bog\,}_{i-2,i-2}(z)]_{(l,h_+;4,h_-;\dots;i-4,h_-)}(R^{Bog}_{i-2,i-2}(z))^{\frac{1}{2}}\|\,.
\end{eqnarray}
%For $k\geq j+2$, we write
%\begin{eqnarray}
%& &\|(R^{Bog}_{k,k}(z))^{\frac{1}{2}}[\Gamma^{Bog\,}_{k,k}(z)]_{(l,m_-; 4,m_-;\dots;j-4,m_-;j-2,m_+)}(R^{Bog}_{k,k}(z))^{\frac{1}{2}}\| \\
%&= &\|(R^{Bog}_{k,k}(z))^{\frac{1}{2}}W_{k,k-2}\,(R^{Bog}_{k-2,k-2}(z))^{\frac{1}{2}} \sum_{l_{k-2}=1}^{+\infty}\Big[(R^{Bog}_{k-2,k-2}(z))^{\frac{1}{2}}[\Gamma^{Bog\,}_{k-2,k-2}(z)]_{(l,m_-;4,m_-;\dots;j-4,m_-;j-2,m_+)}(R^{Bog}_{k-2,k-2}(z))^{\frac{1}{2}} \Big]^{l_{2}}\times \nonumber \\
%& &\quad\quad\quad \times (R^{Bog}_{k-2,k-2}(z))^{\frac{1}{2}} W^*_{k-2,k}(R^{Bog}_{k,k}(z))^{\frac{1}{2}}\|\nonumber\\
%&\leq &\|(R^{Bog}_{k,k}(z))^{\frac{1}{2}}W_{k,k-2}\,(R^{Bog}_{k-2,k-2}(z))^{\frac{1}{2}}\| \sum_{l_{k-2}=1}^{+\infty}\Big[\|(R^{Bog}_{k-2,k-2}(z))^{\frac{1}{2}}[\Gamma^{Bog\,}_{k-2,k-2}(z)]_{(l,m_-;4,m_-;\dots;j-4,m_-;j-2,m_+)}(R^{Bog}_{k-2,k-2}(z))^{\frac{1}{2}}\| \Big]^{l_{2}}\times \nonumber \\
%& &\quad\quad\quad \times \|(R^{Bog}_{k-2,k-2}(z))^{\frac{1}{2}} W^*_{k-2,k}(R^{Bog}_{k,k}(z))^{\frac{1}{2}}\|\nonumber\\
%&\leq &\frac{1}{4(1+a_{\epsilon}-\frac{2b_{\epsilon}}{N-j+2}-\frac{1-c_{\epsilon}}{(N-j+2)^2})}\frac{1}{1-\|(R^{Bog}_{k-2,k-2}(z))^{\frac{1}{2}}[\Gamma^{Bog\,}_{k-2,k-2}(z)]_{(l,m_-;4,m_-;\dots;j-4,m_-;j-2,m_+)}(R^{Bog}_{k-2,k-2}(z))^{\frac{1}{2}}\|}\times \nonumber \label{ineq-RGammaR-bis}\\
%& &\quad \times \Big[\|(R^{Bog}_{k-2,k-2}(z))^{\frac{1}{2}}[\Gamma^{Bog\,}_{k-2,k-2}(z)]_{(l,m_-;4,m_-;\dots;j-4,m_-;j-2,m_+)}(R^{Bog}_{k-2,k-2}(z))^{\frac{1}{2}}\|\Big]
%\end{eqnarray}

With the same iterative procedure exploited in the previous case, we can conclude that
\begin{eqnarray}
& &\|(R^{Bog}_{i,i}(z))^{\frac{1}{2}}[\Gamma^{Bog\,}_{i,i}(z)]_{(r,h_+; r+2,h_-;\dots;i-4,h_-;i-2,h_-)}(R^{Bog}_{i,i}(z))^{\frac{1}{2}}\|\\
&\leq& (Z_{r,\epsilon})^h\,\prod_{f=r+2\,,\, f-r\,\text{even}}^{i}\frac{K_{f,\epsilon}}{(1-Z_{f-2,\epsilon})^2} 
%& &\quad\times \prod_{f=l+2\,,\, f-l\,\text{even}}^{i-2}(1+a_{\epsilon}-\frac{2b_{\epsilon}}{N-f-1}-\frac{1-c_{\epsilon}}{(N-f-1)^2})^{-1}\,.
%\times\\
%& &\quad\quad \times \prod_{f=l+2\,,\, f-l\,\text{even}}^{j-4}(1+a_{\epsilon}-\frac{2b_{\epsilon}}{N-f}-\frac{1-c_{\epsilon}}{(N-f-1)^2})^{-h}
 \end{eqnarray}
\qed
\\

\begin{corollary}\label{truncated}
Under the assumptions of Proposition \ref{lemma-expansion-proof-0},  the following estimate (where for simplicity of the notation $N^{1-\beta}$ is assumed to be an even number)
\begin{eqnarray}
& &\Big\|\sum_{l_{N-2}=0}^{\infty}[(R^{Bog}_{\bold{j}_*\,;\,N-2,N-2}(z))^{\frac{1}{2}}\Gamma^{Bog\,}_{\bold{j}_*\,;\,N-2,N-2}(R^{Bog}_{\bold{j}_*\,;\,N-2,N-2}(z))^{\frac{1}{2}}]^{l_{N-2}}\\
& &\quad - \sum_{l_{N-2}=0}^{\infty}\Big\{(R^{Bog}_{\bold{j}_*\,;\,N-2,N-2}(z))^{\frac{1}{2}}\Big[\sum_{r=N-N^{1-\beta},\,r\, even}^{N-4}[\Gamma^{Bog\,}_{\bold{j}_*\,;\,N-2,N-2}(z)]_{(r,h_-; r+2,h_-;\dots ; N-4,h_-)}|_{h \equiv \infty}\Big](R^{Bog}_{\bold{j}_*\,;\,N-2,N-2}(z))^{\frac{1}{2}}\Big\}^{l_{N-2}}\Big\|\nonumber \\
%&=&\|\sum_{l_{N-2}=1}^{\infty}\sum_{j=0}^{l_{N-2}-1}\mathcal{T}^j(\mathcal{T}-\mathcal{S})\mathcal{S}^{l_{N-2}-j-1}\|\\
%&\leq & \|\mathcal{T}-\mathcal{S}\|\sum_{l_{N-2}=1}^{\infty}l_{N-2}(\frac{4}{5})^{l_{N-2}-1}\\
&\leq &\mathcal{O}(\frac{1}{\sqrt{\epsilon_{\bold{j}_*}}}(\frac{1}{1+c\sqrt{\epsilon_{\bold{j}_*}}})^{N^{1-\beta}})\,
\end{eqnarray}
holds true for $0<\beta <1$ and $\frac{1}{N^{1-\beta}}=o(\sqrt{\epsilon_{\bold{j}_*}})$, where $c$ is a positive constant and the symbol $|_{h \equiv \infty}$ means that $h\equiv \infty $ in the expansion of $\Gamma^{Bog\,}_{\bold{j}_*\,;\,N-2,N-2}(z)$.
\end{corollary}

\noindent
\emph{Proof}

\noindent
Using the estimates in (\ref{gamma-exp-1-0})-(\ref{gamma-exp-2-0}) in Proposition \ref{lemma-expansion-proof-0}, with the help of Remark \ref{prod-control} we estimate 
\begin{eqnarray}
& &\|(R^{Bog}_{\bold{j}_*\,;\,N-2,N-2}(z))^{\frac{1}{2}}\{\Gamma^{Bog\,}_{\bold{j}_*\,;\,N-2,N-2}-\sum_{r=N-N^{1-\beta},\,r\, even}^{N-4}[\Gamma^{Bog\,}_{\bold{j}_*\,;\,N-2,N-2}(z)]_{(r,h_-; r+2,h_-;\dots ; N-4,h_-)}|_{h \equiv \infty}\}(R^{Bog}_{\bold{j}_*\,;\,N-2,N-2}(z))^{\frac{1}{2}}\| \quad\quad\nonumber \\
&=&\|(R^{Bog}_{\bold{j}_*\,;\,N-2,N-2}(z))^{\frac{1}{2}}\sum_{r=2,\,r\, even}^{N-N^{1-\beta}-2}[\Gamma^{Bog\,}_{\bold{j}_*\,;\,N-2,N-2}(z)]_{(r,h_-; r+2,h_-;\dots ; N-4,h_-)}|_{h\equiv \infty}(R^{Bog}_{\bold{j}_*\,;\,N-2,N-2}(z))^{\frac{1}{2}}\|\\
&\leq &\sum_{r=2,\,r\, even}^{N-N^{1-\beta}-2}\,\prod_{f=r+4\,,\, f-r\,\text{even}}^{N-2}\frac{K_{f,\epsilon_{\bold{j}_*}}}{(1-Z_{f-2,\epsilon_{\bold{j}_*}})^2}\\
&\leq &C'\sum_{r=2,\,r\, even}^{N-N^{1-\beta}-2}(\frac{1}{1+c\sqrt{\epsilon_{\bold{j}_*}}})^{N-\frac{C}{\sqrt{\epsilon_{\bold{j}_*}}}-r}\\
&=&\mathcal{O}(\frac{1}{\sqrt{\epsilon_{\bold{j}_*}}}(\frac{1}{1+c\sqrt{\epsilon_{\bold{j}_*}}})^{N^{1-\beta}})
\end{eqnarray}
for some positive constants $c,C,C'$, and where we have assumed  $\frac{1}{N^{1-\beta}}=o(\epsilon_{\bold{j}_*}^{\frac{1}{2}})$ and $\epsilon_{\bold{j}_*}$ sufficiently small. Now, we define 
\begin{equation}
\mathcal{T}:=(R^{Bog}_{\bold{j}_*\,;\,N-2,N-2}(z))^{\frac{1}{2}}\Gamma^{Bog\,}_{\bold{j}_*\,;\,N-2,N-2}(R^{Bog}_{\bold{j}_*\,;\,N-2,N-2}(z))^{\frac{1}{2}}
\end{equation}
and
\begin{equation}
\mathcal{S}:=(R^{Bog}_{\bold{j}_*\,;\,N-2,N-2}(z))^{\frac{1}{2}}\sum_{r=N-N^{1-\beta},\,r\, even}^{N-4}[\Gamma^{Bog\,}_{\bold{j}_*\,;\,N-2,N-2}(z)]_{(r,h_-;r+2,h_-;\dots ; N-4,h_-)}|_{h\equiv \infty}(R^{Bog}_{\bold{j}_*\,;\,N-2,N-2}(z))^{\frac{1}{2}}
\end{equation}
Using Remark  \ref{estimation-proc} we can estimate $\|\mathcal{T}\|\,,\,\|\mathcal{S}\|\leq \frac{4}{5}$. Then, we can write
\begin{eqnarray}
& &\Big\|\sum_{l_{N-2}=0}^{\infty}[(R^{Bog}_{\bold{j}_*\,;\,N-2,N-2}(z))^{\frac{1}{2}}\Gamma^{Bog\,}_{\bold{j}_*\,;\,N-2,N-2}(R^{Bog}_{\bold{j}_*\,;\,N-2,N-2}(z))^{\frac{1}{2}}]^{l_{N-2}}\\
& &\quad - \sum_{l_{N-2}=0}^{\infty}\Big\{(R^{Bog}_{\bold{j}_*\,;\,N-2,N-2}(z))^{\frac{1}{2}}\Big[\sum_{r=N-N^{1-\beta},\,r\, even}^{N-4}[\Gamma^{Bog\,}_{\bold{j}_*\,;\,N-2,N-2}(z)]_{(r,h_-;r+2,h_-;\dots ; N-4,h_-)}|_{h\equiv \infty}\Big](R^{Bog}_{\bold{j}_*\,;\,N-2,N-2}(z))^{\frac{1}{2}}\Big\}^{l_{N-2}}\Big\|\nonumber \\
&=&\|\sum_{l_{N-2}=1}^{\infty}\sum_{j=0}^{l_{N-2}-1}\mathcal{T}^j(\mathcal{T}-\mathcal{S})\mathcal{S}^{l_{N-2}-j-1}\|\\
&\leq & \|\mathcal{T}-\mathcal{S}\|\sum_{l_{N-2}=1}^{\infty}l_{N-2}(\frac{4}{5})^{l_{N-2}-1}\\
&\leq &\mathcal{O}(\frac{1}{\sqrt{\epsilon_{\bold{j}_*}}}(\frac{1}{1+c\sqrt{\epsilon_{\bold{j}_*}}})^{N^{1-\beta}})\,.
\end{eqnarray}
\qed

\noindent
{\bf{Acknowledgements}}

I want to thank D.-A. Deckert, J. Fr\"ohlich, and P. Pickl for stimulating discussions on the contents of this paper. I am indebted to D.-A. Deckert for his contribution  to this project in its earliest stage and for helping with the implementation of crucial numerical simulations. I am indebted to G. Benfatto for his careful reading of this manuscript and for pointing out typos and small mistakes.

\end{document}